%% file: paper.tex
\documentclass[11pt]{report}

\input{paper_definitions.tex}
\usepackage{bbm}
\usepackage{subcaption}
\usepackage{hyperref}
\hypersetup{
    breaklinks=true,  % so long urls are correctly broken across lines
    colorlinks=true,
    urlcolor=urlcolor,
    linkcolor=linkcolor,
    citecolor=citecolor,
    }
\usepackage[square,numbers,sort]{natbib}
\newcommand{\R}{\mathbb{R}}
\newcommand{\C}{\mathbb{C}}
\renewcommand{\i}{\mathrm{i}}
\newcommand{\HP}{\mathcal{H}_P}
\newcommand{\HM}{\mathcal{H}_M}
\newcommand{\UP}{\mathcal{U}_P}
\newcommand{\UM}{\mathcal{U}_M}
\newcommand{\RX}{\mathrm{RX}}
\newcommand{\RZ}{\mathrm{RZ}}
\newcommand{\RZZ}{\mathrm{RZZ}}

\newcommand{\CNOT}{\mathrm{CNOT}}
\newcommand{\X}{\mathrm{X}}
\newcommand{\SX}{\mathrm{\sqrt{X}}}
\newcommand{\SWAP}{\mathrm{SWAP}}
\newcommand{\ket}[1]{\left\vert #1 \right\rangle}

\newcommand{\braket}[1]{\left\langle #1 \right\rangle}
\newcommand{\pmat}[1]{\begin{pmatrix}#1\end{pmatrix}}
\newcommand{\sigmazi}{\sigma_Z^{(i)}}

\title{
    \textsc{\huge Theory and Implementation of the Quantum Approximate Optimization Algorithm} \\
    {\Large A comprehensible introduction and case study using Qiskit and IBM quantum computers}
}
\author{Andreas Sturm \thanks{E-mail: andreas.sturm@iao.fraunhofer.de}
    \\[0.1cm]
    \emph{Fraunhofer IAO, Fraunhofer Institute for Industrial Engineering,}
    \\
    \emph{70569 Stuttgart, Germany}
}
\date{\today}

\begin{document}

\maketitle

\chapter*{Overview}

\section*{Introduction}
The present tutorial aims to provide a comprehensible and easily accessible
introduction into the theory and implementation of the famous Quantum Approximate Optimization Algorithm (QAOA).
We lay our focus on practical aspects and step-by-step guide through the realization of a proof of concept quantum application
based on a real-world use case. In every step we first explain the underlying theory and 
subsequently provide the implementation using IBM's Qiskit. In this way we provide a thorough understanding of the mathematical modelling
and the (quantum) algorithms as well as the equally important knowledge how to properly write the code implementing those theoretical concepts.
As another central aspect of this tutorial we provide extensive experiments on the 27 qubits state-of-the-art quantum computer ibmq\_ehningen.
From the discussion of these experiments we gain an overview on the current status of quantum computers and deduce which problem sizes 
can meaningfully be executed on today's hardware.

\section*{Structure}
This tutorial is divided into four notebooks. In Notebook \ref{chap:notebook-1} 
we introduce the use case that will accompany us all the time and on which we will demonstrate all ideas and concepts.
We present the mathematical modeling of the use case as an optimization problem as well as the associated implementation in Python and Qiskit.
Notebook \ref{chap:notebook-2} is dedicated to the theory and implementation of QAOA. Moreover, we will discuss
the classical optimization that is always present in variational algorithms like QAOA and provide advanced knowledge on
the gate synthesis of the QAOA operators. Subsequently, Notebook \ref{chap:notebook-3} deals with
the topic how QAOA circuits can be optimally executed on real quantum computers. We present a transpilation pipeline that implements this task
and explain how to run experiments on real quantum hardware. Additionally, we demonstrate in this notebook how the experimental results can
be post-processed with the big data tool Pandas. Finally, in Notebook \ref{chap:notebook-4} we present a large variety of
results from experiments on simulators and most of all from the quantum computer ibmq\_ehningen. Among others, we discuss challenges in classical
optimization, that besides the number of required qubits the amount of coupling of the variables in the underlying optimization problem is a central 
factor in judging if a problem is feasible for current quantum hardware, and that the transpilation pipeline and error mitigation techniques have
a great effect on the quality of experimental results. While those techniques allow to get the most out from current quantum computers
our experiments also clearly show today's limits with respect to the problem size and the quality of the results. 

\section*{Code Availability and Technical Requirements}
This tutorial is based and closely follows demonstration notebooks that were developed as a part of the 
project \href{https://websites.fraunhofer.de/sequoia/}{SEQUOIA} within the
\href{https://www.iaf.fraunhofer.de/en/networkers/KQC.html}{Competence Center Quantum Computing Baden-Württemberg}.
The codes shown in this tutorial as well as the codes of the used helper functions are available in the
\href{https://websites.fraunhofer.de/sequoia/use-cases/}{SEQUOIA use case database} or upon request from the author.

All codes were developed with
\begin{itemize}
    \tightlist
    \item \texttt{mthree==1.1.0}
    \item \texttt{numpy==1.19.5}
    \item \texttt{pandas==1.2.1}
    \item \texttt{plotly==4.14.3}
    \item \texttt{qiskit==0.39.4}
    \item \texttt{qiskit\_aer==0.11.1}
    \item \texttt{qiskit\_terra==0.22.2}
\end{itemize}

\section*{Acknowledgement}
The author thanks Christian Tutschku for carefully proofreading this tutorial.

\newpage

\tableofcontents

\chapter[Notebook 1]{} \label{chap:notebook-1}
\input{notebook_1_latex_source.tex}

\chapter[Notebook 2]{} \label{chap:notebook-2}
\input{notebook_2_latex_source.tex}

\chapter[Notebook 3]{} \label{chap:notebook-3}
\input{notebook_3_latex_source.tex}

\chapter[Notebook 4]{} \label{chap:notebook-4}
\input{notebook_4_latex_source.tex}

\appendix

\chapter{Properties of ibmq\_ehningen} \label{app:properties-ehningen}
\input{appendix_ehningen.tex}

\chapter{Codes of Helper Functions} \label{app:codes-helper-functions}
\input{appendix_helper.tex}

\bibliography{literature}

\end{document}

%% file: paper_definitions.tex
\usepackage[breakable]{tcolorbox}
\usepackage{parskip} % Stop auto-indenting (to mimic markdown behaviour)

\usepackage{graphicx}
\usepackage{caption}
\usepackage{float}
\usepackage{xcolor} % Allow colors to be defined
\usepackage{enumerate} % Needed for markdown enumerations to work
\usepackage{geometry} % Used to adjust the document margins
\usepackage{amsmath} % Equations
\usepackage{amssymb} % Equations
\usepackage{textcomp} % defines textquotesingle
\AtBeginDocument{%
    \def\PYZsq{\textquotesingle}% Upright quotes in Pygmentized code
}
\usepackage{upquote} % Upright quotes for verbatim code
\usepackage{eurosym} % defines \euro

\usepackage{iftex}
\ifPDFTeX
    \usepackage[T1]{fontenc}
    \IfFileExists{alphabeta.sty}{
            \usepackage{alphabeta}
        }{
            \usepackage[mathletters]{ucs}
            \usepackage[utf8x]{inputenc}
        }
\else
    \usepackage{fontspec}
    \usepackage{unicode-math}
\fi

\usepackage{fancyvrb} % verbatim replacement that allows latex
\usepackage{grffile} % extends the file name processing of package graphics
\makeatletter % fix for old versions of grffile with XeLaTeX
\@ifpackagelater{grffile}{2019/11/01}
{
}
{
    \def\Gread@@xetex#1{%
    \IfFileExists{"\Gin@base".bb}%
    {\Gread@eps{\Gin@base.bb}}%
    {\Gread@@xetex@aux#1}%
    }
}
\makeatother
\usepackage[Export]{adjustbox} % Used to constrain images to a maximum size
\adjustboxset{max size={0.9\linewidth}{0.9\paperheight}}

\usepackage{titling}
\usepackage{longtable} % longtable support required by pandoc >1.10
\usepackage{booktabs}  % table support for pandoc > 1.12.2
\usepackage{array}     % table support for pandoc >= 2.11.3
\usepackage{calc}      % table minipage width calculation for pandoc >= 2.11.1
\usepackage[inline]{enumitem} % IRkernel/repr support (it uses the enumerate* environment)
\usepackage[normalem]{ulem} % ulem is needed to support strikethroughs (\sout)
\usepackage{mathrsfs}

\definecolor{urlcolor}{rgb}{0,.145,.698}
\definecolor{linkcolor}{rgb}{.71,0.21,0.01}
\definecolor{citecolor}{rgb}{.12,.54,.11}

\definecolor{ansi-black}{HTML}{3E424D}
\definecolor{ansi-black-intense}{HTML}{282C36}
\definecolor{ansi-red}{HTML}{E75C58}
\definecolor{ansi-red-intense}{HTML}{B22B31}
\definecolor{ansi-green}{HTML}{00A250}
\definecolor{ansi-green-intense}{HTML}{007427}
\definecolor{ansi-yellow}{HTML}{DDB62B}
\definecolor{ansi-yellow-intense}{HTML}{B27D12}
\definecolor{ansi-blue}{HTML}{208FFB}
\definecolor{ansi-blue-intense}{HTML}{0065CA}
\definecolor{ansi-magenta}{HTML}{D160C4}
\definecolor{ansi-magenta-intense}{HTML}{A03196}
\definecolor{ansi-cyan}{HTML}{60C6C8}
\definecolor{ansi-cyan-intense}{HTML}{258F8F}
\definecolor{ansi-white}{HTML}{C5C1B4}
\definecolor{ansi-white-intense}{HTML}{A1A6B2}
\definecolor{ansi-default-inverse-fg}{HTML}{FFFFFF}
\definecolor{ansi-default-inverse-bg}{HTML}{000000}

\definecolor{outerrorbackground}{HTML}{FFDFDF}

\providecommand{\tightlist}{%
    \setlength{\itemsep}{0pt}\setlength{\parskip}{0pt}}
\DefineVerbatimEnvironment{Highlighting}{Verbatim}{commandchars=\\\{\}}
\let\Oldtex\TeX
\let\Oldlatex\LaTeX
\renewcommand{\TeX}{\textrm{\Oldtex}}
\renewcommand{\LaTeX}{\textrm{\Oldlatex}}
\title{notebook\_1\_to\_latex}

\makeatletter
\def\PY@reset{\let\PY@it=\relax \let\PY@bf=\relax%
\let\PY@ul=\relax \let\PY@tc=\relax%
\let\PY@bc=\relax \let\PY@ff=\relax}
\def\PY@tok#1{\csname PY@tok@#1\endcsname}
\def\PY@toks#1+{\ifx\relax#1\empty\else%
\PY@tok{#1}\expandafter\PY@toks\fi}
\def\PY@do#1{\PY@bc{\PY@tc{\PY@ul{%
\PY@it{\PY@bf{\PY@ff{#1}}}}}}}
\def\PY#1#2{\PY@reset\PY@toks#1+\relax+\PY@do{#2}}

\@namedef{PY@tok@w}{\def\PY@tc##1{\textcolor[rgb]{0.73,0.73,0.73}{##1}}}
\@namedef{PY@tok@c}{\let\PY@it=\textit\def\PY@tc##1{\textcolor[rgb]{0.24,0.48,0.48}{##1}}}
\@namedef{PY@tok@cp}{\def\PY@tc##1{\textcolor[rgb]{0.61,0.40,0.00}{##1}}}
\@namedef{PY@tok@k}{\let\PY@bf=\textbf\def\PY@tc##1{\textcolor[rgb]{0.00,0.50,0.00}{##1}}}
\@namedef{PY@tok@kp}{\def\PY@tc##1{\textcolor[rgb]{0.00,0.50,0.00}{##1}}}
\@namedef{PY@tok@kt}{\def\PY@tc##1{\textcolor[rgb]{0.69,0.00,0.25}{##1}}}
\@namedef{PY@tok@o}{\def\PY@tc##1{\textcolor[rgb]{0.40,0.40,0.40}{##1}}}
\@namedef{PY@tok@ow}{\let\PY@bf=\textbf\def\PY@tc##1{\textcolor[rgb]{0.67,0.13,1.00}{##1}}}
\@namedef{PY@tok@nb}{\def\PY@tc##1{\textcolor[rgb]{0.00,0.50,0.00}{##1}}}
\@namedef{PY@tok@nf}{\def\PY@tc##1{\textcolor[rgb]{0.00,0.00,1.00}{##1}}}
\@namedef{PY@tok@nc}{\let\PY@bf=\textbf\def\PY@tc##1{\textcolor[rgb]{0.00,0.00,1.00}{##1}}}
\@namedef{PY@tok@nn}{\let\PY@bf=\textbf\def\PY@tc##1{\textcolor[rgb]{0.00,0.00,1.00}{##1}}}
\@namedef{PY@tok@ne}{\let\PY@bf=\textbf\def\PY@tc##1{\textcolor[rgb]{0.80,0.25,0.22}{##1}}}
\@namedef{PY@tok@nv}{\def\PY@tc##1{\textcolor[rgb]{0.10,0.09,0.49}{##1}}}
\@namedef{PY@tok@no}{\def\PY@tc##1{\textcolor[rgb]{0.53,0.00,0.00}{##1}}}
\@namedef{PY@tok@nl}{\def\PY@tc##1{\textcolor[rgb]{0.46,0.46,0.00}{##1}}}
\@namedef{PY@tok@ni}{\let\PY@bf=\textbf\def\PY@tc##1{\textcolor[rgb]{0.44,0.44,0.44}{##1}}}
\@namedef{PY@tok@na}{\def\PY@tc##1{\textcolor[rgb]{0.41,0.47,0.13}{##1}}}
\@namedef{PY@tok@nt}{\let\PY@bf=\textbf\def\PY@tc##1{\textcolor[rgb]{0.00,0.50,0.00}{##1}}}
\@namedef{PY@tok@nd}{\def\PY@tc##1{\textcolor[rgb]{0.67,0.13,1.00}{##1}}}
\@namedef{PY@tok@s}{\def\PY@tc##1{\textcolor[rgb]{0.73,0.13,0.13}{##1}}}
\@namedef{PY@tok@sd}{\let\PY@it=\textit\def\PY@tc##1{\textcolor[rgb]{0.73,0.13,0.13}{##1}}}
\@namedef{PY@tok@si}{\let\PY@bf=\textbf\def\PY@tc##1{\textcolor[rgb]{0.64,0.35,0.47}{##1}}}
\@namedef{PY@tok@se}{\let\PY@bf=\textbf\def\PY@tc##1{\textcolor[rgb]{0.67,0.36,0.12}{##1}}}
\@namedef{PY@tok@sr}{\def\PY@tc##1{\textcolor[rgb]{0.64,0.35,0.47}{##1}}}
\@namedef{PY@tok@ss}{\def\PY@tc##1{\textcolor[rgb]{0.10,0.09,0.49}{##1}}}
\@namedef{PY@tok@sx}{\def\PY@tc##1{\textcolor[rgb]{0.00,0.50,0.00}{##1}}}
\@namedef{PY@tok@m}{\def\PY@tc##1{\textcolor[rgb]{0.40,0.40,0.40}{##1}}}
\@namedef{PY@tok@gh}{\let\PY@bf=\textbf\def\PY@tc##1{\textcolor[rgb]{0.00,0.00,0.50}{##1}}}
\@namedef{PY@tok@gu}{\let\PY@bf=\textbf\def\PY@tc##1{\textcolor[rgb]{0.50,0.00,0.50}{##1}}}
\@namedef{PY@tok@gd}{\def\PY@tc##1{\textcolor[rgb]{0.63,0.00,0.00}{##1}}}
\@namedef{PY@tok@gi}{\def\PY@tc##1{\textcolor[rgb]{0.00,0.52,0.00}{##1}}}
\@namedef{PY@tok@gr}{\def\PY@tc##1{\textcolor[rgb]{0.89,0.00,0.00}{##1}}}
\@namedef{PY@tok@ge}{\let\PY@it=\textit}
\@namedef{PY@tok@gs}{\let\PY@bf=\textbf}
\@namedef{PY@tok@gp}{\let\PY@bf=\textbf\def\PY@tc##1{\textcolor[rgb]{0.00,0.00,0.50}{##1}}}
\@namedef{PY@tok@go}{\def\PY@tc##1{\textcolor[rgb]{0.44,0.44,0.44}{##1}}}
\@namedef{PY@tok@gt}{\def\PY@tc##1{\textcolor[rgb]{0.00,0.27,0.87}{##1}}}
\@namedef{PY@tok@err}{\def\PY@bc##1{{\setlength{\fboxsep}{\string -\fboxrule}\fcolorbox[rgb]{1.00,0.00,0.00}{1,1,1}{\strut ##1}}}}
\@namedef{PY@tok@kc}{\let\PY@bf=\textbf\def\PY@tc##1{\textcolor[rgb]{0.00,0.50,0.00}{##1}}}
\@namedef{PY@tok@kd}{\let\PY@bf=\textbf\def\PY@tc##1{\textcolor[rgb]{0.00,0.50,0.00}{##1}}}
\@namedef{PY@tok@kn}{\let\PY@bf=\textbf\def\PY@tc##1{\textcolor[rgb]{0.00,0.50,0.00}{##1}}}
\@namedef{PY@tok@kr}{\let\PY@bf=\textbf\def\PY@tc##1{\textcolor[rgb]{0.00,0.50,0.00}{##1}}}
\@namedef{PY@tok@bp}{\def\PY@tc##1{\textcolor[rgb]{0.00,0.50,0.00}{##1}}}
\@namedef{PY@tok@fm}{\def\PY@tc##1{\textcolor[rgb]{0.00,0.00,1.00}{##1}}}
\@namedef{PY@tok@vc}{\def\PY@tc##1{\textcolor[rgb]{0.10,0.09,0.49}{##1}}}
\@namedef{PY@tok@vg}{\def\PY@tc##1{\textcolor[rgb]{0.10,0.09,0.49}{##1}}}
\@namedef{PY@tok@vi}{\def\PY@tc##1{\textcolor[rgb]{0.10,0.09,0.49}{##1}}}
\@namedef{PY@tok@vm}{\def\PY@tc##1{\textcolor[rgb]{0.10,0.09,0.49}{##1}}}
\@namedef{PY@tok@sa}{\def\PY@tc##1{\textcolor[rgb]{0.73,0.13,0.13}{##1}}}
\@namedef{PY@tok@sb}{\def\PY@tc##1{\textcolor[rgb]{0.73,0.13,0.13}{##1}}}
\@namedef{PY@tok@sc}{\def\PY@tc##1{\textcolor[rgb]{0.73,0.13,0.13}{##1}}}
\@namedef{PY@tok@dl}{\def\PY@tc##1{\textcolor[rgb]{0.73,0.13,0.13}{##1}}}
\@namedef{PY@tok@s2}{\def\PY@tc##1{\textcolor[rgb]{0.73,0.13,0.13}{##1}}}
\@namedef{PY@tok@sh}{\def\PY@tc##1{\textcolor[rgb]{0.73,0.13,0.13}{##1}}}
\@namedef{PY@tok@s1}{\def\PY@tc##1{\textcolor[rgb]{0.73,0.13,0.13}{##1}}}
\@namedef{PY@tok@mb}{\def\PY@tc##1{\textcolor[rgb]{0.40,0.40,0.40}{##1}}}
\@namedef{PY@tok@mf}{\def\PY@tc##1{\textcolor[rgb]{0.40,0.40,0.40}{##1}}}
\@namedef{PY@tok@mh}{\def\PY@tc##1{\textcolor[rgb]{0.40,0.40,0.40}{##1}}}
\@namedef{PY@tok@mi}{\def\PY@tc##1{\textcolor[rgb]{0.40,0.40,0.40}{##1}}}
\@namedef{PY@tok@il}{\def\PY@tc##1{\textcolor[rgb]{0.40,0.40,0.40}{##1}}}
\@namedef{PY@tok@mo}{\def\PY@tc##1{\textcolor[rgb]{0.40,0.40,0.40}{##1}}}
\@namedef{PY@tok@ch}{\let\PY@it=\textit\def\PY@tc##1{\textcolor[rgb]{0.24,0.48,0.48}{##1}}}
\@namedef{PY@tok@cm}{\let\PY@it=\textit\def\PY@tc##1{\textcolor[rgb]{0.24,0.48,0.48}{##1}}}
\@namedef{PY@tok@cpf}{\let\PY@it=\textit\def\PY@tc##1{\textcolor[rgb]{0.24,0.48,0.48}{##1}}}
\@namedef{PY@tok@c1}{\let\PY@it=\textit\def\PY@tc##1{\textcolor[rgb]{0.24,0.48,0.48}{##1}}}
\@namedef{PY@tok@cs}{\let\PY@it=\textit\def\PY@tc##1{\textcolor[rgb]{0.24,0.48,0.48}{##1}}}

\def\PYZbs{\char`\\}
\def\PYZus{\char`\_}
\def\PYZob{\char`\{}
\def\PYZcb{\char`\}}
\def\PYZca{\char`\^}

\def\PYZlt{\char`\<}
\def\PYZgt{\char`\>}
\def\PYZsh{\char`\#}

\def\PYZdl{\char`\$}
\def\PYZhy{\char`\-}
\def\PYZsq{\char`\'}
\def\PYZdq{\char`\"}

\makeatother

\makeatletter
    \newbox\Wrappedcontinuationbox
    \newbox\Wrappedvisiblespacebox
    \newcommand*\Wrappedvisiblespace {\textcolor{red}{\textvisiblespace}}
    \newcommand*\Wrappedcontinuationsymbol {\textcolor{red}{\llap{\tiny$\m@th\hookrightarrow$}}}
    \newcommand*\Wrappedcontinuationindent {3ex }
    \newcommand*\Wrappedafterbreak {\kern\Wrappedcontinuationindent\copy\Wrappedcontinuationbox}
    \newcommand*\Wrappedbreaksatspecials {%
        \def\PYGZus{\discretionary{\char`\_}{\Wrappedafterbreak}{\char`\_}}%
        \def\PYGZob{\discretionary{}{\Wrappedafterbreak\char`\{}{\char`\{}}%
        \def\PYGZcb{\discretionary{\char`\}}{\Wrappedafterbreak}{\char`\}}}%
        \def\PYGZca{\discretionary{\char`\^}{\Wrappedafterbreak}{\char`\^}}%
        \def\PYGZam{\discretionary{\char`\&}{\Wrappedafterbreak}{\char`\&}}%
        \def\PYGZlt{\discretionary{}{\Wrappedafterbreak\char`\<}{\char`\<}}%
        \def\PYGZgt{\discretionary{\char`\>}{\Wrappedafterbreak}{\char`\>}}%
        \def\PYGZsh{\discretionary{}{\Wrappedafterbreak\char`\#}{\char`\#}}%
        \def\PYGZpc{\discretionary{}{\Wrappedafterbreak\char`\%}{\char`\%}}%
        \def\PYGZdl{\discretionary{}{\Wrappedafterbreak\char`\$}{\char`\$}}%
        \def\PYGZhy{\discretionary{\char`\-}{\Wrappedafterbreak}{\char`\-}}%
        \def\PYGZsq{\discretionary{}{\Wrappedafterbreak\textquotesingle}{\textquotesingle}}%
        \def\PYGZdq{\discretionary{}{\Wrappedafterbreak\char`\"}{\char`\"}}%
        \def\PYGZti{\discretionary{\char`\~}{\Wrappedafterbreak}{\char`\~}}%
    }
    \newcommand*\Wrappedbreaksatpunct {%
        \lccode`\~`\.\lowercase{\def~}{\discretionary{\hbox{\char`\.}}{\Wrappedafterbreak}{\hbox{\char`\.}}}%
        \lccode`\~`\,\lowercase{\def~}{\discretionary{\hbox{\char`\,}}{\Wrappedafterbreak}{\hbox{\char`\,}}}%
        \lccode`\~`\;\lowercase{\def~}{\discretionary{\hbox{\char`\;}}{\Wrappedafterbreak}{\hbox{\char`\;}}}%
        \lccode`\~`\:\lowercase{\def~}{\discretionary{\hbox{\char`\:}}{\Wrappedafterbreak}{\hbox{\char`\:}}}%
        \lccode`\~`\?\lowercase{\def~}{\discretionary{\hbox{\char`\?}}{\Wrappedafterbreak}{\hbox{\char`\?}}}%
        \lccode`\~`\!\lowercase{\def~}{\discretionary{\hbox{\char`\!}}{\Wrappedafterbreak}{\hbox{\char`\!}}}%
        \lccode`\~`\/\lowercase{\def~}{\discretionary{\hbox{\char`\/}}{\Wrappedafterbreak}{\hbox{\char`\/}}}%
        \catcode`\.\active
        \catcode`\,\active
        \catcode`\;\active
        \catcode`\:\active
        \catcode`\?\active
        \catcode`\!\active
        \catcode`\/\active
        \lccode`\~`\~
    }
\makeatother

\let\OriginalVerbatim=\Verbatim
\makeatletter
\renewcommand{\Verbatim}[1][1]{%
    \sbox\Wrappedcontinuationbox {\Wrappedcontinuationsymbol}%
    \sbox\Wrappedvisiblespacebox {\FV@SetupFont\Wrappedvisiblespace}%
    \def\FancyVerbFormatLine ##1{\hsize\linewidth
        \vtop{\raggedright\hyphenpenalty\z@\exhyphenpenalty\z@
            \doublehyphendemerits\z@\finalhyphendemerits\z@
            \strut ##1\strut}%
    }%
    \def\FV@Space {%
        \nobreak\hskip\z@ plus\fontdimen3\font minus\fontdimen4\font
        \discretionary{\copy\Wrappedvisiblespacebox}{\Wrappedafterbreak}
        {\kern\fontdimen2\font}%
    }%

    \Wrappedbreaksatspecials
    \OriginalVerbatim[#1,codes*=\Wrappedbreaksatpunct]%
}
\makeatother

\definecolor{incolor}{HTML}{303F9F}
\definecolor{outcolor}{HTML}{D84315}
\definecolor{cellborder}{HTML}{CFCFCF}
\definecolor{cellbackground}{HTML}{F7F7F7}

\makeatletter
\newcommand{\boxspacing}{\kern\kvtcb@left@rule\kern\kvtcb@boxsep}
\makeatother
\newcommand{\prompt}[4]{
    {\ttfamily\llap{{\color{#2}[#3]:\hspace{3pt}#4}}\vspace{-\baselineskip}}
}

%% file: notebook_1_latex_source.tex
\hypertarget{real-world-problem}{%
\section{Real World Problem}\label{sec:real-world-problem}}

    The project \textbf{LamA - Laden am Arbeitsplatz} (english: charging at
work) is a joint project led by Fraunhofer IAO and funded by the federal
government of Germany. Among others the goal is to build up charging
infrastructure for electric vehicles at 37 institutes of the Fraunhofer
society across Germany so that employees can charge their electric cars at work \cite{LamaWeb,IaoWeb,IegWeb}.

    In this tutorial we aim to provide an optimal charging schedule for
this charging infrastructure. Clearly, such an optimization can be
implemented with respect to different aspects, e.g.~by involving weather
predictions we could optimize the schedule such that as much clean
energy (sun, wind, etc.) as possible is used. Or, considering that the
price of energy from the public electricity grid varies over time and the
electric vehicles are charged during working hours, i.e.~around a fairly
long period of about 8 hours, we could optimize the schedule such that
it takes energy from the grid when it is the cheapest. Having in mind
that quantum computing is currently in the NISQ (noisy intermediate
scale quantum) era we decided for this tutorial to consider the aspect of
\textbf{minimizing the peak load} that is taken from the eletricity
grid. This reduces costs, relieves the public electricity grid and -- as
we will see -- can be reduced to a meaningful proof of concept problem
that can be executed on today's available quantum computers.

    \hypertarget{proof-of-concept-poc-model}{%
\section{Proof of Concept Model}\label{sec:poc-model}}

    Let us begin by introducing the proof of concept (POC) model that we will consider in this
tutorial. It is an \textbf{optimization task} where we are \textbf{given}
\begin{itemize}
    \tightlist
    \item the number of cars that have to be charged,
    \item the arrival and departure times of these cars, and
    \item the required energies they need to charge.
\end{itemize}

    Our \textbf{aim} is to
\begin{itemize}
\tightlist
\item minimize the peak load taken from the electricity grid,
\item meet the time restrictions imposed by the arrival and departure times, and
\item charge the correct amount of energies.
\end{itemize}

    In order to make this optimization task feasible for a NISQ computer we
make the following \textbf{simplifications}:

\begin{itemize}
\tightlist
\item
  We work with discrete time slots and
\item
  we assume that we can only charge on discrete loading levels.
\end{itemize}

    Let us illustrate this rather abstract optimization task with a simple toy
example.

    \hypertarget{example}{%
\subsection{Toy Example}\label{sec:notebook-1-example}}

\begin{figure}
    \begin{center}
        \includegraphics{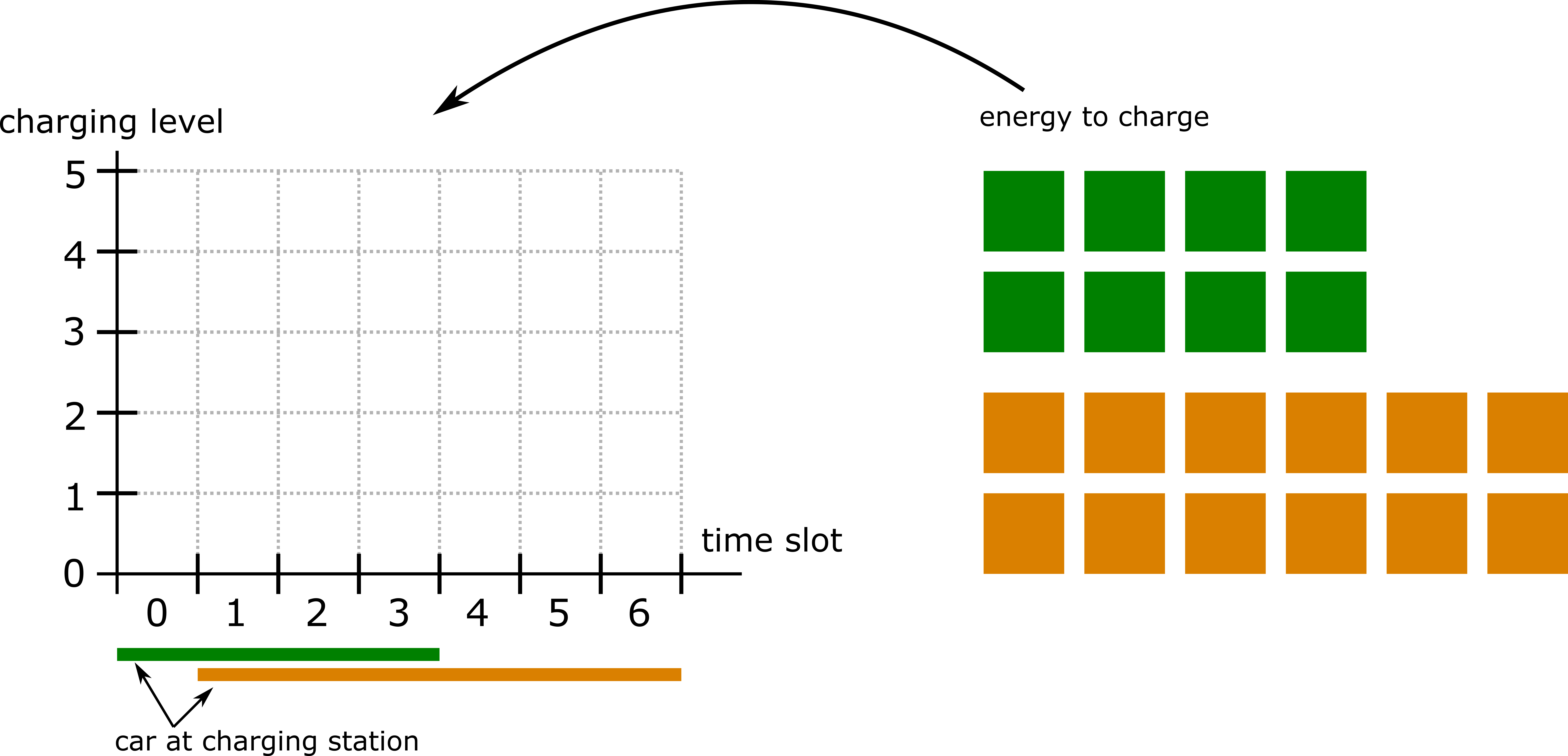}
        \caption{A simple example with one charging station and two cars.}
        \label{fig:notebook-1-poc-example}
    \end{center}
\end{figure}

    We consider a charging station with 6 charging levels (i.e.~levels
0, 1, \ldots, 5) and 7 available time slots (i.e.~slots
0, 1, \ldots, 6). Moreover, let us take two cars (named
car\_green and car\_orange), where
\begin{itemize}
    \tightlist
    \item car\_green is at the charging station at time slots 0, \ldots, 3 and needs to charge 8
    energy units, and
    \item car\_orange is at the charging station at time slots
    1, \ldots, 6 and needs to charge 12 energy units.
\end{itemize}
A visualization of the example is given in Figure \ref{fig:notebook-1-poc-example}.

\begin{figure}
    \begin{center}
        \begin{subfigure}{0.33\textwidth}
            \centering
            \includegraphics{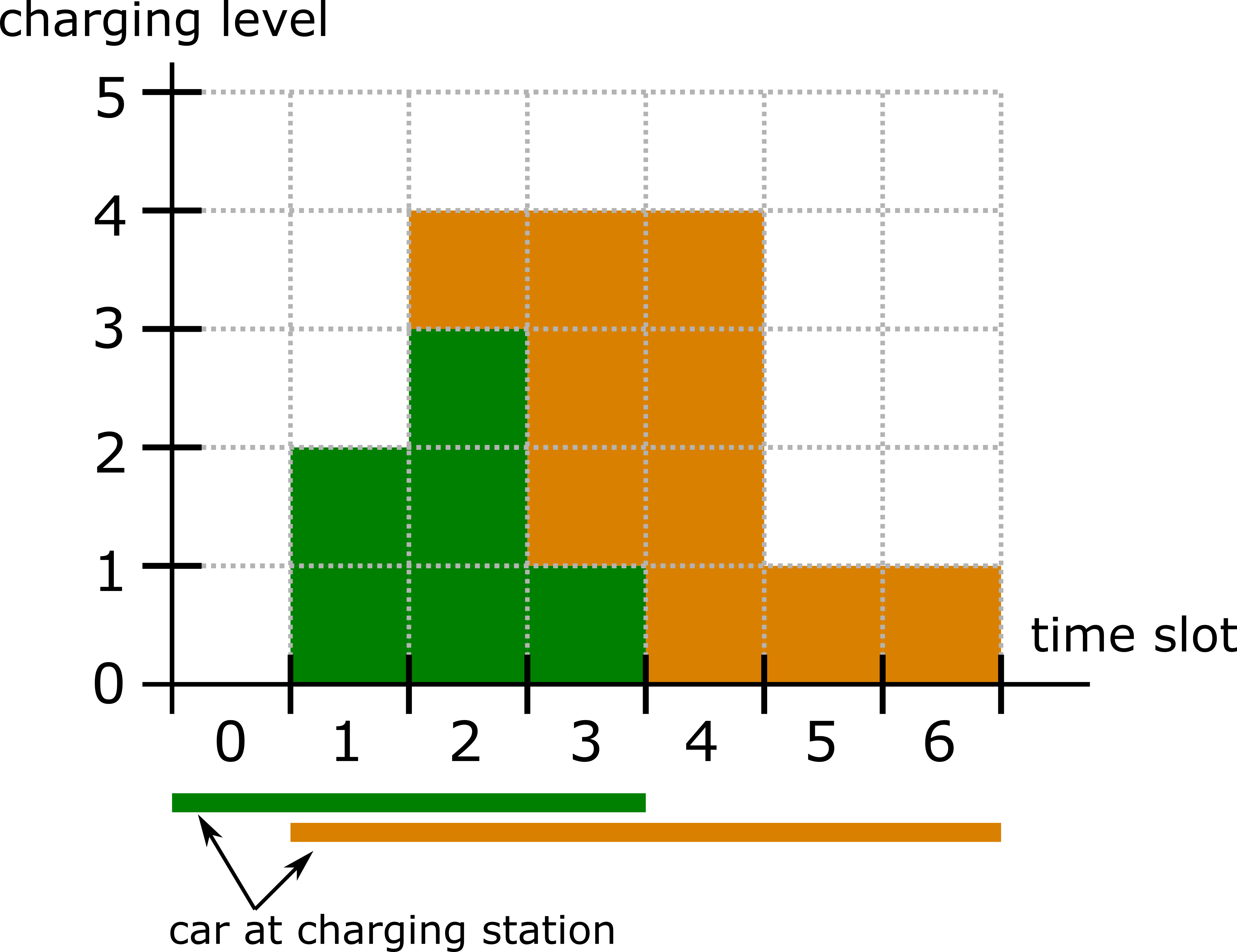}
            \caption{A non-feasible schedule because the energy constraint is violated.}
            \label{fig:notebook-1-poc-example-non-feasible}
        \end{subfigure}
        \qquad
        \begin{subfigure}{0.33\textwidth}
            \centering
            \includegraphics{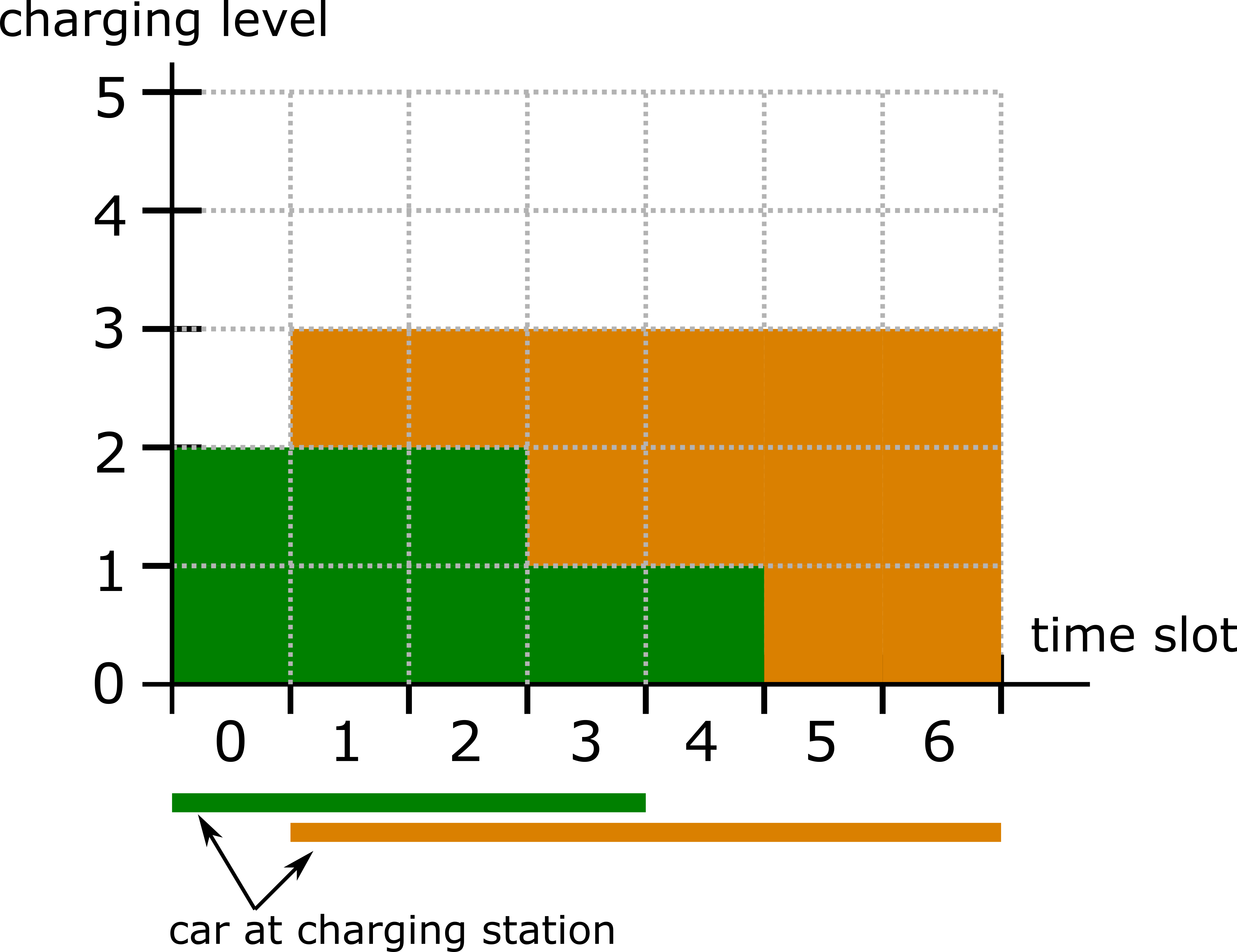}
            \caption{A non-feasible schedule because the time constraint is violated.}
            \label{fig:notebook-1-poc-example-non-feasible-2}
        \end{subfigure}
        \\[0.5cm]
        \begin{subfigure}{0.33\textwidth}
            \centering
            \includegraphics{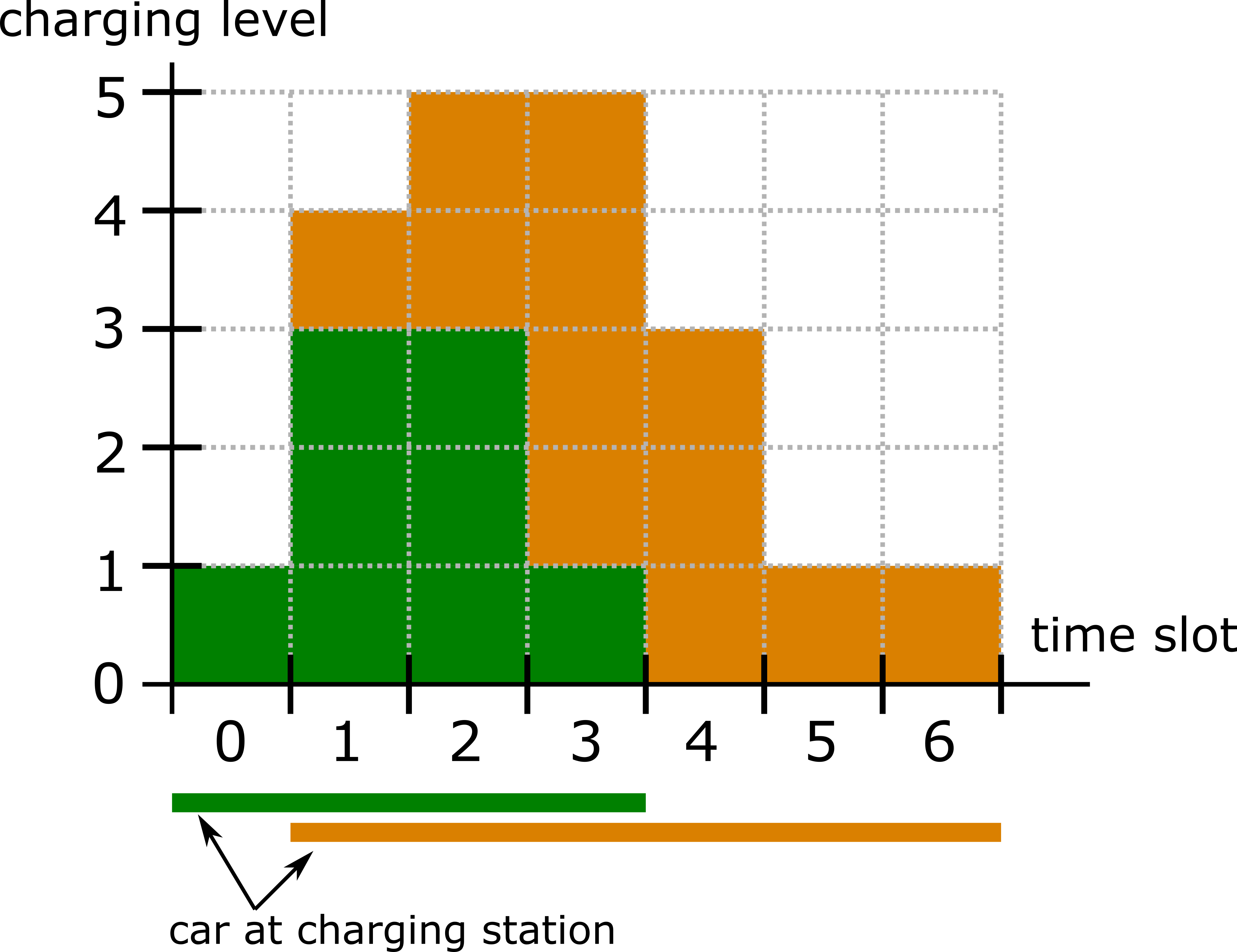}
            \caption{A non-optimal schedule because the peak load is not as low as possible.}
            \label{fig:notebook-1-poc-example-non-optimal}
        \end{subfigure}
        \qquad
        \begin{subfigure}{0.33\textwidth}
            \centering
            \includegraphics{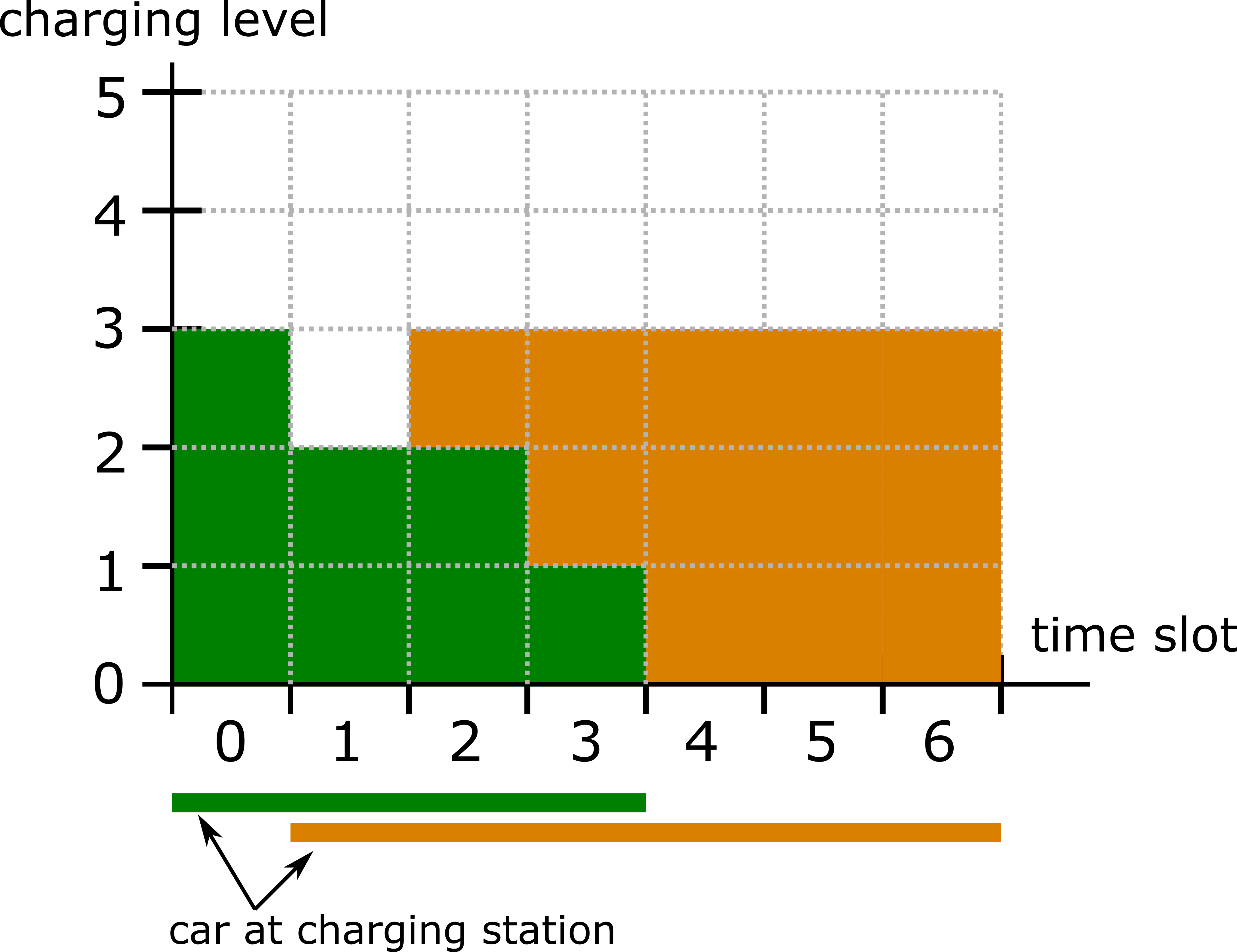}
            \caption{An optimal schedule.\\\phantom{a}\\\phantom{a}}
            \label{fig:notebook-1-poc-example-optimal}
        \end{subfigure}
    \end{center}
    \caption{Possible outcomes of an optimization process for 
    the situation described in Figure \ref{fig:notebook-1-poc-example}}
    \label{fig:notebook-1-poc-example-possible-situations}
\end{figure}

    An optimization process for the above situation could for example yield the 
charging schedules depicted in Figure \ref{fig:notebook-1-poc-example-possible-situations}.

    Our next step is to find a formal mathematical description for our POC
model.

    \hypertarget{mathematical-description-of-poc-model}{%
\section{Mathematical Description of POC
Model}\label{sec:mathematical-description-poc-model}}

    \hypertarget{notation}{%
\subsection{Notation}\label{sec:notebook-1-notation}}

    In the following we use the notation: 
\begin{itemize}
    \tightlist
    \item $T \in \mathbb{N}$: Number of time slots \(\rightarrow\) we have time slots \(0, 1, \dots, T-1\).
    \item $L \in \mathbb{N}$: Number of charging levels \(\rightarrow\) we have charging levels \(0, 1, \dots, L-1\).
    \item $K \in \mathbb{N}$: Number of cars \(\rightarrow\) we have $K$ cars that we label as \(0, 1, \dots, K-1\).
\end{itemize}
Moreover, for every car \(k \in \{0, \dots, K-1\}\) we define:
\begin{itemize}
    \tightlist
    \item \(t_k^a \in \{0, \dots, T-1\}\): Arrival time of car \(k\).
    \item \(t_k^d \in \{0, \dots, T-1\}\): Departure time of car \(k\).
    \item \(e_k \in \mathbb{N}\): Required amount of energy.
\end{itemize}
Finally, for every car \(k \in \{0, \dots, K-1\}\) and for every time slot \(t \in \{0, \dots, T-1\}\)
we define:
\begin{itemize}
    \tightlist
    \item \(p_k^t \in \{0, \dots, L-1\}\): Charging level on which car \(k\) charges at time slot \(t\)
    (i.e.~the amount of energy packages it draws from the electricity grid on time slot $t$).
\end{itemize}

    For a shorter notation we introduce the vectors
\begin{itemize}
    \tightlist
    \item \(\vec{p}_k\) = \(\pmat{p_k^0 \\ \vdots \\ p_k^{T-1}} \in \{0, \dots, L-1\}^T\) ,
    where \(k \in \{0, \dots, K-1\}\), 
    \vspace{0.3cm}
    \item \(\vec{p}\) = \(\pmat{\vec{p}_0 \\ \vdots \\ \vec{p}_{K-1}} \in \{0, \dots, L-1\}^{K T}\) . 
\end{itemize}
The entries of \(\vec{p}_k\) form the charging curve for car \(k\), see Figure \ref{fig:notebook-1-poc-example-possible-situations}.

Last, we define the vector \(\vec{p}_\text{sum}\) so that its
\(t\)-th component is the sum of the charging levels of all cars for time slot
\(t\):
\begin{equation*}
    \vec{p}_\text{sum}
    = \vec{p}_0 + \vec{p}_1 + \dots + \vec{p}_{K-1}
    = \pmat{
        p_0^0 + p_1^0 + \dots p_{K-1}^0
        \\
        \vdots
        \\[0.07cm]
        p_0^{T-1} + p_1^{T-1} + \dots p_{K-1}^{T-1}
    } \ .
\end{equation*}
It is easy to see that we have
\begin{equation}
    \vec{p}_\text{sum}
    = \bigl(\underbrace{I_T \; I_T \; \cdots \; I_T}_{K \text{ times}}\bigr) \vec{p} \ ,
    \label{eq:p-sum-1}
\end{equation}
where $I_T$ is the $T \times T$ identity matrix. 

With this notation at hand we can set up a cost function that assigns a cost to a charging schedule.

    \hypertarget{cost-function}{%
\subsection{Cost Function}\label{sec:notebook-1-cost-function}}

Note that the summation of all components of the vector \(\vec{p}_\text{sum}\) gives the total energy drawn from the electricity grid. 
Further, observe that for the same total energy the smaller the \(\ell_2\)-norm \(\|\vec{p}_\text{sum}\|_2\) of \(\vec{p}_\text{sum}\) the smaller the peak load. E.g.

\[
\left\| \pmat{1 \\ 2 \\ 3} \right\|_2^2
= 1^2 + 2^2 + 3^2 
= 14,
\qquad
\left\| \pmat{2 \\ 2 \\ 2} \right\|_2^2
= 2^2 + 2^2 + 2^2 
= 12.
\]

    Moreover, note that
\begin{equation*}
    \|\vec{p}_\text{sum}\|_2^2
    = \vec{p}_\text{sum}^{\; t} \vec{p}_\text{sum}
    = \vec{p}^{\; t} \! A \vec{p} \ , 
    \qquad
    A = \mathbbm{1}_K^t \mathbbm{1}_K \otimes I_T \ ,
\end{equation*}
where we used \eqref{eq:p-sum-1} and where $\mathbbm{1}_K = (1, 1 \dots, 1)^t \in \R^K$.

    So, we can use the following \textbf{cost function} for the minimization
of the peak load:

\begin{equation}
    f_1(\vec{p \, })
    = \vec{p}^{\ t} \! A \vec{p} \ .
    \label{eq:f-1}
\end{equation}

    Clearly, only minimizing \(f_1\) would yield
\(\vec{p} = (0, 0, \dots, 0)^t\), which means that no charging would take
place at all. Thus, we have to incorporate constraints that enforce the
charging of the correct amount of energy at valid time slots.

    \hypertarget{constraints}{%
\subsection{Constraints}\label{sec:notebook-1-constraints}}

    For every \(k \in \{0, \dots, K-1\}\) we define

\[
C_k = \bigl( 0 \; \dots 0 \; \underbrace{1}_{t_k^a} \; 1 \dots 1 \; \underbrace{1}_{t_k^d} \; 0 \dots 0 \bigr) \ ,
\]

    and collect all \(C_k\) in a matrix \(C\):

\[
C = 
\pmat{
    C_0 &     &        &     
    \\
        & C_1 &        &     
    \\
        &     & \ddots &     
    \\
        &     &        & C_{K-1}
} \ .
\]

    Then, the \textbf{constraints} (charge at valid times the right amount
of energy) are given by

\[
C \vec{p} = \vec{e} \ ,
\qquad
\vec{e} = \pmat{e_0 \\ \vdots \\ e_{K-1}},
\]

where \(e_k\) is the amount of energy car \(k\) needs to charge, see Sections \ref{sec:notebook-1-cost-function} and \ref{sec:notebook-1-constraints}.

    \hypertarget{minimization-problem}{%
\subsection{Minimization Problem}\label{minimization-problem}}

Putting all together we have a \textbf{quadratic constrained integer
optimization} problem that reads

\begin{equation}
\left\{ \; \;
\begin{aligned}
    \min_{\vec{p} \in \{0, \dots, L-1\}^{KT}} f_1(\vec{p}\, ) \ ,
    \\[0.2cm]
    \text{such that } C \vec{p} = \vec{e} \ .
\end{aligned}
\label{eq:qcio}
\tag{QCIO}
\right.
\end{equation}

    Next, we want to implement the logic of \eqref{eq:qcio} in \textbf{Python} and
\textbf{Qiskit}.

    \hypertarget{implementation-of-poc-model}{%
\section{Implementation of POC
Model}\label{implementation-of-poc-model}}

    \hypertarget{part-1-python}{%
\subsection{Part 1: Python}\label{sec:implementation-python}}

    First, we write a class that holds the data for a car that should be
charged.

\begin{tcolorbox}[breakable, size=fbox, boxrule=1pt, pad at break*=1mm,colback=cellbackground, colframe=cellborder]
\prompt{In}{incolor}{2}{\boxspacing}
\begin{Verbatim}[commandchars=\\\{\}]
\PY{k+kn}{from} \PY{n+nn}{typing} \PY{k+kn}{import} \PY{n}{List} 

\PY{k}{class} \PY{n+nc}{Car}\PY{p}{:}
    \PY{k}{def} \PY{n+nf+fm}{\PYZus{}\PYZus{}init\PYZus{}\PYZus{}}\PY{p}{(}
        \PY{n+nb+bp}{self}\PY{p}{,}
        \PY{n}{car\PYZus{}id}\PY{p}{:} \PY{n+nb}{str}\PY{p}{,} \PY{c+c1}{\PYZsh{} an arbitrary name for the car}
        \PY{n}{time\PYZus{}slots\PYZus{}at\PYZus{}charging\PYZus{}unit}\PY{p}{:} \PY{n}{List}\PY{p}{[}\PY{n+nb}{int}\PY{p}{]}\PY{p}{,} \PY{c+c1}{\PYZsh{} time slots when the}
                                                \PY{c+c1}{\PYZsh{} car is at the charging unit}
        \PY{n}{required\PYZus{}energy}\PY{p}{:} \PY{n+nb}{int} \PY{c+c1}{\PYZsh{} energy units that should be charged}
    \PY{p}{)} \PY{o}{\PYZhy{}}\PY{o}{\PYZgt{}} \PY{k+kc}{None}\PY{p}{:}
        \PY{n+nb+bp}{self}\PY{o}{.}\PY{n}{car\PYZus{}id} \PY{o}{=} \PY{n}{car\PYZus{}id}
        \PY{n+nb+bp}{self}\PY{o}{.}\PY{n}{time\PYZus{}slots\PYZus{}at\PYZus{}charging\PYZus{}unit} \PY{o}{=} \PY{n}{time\PYZus{}slots\PYZus{}at\PYZus{}charging\PYZus{}unit}
        \PY{n+nb+bp}{self}\PY{o}{.}\PY{n}{required\PYZus{}energy} \PY{o}{=} \PY{n}{required\PYZus{}energy}

    \PY{k}{def} \PY{n+nf+fm}{\PYZus{}\PYZus{}str\PYZus{}\PYZus{}}\PY{p}{(}\PY{n+nb+bp}{self}\PY{p}{)} \PY{o}{\PYZhy{}}\PY{o}{\PYZgt{}} \PY{n+nb}{str}\PY{p}{:}
        \PY{k}{return} \PY{l+s+sa}{f}\PY{l+s+s2}{\PYZdq{}}\PY{l+s+s2}{Car }\PY{l+s+s2}{\PYZsq{}}\PY{l+s+si}{\PYZob{}}\PY{n+nb+bp}{self}\PY{o}{.}\PY{n}{car\PYZus{}id}\PY{l+s+si}{\PYZcb{}}\PY{l+s+s2}{\PYZsq{}}\PY{l+s+s2}{:}\PY{l+s+se}{\PYZbs{}n}\PY{l+s+s2}{\PYZdq{}} \PYZbs{}
            \PY{l+s+s2}{\PYZdq{}}\PY{l+s+s2}{  at charging station at time slots }\PY{l+s+s2}{\PYZdq{}} \PYZbs{}
            \PY{l+s+sa}{f}\PY{l+s+s2}{\PYZdq{}}\PY{l+s+si}{\PYZob{}}\PY{n+nb+bp}{self}\PY{o}{.}\PY{n}{time\PYZus{}slots\PYZus{}at\PYZus{}charging\PYZus{}unit}\PY{l+s+si}{\PYZcb{}}\PY{l+s+se}{\PYZbs{}n}\PY{l+s+s2}{\PYZdq{}} \PYZbs{}
            \PY{l+s+s2}{\PYZdq{}}\PY{l+s+s2}{  requires }\PY{l+s+s2}{\PYZdq{}} \PYZbs{}
            \PY{l+s+sa}{f}\PY{l+s+s2}{\PYZdq{}}\PY{l+s+si}{\PYZob{}}\PY{n+nb+bp}{self}\PY{o}{.}\PY{n}{required\PYZus{}energy}\PY{l+s+si}{\PYZcb{}}\PY{l+s+s2}{ energy units}\PY{l+s+s2}{\PYZdq{}}
\end{Verbatim}
\end{tcolorbox}

    Next, we write a class to hold the data for a charging unit and with the
possibility to register cars to charge at it. Moreover, to keep things
simple we give this class the task to generate the matrices of the
mathematical formulation of our POC model, see Section \ref{sec:mathematical-description-poc-model}.

\begin{tcolorbox}[breakable, size=fbox, boxrule=1pt, pad at break*=1mm,colback=cellbackground, colframe=cellborder]
\prompt{In}{incolor}{6}{\boxspacing}
\begin{Verbatim}[commandchars=\\\{\}]
\PY{k+kn}{import} \PY{n+nn}{numpy} \PY{k}{as} \PY{n+nn}{np}

\PY{k}{class} \PY{n+nc}{ChargingUnit}\PY{p}{:}
    \PY{k}{def} \PY{n+nf+fm}{\PYZus{}\PYZus{}init\PYZus{}\PYZus{}}\PY{p}{(}
        \PY{n+nb+bp}{self}\PY{p}{,}
        \PY{n}{charging\PYZus{}unit\PYZus{}id}\PY{p}{:} \PY{n+nb}{str}\PY{p}{,} \PY{c+c1}{\PYZsh{} an arbitrary name for the charging unit}
        \PY{n}{number\PYZus{}charging\PYZus{}levels}\PY{p}{:} \PY{n+nb}{int}\PY{p}{,}
        \PY{n}{number\PYZus{}time\PYZus{}slots}\PY{p}{:} \PY{n+nb}{int}\PY{p}{,}
    \PY{p}{)} \PY{o}{\PYZhy{}}\PY{o}{\PYZgt{}} \PY{k+kc}{None}\PY{p}{:}
        \PY{n+nb+bp}{self}\PY{o}{.}\PY{n}{charging\PYZus{}unit\PYZus{}id} \PY{o}{=} \PY{n}{charging\PYZus{}unit\PYZus{}id}
        \PY{n+nb+bp}{self}\PY{o}{.}\PY{n}{number\PYZus{}charging\PYZus{}levels} \PY{o}{=} \PY{n}{number\PYZus{}charging\PYZus{}levels}
        \PY{n+nb+bp}{self}\PY{o}{.}\PY{n}{number\PYZus{}time\PYZus{}slots} \PY{o}{=} \PY{n}{number\PYZus{}time\PYZus{}slots}
        \PY{n+nb+bp}{self}\PY{o}{.}\PY{n}{cars\PYZus{}to\PYZus{}charge} \PY{o}{=} \PY{p}{[}\PY{p}{]}

    \PY{k}{def} \PY{n+nf+fm}{\PYZus{}\PYZus{}str\PYZus{}\PYZus{}}\PY{p}{(}\PY{n+nb+bp}{self}\PY{p}{)} \PY{o}{\PYZhy{}}\PY{o}{\PYZgt{}} \PY{n+nb}{str}\PY{p}{:}
        \PY{n}{info\PYZus{}cars\PYZus{}registered} \PY{o}{=} \PY{l+s+s2}{\PYZdq{}}\PY{l+s+s2}{\PYZdq{}}
        \PY{k}{for} \PY{n}{car} \PY{o+ow}{in} \PY{n+nb+bp}{self}\PY{o}{.}\PY{n}{cars\PYZus{}to\PYZus{}charge}\PY{p}{:}
            \PY{n}{info\PYZus{}cars\PYZus{}registered} \PY{o}{=} \PY{n}{info\PYZus{}cars\PYZus{}registered} \PY{o}{+} \PY{l+s+s2}{\PYZdq{}}\PY{l+s+s2}{ }\PY{l+s+s2}{\PYZdq{}} \PY{o}{+} \PY{n}{car}\PY{o}{.}\PY{n}{car\PYZus{}id}
        \PY{k}{return} \PY{l+s+s2}{\PYZdq{}}\PY{l+s+s2}{Charging unit with}\PY{l+s+se}{\PYZbs{}n}\PY{l+s+s2}{\PYZdq{}} \PYZbs{}
            \PY{l+s+s2}{\PYZdq{}}\PY{l+s+s2}{  charging levels: }\PY{l+s+s2}{\PYZdq{}} \PYZbs{}
            \PY{l+s+sa}{f}\PY{l+s+s2}{\PYZdq{}}\PY{l+s+si}{\PYZob{}}\PY{n+nb}{list}\PY{p}{(}\PY{n+nb}{range}\PY{p}{(}\PY{n+nb+bp}{self}\PY{o}{.}\PY{n}{number\PYZus{}charging\PYZus{}levels}\PY{p}{)}\PY{p}{)}\PY{p}{[}\PY{l+m+mi}{1}\PY{p}{:}\PY{o}{\PYZhy{}}\PY{l+m+mi}{1}\PY{p}{]}\PY{l+s+si}{\PYZcb{}}\PY{l+s+se}{\PYZbs{}n}\PY{l+s+s2}{\PYZdq{}} \PYZbs{}
            \PY{l+s+s2}{\PYZdq{}}\PY{l+s+s2}{  time slots: }\PY{l+s+s2}{\PYZdq{}} \PYZbs{}
            \PY{l+s+sa}{f}\PY{l+s+s2}{\PYZdq{}}\PY{l+s+si}{\PYZob{}}\PY{n+nb}{list}\PY{p}{(}\PY{n+nb}{range}\PY{p}{(}\PY{n+nb+bp}{self}\PY{o}{.}\PY{n}{number\PYZus{}time\PYZus{}slots}\PY{p}{)}\PY{p}{)}\PY{p}{[}\PY{l+m+mi}{1}\PY{p}{:}\PY{o}{\PYZhy{}}\PY{l+m+mi}{1}\PY{p}{]}\PY{l+s+si}{\PYZcb{}}\PY{l+s+se}{\PYZbs{}n}\PY{l+s+s2}{\PYZdq{}} \PYZbs{}
            \PY{l+s+s2}{\PYZdq{}}\PY{l+s+s2}{  cars to charge:}\PY{l+s+s2}{\PYZdq{}} \PYZbs{}
            \PY{o}{+} \PY{n}{info\PYZus{}cars\PYZus{}registered}

    \PY{k}{def} \PY{n+nf}{register\PYZus{}car\PYZus{}for\PYZus{}charging}\PY{p}{(}\PY{n+nb+bp}{self}\PY{p}{,} \PY{n}{car}\PY{p}{:} \PY{n}{Car}\PY{p}{)} \PY{o}{\PYZhy{}}\PY{o}{\PYZgt{}} \PY{k+kc}{None}\PY{p}{:}
        \PY{k}{if} \PY{n+nb}{max}\PY{p}{(}\PY{n}{car}\PY{o}{.}\PY{n}{time\PYZus{}slots\PYZus{}at\PYZus{}charging\PYZus{}unit}\PY{p}{)} \PY{o}{\PYZgt{}} \PY{n+nb+bp}{self}\PY{o}{.}\PY{n}{number\PYZus{}time\PYZus{}slots} \PY{o}{\PYZhy{}} \PY{l+m+mi}{1}\PY{p}{:}
            \PY{k}{raise} \PY{n+ne}{ValueError}\PY{p}{(}\PY{l+s+s2}{\PYZdq{}}\PY{l+s+s2}{From car required time slots not compatible }\PY{l+s+s2}{\PYZdq{}}
                \PY{l+s+s2}{\PYZdq{}}\PY{l+s+s2}{ with charging unit.}\PY{l+s+s2}{\PYZdq{}}\PY{p}{)}
        \PY{n+nb+bp}{self}\PY{o}{.}\PY{n}{cars\PYZus{}to\PYZus{}charge}\PY{o}{.}\PY{n}{append}\PY{p}{(}\PY{n}{car}\PY{p}{)}

    \PY{k}{def} \PY{n+nf}{reset\PYZus{}cars\PYZus{}for\PYZus{}charging}\PY{p}{(}\PY{n+nb+bp}{self}\PY{p}{)} \PY{o}{\PYZhy{}}\PY{o}{\PYZgt{}} \PY{k+kc}{None}\PY{p}{:}
        \PY{n+nb+bp}{self}\PY{o}{.}\PY{n}{cars\PYZus{}to\PYZus{}charge} \PY{o}{=} \PY{p}{[}\PY{p}{]}

    \PY{k}{def} \PY{n+nf}{generate\PYZus{}constraint\PYZus{}matrix}\PY{p}{(}\PY{n+nb+bp}{self}\PY{p}{)} \PY{o}{\PYZhy{}}\PY{o}{\PYZgt{}} \PY{n}{np}\PY{o}{.}\PY{n}{ndarray}\PY{p}{:}
        \PY{l+s+sd}{\PYZdq{}\PYZdq{}\PYZdq{}Matrix with ones for times when car is at charging station}
\PY{l+s+sd}{         and with zeros if car is not at charging station\PYZdq{}\PYZdq{}\PYZdq{}}
        \PY{n}{number\PYZus{}cars\PYZus{}to\PYZus{}charge} \PY{o}{=} \PY{n+nb}{len}\PY{p}{(}\PY{n+nb+bp}{self}\PY{o}{.}\PY{n}{cars\PYZus{}to\PYZus{}charge}\PY{p}{)}
        \PY{n}{constraint\PYZus{}matrix} \PY{o}{=} \PY{n}{np}\PY{o}{.}\PY{n}{zeros}\PY{p}{(}
            \PY{p}{(}\PY{n}{number\PYZus{}cars\PYZus{}to\PYZus{}charge}\PY{p}{,} 
                \PY{n}{number\PYZus{}cars\PYZus{}to\PYZus{}charge}\PY{o}{*}\PY{n+nb+bp}{self}\PY{o}{.}\PY{n}{number\PYZus{}time\PYZus{}slots}\PY{p}{)}\PY{p}{)}
        \PY{k}{for} \PY{n}{row\PYZus{}index} \PY{o+ow}{in} \PY{n+nb}{range}\PY{p}{(}\PY{l+m+mi}{0}\PY{p}{,} \PY{n}{number\PYZus{}cars\PYZus{}to\PYZus{}charge}\PY{p}{)}\PY{p}{:}
            \PY{n}{offset} \PY{o}{=} \PY{n}{row\PYZus{}index}\PY{o}{*}\PY{n+nb+bp}{self}\PY{o}{.}\PY{n}{number\PYZus{}time\PYZus{}slots}
            \PY{n}{cols} \PY{o}{=} \PY{n}{np}\PY{o}{.}\PY{n}{array}\PY{p}{(}
                \PY{n+nb+bp}{self}\PY{o}{.}\PY{n}{cars\PYZus{}to\PYZus{}charge}\PY{p}{[}\PY{n}{row\PYZus{}index}\PY{p}{]}\PY{o}{.}\PY{n}{time\PYZus{}slots\PYZus{}at\PYZus{}charging\PYZus{}unit}\PY{p}{)}
            \PY{n}{constraint\PYZus{}matrix}\PY{p}{[}\PY{n}{row\PYZus{}index}\PY{p}{,} \PY{n}{offset}\PY{o}{+}\PY{n}{cols}\PY{p}{]} \PY{o}{=} \PY{l+m+mi}{1}
        \PY{k}{return} \PY{n}{constraint\PYZus{}matrix}

    \PY{k}{def} \PY{n+nf}{generate\PYZus{}constraint\PYZus{}rhs}\PY{p}{(}\PY{n+nb+bp}{self}\PY{p}{)} \PY{o}{\PYZhy{}}\PY{o}{\PYZgt{}} \PY{n}{np}\PY{o}{.}\PY{n}{ndarray}\PY{p}{:}
        \PY{l+s+sd}{\PYZdq{}\PYZdq{}\PYZdq{}Vector with required energy as entries\PYZdq{}\PYZdq{}\PYZdq{}}
        \PY{n}{number\PYZus{}cars\PYZus{}to\PYZus{}charge} \PY{o}{=} \PY{n+nb}{len}\PY{p}{(}\PY{n+nb+bp}{self}\PY{o}{.}\PY{n}{cars\PYZus{}to\PYZus{}charge}\PY{p}{)}
        \PY{n}{constraint\PYZus{}rhs} \PY{o}{=} \PY{n}{np}\PY{o}{.}\PY{n}{zeros}\PY{p}{(}\PY{p}{(}\PY{n}{number\PYZus{}cars\PYZus{}to\PYZus{}charge}\PY{p}{,} \PY{l+m+mi}{1}\PY{p}{)}\PY{p}{)}
        \PY{k}{for} \PY{n}{row\PYZus{}index} \PY{o+ow}{in} \PY{n+nb}{range}\PY{p}{(}\PY{l+m+mi}{0}\PY{p}{,} \PY{n}{number\PYZus{}cars\PYZus{}to\PYZus{}charge}\PY{p}{)}\PY{p}{:}
            \PY{n}{constraint\PYZus{}rhs}\PY{p}{[}\PY{n}{row\PYZus{}index}\PY{p}{]} \PY{o}{=} \PYZbs{}
                \PY{n+nb+bp}{self}\PY{o}{.}\PY{n}{cars\PYZus{}to\PYZus{}charge}\PY{p}{[}\PY{n}{row\PYZus{}index}\PY{p}{]}\PY{o}{.}\PY{n}{required\PYZus{}energy}
        \PY{k}{return} \PY{n}{constraint\PYZus{}rhs}

    \PY{k}{def} \PY{n+nf}{generate\PYZus{}cost\PYZus{}matrix}\PY{p}{(}\PY{n+nb+bp}{self}\PY{p}{)} \PY{o}{\PYZhy{}}\PY{o}{\PYZgt{}} \PY{n}{np}\PY{o}{.}\PY{n}{ndarray}\PY{p}{:}
        \PY{n}{number\PYZus{}cars\PYZus{}to\PYZus{}charge} \PY{o}{=} \PY{n+nb}{len}\PY{p}{(}\PY{n+nb+bp}{self}\PY{o}{.}\PY{n}{cars\PYZus{}to\PYZus{}charge}\PY{p}{)}
        \PY{k}{return} \PY{n}{np}\PY{o}{.}\PY{n}{kron}\PY{p}{(}
            \PY{n}{np}\PY{o}{.}\PY{n}{ones}\PY{p}{(}\PY{p}{(}\PY{n}{number\PYZus{}cars\PYZus{}to\PYZus{}charge}\PY{p}{,} \PY{l+m+mi}{1}\PY{p}{)}\PY{p}{)} \PYZbs{}
                \PY{o}{@} \PY{n}{np}\PY{o}{.}\PY{n}{ones}\PY{p}{(}\PY{p}{(}\PY{l+m+mi}{1}\PY{p}{,} \PY{n}{number\PYZus{}cars\PYZus{}to\PYZus{}charge}\PY{p}{)}\PY{p}{)}\PY{p}{,}
            \PY{n}{np}\PY{o}{.}\PY{n}{eye}\PY{p}{(}\PY{n+nb+bp}{self}\PY{o}{.}\PY{n}{number\PYZus{}time\PYZus{}slots}\PY{p}{)}\PY{p}{)}
\end{Verbatim}
\end{tcolorbox}

\hypertarget{example}{%
\subsubsection{Toy Example: Implementation in Python}}

    Let's instantiate the objects for our upper example:

    \begin{tcolorbox}[breakable, size=fbox, boxrule=1pt, pad at break*=1mm,colback=cellbackground, colframe=cellborder]
\prompt{In}{incolor}{3}{\boxspacing}
\begin{Verbatim}[commandchars=\\\{\}]
\PY{n}{car\PYZus{}green} \PY{o}{=} \PY{n}{Car}\PY{p}{(}
    \PY{n}{car\PYZus{}id}\PY{o}{=}\PY{l+s+s2}{\PYZdq{}}\PY{l+s+s2}{car\PYZus{}green}\PY{l+s+s2}{\PYZdq{}}\PY{p}{,}
    \PY{n}{time\PYZus{}slots\PYZus{}at\PYZus{}charging\PYZus{}unit}\PY{o}{=}\PY{p}{[}\PY{l+m+mi}{0}\PY{p}{,} \PY{l+m+mi}{1}\PY{p}{,} \PY{l+m+mi}{2}\PY{p}{,} \PY{l+m+mi}{3}\PY{p}{]}\PY{p}{,}
    \PY{n}{required\PYZus{}energy}\PY{o}{=}\PY{l+m+mi}{8}\PY{p}{)}

\PY{n}{car\PYZus{}orange} \PY{o}{=} \PY{n}{Car}\PY{p}{(}
    \PY{n}{car\PYZus{}id}\PY{o}{=}\PY{l+s+s2}{\PYZdq{}}\PY{l+s+s2}{car\PYZus{}orange}\PY{l+s+s2}{\PYZdq{}}\PY{p}{,}
    \PY{n}{time\PYZus{}slots\PYZus{}at\PYZus{}charging\PYZus{}unit}\PY{o}{=}\PY{p}{[}\PY{l+m+mi}{1}\PY{p}{,} \PY{l+m+mi}{2}\PY{p}{,} \PY{l+m+mi}{3}\PY{p}{,} \PY{l+m+mi}{4}\PY{p}{,} \PY{l+m+mi}{5}\PY{p}{,}  \PY{l+m+mi}{6}\PY{p}{]}\PY{p}{,}
    \PY{n}{required\PYZus{}energy}\PY{o}{=}\PY{l+m+mi}{12}\PY{p}{)}
\end{Verbatim}
\end{tcolorbox}

    \begin{tcolorbox}[breakable, size=fbox, boxrule=1pt, pad at break*=1mm,colback=cellbackground, colframe=cellborder]
\prompt{In}{incolor}{4}{\boxspacing}
\begin{Verbatim}[commandchars=\\\{\}]
\PY{n+nb}{print}\PY{p}{(}\PY{n}{car\PYZus{}green}\PY{p}{)}
\PY{n+nb}{print}\PY{p}{(}\PY{n}{car\PYZus{}orange}\PY{p}{)}
\end{Verbatim}
\end{tcolorbox}

    \begin{Verbatim}[commandchars=\\\{\}]
Car 'car\_green':
  at charging station at time slots [0, 1, 2, 3]
  requires 8 energy units
Car 'car\_orange':
  at charging station at time slots [1, 2, 3, 4, 5, 6]
  requires 12 energy units
    \end{Verbatim}

    \begin{tcolorbox}[breakable, size=fbox, boxrule=1pt, pad at break*=1mm,colback=cellbackground, colframe=cellborder]
\prompt{In}{incolor}{5}{\boxspacing}
\begin{Verbatim}[commandchars=\\\{\}]
\PY{n}{charging\PYZus{}unit} \PY{o}{=} \PY{n}{ChargingUnit}\PY{p}{(}
    \PY{n}{charging\PYZus{}unit\PYZus{}id}\PY{o}{=}\PY{l+s+s2}{\PYZdq{}}\PY{l+s+s2}{charging\PYZus{}unit}\PY{l+s+s2}{\PYZdq{}}\PY{p}{,}
    \PY{n}{number\PYZus{}charging\PYZus{}levels}\PY{o}{=}\PY{l+m+mi}{6}\PY{p}{,}
    \PY{n}{number\PYZus{}time\PYZus{}slots}\PY{o}{=}\PY{l+m+mi}{7}\PY{p}{)}
\end{Verbatim}
\end{tcolorbox}

    \begin{tcolorbox}[breakable, size=fbox, boxrule=1pt, pad at break*=1mm,colback=cellbackground, colframe=cellborder]
\prompt{In}{incolor}{6}{\boxspacing}
\begin{Verbatim}[commandchars=\\\{\}]
\PY{n+nb}{print}\PY{p}{(}\PY{n}{charging\PYZus{}unit}\PY{p}{)}
\end{Verbatim}
\end{tcolorbox}

    \begin{Verbatim}[commandchars=\\\{\}]
Charging unit with
  charging levels: 0, 1, 2, 3, 4, 5
  time slots: 0, 1, 2, 3, 4, 5, 6
  cars to charge:
    \end{Verbatim}

    \begin{tcolorbox}[breakable, size=fbox, boxrule=1pt, pad at break*=1mm,colback=cellbackground, colframe=cellborder]
\prompt{In}{incolor}{7}{\boxspacing}
\begin{Verbatim}[commandchars=\\\{\}]
\PY{n}{charging\PYZus{}unit}\PY{o}{.}\PY{n}{register\PYZus{}car\PYZus{}for\PYZus{}charging}\PY{p}{(}\PY{n}{car\PYZus{}green}\PY{p}{)}
\PY{n}{charging\PYZus{}unit}\PY{o}{.}\PY{n}{register\PYZus{}car\PYZus{}for\PYZus{}charging}\PY{p}{(}\PY{n}{car\PYZus{}orange}\PY{p}{)}
\PY{n+nb}{print}\PY{p}{(}\PY{n}{charging\PYZus{}unit}\PY{p}{)}
\end{Verbatim}
\end{tcolorbox}

    \begin{Verbatim}[commandchars=\\\{\}]
Charging unit with
  charging levels: 0, 1, 2, 3, 4, 5
  time slots: 0, 1, 2, 3, 4, 5, 6
  cars to charge: car\_green car\_orange
    \end{Verbatim}

    Now, let's get the cost matrix \(A\), the constraint matrix \(C\), and the constraint right-hand side (RHS) vector \(\vec{e}\):

    \begin{tcolorbox}[breakable, size=fbox, boxrule=1pt, pad at break*=1mm,colback=cellbackground, colframe=cellborder]
\prompt{In}{incolor}{8}{\boxspacing}
\begin{Verbatim}[commandchars=\\\{\}]
\PY{n}{A} \PY{o}{=} \PY{n}{charging\PYZus{}unit}\PY{o}{.}\PY{n}{generate\PYZus{}cost\PYZus{}matrix}\PY{p}{(}\PY{p}{)}
\PY{n+nb}{print}\PY{p}{(}\PY{l+s+sa}{f}\PY{l+s+s2}{\PYZdq{}}\PY{l+s+s2}{A =}\PY{l+s+se}{\PYZbs{}n}\PY{l+s+si}{\PYZob{}}\PY{n}{A}\PY{l+s+si}{\PYZcb{}}\PY{l+s+s2}{\PYZdq{}}\PY{p}{)}
\end{Verbatim}
\end{tcolorbox}

    \begin{Verbatim}[commandchars=\\\{\}]
A =
[[1. 0. 0. 0. 0. 0. 0. 1. 0. 0. 0. 0. 0. 0.]
 [0. 1. 0. 0. 0. 0. 0. 0. 1. 0. 0. 0. 0. 0.]
 [0. 0. 1. 0. 0. 0. 0. 0. 0. 1. 0. 0. 0. 0.]
 [0. 0. 0. 1. 0. 0. 0. 0. 0. 0. 1. 0. 0. 0.]
 [0. 0. 0. 0. 1. 0. 0. 0. 0. 0. 0. 1. 0. 0.]
 [0. 0. 0. 0. 0. 1. 0. 0. 0. 0. 0. 0. 1. 0.]
 [0. 0. 0. 0. 0. 0. 1. 0. 0. 0. 0. 0. 0. 1.]
 [1. 0. 0. 0. 0. 0. 0. 1. 0. 0. 0. 0. 0. 0.]
 [0. 1. 0. 0. 0. 0. 0. 0. 1. 0. 0. 0. 0. 0.]
 [0. 0. 1. 0. 0. 0. 0. 0. 0. 1. 0. 0. 0. 0.]
 [0. 0. 0. 1. 0. 0. 0. 0. 0. 0. 1. 0. 0. 0.]
 [0. 0. 0. 0. 1. 0. 0. 0. 0. 0. 0. 1. 0. 0.]
 [0. 0. 0. 0. 0. 1. 0. 0. 0. 0. 0. 0. 1. 0.]
 [0. 0. 0. 0. 0. 0. 1. 0. 0. 0. 0. 0. 0. 1.]]
    \end{Verbatim}

    \begin{tcolorbox}[breakable, size=fbox, boxrule=1pt, pad at break*=1mm,colback=cellbackground, colframe=cellborder]
\prompt{In}{incolor}{9}{\boxspacing}
\begin{Verbatim}[commandchars=\\\{\}]
\PY{n}{C} \PY{o}{=} \PY{n}{charging\PYZus{}unit}\PY{o}{.}\PY{n}{generate\PYZus{}constraint\PYZus{}matrix}\PY{p}{(}\PY{p}{)}
\PY{n}{e} \PY{o}{=} \PY{n}{charging\PYZus{}unit}\PY{o}{.}\PY{n}{generate\PYZus{}constraint\PYZus{}rhs}\PY{p}{(}\PY{p}{)}
\PY{n+nb}{print}\PY{p}{(}\PY{l+s+sa}{f}\PY{l+s+s2}{\PYZdq{}}\PY{l+s+s2}{C =}\PY{l+s+se}{\PYZbs{}n}\PY{l+s+si}{\PYZob{}}\PY{n}{C}\PY{l+s+si}{\PYZcb{}}\PY{l+s+se}{\PYZbs{}n}\PY{l+s+s2}{\PYZdq{}}\PY{p}{)}
\PY{n+nb}{print}\PY{p}{(}\PY{l+s+sa}{f}\PY{l+s+s2}{\PYZdq{}}\PY{l+s+s2}{e =}\PY{l+s+se}{\PYZbs{}n}\PY{l+s+si}{\PYZob{}}\PY{n}{e}\PY{l+s+si}{\PYZcb{}}\PY{l+s+s2}{\PYZdq{}}\PY{p}{)}
\end{Verbatim}
\end{tcolorbox}

    \begin{Verbatim}[commandchars=\\\{\}]
C =
[[1. 1. 1. 1. 0. 0. 0. 0. 0. 0. 0. 0. 0. 0.]
 [0. 0. 0. 0. 0. 0. 0. 0. 1. 1. 1. 1. 1. 1.]]

e =
[[ 8.]
 [12.]]
    \end{Verbatim}

    Next, we move to Qiskit, where we can easily implement the minimization problem.

    \hypertarget{part-2-qiskit}{%
\subsection{Part 2: Qiskit}\label{sec:notebook-1-part-2-qiskit}}

    The module \href{https://qiskit.org/documentation/optimization/apidocs/qiskit_optimization.html}{\texttt{qiskit\_optimization}} contains the class
\href{https://qiskit.org/documentation/optimization/stubs/qiskit_optimization.QuadraticProgram.html#qiskit_optimization.QuadraticProgram}{\texttt{QuadraticProgram}}
which is used to represent quadratic optimization problems. In the
following code we write a function that generates a
\texttt{QuadraticProgram} instance according to \eqref{eq:qcio}, where the
necessary data stem from a \texttt{ChargingUnit} object.

    \emph{Remark:} For further information about \texttt{qiskit\_optimization} and how it is used see e.g. \cite{QiskitOptimizationWeb}.

\begin{tcolorbox}[breakable, size=fbox, boxrule=1pt, pad at break*=1mm,colback=cellbackground, colframe=cellborder]
\prompt{In}{incolor}{18}{\boxspacing}
\begin{Verbatim}[commandchars=\\\{\}]
\PY{k+kn}{from} \PY{n+nn}{qiskit\PYZus{}optimization} \PY{k+kn}{import} \PY{n}{QuadraticProgram}

\PY{k}{def} \PY{n+nf}{generate\PYZus{}qcio}\PY{p}{(}
    \PY{n}{charging\PYZus{}unit}\PY{p}{:} \PY{n}{ChargingUnit}\PY{p}{,}
    \PY{n}{name}\PY{p}{:} \PY{n+nb}{str}\PY{o}{=}\PY{k+kc}{None}
\PY{p}{)} \PY{o}{\PYZhy{}}\PY{o}{\PYZgt{}} \PY{n}{QuadraticProgram}\PY{p}{:}
    \PY{k}{if} \PY{n}{name} \PY{o+ow}{is} \PY{k+kc}{None}\PY{p}{:}
        \PY{n}{name} \PY{o}{=} \PY{l+s+s2}{\PYZdq{}}\PY{l+s+s2}{\PYZdq{}}
    \PY{n}{qcio} \PY{o}{=} \PY{n}{QuadraticProgram}\PY{p}{(}\PY{n}{name}\PY{p}{)}

    \PY{k}{for} \PY{n}{car} \PY{o+ow}{in} \PY{n}{charging\PYZus{}unit}\PY{o}{.}\PY{n}{cars\PYZus{}to\PYZus{}charge}\PY{p}{:}
        \PY{n}{qcio}\PY{o}{.}\PY{n}{integer\PYZus{}var\PYZus{}list}\PY{p}{(}
            \PY{n}{keys}\PY{o}{=}\PY{p}{[}\PY{l+s+sa}{f}\PY{l+s+s2}{\PYZdq{}}\PY{l+s+si}{\PYZob{}}\PY{n}{car}\PY{o}{.}\PY{n}{car\PYZus{}id}\PY{l+s+si}{\PYZcb{}}\PY{l+s+s2}{\PYZus{}t}\PY{l+s+si}{\PYZob{}}\PY{n}{t}\PY{l+s+si}{\PYZcb{}}\PY{l+s+s2}{\PYZdq{}} \PYZbs{}
                    \PY{k}{for} \PY{n}{t} \PY{o+ow}{in} \PY{n+nb}{range}\PY{p}{(}\PY{l+m+mi}{0}\PY{p}{,} \PY{n}{charging\PYZus{}unit}\PY{o}{.}\PY{n}{number\PYZus{}time\PYZus{}slots}\PY{p}{)}\PY{p}{]}\PY{p}{,}
            \PY{n}{lowerbound}\PY{o}{=}\PY{l+m+mi}{0}\PY{p}{,}
            \PY{n}{upperbound}\PY{o}{=}\PY{n}{charging\PYZus{}unit}\PY{o}{.}\PY{n}{number\PYZus{}charging\PYZus{}levels}\PY{o}{\PYZhy{}}\PY{l+m+mi}{1}\PY{p}{,}
            \PY{n}{name}\PY{o}{=}\PY{l+s+s2}{\PYZdq{}}\PY{l+s+s2}{p_}\PY{l+s+s2}{\PYZdq{}}\PY{p}{)}

    \PY{n}{constraint\PYZus{}matrix} \PY{o}{=} \PY{n}{charging\PYZus{}unit}\PY{o}{.}\PY{n}{generate\PYZus{}constraint\PYZus{}matrix}\PY{p}{(}\PY{p}{)}
    \PY{n}{constraint\PYZus{}rhs} \PY{o}{=} \PY{n}{charging\PYZus{}unit}\PY{o}{.}\PY{n}{generate\PYZus{}constraint\PYZus{}rhs}\PY{p}{(}\PY{p}{)}
    \PY{k}{for} \PY{n}{row\PYZus{}index} \PY{o+ow}{in} \PY{n+nb}{range}\PY{p}{(}\PY{l+m+mi}{0}\PY{p}{,} \PY{n}{constraint\PYZus{}matrix}\PY{o}{.}\PY{n}{shape}\PY{p}{[}\PY{l+m+mi}{0}\PY{p}{]}\PY{p}{)}\PY{p}{:}
        \PY{n}{qcio}\PY{o}{.}\PY{n}{linear\PYZus{}constraint}\PY{p}{(}
            \PY{n}{linear}\PY{o}{=}\PY{n}{constraint\PYZus{}matrix}\PY{p}{[}\PY{n}{row\PYZus{}index}\PY{p}{,} \PY{p}{:}\PY{p}{]}\PY{p}{,}
            \PY{n}{rhs}\PY{o}{=}\PY{n}{constraint\PYZus{}rhs}\PY{p}{[}\PY{n}{row\PYZus{}index}\PY{p}{]}\PY{p}{[}\PY{l+m+mi}{0}\PY{p}{]}\PY{p}{,}
            \PY{n}{sense}\PY{o}{=}\PY{l+s+s2}{\PYZdq{}}\PY{l+s+s2}{==}\PY{l+s+s2}{\PYZdq{}}\PY{p}{,}
            \PY{n}{name}\PY{o}{=}\PY{l+s+sa}{f}\PY{l+s+s2}{\PYZdq{}}\PY{l+s+s2}{charge\PYZus{}correct\PYZus{}energy\PYZus{}for\PYZus{}}\PY{l+s+si}{\PYZob{}}\PY{n}{charging\PYZus{}unit}\PY{o}{.}\PY{n}{cars\PYZus{}to\PYZus{}charge}\PY{p}{[}\PY{n}{row\PYZus{}index}\PY{p}{]}\PY{o}{.}\PY{n}{car\PYZus{}id}\PY{l+s+si}{\PYZcb{}}\PY{l+s+s2}{\PYZdq{}}\PY{p}{)}

    \PY{n}{cost\PYZus{}matrix} \PY{o}{=} \PY{n}{charging\PYZus{}unit}\PY{o}{.}\PY{n}{generate\PYZus{}cost\PYZus{}matrix}\PY{p}{(}\PY{p}{)}
    \PY{n}{qcio}\PY{o}{.}\PY{n}{minimize}\PY{p}{(}\PY{n}{quadratic}\PY{o}{=}\PY{n}{cost\PYZus{}matrix}\PY{p}{)}

    \PY{k}{return} \PY{n}{qcio}
\end{Verbatim}
\end{tcolorbox}

    \hypertarget{example}{%
\subsubsection{Toy Example: Implementation in Qiskit}\label{sec:notebook-1-example-qiskit}}

    For our upper \texttt{charging\_unit} we get:

    \begin{tcolorbox}[breakable, size=fbox, boxrule=1pt, pad at break*=1mm,colback=cellbackground, colframe=cellborder]
\prompt{In}{incolor}{11}{\boxspacing}
\begin{Verbatim}[commandchars=\\\{\}]
\PY{n}{qcio} \PY{o}{=} \PY{n}{generate\PYZus{}qcio}\PY{p}{(}\PY{n}{charging\PYZus{}unit}\PY{p}{,} \PY{n}{name}\PY{o}{=}\PY{l+s+s2}{\PYZdq{}}\PY{l+s+s2}{QCIO}\PY{l+s+s2}{\PYZdq{}}\PY{p}{)}
\end{Verbatim}
\end{tcolorbox}

    \begin{tcolorbox}[breakable, size=fbox, boxrule=1pt, pad at break*=1mm,colback=cellbackground, colframe=cellborder]
\prompt{In}{incolor}{12}{\boxspacing}
\begin{Verbatim}[commandchars=\\\{\}]
\PY{n+nb}{print}\PY{p}{(}\PY{n}{qcio}\PY{o}{.}\PY{n}{prettyprint}\PY{p}{(}\PY{p}{)}\PY{p}{)}
\end{Verbatim}
\end{tcolorbox}

    \begin{Verbatim}[commandchars=\\\{\}]
Problem name: QCIO

Minimize
  p_car\_green\_t0\^{}2 + 2*p_car\_green\_t0*p_car\_orange\_t0
  + p_car\_green\_t1\^{}2 + 2*p_car\_green\_t1*p_car\_orange\_t1
  + p_car\_green\_t2\^{}2 + 2*p_car\_green\_t2*p_car\_orange\_t2
  + p_car\_green\_t3\^{}2 + 2*p_car\_green\_t3*p_car\_orange\_t3
  + p_car\_green\_t4\^{}2 + 2*p_car\_green\_t4*p_car\_orange\_t4
  + p_car\_green\_t5\^{}2 + 2*p_car\_green\_t5*p_car\_orange\_t5
  + p_car\_green\_t6\^{}2 + 2*p_car\_green\_t6*p_car\_orange\_t6
  + p_car\_orange\_t0\^{}2 + p_car\_orange\_t1\^{}2 + p_car\_orange\_t2\^{}2
  + p_car\_orange\_t3\^{}2 + p_car\_orange\_t4\^{}2 + p_car\_orange\_t5\^{}2
  + p_car\_orange\_t6\^{}2

Subject to
  Linear constraints (2)
    p_car\_green\_t0 + p_car\_green\_t1 + p_car\_green\_t2
    + p_car\_green\_t3 == 8  'charge\_correct\_energy\_for\_car\_green'
    p_car\_orange\_t1 + p_car\_orange\_t2 + p_car\_orange\_t3
    + p_car\_orange\_t4 + p_car\_orange\_t5 + p_car\_orange\_t6
    == 12  'charge\_correct\_energy\_for\_car\_orange'

  Integer variables (14)
    0 <= p_car\_green\_t0 <= 5
    0 <= p_car\_green\_t1 <= 5
    0 <= p_car\_green\_t2 <= 5
    0 <= p_car\_green\_t3 <= 5
    0 <= p_car\_green\_t4 <= 5
    0 <= p_car\_green\_t5 <= 5
    0 <= p_car\_green\_t6 <= 5
    0 <= p_car\_orange\_t0 <= 5
    0 <= p_car\_orange\_t1 <= 5
    0 <= p_car\_orange\_t2 <= 5
    0 <= p_car\_orange\_t3 <= 5
    0 <= p_car\_orange\_t4 <= 5
    0 <= p_car\_orange\_t5 <= 5
    0 <= p_car\_orange\_t6 <= 5

    \end{Verbatim}

    \hypertarget{solve-poc-model-with-classical-solver}{%
\section{Solve POC Model with Classical
Solver}\label{solve-poc-model-with-classical-solver}}

    In order to assess our later quantum algorithm it is of great
advantage to know the exact solution of \eqref{eq:qcio}. For small POC examples
such as we are considering here this solution can be computed by a
classical solver. Qiskit provides two such solvers
(\texttt{CplexOptimizer} and \texttt{GurobiOptimizer}) in
\href{https://qiskit.org/documentation/optimization/apidocs/qiskit_optimization.algorithms.html}{\texttt{qiskit\_optimization.algorithms}}.

\emph{Remark:} Further information is available at \cite{QiskitOptimizationWeb2}.

    \hypertarget{example}{%
\subsection{Toy Example: Classical Solver}\label{sec:notebook-1-example-classical-solver}}

    We solve \texttt{qcio} from above with \texttt{CplexOptimizer}.

\begin{tcolorbox}[breakable, size=fbox, boxrule=1pt, pad at break*=1mm,colback=cellbackground, colframe=cellborder]
\prompt{In}{incolor}{21}{\boxspacing}
\begin{Verbatim}[commandchars=\\\{\}]
\PY{k+kn}{from} \PY{n+nn}{qiskit\PYZus{}optimization}\PY{n+nn}{.}\PY{n+nn}{algorithms} \PY{k+kn}{import} \PY{n}{CplexOptimizer}

\PY{n}{cplex\PYZus{}optimizer} \PY{o}{=} \PY{n}{CplexOptimizer}\PY{p}{(}\PY{p}{)}

\PY{n}{qcio\PYZus{}minimization\PYZus{}result} \PY{o}{=} \PY{n}{cplex\PYZus{}optimizer}\PY{o}{.}\PY{n}{solve}\PY{p}{(}\PY{n}{qcio}\PY{p}{)}

\PY{n+nb}{print}\PY{p}{(}\PY{l+s+s2}{\PYZdq{}}\PY{l+s+s2}{minimum point: p\PYZus{}min = }\PY{l+s+s2}{\PYZdq{}}\PY{p}{,} \PY{n}{qcio\PYZus{}minimization\PYZus{}result}\PY{o}{.}\PY{n}{x}\PY{p}{)}
\PY{n+nb}{print}\PY{p}{(}\PY{l+s+s2}{\PYZdq{}}\PY{l+s+s2}{minimum value: f\PYZus{}1(p\PYZus{}min) = }\PY{l+s+s2}{\PYZdq{}}\PY{p}{,} \PY{n}{qcio\PYZus{}minimization\PYZus{}result}\PY{o}{.}\PY{n}{fval}\PY{p}{)}
\end{Verbatim}
\end{tcolorbox}

    \begin{Verbatim}[commandchars=\\\{\}]
minimum point: p\_min =  [3. 2. 0. 3. 0. 0. 0. 0. 0. 3. 0. 3. 3. 3.]
minimum value: f\_1(p\_min) =  58.0
    \end{Verbatim}

    The array \texttt{qcio\_minimization\_result.x} corresponds to a
solution
\(\vec{p}_\mathrm{min} = \pmat{\vec{p}_{0, \mathrm{min}} \\ \vec{p}_{1, \mathrm{min}}}\)
of \eqref{eq:qcio}, where \(\vec{p}_{0, \mathrm{min}}\) and
\(\vec{p}_{1, \mathrm{min}}\) correspond to solutions for car\_green and
car\_orange, respectively.

\begin{tcolorbox}[breakable, size=fbox, boxrule=1pt, pad at break*=1mm,colback=cellbackground, colframe=cellborder]
\prompt{In}{incolor}{24}{\boxspacing}
\begin{Verbatim}[commandchars=\\\{\}]
\PY{n+nb}{print}\PY{p}{(}\PY{l+s+s2}{\PYZdq{}}\PY{l+s+s2}{minimum point for car\PYZus{}green: p\PYZus{}0,min = }\PY{l+s+s2}{\PYZdq{}}
        \PY{l+s+sa}{f}\PY{l+s+s2}{\PYZdq{}}\PY{l+s+si}{\PYZob{}}\PY{n}{qcio\PYZus{}minimization\PYZus{}result}\PY{o}{.}\PY{n}{x}\PY{p}{[}\PY{l+m+mi}{0}\PY{p}{:}\PY{n}{charging\PYZus{}unit}\PY{o}{.}\PY{n}{number\PYZus{}time\PYZus{}slots}\PY{p}{]}\PY{l+s+si}{\PYZcb{}}\PY{l+s+s2}{\PYZdq{}}\PY{p}{)}
\PY{n+nb}{print}\PY{p}{(}\PY{l+s+s2}{\PYZdq{}}\PY{l+s+s2}{minimum point for car\PYZus{}orange: p\PYZus{}1,min = }\PY{l+s+s2}{\PYZdq{}}
        \PY{l+s+sa}{f}\PY{l+s+s2}{\PYZdq{}}\PY{l+s+si}{\PYZob{}}\PY{n}{qcio\PYZus{}minimization\PYZus{}result}\PY{o}{.}\PY{n}{x}\PY{p}{[}\PY{n}{charging\PYZus{}unit}\PY{o}{.}\PY{n}{number\PYZus{}time\PYZus{}slots}\PY{p}{:}\PY{p}{]}\PY{l+s+si}{\PYZcb{}}\PY{l+s+s2}{\PYZdq{}}\PY{p}{)}
\end{Verbatim}
\end{tcolorbox}

    \begin{Verbatim}[commandchars=\\\{\}]
minimum point for car\_green: p\_0,min =  [3. 2. 0. 3. 0. 0. 0.]
minimum point for car\_orange: p\_1,min =  [0. 0. 3. 0. 3. 3. 3.]
    \end{Verbatim}

    We can plot the solution with the provided function
\texttt{plot\_charging\_schedule}

\begin{tcolorbox}[breakable, size=fbox, boxrule=1pt, pad at break*=1mm,colback=cellbackground, colframe=cellborder]
\prompt{In}{incolor}{25}{\boxspacing}
\begin{Verbatim}[commandchars=\\\{\}]
\PY{k+kn}{from} \PY{n+nn}{utils} \PY{k+kn}{import} \PY{n}{plot\PYZus{}charging\PYZus{}schedule}

\PY{n}{fig} \PY{o}{=} \PY{n}{plot\PYZus{}charging\PYZus{}schedule}\PY{p}{(}
    \PY{n}{charging\PYZus{}unit}\PY{p}{,} \PY{n}{qcio\PYZus{}minimization\PYZus{}result}\PY{o}{.}\PY{n}{x}\PY{p}{,} \PY{n}{marker\PYZus{}size}\PY{o}{=}\PY{l+m+mi}{20}\PY{p}{)}
\PY{n}{fig}\PY{o}{.}\PY{n}{update\PYZus{}layout}\PY{p}{(}\PY{n}{width}\PY{o}{=}\PY{l+m+mi}{400}\PY{p}{,} \PY{n}{height}\PY{o}{=}\PY{l+m+mi}{300}\PY{p}{)}
\PY{n}{fig}\PY{o}{.}\PY{n}{show}\PY{p}{(}\PY{p}{)}
\end{Verbatim}
\end{tcolorbox}

\includegraphics[width=7cm]{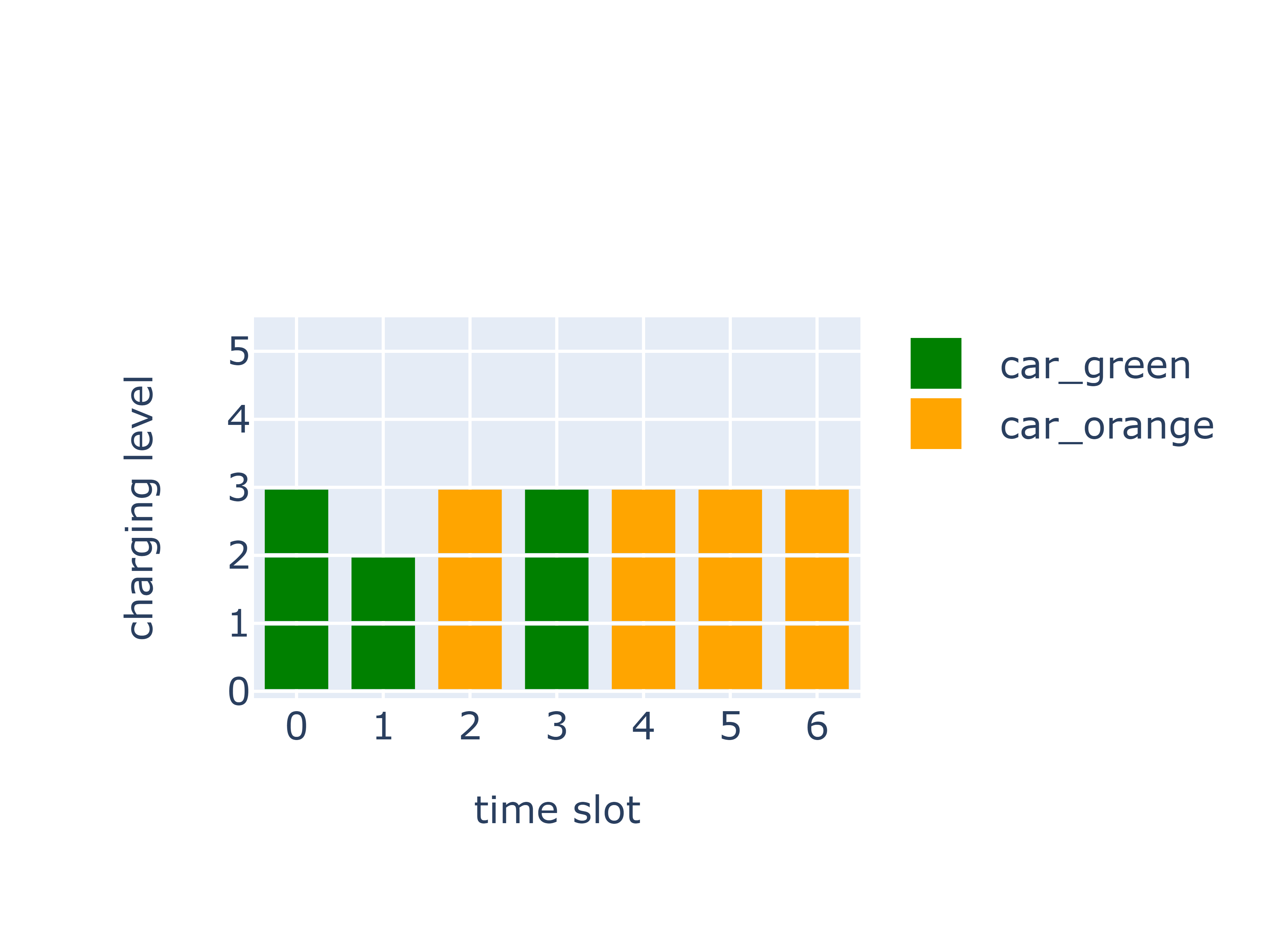}
    
Recalling our toy example, see Section \ref{sec:notebook-1-example} and Figure \ref{fig:notebook-1-poc-example}, we see that this is indeed an optimal solution.

    \hypertarget{convert-qcio-to-a-qubo}{%
\section{Convert QCIO to a QUBO}\label{sec:qcio-to-qubo}}

    Later we want to solve our optimization problem \eqref{eq:qcio} with the quantum
algorithm QAOA. This algorithm requires our optimization problem in a different form,
which will derive in this section.

    Recall that \eqref{eq:qcio}  was given by 
\[
\min_{\vec{p\, } \in \{0, \dots, L\}^{KT}} f_1(\vec{p \, })
\qquad
\text{such that } C \vec{p} = \vec{e} \ ,
\qquad
\text{where } f_1(\vec{p \, }) = \vec{p \, }^t \! A \vec{p} \ .
\]

    In the sequence we transform \eqref{eq:qcio} by the following two steps:
\begin{itemize}[leftmargin=50pt]
    \item[Step 1:] Convert hard constraints to soft constraints. \\
        This means include $C \vec{p} = \vec{e}$ into the cost function.
    \item[Step 2:] Binary encoding of integer variables $\vec{p}$. \\
        This means to transform the problem such that we have binary variables $\vec{b}$.
\end{itemize}

    \hypertarget{convert-hard-to-soft-constraints}{%
\subsection{Convert Hard to Soft Constraints}\label{sec:hard-to-soft-constraints}}

    For a \textbf{penalty parameter} \(\varrho \ge 0\) we define

\[
f_2(\vec{p} \, ; \varrho)
= f_1(\vec{p}\, ) + \varrho \|C \vec{p} - \vec{e} \, \|_2^2 \ .
\]

    Note that \(f_2\) is also a quadratic cost function with

\begin{align*}
f_2(\vec{p} \, ; \varrho) = \vec{p \, }^t \! \hat{A}_\varrho \vec{p} + \hat{L}_\varrho \vec{p} + \hat{c}_\varrho \ ,
\end{align*}
where
\begin{align*}
\hat{A}_\varrho = A + \varrho C^t C \ ,
\qquad
\hat{L}_\varrho = - 2 \varrho \vec{e \, }^t C \ ,
\qquad
\hat{c}_\varrho = \varrho \|\vec{e} \, \|_2^2 \ .
\end{align*}

    Now, we have (for a fixed penalty parameter \(\varrho\)) a
\textbf{quadratic unconstrained integer optimization} problem

\begin{equation}
    \min_{\vec{p} \in \{0, \dots, L\}^{KT}} f_2(\vec{p} \, ; \varrho) \ .
    \label{eq:quio}
    \tag{QUIO}
\end{equation}

    It is important to note that if \(\varrho\) is chosen large enough then
the solution of \eqref{eq:quio} is also a solution of \eqref{eq:qcio}. On the other hand,
this means that choosing \(\varrho\) too small can lead to
\textbf{unfeasible} solutions, i.e.~solutions \(\vec{p}\) of \eqref{eq:quio} that
do not satisfy the constraint \(C \vec{p} = \vec{e}\) of \eqref{eq:qcio}.

    \hypertarget{binary-encoding}{%
\subsection{Binary Encoding}\label{sec:binary-encoding}}

    A \textbf{binary encoding} is given by a transformation matrix \(B\)
such that \(\vec{p} = B \vec{b}\), where the coefficients \(b_i\) of
\(\vec{b}\) are binary, i.e.~\(b_i \in \{0, 1\}\).

    By substituting \(\vec{p} = B \vec{b}\) in (QUIO) we get a
\textbf{quadratic unconstrained binary optimization} problem

\begin{equation}
    \min_{\vec{b} \in \{0, 1\}^{\widetilde{N}}} f_3(\vec{b} \, ; \varrho) \ ,
    \label{eq:qubo}
    \tag{QUBO}
\end{equation}

    where 
\[
f_3(\vec{b} \, ; \varrho)
= \vec{b \, }^t \! \widetilde{A}_\varrho \vec{b} + \widetilde{L}_\varrho \vec{b} + \widetilde{c}_ \varrho,
\qquad
\widetilde{A}_\varrho = B^t \hat{A}_\varrho B \ , \quad
\widetilde{L}_\varrho = \hat{L}_\varrho B, \quad
\widetilde{c}_\varrho = \hat{c}_\varrho \ .
\]

    In the next lines we give a simple example of a binary encoding and
refer to the literature for advanced encodings.

    \hypertarget{example}{%
\subsubsection{Toy example: fixed width binary encoding}\label{sec:example-binary-encoding}}

    Let \(w=3\) be a fixed encoding width. Then, we can represent every
component \(p_i \in \{0, 1, \dots, 5\}\) of our vector \(\vec{p}\) by

    \[
p_i
= b_{i, 0} \cdot 2^0 + b_{i, 1} \cdot 2^1 + b_{i, 2} \cdot 2^2
= \underbrace{\pmat{2^0 \; 2^1 \; 2^2}}_{\widetilde{B}} \underbrace{\pmat{b_{i,0} \\ b_{i,1} \\ b_{i,2}}}_{\vec{b_{i}}},
\quad 
\vec{b_{i}} \in \{0, 1\}^3.
\]

    Using this, we can write

\[
\vec{p}
= \pmat{p_0 \\ \vdots \\ p_{N-1}}
= \underbrace{
  \pmat{
    \widetilde{B} &           &        &           &
    \\
              & \widetilde{B} &        &           &
    \\
              &           & \ddots &           &
    \\
              &           &        & \widetilde{B} &
}}_{B}
\underbrace{\pmat{\vec{b}_0 \\ \vdots \\ \vec{b}_{N-1}}}_{\vec{b}},
\qquad
N = K T.
\]

    Note that the dimension increases: \(\vec{p}\) has \(N = K T\) entries,
whereas \(\vec{b}\) has \(\widetilde{N} = w K T\) entries.

    \hypertarget{other-binary-encodings}{%
\subsubsection{Other binary encodings}\label{other-binary-encodings}}

    Further information on encondings and more examples can be found in
\cite{KarP19}. Also the \textbf{bounded-coefficient encoding}, which
\textbf{Qiskit uses by default}, is proposed in this paper (see the
documention of the class
\href{https://qiskit.org/documentation/optimization/stubs/qiskit_optimization.converters.IntegerToBinary.html#qiskit_optimization.converters.IntegerToBinary}{\texttt{qiskit\_optimization.converters.IntegerToBinary}}).

    \hypertarget{remark}{%
\subsubsection{Remarks}\label{remark}}

    1) It is easy to prove that the matrix \(\widetilde{A}_\varrho\) in \eqref{eq:qubo} can
always be transformed such that it is an upper triangular matrix:

\[
\widetilde{A}_\varrho
= \left( \begin{array}{rrrrr}
    * & * & \dots & * & * 
    \\
    0 & * & \dots & * & *
    \\
    \vdots &  & \ddots & & \vdots
    \\
    0 & 0 &\dots & * & *
    \\
    0 & 0 &\dots & 0 & *
\end{array} \right) \ .
\]

    This also holds true for \(A\) and \(\hat{A}_\varrho\). We note this
here because Qiskit will save the matrices in such a way.

2) In this tutorial we consider the original form of QAOA, i.e. the Quantum Approximate Optimization Algorithm.
An adaption of this algorithm is the Quantum Alternating Operator Ansatz \cite{HadWORVB19} (often also
abbreviated with QAOA), where hard constraints are handled differently and thus Step 1 is not needed. Many examples can
be found in \cite{HadWORVB19} and also in \cite{KosBTS23}.

    \hypertarget{implementation}{%
\subsection{Implementation}\label{implementation}}

    Next, we write a class \texttt{Converter} the implements the upper
transformations. For this we use the following Qiskit classes: 
\begin{itemize}
    \tightlist
    \item \href{https://qiskit.org/documentation/optimization/stubs/qiskit_optimization.converters.QuadraticProgramConverter.html#qiskit_optimization.converters.QuadraticProgramConverter}{\texttt{QuadraticProgramConverter}}: This is the abstract class for
    converters of quadratic programs in Qiskit. It enforces that we
    implement the methods \texttt{convert} and \texttt{interpret} in our
    class \texttt{Converter}.
    \item \href{https://qiskit.org/documentation/optimization/stubs/qiskit_optimization.converters.LinearEqualityToPenalty.html#qiskit_optimization.converters.LinearEqualityToPenalty}{\texttt{LinearEqualityToPenalty}}: For Step 1.
    \item \href{https://qiskit.org/documentation/optimization/stubs/qiskit_optimization.converters.IntegerToBinary.html#qiskit_optimization.converters.IntegerToBinary}{\texttt{IntegerToBinary}}: For Step 2.
\end{itemize}

\begin{tcolorbox}[breakable, size=fbox, boxrule=1pt, pad at break*=1mm,colback=cellbackground, colframe=cellborder]
\prompt{In}{incolor}{30}{\boxspacing}
\begin{Verbatim}[commandchars=\\\{\}]
\PY{k+kn}{from} \PY{n+nn}{typing} \PY{k+kn}{import} \PY{n}{Union}
\PY{k+kn}{from} \PY{n+nn}{qiskit\PYZus{}optimization}\PY{n+nn}{.}\PY{n+nn}{converters} \PY{k+kn}{import} \PY{n}{QuadraticProgramConverter}\PY{p}{,} \PYZbs{}
    \PY{n}{LinearEqualityToPenalty}\PY{p}{,} \PY{n}{IntegerToBinary}

\PY{k}{class} \PY{n+nc}{Converter}\PY{p}{(}\PY{n}{QuadraticProgramConverter}\PY{p}{)}\PY{p}{:}
    \PY{k}{def} \PY{n+nf+fm}{\PYZus{}\PYZus{}init\PYZus{}\PYZus{}}\PY{p}{(}
        \PY{n+nb+bp}{self}\PY{p}{,} 
        \PY{n}{penalty}\PY{p}{:} \PY{n+nb}{float}\PY{o}{=}\PY{k+kc}{None} \PY{c+c1}{\PYZsh{} the penalty parameter for step 1}
    \PY{p}{)} \PY{o}{\PYZhy{}}\PY{o}{\PYZgt{}} \PY{k+kc}{None}\PY{p}{:}
        \PY{n+nb}{super}\PY{p}{(}\PY{p}{)}\PY{o}{.}\PY{n+nf+fm}{\PYZus{}\PYZus{}init\PYZus{}\PYZus{}}\PY{p}{(}\PY{p}{)}
        \PY{n+nb+bp}{self}\PY{o}{.}\PY{n}{\PYZus{}penalty} \PY{o}{=} \PY{n}{penalty}
        \PY{n+nb+bp}{self}\PY{o}{.}\PY{n}{linear\PYZus{}equality\PYZus{}to\PYZus{}penalty\PYZus{}converter} \PY{o}{=} \PYZbs{}
            \PY{n}{LinearEqualityToPenalty}\PY{p}{(}\PY{n}{penalty}\PY{p}{)}
        \PY{n+nb+bp}{self}\PY{o}{.}\PY{n}{integer\PYZus{}to\PYZus{}binary\PYZus{}converter} \PY{o}{=} \PY{n}{IntegerToBinary}\PY{p}{(}\PY{p}{)}

    \PY{k}{def} \PY{n+nf}{convert}\PY{p}{(}\PY{n+nb+bp}{self}\PY{p}{,} \PY{n}{quadratic\PYZus{}program}\PY{p}{:} \PY{n}{QuadraticProgram}\PY{p}{)} \PY{o}{\PYZhy{}}\PY{o}{\PYZgt{}} \PY{n}{QuadraticProgram}\PY{p}{:}
        \PY{k}{return} \PY{n+nb+bp}{self}\PY{o}{.}\PY{n}{integer\PYZus{}to\PYZus{}binary\PYZus{}converter}\PY{o}{.}\PY{n}{convert}\PY{p}{(}
            \PY{n+nb+bp}{self}\PY{o}{.}\PY{n}{linear\PYZus{}equality\PYZus{}to\PYZus{}penalty\PYZus{}converter}\PY{o}{.}\PY{n}{convert}\PY{p}{(}\PY{n}{quadratic\PYZus{}program}\PY{p}{)}\PY{p}{)}
    
    \PY{k}{def} \PY{n+nf}{interpret}\PY{p}{(}\PY{n+nb+bp}{self}\PY{p}{,} \PY{n}{x}\PY{p}{:} \PY{n}{Union}\PY{p}{[}\PY{n}{np}\PY{o}{.}\PY{n}{ndarray}\PY{p}{,} \PY{n}{List}\PY{p}{[}\PY{n+nb}{float}\PY{p}{]}\PY{p}{]}\PY{p}{)} \PY{o}{\PYZhy{}}\PY{o}{\PYZgt{}} \PY{n}{np}\PY{o}{.}\PY{n}{ndarray}\PY{p}{:}
        \PY{k}{return} \PY{n+nb+bp}{self}\PY{o}{.}\PY{n}{linear\PYZus{}equality\PYZus{}to\PYZus{}penalty\PYZus{}converter}\PY{o}{.}\PY{n}{interpret}\PY{p}{(}
            \PY{n+nb+bp}{self}\PY{o}{.}\PY{n}{integer\PYZus{}to\PYZus{}binary\PYZus{}converter}\PY{o}{.}\PY{n}{interpret}\PY{p}{(}\PY{n}{x}\PY{p}{)}\PY{p}{)}
\end{Verbatim}
\end{tcolorbox}

    \hypertarget{example}{%
\subsection{Toy Example: Convert to QUBO}\label{sec:notebook-1-example-convert-to-qubo}}

    Let us use our \texttt{Converter} to convert \texttt{qcio}.

    \begin{tcolorbox}[breakable, size=fbox, boxrule=1pt, pad at break*=1mm,colback=cellbackground, colframe=cellborder]
\prompt{In}{incolor}{17}{\boxspacing}
\begin{Verbatim}[commandchars=\\\{\}]
\PY{c+c1}{\PYZsh{} Note: penalty \PYZlt{}= 5.0 will give a non\PYZhy{}feasible solution,}
\PY{c+c1}{\PYZsh{} penalty \PYZgt 5.0 will give a feasible solution.}
\PY{n}{converter} \PY{o}{=} \PY{n}{Converter}\PY{p}{(}\PY{n}{penalty}\PY{o}{=}\PY{l+m+mf}{5.1}\PY{p}{)}
\PY{n}{qubo} \PY{o}{=} \PY{n}{converter}\PY{o}{.}\PY{n}{convert}\PY{p}{(}\PY{n}{qcio}\PY{p}{)}
\PY{n}{qubo}\PY{o}{.}\PY{n}{name} \PY{o}{=} \PY{l+s+s2}{\PYZdq{}}\PY{l+s+s2}{QUBO}\PY{l+s+s2}{\PYZdq{}}
\PY{n+nb}{print}\PY{p}{(}\PY{n}{qubo}\PY{o}{.}\PY{n}{prettyprint}\PY{p}{(}\PY{p}{)}\PY{p}{)}
\end{Verbatim}
\end{tcolorbox}

    \begin{Verbatim}[commandchars=\\\{\}]
Problem name: QUBO

Minimize
  6.1*p_car\_green\_t0@0\^{}2 + 24.4*p_car\_green\_t0@0*p_car\_green\_t0@1
  + 24.4*p_car\_green\_t0@0*p_car\_green\_t0@2
  + 10.2*p_car\_green\_t0@0*p_car\_green\_t1@0
  + 20.4*p_car\_green\_t0@0*p_car\_green\_t1@1
  + 20.4*p_car\_green\_t0@0*p_car\_green\_t1@2
  + 10.2*p_car\_green\_t0@0*p_car\_green\_t2@0
  + 20.4*p_car\_green\_t0@0*p_car\_green\_t2@1
  + 20.4*p_car\_green\_t0@0*p_car\_green\_t2@2
  + 10.2*p_car\_green\_t0@0*p_car\_green\_t3@0
  + 20.4*p_car\_green\_t0@0*p_car\_green\_t3@1
  + 20.4*p_car\_green\_t0@0*p_car\_green\_t3@2
  + 2*p_car\_green\_t0@0*p_car\_orange\_t0@0
  + 4*p_car\_green\_t0@0*p_car\_orange\_t0@1
  + 4*p_car\_green\_t0@0*p_car\_orange\_t0@2 + 24.4*p_car\_green\_t0@1\^{}2
  + 48.8*p_car\_green\_t0@1*p_car\_green\_t0@2
  ...

Subject to
  No constraints

  Binary variables (42)
    p_car\_green\_t0@0 p_car\_green\_t0@1 p_car\_green\_t0@2
    p_car\_green\_t1@0 p_car\_green\_t1@1 p_car\_green\_t1@2
    ...
    \end{Verbatim}

    We can verify that the number of binary variables has grown in
comparison with the number of integer variables:

\begin{tcolorbox}[breakable, size=fbox, boxrule=1pt, pad at break*=1mm,colback=cellbackground, colframe=cellborder]
\prompt{In}{incolor}{33}{\boxspacing}
\begin{Verbatim}[commandchars=\\\{\}]
\PY{n}{number\PYZus{}integer\PYZus{}variables} \PY{o}{=} \PY{n}{qcio}\PY{o}{.}\PY{n}{get\PYZus{}num\PYZus{}integer\PYZus{}vars}\PY{p}{(}\PY{p}{)}
\PY{n+nb}{print}\PY{p}{(}\PY{l+s+sa}{f}\PY{l+s+s2}{\PYZdq{}}\PY{l+s+s2}{Number integer variables: }\PY{l+s+si}{\PYZob{}}\PY{n}{number\PYZus{}integer\PYZus{}variables}\PY{l+s+si}{\PYZcb{}}\PY{l+s+s2}{\PYZdq{}}\PY{p}{)}
        
\PY{n}{number\PYZus{}binary\PYZus{}variables} \PY{o}{=} \PY{n}{qubo}\PY{o}{.}\PY{n}{get\PYZus{}num\PYZus{}binary\PYZus{}vars}\PY{p}{(}\PY{p}{)}
\PY{n+nb}{print}\PY{p}{(}\PY{l+s+sa}{f}\PY{l+s+s2}{\PYZdq{}}\PY{l+s+s2}{Number binary variables: }\PY{l+s+si}{\PYZob{}}\PY{n}{number\PYZus{}binary\PYZus{}variables}\PY{l+s+si}{\PYZcb{}}\PY{l+s+s2}{\PYZdq{}}\PY{p}{)}
\end{Verbatim}
\end{tcolorbox}

    \begin{Verbatim}[commandchars=\\\{\}]
Number integer variables:  14
Number binary variables:  42
    \end{Verbatim}

    As a last step let's retrieve the matrix \(\tilde{A}\) from
\texttt{qubo} and make a plot of its sparsity pattern.

\begin{tcolorbox}[breakable, size=fbox, boxrule=1pt, pad at break*=1mm,colback=cellbackground, colframe=cellborder]
\prompt{In}{incolor}{34}{\boxspacing}
\begin{Verbatim}[commandchars=\\\{\}]
\PY{n}{A\PYZus{}tilde} \PY{o}{=} \PY{n}{qubo}\PY{o}{.}\PY{n}{objective}\PY{o}{.}\PY{n}{quadratic}\PY{o}{.}\PY{n}{to\PYZus{}array}\PY{p}{(}\PY{p}{)}
\PY{n+nb}{print}\PY{p}{(}\PY{l+s+sa}{f}\PY{l+s+s2}{\PYZdq{}}\PY{l+s+s2}{Dimension: }\PY{l+s+si}{\PYZob{}}\PY{n}{A\PYZus{}tilde}\PY{o}{.}\PY{n}{shape}\PY{l+s+si}{\PYZcb{}}\PY{l+s+s2}{\PYZdq{}}\PY{p}{)}
\end{Verbatim}
\end{tcolorbox}

    \begin{Verbatim}[commandchars=\\\{\}]
Dimension:  (42, 42)
    \end{Verbatim}

\begin{tcolorbox}[breakable, size=fbox, boxrule=1pt, pad at break*=1mm,colback=cellbackground, colframe=cellborder]
\prompt{In}{incolor}{51}{\boxspacing}
\begin{Verbatim}[commandchars=\\\{\}]
\PY{k+kn}{import} \PY{n+nn}{matplotlib}\PY{n+nn}{.}\PY{n+nn}{pyplot} \PY{k}{as} \PY{n+nn}{plt}
\PY{n}{fig} \PY{o}{=} \PY{n}{plt}\PY{o}{.}\PY{n}{figure}\PY{p}{(}\PY{p}{)}
\PY{n}{ax} \PY{o}{=} \PY{n}{fig}\PY{o}{.}\PY{n}{add\PYZus{}subplot}\PY{p}{(}\PY{l+m+mi}{1}\PY{p}{,} \PY{l+m+mi}{1}\PY{p}{,} \PY{l+m+mi}{1}\PY{p}{)}
\PY{n}{ax}\PY{o}{.}\PY{n}{spy}\PY{p}{(}\PY{n}{A\PYZus{}tilde}\PY{p}{,} \PY{n}{markersize}\PY{o}{=}\PY{l+m+mi}{3}\PY{p}{)}
\end{Verbatim}
\end{tcolorbox}

\includegraphics[width=8cm]{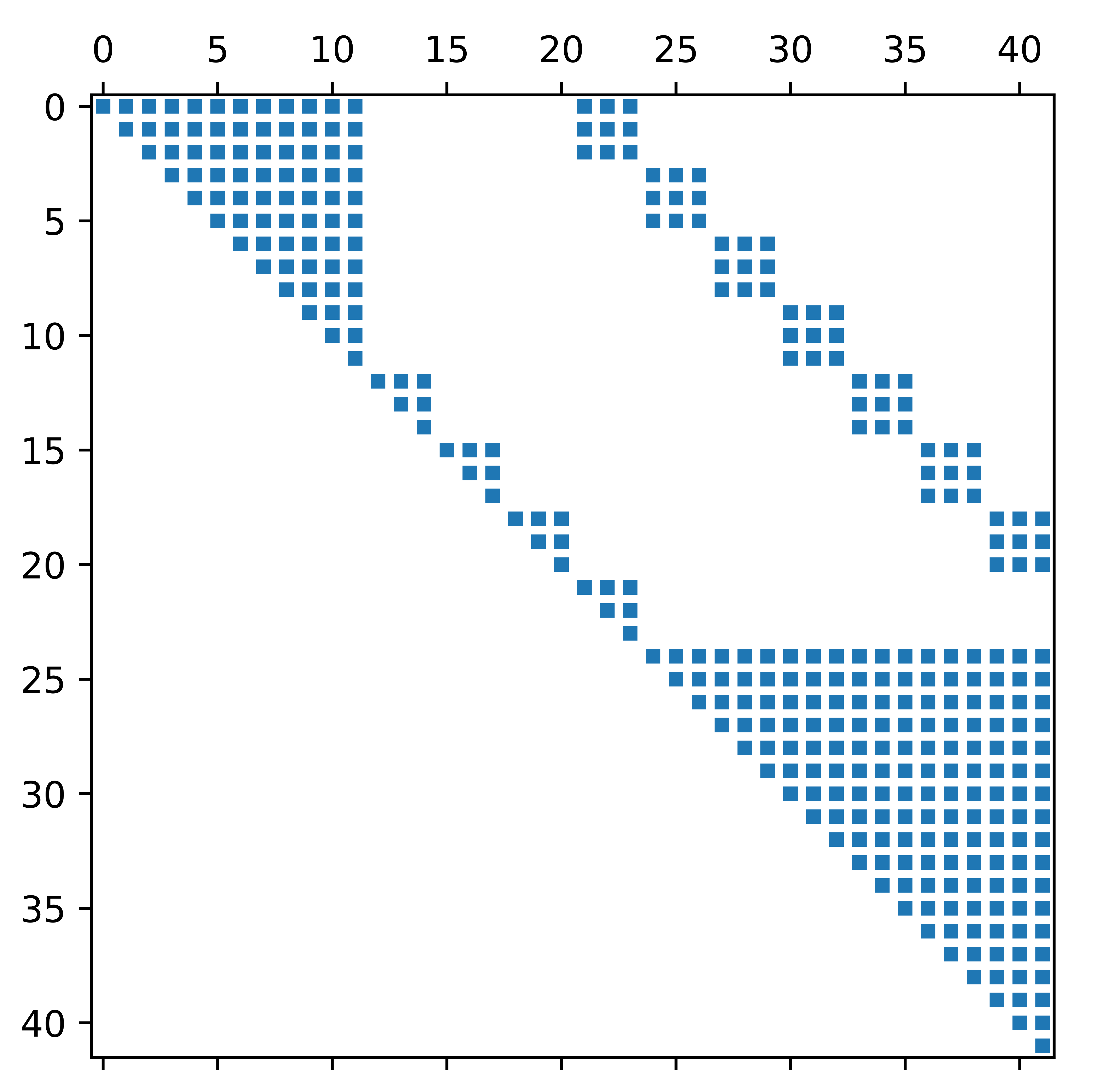}
    
    In Notebook \ref{chap:notebook-4} we will discuss the meaning and the importance of the
sparsity pattern.

    \hypertarget{solve-with-a-classical-solver}{%
\subsection{Solve with a Classical Solver}\label{sec:solve-with-a-classical-solver}}

    Analog to \eqref{eq:qcio} we can use a classical solver to solve \eqref{eq:qubo}. In the
next cells we do this for our \texttt{qubo} from above.

    \begin{tcolorbox}[breakable, size=fbox, boxrule=1pt, pad at break*=1mm,colback=cellbackground, colframe=cellborder]
\prompt{In}{incolor}{21}{\boxspacing}
\begin{Verbatim}[commandchars=\\\{\}]
\PY{n}{qubo\PYZus{}minimization\PYZus{}result} \PY{o}{=} \PY{n}{cplex\PYZus{}optimizer}\PY{o}{.}\PY{n}{solve}\PY{p}{(}\PY{n}{qubo}\PY{p}{)}

\PY{n+nb}{print}\PY{p}{(}\PY{l+s+s2}{\PYZdq{}}\PY{l+s+s2}{minimum point (binary): b\PYZus{}min = }\PY{l+s+s2}{\PYZdq{}}\PY{p}{,} \PY{n}{qubo\PYZus{}minimization\PYZus{}result}\PY{o}{.}\PY{n}{x}\PY{p}{)}
\PY{n+nb}{print}\PY{p}{(}\PY{l+s+s2}{\PYZdq{}}\PY{l+s+s2}{minimum value: f\PYZus{}3(b\PYZus{}min) = }\PY{l+s+s2}{\PYZdq{}}\PY{p}{,} \PY{n}{qubo\PYZus{}minimization\PYZus{}result}\PY{o}{.}\PY{n}{fval}\PY{p}{)}
\end{Verbatim}
\end{tcolorbox}

    \begin{Verbatim}[commandchars=\\\{\}]
minimum point (binary): b\_min =  [1. 0. 1. 1. 0. 0. 1. 0. 1. 1. 0. 0. 0. 0. 0. 0. 0. 0. 0. 0. 0. 0. 0. 0. 0. 0. 1. 0. 0. 0. 0. 0. 1. 0. 1. 0. 1. 1. 0. 1. 1. 0.]
minimum value: f\_3(b\_min) =  57.999999999999886
    \end{Verbatim}

    The array \texttt{qubo\_minimization\_result.x} corresponds to a
solution \(\vec{b}_\text{min}\) of \eqref{eq:qubo}. With the method
\texttt{interpret} of \texttt{converter} we can transform the binary
vector \(\vec{b}_\text{min}\) to the integer vector
\(\vec{p}_\text{min}\) (which is a feasible solution of \eqref{eq:qcio} if
\(\varrho\) is chosen large enough).

    \begin{tcolorbox}[breakable, size=fbox, boxrule=1pt, pad at break*=1mm,colback=cellbackground, colframe=cellborder]
\prompt{In}{incolor}{22}{\boxspacing}
\begin{Verbatim}[commandchars=\\\{\}]
\PY{n}{b\PYZus{}min} \PY{o}{=} \PY{n}{qubo\PYZus{}minimization\PYZus{}result}\PY{o}{.}\PY{n}{x}
\PY{n}{p\PYZus{}min} \PY{o}{=} \PY{n}{converter}\PY{o}{.}\PY{n}{interpret}\PY{p}{(}\PY{n}{b\PYZus{}min}\PY{p}{)}

\PY{n+nb}{print}\PY{p}{(}\PY{l+s+s2}{\PYZdq{}}\PY{l+s+s2}{minimum point (integer): p\PYZus{}min = }\PY{l+s+s2}{\PYZdq{}}\PY{p}{,} \PY{n}{p\PYZus{}min}\PY{p}{)}
\PY{n+nb}{print}\PY{p}{(}\PY{l+s+s2}{\PYZdq{}}\PY{l+s+s2}{minimum value: f\PYZus{}1(p\PYZus{}min) = }\PY{l+s+s2}{\PYZdq{}}\PY{p}{,} \PY{n}{qcio}\PY{o}{.}\PY{n}{objective}\PY{o}{.}\PY{n}{evaluate}\PY{p}{(}\PY{n}{p\PYZus{}min}\PY{p}{)}\PY{p}{)}
\PY{n+nb}{print}\PY{p}{(}\PY{l+s+s2}{\PYZdq{}}\PY{l+s+s2}{minimum point feasible = }\PY{l+s+s2}{\PYZdq{}}\PY{p}{,} \PY{n}{qcio}\PY{o}{.}\PY{n}{is\PYZus{}feasible}\PY{p}{(}\PY{n}{p\PYZus{}min}\PY{p}{)}\PY{p}{)}
\end{Verbatim}
\end{tcolorbox}

    \begin{Verbatim}[commandchars=\\\{\}]
minimum point (integer): p\_min =  [3. 1. 3. 1. 0. 0. 0. 0. 2. 0. 2. 2. 3. 3.]
minimum value: f\_1(p\_min) =  58.0
minimum point feasible =  True
    \end{Verbatim}

    Last, let's plot the solution:

    \begin{tcolorbox}[breakable, size=fbox, boxrule=1pt, pad at break*=1mm,colback=cellbackground, colframe=cellborder]
\prompt{In}{incolor}{23}{\boxspacing}
\begin{Verbatim}[commandchars=\\\{\}]
\PY{n}{fig} \PY{o}{=} \PY{n}{plot\PYZus{}charging\PYZus{}schedule}\PY{p}{(}\PY{n}{charging\PYZus{}unit}\PY{p}{,} \PY{n}{p\PYZus{}min}\PY{p}{,} \PY{n}{marker\PYZus{}size}\PY{o}{=}\PY{l+m+mi}{20}\PY{p}{)}
\PY{n}{fig}\PY{o}{.}\PY{n}{update\PYZus{}layout}\PY{p}{(}\PY{n}{width}\PY{o}{=}\PY{l+m+mi}{400}\PY{p}{,} \PY{n}{height}\PY{o}{=}\PY{l+m+mi}{300}\PY{p}{)}
\PY{n}{fig}\PY{o}{.}\PY{n}{show}\PY{p}{(}\PY{p}{)}
\end{Verbatim}
\end{tcolorbox}

\includegraphics[width=7cm]{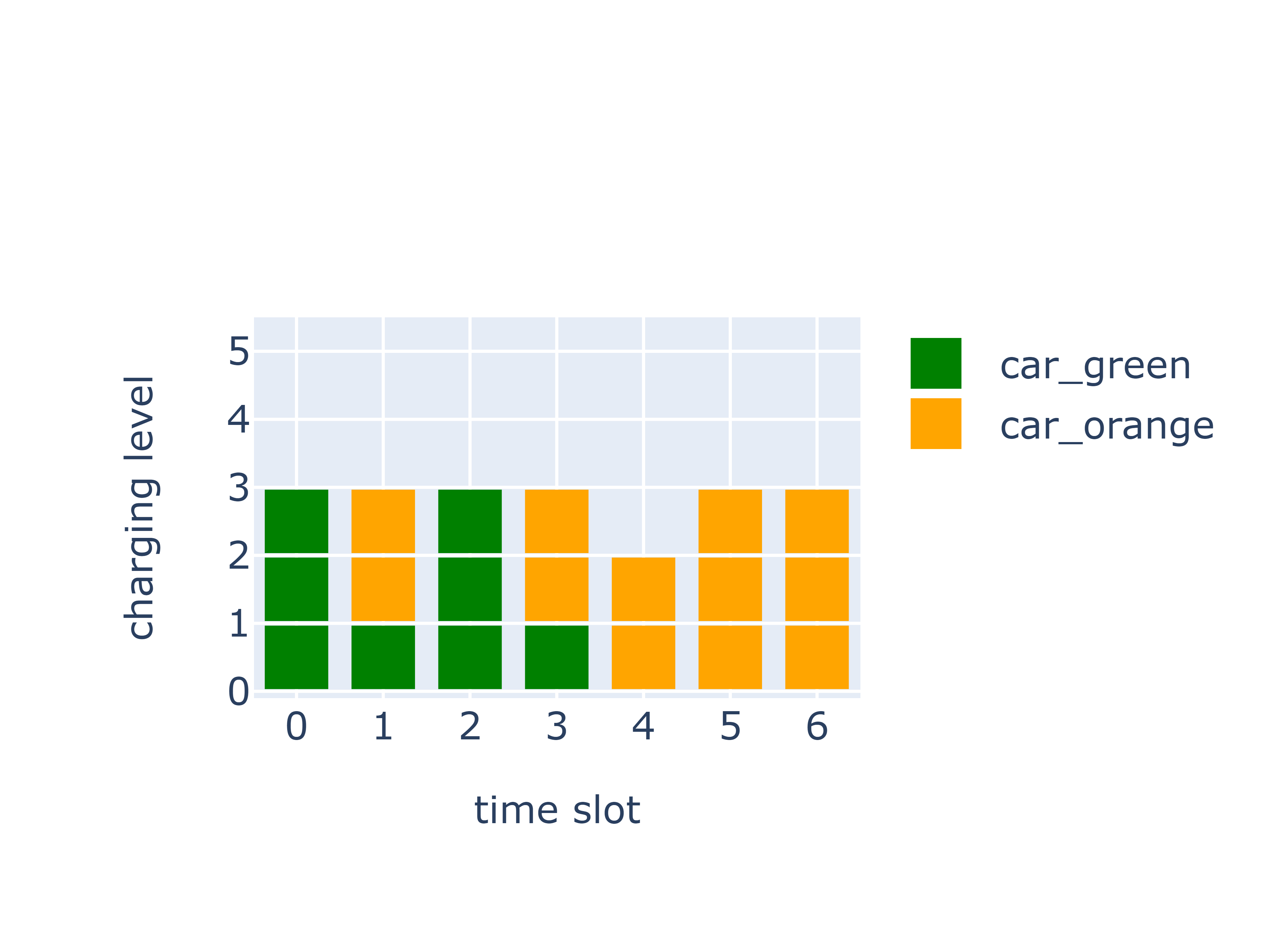}

%% file: notebook_2_latex_source.tex
\hypertarget{introduction}{%
\section{Introduction}\label{sec:notebook-2-introduction}}

    In Notebook \ref{chap:notebook-1} we have seen how to model and implement (a proof of
concept version of) a real-world problem on optimizing the charging
schedules for electric vehicles. Moreover, we showed how to transform
and implement this problem as a QUBO which is the required
form for the \textbf{Quantum Approximate Optimization Algorithm (QAOA)} \cite{FarGG14} .

    In this notebook we will introduce QAOA, explain how a QAOA circuit can
be derived from a QUBO and how it can be implemented in Qiskit.

    We will demonstrate these points with a small example from the use case
introduced in Notebook \ref{chap:notebook-1}.

    \hypertarget{example-for-this-notebook}{%
\section{Example for this Notebook}\label{sec:notebook-2-example}}

\begin{figure}
    \begin{center}
        \includegraphics[width=7cm]{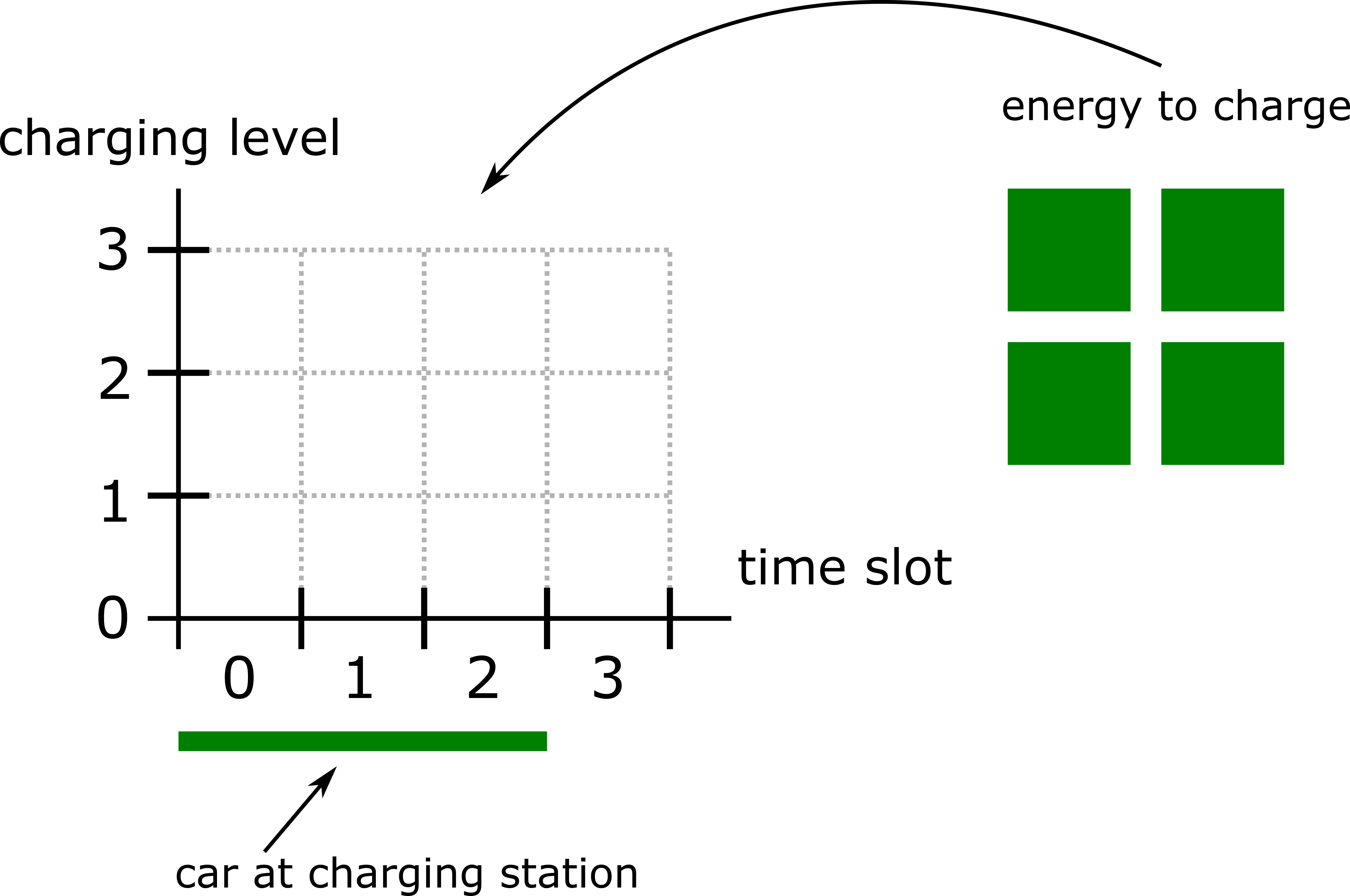}
        \caption{Example for this notebook.}
        \label{fig:notebook-2-example}
    \end{center}
\end{figure}

    Our example for this whole notebook is a \textbf{charging station} with
\textbf{4 charging levels} and \textbf{4 available time slots}.
\textbf{One car} is at the charging station at \textbf{time slots 0, 1
and 2}, and needs to \textbf{charge 4 energy units}. The situation is
depicted in Figure \ref{fig:notebook-2-example}.

    The following cell implements the upper situation based on our codes
from Notebook \ref{chap:notebook-1}.

    \begin{tcolorbox}[breakable, size=fbox, boxrule=1pt, pad at break*=1mm,colback=cellbackground, colframe=cellborder]
\prompt{In}{incolor}{1}{\boxspacing}
\begin{Verbatim}[commandchars=\\\{\}]
\PY{k+kn}{from} \PY{n+nn}{codes\PYZus{}notebook\PYZus{}1} \PY{k+kn}{import} \PY{n}{generate\PYZus{}example}
\PY{n}{charging\PYZus{}unit}\PY{p}{,} \PY{n}{car\PYZus{}green}\PY{p}{,} \PY{n}{qcio}\PY{p}{,} \PY{n}{converter}\PY{p}{,} \PY{n}{qubo}\PY{p}{,} \PYZbs{}
    \PY{n}{number\PYZus{}binary\PYZus{}variables}\PY{p}{,} \PY{n}{qubo\PYZus{}minimization\PYZus{}result} \PY{o}{=} \PY{n}{generate\PYZus{}example}\PY{p}{(}\PY{p}{)}
\end{Verbatim}
\end{tcolorbox}

    \begin{tcolorbox}[breakable, size=fbox, boxrule=1pt, pad at break*=1mm,colback=cellbackground, colframe=cellborder]
\prompt{In}{incolor}{2}{\boxspacing}
\begin{Verbatim}[commandchars=\\\{\}]
\PY{n+nb}{print}\PY{p}{(}\PY{n}{charging\PYZus{}unit}\PY{p}{)}
\PY{n+nb}{print}\PY{p}{(}\PY{n}{car\PYZus{}green}\PY{p}{)}
\end{Verbatim}
\end{tcolorbox}

    \begin{Verbatim}[commandchars=\\\{\}]
Charging unit with
  charging levels: 0, 1, 2, 3
  time slots: 0, 1, 2, 3
  cars to charge: car\_green
Car 'car\_green':
  at charging station at time slots [0, 1, 2]
  requires 4 energy units
    \end{Verbatim}

    Let us compute an exact solution so that we have a reference to compare
our quantum solution with.

    \begin{tcolorbox}[breakable, size=fbox, boxrule=1pt, pad at break*=1mm,colback=cellbackground, colframe=cellborder]
\prompt{In}{incolor}{3}{\boxspacing}
\begin{Verbatim}[commandchars=\\\{\}]
\PY{n}{b\PYZus{}min} \PY{o}{=} \PY{n}{qubo\PYZus{}minimization\PYZus{}result}\PY{o}{.}\PY{n}{x}
\PY{n}{f\PYZus{}3\PYZus{}min} \PY{o}{=} \PY{n}{qubo\PYZus{}minimization\PYZus{}result}\PY{o}{.}\PY{n}{fval}
\PY{n+nb}{print}\PY{p}{(}\PY{l+s+s2}{\PYZdq{}}\PY{l+s+s2}{minimum point (binary): b\PYZus{}min = }\PY{l+s+s2}{\PYZdq{}}\PY{p}{,} \PY{n}{b\PYZus{}min}\PY{p}{)}
\PY{n+nb}{print}\PY{p}{(}\PY{l+s+s2}{\PYZdq{}}\PY{l+s+s2}{minimum value: f\PYZus{}3(b\PYZus{}min) = }\PY{l+s+s2}{\PYZdq{}}\PY{p}{,} \PY{n}{f\PYZus{}3\PYZus{}min}\PY{p}{)}
\end{Verbatim}
\end{tcolorbox}

    \begin{Verbatim}[commandchars=\\\{\}]
minimum point (binary): b\_min =  [0. 1. 1. 0. 1. 0. 0. 0.]
minimum value: f\_3(b\_min) =  5.999999999999993
    \end{Verbatim}

    \begin{tcolorbox}[breakable, size=fbox, boxrule=1pt, pad at break*=1mm,colback=cellbackground, colframe=cellborder]
\prompt{In}{incolor}{4}{\boxspacing}
\begin{Verbatim}[commandchars=\\\{\}]
\PY{n}{p\PYZus{}min} \PY{o}{=} \PY{n}{converter}\PY{o}{.}\PY{n}{interpret}\PY{p}{(}\PY{n}{b\PYZus{}min}\PY{p}{)}
\PY{n}{f\PYZus{}1\PYZus{}min} \PY{o}{=} \PY{n}{qcio}\PY{o}{.}\PY{n}{objective}\PY{o}{.}\PY{n}{evaluate}\PY{p}{(}\PY{n}{p\PYZus{}min}\PY{p}{)}
\PY{n}{p\PYZus{}min\PYZus{}feasible} \PY{o}{=} \PY{n}{qcio}\PY{o}{.}\PY{n}{is\PYZus{}feasible}\PY{p}{(}\PY{n}{p\PYZus{}min}\PY{p}{)}

\PY{n+nb}{print}\PY{p}{(}\PY{l+s+s2}{\PYZdq{}}\PY{l+s+s2}{minimum point (integer): p\PYZus{}min = }\PY{l+s+s2}{\PYZdq{}}\PY{p}{,} \PY{n}{p\PYZus{}min}\PY{p}{)}
\PY{n+nb}{print}\PY{p}{(}\PY{l+s+s2}{\PYZdq{}}\PY{l+s+s2}{minimum value: f\PYZus{}1(p\PYZus{}min) = }\PY{l+s+s2}{\PYZdq{}}\PY{p}{,} \PY{n}{f\PYZus{}1\PYZus{}min}\PY{p}{)}
\PY{n+nb}{print}\PY{p}{(}\PY{l+s+s2}{\PYZdq{}}\PY{l+s+s2}{minimum point feasible = }\PY{l+s+s2}{\PYZdq{}}\PY{p}{,} \PY{n}{p\PYZus{}min\PYZus{}feasible}\PY{p}{)}
\end{Verbatim}
\end{tcolorbox}

    \begin{Verbatim}[commandchars=\\\{\}]
minimum point (integer): p\_min =  [2. 1. 1. 0.]
minimum value: f\_1(p\_min) =  6.0
minimum point feasible =  True
    \end{Verbatim}

    \begin{tcolorbox}[breakable, size=fbox, boxrule=1pt, pad at break*=1mm,colback=cellbackground, colframe=cellborder]
\prompt{In}{incolor}{5}{\boxspacing}
\begin{Verbatim}[commandchars=\\\{\}]
\PY{k+kn}{from} \PY{n+nn}{utils} \PY{k+kn}{import} \PY{n}{plot\PYZus{}charging\PYZus{}schedule}
\PY{n}{fig} \PY{o}{=} \PY{n}{plot\PYZus{}charging\PYZus{}schedule}\PY{p}{(}
    \PY{n}{charging\PYZus{}unit}\PY{p}{,} \PY{n}{p\PYZus{}min}\PY{p}{,} \PY{n}{marker\PYZus{}size}\PY{o}{=}\PY{l+m+mi}{30}\PY{p}{)}
\PY{n}{fig}\PY{o}{.}\PY{n}{update\PYZus{}layout}\PY{p}{(}\PY{n}{width}\PY{o}{=}\PY{l+m+mi}{350}\PY{p}{,} \PY{n}{height}\PY{o}{=}\PY{l+m+mi}{300}\PY{p}{)}
\PY{n}{fig}\PY{o}{.}\PY{n}{show}\PY{p}{(}\PY{p}{)}
\end{Verbatim}
\end{tcolorbox}

\includegraphics[width=7cm]{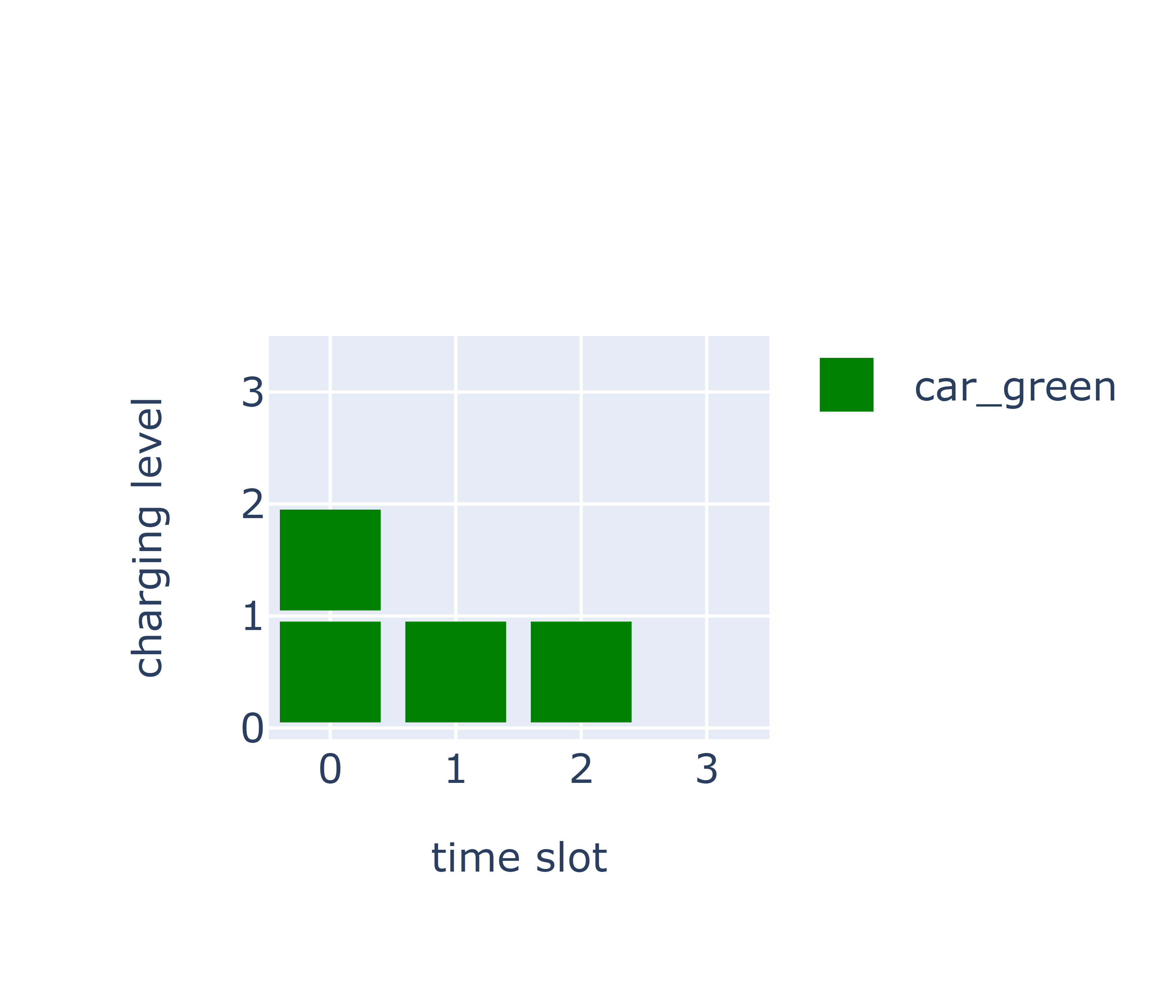}

    \hypertarget{revision-qubo}{%
\section{Revision: QUBO}\label{sec:revision-qubo}}

    In Notebook \ref{chap:notebook-1} we discussed how a QUBO can be derived from an
optimization use case. For this notebook we move the use case more in
the background and write a QUBO in the general form

\begin{equation}
    \min_{\vec{b} \in \{0, 1\}^n} f_3(\vec{b}),
    \qquad
    f_3(\vec{b}) = \vec{b}^{\, t} A \vec{b} + L \vec{b} + c \ ,
    \qquad
    A \in \R^{n \times n}, \
    L \in \R^{n \times 1}, \
    c \in \R.
    \label{eq:qubo2}
    \tag{QUBO}
\end{equation}
From now on, if we refer to \eqref{eq:qubo2} we mean the problem above.

    A first step to solve \eqref{eq:qubo2} with QAOA is to transform it into an Ising
Hamiltonian \cite{Gue05,IsingWeb}.

    \hypertarget{convert-qubo-to-ising-hamiltonian}{%
\section{Convert QUBO to Ising
Hamiltonian}\label{sec:qubo-to-ising}}

    \hypertarget{theory}{%
\subsection{Theory}\label{theory}}

    We write the QUBO cost function \(f_3\) as

\[
f_3(\vec{b})
= \sum_{i=0}^{n-1} \sum_{j>i}^{n-1} a_{ij} b_i b_j
+ \sum_{i=0}^{n-1} l_i b_i 
+ c \ ,
\]

where \(a_{ij}\) and \(l_i\) are the entries of \(A\) and \(L\),
respectively. Subsequently, we replace 

\begin{equation*}
    b_i \leftrightarrow 
    \tfrac12 \bigl( I^{\otimes n} - \sigmazi \bigr),
\end{equation*}
where
\begin{equation*}
I^{\otimes n} = \underbrace{I \otimes \dots \otimes I}_{n \text{ times}} \, ,
\qquad
I = \pmat{1 & 0 \\ 0 & 1} \ ,
\end{equation*}
and
\begin{equation*}
\sigma_Z^{(i)} = I \otimes \dots \otimes I \otimes \underbrace{Z}_{\text{position } i} \otimes \, I \otimes \dots \otimes I \, ,
\qquad
Z = \pmat{1 & 0 \\ 0 & -1} \ .
\end{equation*}

    This gives the \textbf{cost (or problem) Hamiltonian} \(\HP\), which is
of the form

\begin{equation}
\HP = \sum_{i=0}^{n-1} \sum_{j>i}^{n-1} h_{ij} \sigma_Z^{(i)} \sigma_Z^{(j)}
+ \sum_{i=0}^{n-1} h_i' \sigma_Z^{(i)}
+ h'' I^{\otimes n} \ ,
\label{eq:HP}
\end{equation}

    where the coefficients \(h_{ij}, h_i'\) and \(h''\) can be computed from
\(a_{ij}\), \(l_i\) and \(c\).

    The following theorem establishes the connection between the QUBO cost
function \(f_3\) and the cost Hamiltonian \(\HP\).

    \textbf{Theorem} For a quantum state \(\ket{\psi}\) with amplitudes
\(\lambda_{b}\), i.e.

\[
\ket{\psi}
= \sum_{\vec{b} \in \{0, 1\}^n}
\lambda_{b} |b\rangle \ ,
\qquad
\lambda_{b} \in \C \ ,
\
\sum_{\vec{b} \in \{0, 1\}^n}
|\lambda_{b}|^2
= 1 \ ,
\]

we have
\begin{equation}
    \langle \psi | \HP | \psi \rangle
    = \sum_{\vec{b} \in \{0, 1\}^n}
    |\lambda_{b}|^2 f_3(\vec{b}) \ .
    \label{eq:HP_f3}
\end{equation}

    \emph{Remark 1}: By writing \(b\) instead of \(\vec{b}\) we mean the
bitstring \(b = b_0 b_1 \dots b_{n-1}\) associated with the vector
\(\vec{b} = (b_0, b_1, \dots, b_{n-1})^t \).

    \emph{Remark 2:} Since this tutorial is concerned with practical
aspects of quantum computing we leave out the mathematical details and
derivations. However, it is a good exercise to do the computation QUBO
\(\to\) Ising and to prove the theorem.

    \hypertarget{implementation}{%
\subsection{Implementation}\label{sec:implementation-qubo-to-ising}}

    In Notebook \ref{chap:notebook-1} we already got to know the class
\texttt{QuadraticProgram}. In particular, \texttt{qubo} is an object of
this class. Now, we need this class' method \texttt{to\_ising} since it
implements the transformation discussed above. However, note that
\texttt{to\_ising} splits \(\HP\) into two parts, namely
\begin{itemize}
    \item \texttt{ising} that contains the first two terms of the RHS of \eqref{eq:HP}, and
    \item \texttt{ising\_offset} which contains the third term of the RHS of \eqref{eq:HP}.
\end{itemize}
    \begin{tcolorbox}[breakable, size=fbox, boxrule=1pt, pad at break*=1mm,colback=cellbackground, colframe=cellborder]
\prompt{In}{incolor}{6}{\boxspacing}
\begin{Verbatim}[commandchars=\\\{\}]
\PY{n}{ising}\PY{p}{,} \PY{n}{ising\PYZus{}offset} \PY{o}{=} \PY{n}{qubo}\PY{o}{.}\PY{n}{to\PYZus{}ising}\PY{p}{(}\PY{p}{)}
\end{Verbatim}
\end{tcolorbox}

    \begin{tcolorbox}[breakable, size=fbox, boxrule=1pt, pad at break*=1mm,colback=cellbackground, colframe=cellborder]
\prompt{In}{incolor}{7}{\boxspacing}
\begin{Verbatim}[commandchars=\\\{\}]
\PY{n+nb}{print}\PY{p}{(}\PY{l+s+s2}{\PYZdq{}}\PY{l+s+s2}{ising: }\PY{l+s+s2}{\PYZdq{}}\PY{p}{,} \PY{n}{ising}\PY{p}{)}
\end{Verbatim}
\end{tcolorbox}

    \begin{Verbatim}[commandchars=\\\{\}]
ising:  -3.3 * IIIIIIIZ
- 6.599999999999998 * IIIIIIZI
- 3.3000000000000007 * IIIIIZII
- 6.6 * IIIIZIII
- 3.2999999999999994 * IIIZIIII
- 6.599999999999999 * IIZIIIII
+ 4.6 * IIIIIIZZ
+ 1.8 * IIIIIZIZ
+ 3.6 * IIIIIZZI
+ 3.6 * IIIIZIIZ
+ 7.2 * IIIIZIZI
+ 4.6 * IIIIZZII
+ 1.8 * IIIZIIIZ
+ 3.6 * IIIZIIZI
+ 1.8 * IIIZIZII
+ 3.6 * IIIZZIII
+ 3.6 * IIZIIIIZ
+ 7.2 * IIZIIIZI
+ 3.6 * IIZIIZII
+ 7.2 * IIZIZIII
+ 4.6 * IIZZIIII
- 1.5 * IZIIIIII
+ 1.0 * ZZIIIIII
- 3.0 * ZIIIIIII
    \end{Verbatim}

    \begin{tcolorbox}[breakable, size=fbox, boxrule=1pt, pad at break*=1mm,colback=cellbackground, colframe=cellborder]
\prompt{In}{incolor}{8}{\boxspacing}
\begin{Verbatim}[commandchars=\\\{\}]
\PY{n+nb}{print}\PY{p}{(}\PY{l+s+s2}{\PYZdq{}}\PY{l+s+s2}{ising\PYZus{}offset:}\PY{l+s+s2}{\PYZdq{}}\PY{p}{,} \PY{n}{ising\PYZus{}offset}\PY{p}{)}
\end{Verbatim}
\end{tcolorbox}

    \begin{Verbatim}[commandchars=\\\{\}]
ising\_offset: 28.400000000000013
    \end{Verbatim}

    Let's verify Equation \eqref{eq:HP_f3} with the following state

\[
\ket{\psi} = 0.7 \ket{00011100} + 0.2\i \ket{10101010} + (0.6 + \sqrt{0.11}\i)\ket{11110000} \ .
\]

    For this purpose we use the following classes from
\href{https://qiskit.org/documentation/apidoc/opflow.html}{\texttt{qiskit.opflow}}:
\begin{itemize}
    \item \href{https://qiskit.org/documentation/stubs/qiskit.opflow.state_fns.DictStateFn.html#qiskit.opflow.state_fns.DictStateFn}{\texttt{DictStateFn}} to represent \(\ket{\psi}\) and
    \item \href{https://qiskit.org/documentation/stubs/qiskit.opflow.state_fns.OperatorStateFn.html#qiskit.opflow.state_fns.OperatorStateFn}{\texttt{OperatorStateFn}} with parameter \texttt{is\_measurement=True} to
    represent the observable \(\HP\) and to calculate the expecation value
    \(\braket{\psi | \HP | \psi}\) .
\end{itemize}

\emph{Remark:} For a tutorial on \texttt{qiskit.opflow} see \cite{QiskitOpflowWeb}.

\begin{tcolorbox}[breakable, size=fbox, boxrule=1pt, pad at break*=1mm,colback=cellbackground, colframe=cellborder]
\prompt{In}{incolor}{9}{\boxspacing}
\begin{Verbatim}[commandchars=\\\{\}]
\PY{k+kn}{import} \PY{n+nn}{numpy} \PY{k}{as} \PY{n+nn}{np}
\PY{k+kn}{from} \PY{n+nn}{qiskit}\PY{n+nn}{.}\PY{n+nn}{opflow} \PY{k+kn}{import} \PY{n}{DictStateFn}\PY{p}{,} \PY{n}{OperatorStateFn}

\PY{n}{psi} \PY{o}{=} \PY{n}{DictStateFn}\PY{p}{(}\PY{p}{\PYZob{}}\PY{l+s+s2}{\PYZdq{}}\PY{l+s+s2}{00011100}\PY{l+s+s2}{\PYZdq{}}\PY{p}{:} \PY{l+m+mf}{0.7}\PY{p}{,}
                   \PY{l+s+s2}{\PYZdq{}}\PY{l+s+s2}{10101010}\PY{l+s+s2}{\PYZdq{}}\PY{p}{:} \PY{l+m+mf}{0.2}\PY{n}{j}\PY{p}{,}
                   \PY{l+s+s2}{\PYZdq{}}\PY{l+s+s2}{11110000}\PY{l+s+s2}{\PYZdq{}}\PY{p}{:} \PY{l+m+mf}{0.6}\PY{o}{+}\PY{l+m+mi}{1}\PY{n}{j}\PY{o}{*}\PY{n}{np}\PY{o}{.}\PY{n}{sqrt}\PY{p}{(}\PY{l+m+mf}{0.11}\PY{p}{)}\PY{p}{\PYZcb{}}\PY{p}{)}

\PY{c+c1}{\PYZsh{} Note that observable does not contain ising\PYZus{}offset}
\PY{n}{observable} \PY{o}{=} \PY{n}{OperatorStateFn}\PY{p}{(}\PY{n}{ising}\PY{p}{,} \PY{n}{is\PYZus{}measurement}\PY{o}{=}\PY{k+kc}{True}\PY{p}{)} 

\PY{n}{expectation\PYZus{}wo\PYZus{}offset} \PY{o}{=} \PY{n}{np}\PY{o}{.}\PY{n}{real}\PY{p}{(}\PY{n}{observable}\PY{o}{.}\PY{n}{eval}\PY{p}{(}\PY{n}{psi}\PY{p}{)}\PY{p}{)}
\PY{n}{expectation} \PY{o}{=} \PY{n}{expectation\PYZus{}wo\PYZus{}offset} \PY{o}{+} \PY{n}{ising\PYZus{}offset}

\PY{n+nb}{print}\PY{p}{(}\PY{l+s+s2}{\PYZdq{}}\PY{l+s+s2}{\PYZlt{}psi | H\PYZus{}P | psi\PYZgt{} = }\PY{l+s+s2}{\PYZdq{}}\PY{p}{,} \PY{n}{expectation}\PY{p}{)}
\end{Verbatim}
\end{tcolorbox}

    \begin{Verbatim}[commandchars=\\\{\}]
<psi | H\_P | psi> =  16.268000000000015
    \end{Verbatim}

    Before we calculate the RHS of \eqref{eq:HP_f3} we recall that Qiskit uses a
different ordering of the qubits than most textbooks and different to
the default ordering in Python. In fact, we have \[
\text{ordering textbooks/Python: } \ket{b_0 \dots b_{n-1}}
\qquad
\leftrightarrow
\qquad
\text{ordering Qiskit: } \ket{b_{n-1} \dots b_0}
\]

    It is important to note that methods like
\texttt{QuadraticObjective.evaluate} or
\texttt{QuadraticProgramConverter.interpret} expect their input in the
\textbf{textbook/Python ordering}.

    With this knowledge we can implement the RHS of \eqref{eq:HP_f3}:

    \begin{tcolorbox}[breakable, size=fbox, boxrule=1pt, pad at break*=1mm,colback=cellbackground, colframe=cellborder]
\prompt{In}{incolor}{10}{\boxspacing}
\begin{Verbatim}[commandchars=\\\{\}]
\PY{n}{result} \PY{o}{=} \PY{l+m+mi}{0}
\PY{k}{for} \PY{n}{bitstring}\PY{p}{,} \PY{n}{amplitude} \PY{o+ow}{in} \PY{n}{psi}\PY{o}{.}\PY{n}{primitive}\PY{o}{.}\PY{n}{items}\PY{p}{(}\PY{p}{)}\PY{p}{:}
    \PY{c+c1}{\PYZsh{} Convert string to array and CHANGE ORDERRING}
    \PY{n}{bitarray} \PY{o}{=} \PY{n}{np}\PY{o}{.}\PY{n}{fromiter}\PY{p}{(}\PY{n}{bitstring}\PY{p}{,} \PY{n+nb}{int}\PY{p}{)}\PY{p}{[}\PY{p}{:}\PY{p}{:}\PY{o}{\PYZhy{}}\PY{l+m+mi}{1}\PY{p}{]}
    \PY{n}{result} \PY{o}{+}\PY{o}{=}  \PY{p}{(}\PY{n}{np}\PY{o}{.}\PY{n}{abs}\PY{p}{(}\PY{n}{amplitude}\PY{p}{)}\PY{o}{*}\PY{o}{*}\PY{l+m+mi}{2}\PY{p}{)} \PY{o}{*} \PY{n}{qubo}\PY{o}{.}\PY{n}{objective}\PY{o}{.}\PY{n}{evaluate}\PY{p}{(}\PY{n}{bitarray}\PY{p}{)}
\PY{n+nb}{print}\PY{p}{(}\PY{l+s+s2}{\PYZdq{}}\PY{l+s+s2}{sum\PYZus{}b |lambda\PYZus{}b|\PYZca{}2 f\PYZus{}3(b) = }\PY{l+s+s2}{\PYZdq{}}\PY{p}{,} \PY{n}{result}\PY{p}{)}
\end{Verbatim}
\end{tcolorbox}

    \begin{Verbatim}[commandchars=\\\{\}]
sum\_b |lambda\_b|\^{}2 f\_3(b) =  16.267999999999994
    \end{Verbatim}

    \hypertarget{qaoa-theory}{%
\section{QAOA: Theory}\label{sec:qaoa-theory}}

    QAOA aims to construct a quantum state
\begin{equation*}
    \ket{\psi} 
    = \sum\limits_{\vec{b} \in \{0, 1\}^n} \lambda_b \ket{b}
\end{equation*}
that has amplitudes \(\lambda_{b}\) with large
absolute value for those basis states \(\ket{b}\) where \(f_3(\vec{b})\)
is small (or even better where \(f_3\) is minimal). \textbf{Measuring} such a
quantum state will result with \textbf{high probability} in a bitstring
\(\vec{b}^\ast\) that is a \textbf{(nearly) optimal solution} of \eqref{eq:qubo2}.

    Equation \eqref{eq:HP_f3} describes that such a quantum state can be obtained by
minimizing the expectation value \(\braket{\psi | \HP | \psi}\). In
fact, for
\begin{equation*}
    \ket{\psi_{\text{min}}} = 
    \sum\limits_{\vec{b} \in \{0, 1\}^n} \lambda_{b, \text{min}} \ket{b}
    \qquad
    \text{with}
    \qquad
    \ket{\psi_{\text{min}}} 
    = \underset{\, \ket{\psi} \in \mathbb{H}^{\otimes n}}{\mathrm{argmin}} \braket{\psi | \HP | \psi}
\end{equation*}
we have that every basis state \(\ket{b}\) with
\(\lambda_{b, \text{min}} \neq 0\) is a minimum of \(f_3\). Here
\(\mathbb{H}^{\otimes n}\) is the space of all \(n\) qubit states.

    In order to construct an \textbf{approximation} of such an optimal state
QAOA starts at the \textbf{uniform superposition}

\[
\ket{+}^{\otimes n} = \underbrace{\ket{+} \otimes \dots \otimes \ket{+}}_{n \text{ times}} \ ,
\qquad
\ket{+} = \tfrac{1}{\sqrt2} \bigl(\ket{0} + \ket{1}\bigr) \ ,
\]

    and then alternatingly applies \(p\) times a \textbf{phase operator}
\(\UP\) and a \textbf{mixing operator} \(\UM\):

\begin{equation}
    | \psi_\mathrm{QAOA}(\vec{\beta}, \vec{\gamma}) \rangle
    = \UM(\beta_{p-1})
    \UP(\gamma_{p-1})
    \cdots
    \UM(\beta_0)
    \UP(\gamma_0)
    |+\rangle^{\otimes n}
    \ .
    \label{eq:qaoa1}
    \tag{QAOA-1}
\end{equation}

    Here, \(\vec{\beta} = (\beta_0, \dots, \beta_{p-1})^t \in [0, \pi]^p\)
and \(\vec{\gamma} = (\gamma_0, \dots,\gamma_{p-1})^t \in \R^p\) are
parameters, and the phase and mixing operator are given by
\begin{equation*}
    \UP(\gamma) = \mathrm{exp}(- \mathrm{i} \gamma \HP)
    \quad \text{and} \quad
    \UM(\beta) =  \mathrm{exp}(- \mathrm{i} \beta \HM),
\end{equation*}
respectively. The \textbf{mixer} \(\HM\) is defined by

\[
\HM = \sum_{i=0}^{n-1} \sigma_X^{(i)},
\qquad
\sigma_X^{(i)} = I \otimes \dots \otimes I \otimes \underbrace{X}_{\text{place } i} \otimes \ I \otimes \dots \otimes I,
\qquad
X = \pmat{0 & 1 \\ 1 & 0}.
\]

    Following the reasoning above the parameters \(\vec\beta\) and
\(\vec\gamma\) in \eqref{eq:qaoa1} should be chosen such that the expectation
value

\begin{equation}
    e(\vec{\beta}, \vec{\gamma})
    =
    \langle \psi_\mathrm{QAOA}(\vec{\beta}, \vec{\gamma}) | \HP |\psi_\mathrm{QAOA}(\vec{\beta}, \vec{\gamma}) \rangle
    \label{eq:qaoa2}
    \tag{QAOA-2}
\end{equation}
is minimized, i.e.

\begin{equation}
    (\vec{\beta}^\ast, \vec{\gamma}^\ast)
    = \underset{\vec{\beta}, \vec{\gamma}}{\mathrm{argmin}} \ e(\vec{\beta}, \vec{\gamma})
    \ .
    \label{eq:qaoa3}
    \tag{QAOA-3}
\end{equation}

    Then, measuring
\(| \psi_\mathrm{QAOA}(\vec{\beta}^\ast, \vec{\gamma}^\ast) \rangle\)
should result in a bitstring \(\vec{b}^\ast\) that is a good
approximation of the minimum of \(f_3\), i.e. 
\[
f_3(\vec{b}^\ast) \approx \min_{\vec{b} \in \{0, 1\}^n} f_3(\vec{b}) \ .
\]

    \hypertarget{classical-optimizers}{%
\subsection{Classical Optimizers}\label{classical-optimizers}}

    The minimization task \eqref{eq:qaoa3} can be carried out by a classical
optimizer (as e.g.~provided in \href{https://qiskit.org/documentation/stubs/qiskit.algorithms.optimizers.html}{\texttt{qiskit.algorithms.optimizers}}).

    Very roughly speaking such an optimizer works in the following way:
Starting from initial guesses \(\vec{\beta}^{(0)}\) and
\(\vec{\gamma}^{(0)}\) it computes iteratively a next set of parameters
\(\vec{\beta}^{(j+1)}\) and \(\vec{\gamma}^{(j+1)}\) with
\(e(\vec{\beta}^{(j+1)}, \vec{\gamma}^{(j+1)}) \le e(\vec{\beta}^{(j)}, \vec{\gamma}^{(j)})\),
\(j=0, 1, \dots\) . In this way \(\vec{\beta}^{(j+1)}\) and
\(\vec{\gamma}^{(j+1)}\) approach \(\vec{\beta}^\ast\) and
\(\vec{\gamma}^\ast\) of \eqref{eq:qaoa3} with every iteration. Clearly, in
order to do the step
\(\vec{\beta}^{(j)}, \vec{\gamma}^{(j)} \curvearrowright \vec{\beta}^{(j+1)}, \vec{\gamma}^{(j+1)}\)
the value of \(e\) in \((\vec{\beta}^{(j)}, \vec{\gamma}^{(j)})\) and
(usually) the values of
\(e(\vec{\beta}^{(j)} + \vec{h}_{\beta}, \vec{\gamma}^{(j)} + \vec{h}_{\gamma})\)
for some \(\vec{h}_{\beta}, \vec{h}_{\gamma}\) (to compute a gradient) are
needed.

    We see that this is a \textbf{hybrid workflow}, where for a given set of
parameters the quantum state in \eqref{eq:qaoa1} is constructed on a quantum
computer and the evaluation \eqref{eq:qaoa2} is computed from measurements of
this quantum state. From this evaluation (and usually evaluations of
small variations of the given parameters to obtain a gradient) a
classical optimizer proposes a next set of parameters and the process starts from the beginning. 
In this way \eqref{eq:qaoa3} is iteratively approximated.

    We note that this optimization is usually a difficult task and for many
applications, initial guesses, hyperparameter choices, etc. the
classical optimizer gets stuck in a \textbf{local minimum} and is not
able to converge to the \textbf{global minimum} given by the optimal
parameters \(\vec{\beta}^\ast\) and \(\vec{\gamma}^\ast\). Another issue
are \textbf{barren plateaus} where the gradient vanishes exponentially.
We refer to \cite{CerEtAl21,McCEtAl18} for more insights into these problems.

\emph{Remark 1:} We usually omit the dependence on \(\vec{\beta}\) and \(\vec{\gamma}\)
and just write
\(\ket{\psi_\mathrm{QAOA}} = | \psi_\mathrm{QAOA}(\vec{\beta}, \vec{\gamma}) \rangle\).

\emph{Remark 2:} Note that $\UP(\gamma)$, $\UM(\beta)$, $\HP$, and $\HM$ are $2^n \times 2^n$ matrices
whereas $A$ is a $n \times n$ matrix. Also note that if \(f_3\) depends on
\(\varrho\) also $\HP$, $\UP(\gamma)$ and $\ket{\psi_\mathrm{QAOA}}$ depend
on \(\varrho\).

    \hypertarget{qaoa-implementation}{%
\section{QAOA: Implementation}\label{sec:qaoa-implementation}}

    A quantum circuit that implements \eqref{eq:qaoa1} can be obtained from the
class \href{https://qiskit.org/documentation/stubs/qiskit.circuit.library.QAOAAnsatz.html?highlight=qaoaansatz#}{\texttt{QAOAAnsatz}}.

    In the next cell we use this class to get a QAOA circuit with \(p=2\)
for our cost Hamiltonian \texttt{ising} from above.

    \begin{tcolorbox}[breakable, size=fbox, boxrule=1pt, pad at break*=1mm,colback=cellbackground, colframe=cellborder]
\prompt{In}{incolor}{11}{\boxspacing}
\begin{Verbatim}[commandchars=\\\{\}]
\PY{k+kn}{from} \PY{n+nn}{qiskit}\PY{n+nn}{.}\PY{n+nn}{circuit}\PY{n+nn}{.}\PY{n+nn}{library} \PY{k+kn}{import} \PY{n}{QAOAAnsatz}

\PY{n}{qaoa\PYZus{}reps} \PY{o}{=} \PY{l+m+mi}{2} \PY{c+c1}{\PYZsh{} this corresponds to the parameter p in (QAOA)}
\PY{n}{qaoa\PYZus{}ansatz} \PY{o}{=} \PY{n}{QAOAAnsatz}\PY{p}{(}\PY{n}{cost\PYZus{}operator}\PY{o}{=}\PY{n}{ising}\PY{p}{,} \PY{n}{reps}\PY{o}{=}\PY{n}{qaoa\PYZus{}reps}\PY{p}{,} \PY{n}{name}\PY{o}{=}\PY{l+s+s1}{\PYZsq{}}\PY{l+s+s1}{qaoa}\PY{l+s+s1}{\PYZsq{}}\PY{p}{)}
\PY{n}{qaoa\PYZus{}ansatz}\PY{o}{.}\PY{n}{measure\PYZus{}active}\PY{p}{(}\PY{p}{)}
\end{Verbatim}
\end{tcolorbox}

    \begin{tcolorbox}[breakable, size=fbox, boxrule=1pt, pad at break*=1mm,colback=cellbackground, colframe=cellborder]
\prompt{In}{incolor}{12}{\boxspacing}
\begin{Verbatim}[commandchars=\\\{\}]
\PY{n}{qaoa\PYZus{}ansatz}\PY{o}{.}\PY{n}{decompose}\PY{p}{(}\PY{n}{reps}\PY{o}{=}\PY{l+m+mi}{1}\PY{p}{)}\PY{o}{.}\PY{n}{draw}\PY{p}{(}\PY{n}{scale}\PY{o}{=}\PY{l+m+mf}{0.5}\PY{p}{,} \PY{n}{fold}\PY{o}{=}\PY{o}{\PYZhy{}}\PY{l+m+mi}{1}\PY{p}{)}
\end{Verbatim}
\end{tcolorbox}
 
\prompt{Out}{outcolor}{12}{}
    
\begin{center}
\adjustimage{width=21cm,max size={0.95\linewidth}{0.9\paperheight}}{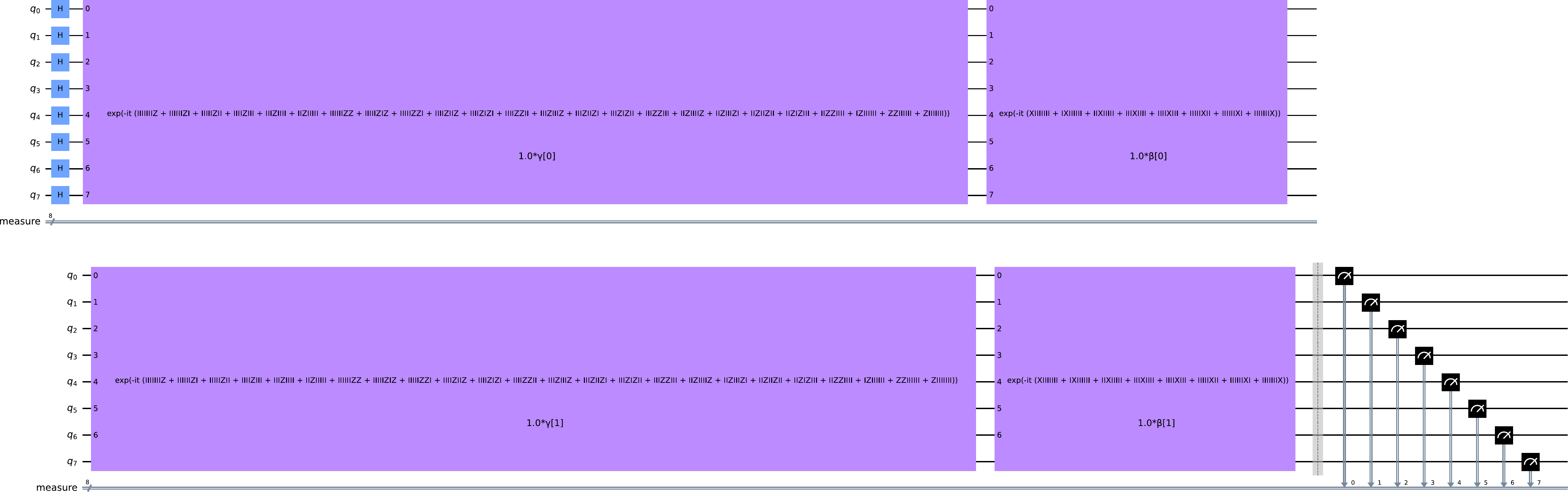}
\end{center}
{ \hspace*{\fill} \\}

    Note in which order \texttt{qaoa\_ansatz} expects its parameters:

    \begin{tcolorbox}[breakable, size=fbox, boxrule=1pt, pad at break*=1mm,colback=cellbackground, colframe=cellborder]
\prompt{In}{incolor}{13}{\boxspacing}
\begin{Verbatim}[commandchars=\\\{\}]
\PY{n}{qaoa\PYZus{}ansatz}\PY{o}{.}\PY{n}{parameters}
\end{Verbatim}
\end{tcolorbox}

            \begin{tcolorbox}[breakable, size=fbox, boxrule=.5pt, pad at break*=1mm, opacityfill=0]
\prompt{Out}{outcolor}{13}{\boxspacing}
\begin{Verbatim}[commandchars=\\\{\}]
ParameterView([ParameterVectorElement(β[0]), ParameterVectorElement(β[1]),
ParameterVectorElement(γ[0]), ParameterVectorElement(γ[1])])
\end{Verbatim}
\end{tcolorbox}
        
    Let us execute the circuit on a simulator for some arbitrarily chosen
parameters and visualize the result.

\begin{tcolorbox}[breakable, size=fbox, boxrule=1pt, pad at break*=1mm,colback=cellbackground, colframe=cellborder]
\prompt{In}{incolor}{28}{\boxspacing}
\begin{Verbatim}[commandchars=\\\{\}]
\PY{n}{betas} \PY{o}{=} \PY{p}{[}\PY{l+m+mf}{1.23}\PY{p}{,} \PY{l+m+mf}{2.31}\PY{p}{]}
\PY{n}{gammas} \PY{o}{=} \PY{p}{[}\PY{l+m+mf}{3.21}\PY{p}{,} \PY{l+m+mf}{4.32}\PY{p}{]}

\PY{c+c1}{\PYZsh{} See above the ordering that qaoa\PYZus{}ansatz expects}
\PY{n}{parameter\PYZus{}values} \PY{o}{=} \PY{p}{[}\PY{o}{*}\PY{n}{betas}\PY{p}{,} \PY{o}{*}\PY{n}{gammas}\PY{p}{]} 
\PY{n}{parameters} \PY{o}{=} \PY{n}{qaoa\PYZus{}ansatz}\PY{o}{.}\PY{n}{parameters}
\PY{n}{parameter\PYZus{}bindings} \PY{o}{=} \PY{n+nb}{dict}\PY{p}{(}\PY{n+nb}{zip}\PY{p}{(}\PY{n}{parameters}\PY{p}{,} \PY{n}{parameter\PYZus{}values}\PY{p}{)}\PY{p}{)}

\PY{n}{qaoa\PYZus{}with\PYZus{}parameters} \PY{o}{=} \PY{n}{qaoa\PYZus{}ansatz}\PY{o}{.}\PY{n}{bind\PYZus{}parameters}\PY{p}{(}\PY{n}{parameter\PYZus{}bindings}\PY{p}{)}
\end{Verbatim}
\end{tcolorbox}

\begin{tcolorbox}[breakable, size=fbox, boxrule=1pt, pad at break*=1mm,colback=cellbackground, colframe=cellborder]
\prompt{In}{incolor}{15}{\boxspacing}
\begin{Verbatim}[commandchars=\\\{\}]
\PY{n}{qaoa\PYZus{}with\PYZus{}parameters}\PY{o}{.}\PY{n}{decompose}\PY{p}{(}\PY{n}{reps}\PY{o}{=}\PY{l+m+mi}{1}\PY{p}{)}\PY{o}{.}\PY{n}{draw}\PY{p}{(}\PY{n}{scale}\PY{o}{=}\PY{l+m+mf}{0.5}\PY{p}{,} \PY{n}{fold}\PY{o}{=}\PY{o}{\PYZhy{}}\PY{l+m+mi}{1}\PY{p}{)}
\end{Verbatim}
\end{tcolorbox}

\prompt{Out}{outcolor}{15}{}
    
\begin{center}
\adjustimage{width=21cm,max size={0.95\linewidth}{0.9\paperheight}}{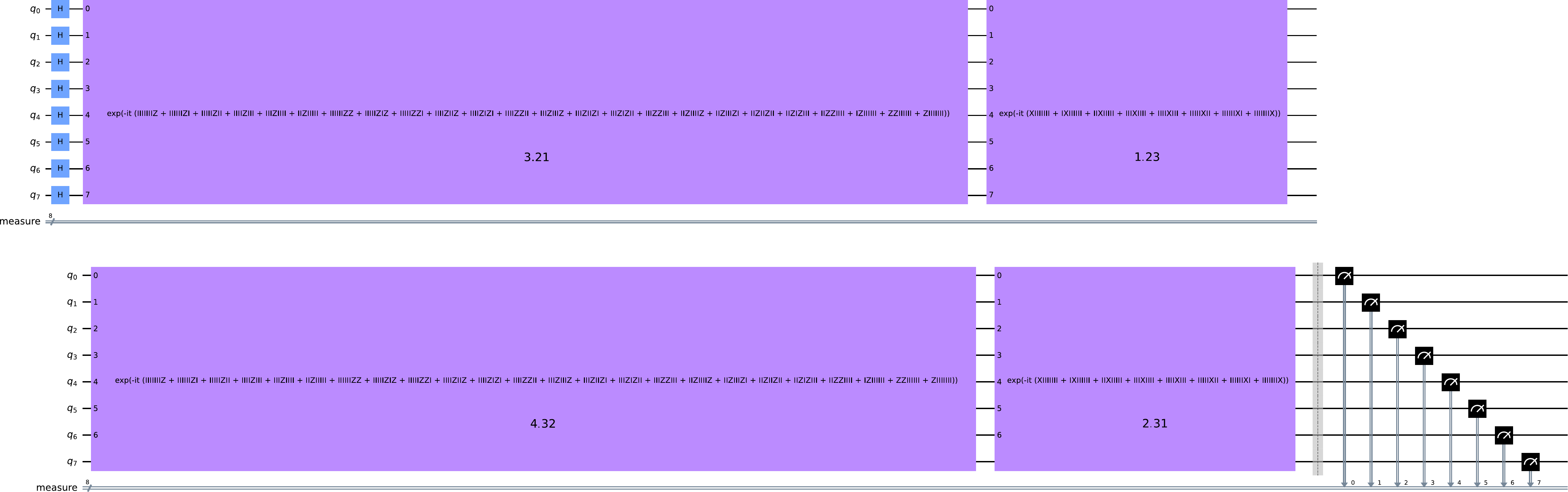}
\end{center}
{ \hspace*{\fill} \\}

    \begin{tcolorbox}[breakable, size=fbox, boxrule=1pt, pad at break*=1mm,colback=cellbackground, colframe=cellborder]
\prompt{In}{incolor}{16}{\boxspacing}
\begin{Verbatim}[commandchars=\\\{\}]
\PY{k+kn}{from} \PY{n+nn}{qiskit\PYZus{}aer} \PY{k+kn}{import} \PY{n}{AerSimulator}

\PY{n}{number\PYZus{}of\PYZus{}shots} \PY{o}{=} \PY{l+m+mi}{8000}
\PY{n}{aer\PYZus{}simulator} \PY{o}{=} \PY{n}{AerSimulator}\PY{p}{(}\PY{n}{method}\PY{o}{=}\PY{l+s+s2}{\PYZdq{}}\PY{l+s+s2}{statevector}\PY{l+s+s2}{\PYZdq{}}\PY{p}{,} \PY{n}{shots}\PY{o}{=}\PY{n}{number\PYZus{}of\PYZus{}shots}\PY{p}{)}

\PY{c+c1}{\PYZsh{} Decompose the circuit such that the gates can be simulated with the}
\PY{c+c1}{\PYZsh{} AerSimulator}
\PY{n}{qaoa\PYZus{}with\PYZus{}parameters\PYZus{}decomposed} \PY{o}{=} \PY{n}{qaoa\PYZus{}with\PYZus{}parameters}\PY{o}{.}\PY{n}{decompose}\PY{p}{(}\PY{n}{reps}\PY{o}{=}\PY{l+m+mi}{3}\PY{p}{)}

\PY{n}{result} \PY{o}{=} \PY{n}{aer\PYZus{}simulator}\PY{o}{.}\PY{n}{run}\PY{p}{(}\PY{n}{qaoa\PYZus{}with\PYZus{}parameters\PYZus{}decomposed}\PY{p}{)}\PY{o}{.}\PY{n}{result}\PY{p}{(}\PY{p}{)}
\end{Verbatim}
\end{tcolorbox}

    \begin{tcolorbox}[breakable, size=fbox, boxrule=1pt, pad at break*=1mm,colback=cellbackground, colframe=cellborder]
\prompt{In}{incolor}{17}{\boxspacing}
\begin{Verbatim}[commandchars=\\\{\}]
\PY{k+kn}{from} \PY{n+nn}{qiskit}\PY{n+nn}{.}\PY{n+nn}{visualization} \PY{k+kn}{import} \PY{n}{plot\PYZus{}histogram}

\PY{n}{counts} \PY{o}{=} \PY{n}{result}\PY{o}{.}\PY{n}{get\PYZus{}counts}\PY{p}{(}\PY{p}{)}

\PY{c+c1}{\PYZsh{} Plot the 40 bitstrings with the highest count.}
\PY{n}{plot\PYZus{}histogram}\PY{p}{(}
    \PY{n}{counts}\PY{p}{,}
    \PY{n}{number\PYZus{}to\PYZus{}keep}\PY{o}{=}\PY{l+m+mi}{40}\PY{p}{,}
    \PY{n}{sort}\PY{o}{=}\PY{l+s+s2}{\PYZdq{}}\PY{l+s+s2}{value\PYZus{}desc}\PY{l+s+s2}{\PYZdq{}}\PY{p}{,}
    \PY{n}{bar\PYZus{}labels}\PY{o}{=}\PY{k+kc}{False}\PY{p}{,}
    \PY{n}{figsize}\PY{o}{=}\PY{p}{(}\PY{l+m+mi}{10}\PY{p}{,}\PY{l+m+mi}{5}\PY{p}{)}\PY{p}{)}
\end{Verbatim}
\end{tcolorbox}

\prompt{Out}{outcolor}{17}{}
    
    \begin{center}
    \adjustimage{max size={0.9\linewidth}{0.9\paperheight}}{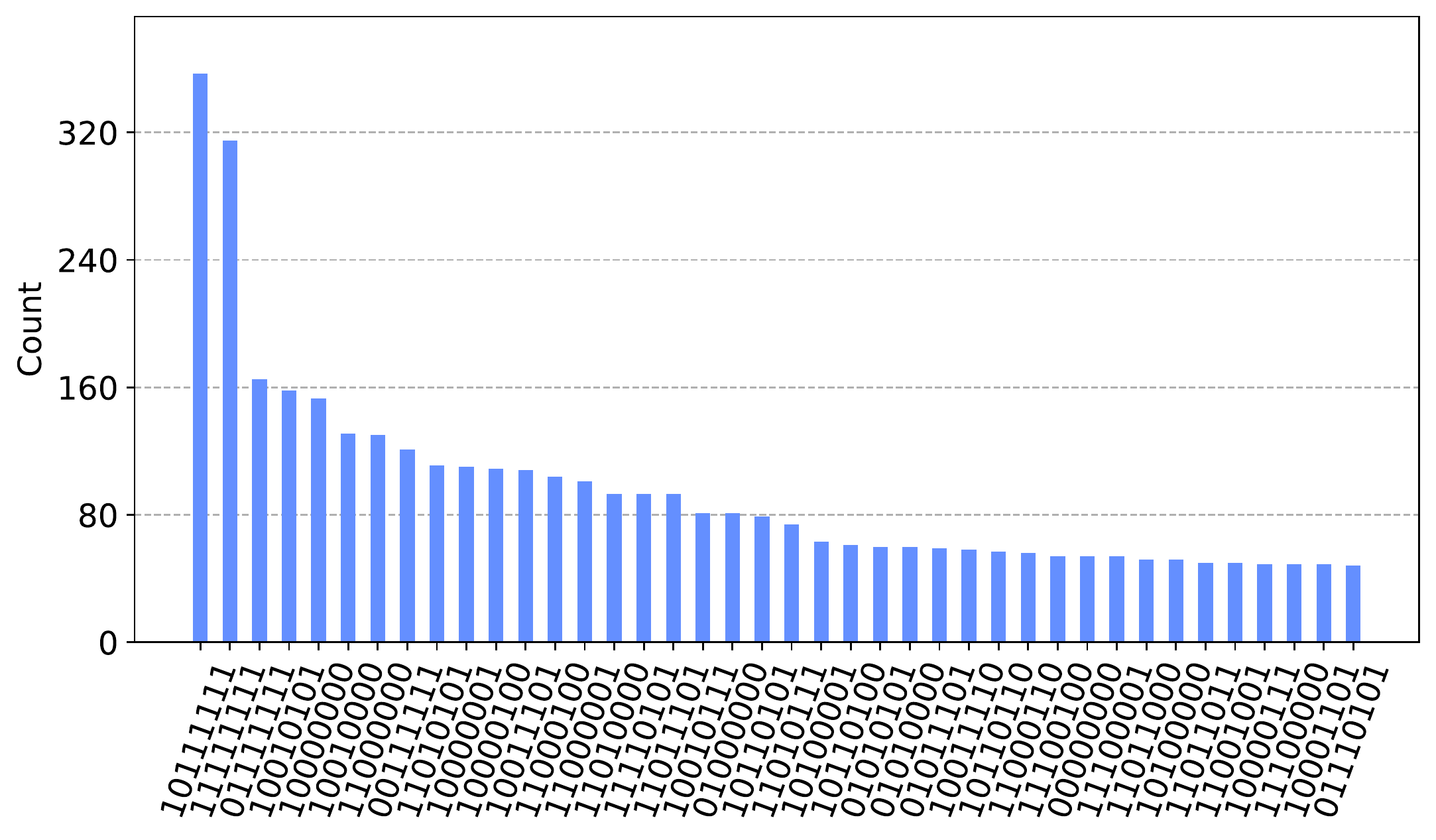}
    \end{center}
    { \hspace*{\fill} \\}

    Next, let us calculate the expectation value for our result and for the
bitstring with the highest count/probability.

\begin{tcolorbox}[breakable, size=fbox, boxrule=1pt, pad at break*=1mm,colback=cellbackground, colframe=cellborder]
\prompt{In}{incolor}{19}{\boxspacing}
\begin{Verbatim}[commandchars=\\\{\}]
\PY{n}{sqrt\PYZus{}probabilities} \PY{o}{=} \PY{p}{\PYZob{}}\PY{n}{bitstring}\PY{p}{:} \PY{n}{np}\PY{o}{.}\PY{n}{sqrt}\PY{p}{(}\PY{n}{count}\PY{o}{/}\PY{n}{number\PYZus{}of\PYZus{}shots}\PY{p}{)} \PYZbs{}
                    \PY{k}{  for} \PY{n}{bitstring}\PY{p}{,} \PY{n}{count} \PY{o+ow}{in} \PY{n}{counts}\PY{o}{.}\PY{n}{items}\PY{p}{(}\PY{p}{)}\PY{p}{\PYZcb{}}

\PY{n}{expectation} \PY{o}{=} \PY{n}{np}\PY{o}{.}\PY{n}{real}\PY{p}{(}\PY{n}{observable}\PY{o}{.}\PY{n}{eval}\PY{p}{(}\PY{n}{sqrt\PYZus{}probabilities}\PY{p}{)}\PY{p}{)} \PYZbs{}
              \PY{o}{+} \PY{n}{ising\PYZus{}offset}

\PY{n+nb}{print}\PY{p}{(}\PY{l+s+sa}{f}\PY{l+s+s2}{\PYZdq{}}\PY{l+s+s2}{e(}\PY{l+s+si}{\PYZob{}}\PY{n}{betas}\PY{l+s+si}{\PYZcb{}}\PY{l+s+s2}{, }\PY{l+s+si}{\PYZob{}}\PY{n}{gammas}\PY{l+s+si}{\PYZcb{}}\PY{l+s+s2}{) = }\PY{l+s+s2}{\PYZdq{}}\PY{p}{,} \PY{n}{expectation}\PY{p}{)}
\end{Verbatim}
\end{tcolorbox}

    \begin{Verbatim}[commandchars=\\\{\}]
e([1.23, 2.31], [3.21, 4.32]) =  39.16945000000001
    \end{Verbatim}

\begin{tcolorbox}[breakable, size=fbox, boxrule=1pt, pad at break*=1mm,colback=cellbackground, colframe=cellborder]
\prompt{In}{incolor}{20}{\boxspacing}
\begin{Verbatim}[commandchars=\\\{\}]
\PY{n}{bitstring\PYZus{}highest\PYZus{}prob} \PY{o}{=} \PY{n+nb}{max}\PY{p}{(}
    \PY{n}{sqrt\PYZus{}probabilities}\PY{p}{,} \PY{n}{key}\PY{o}{=}\PY{k}{lambda} \PY{n}{key}\PY{p}{:} \PY{n}{sqrt\PYZus{}probabilities}\PY{p}{[}\PY{n}{key}\PY{p}{]}\PY{p}{)}

\PY{n+nb}{print}\PY{p}{(}\PY{l+s+s2}{\PYZdq{}}\PY{l+s+s2}{bitstring with highest probability:}\PY{l+s+s2}{\PYZdq{}}\PY{p}{,} \PY{n}{bitstring\PYZus{}highest\PYZus{}prob}\PY{p}{)}
\end{Verbatim}
\end{tcolorbox}

    \begin{Verbatim}[commandchars=\\\{\}]
bitstring with highest probability: 10111111
    \end{Verbatim}

    \begin{tcolorbox}[breakable, size=fbox, boxrule=1pt, pad at break*=1mm,colback=cellbackground, colframe=cellborder]
\prompt{In}{incolor}{20}{\boxspacing}
\begin{Verbatim}[commandchars=\\\{\}]
\PY{n}{expectation} \PY{o}{=} \PY{n}{np}\PY{o}{.}\PY{n}{real}\PY{p}{(}\PY{n}{observable}\PY{o}{.}\PY{n}{eval}\PY{p}{(}\PY{n}{bitstring\PYZus{}highest\PYZus{}prob}\PY{p}{)}\PY{p}{)} \PYZbs{}
              \PY{o}{+} \PY{n}{ising\PYZus{}offset}

\PY{n+nb}{print}\PY{p}{(}\PY{l+s+sa}{f}\PY{l+s+s2}{\PYZdq{}}\PY{l+s+s2}{\PYZlt{}}\PY{l+s+si}{\PYZob{}}\PY{n}{bitstring\PYZus{}highest\PYZus{}prob}\PY{l+s+si}{\PYZcb{}}\PY{l+s+s2}{ | H\PYZus{}P | }\PY{l+s+si}{\PYZob{}}\PY{n}{bitstring\PYZus{}highest\PYZus{}prob}\PY{l+s+si}{\PYZcb{}}\PY{l+s+s2}{\PYZgt{} = }\PY{l+s+s2}{\PYZdq{}}
      \PY{l+s+sa}{f}\PY{l+s+s2}{\PYZdq{}}\PY{l+s+si}{\PYZob{}}\PY{n}{expectation}\PY{l+s+si}{\PYZcb{}}\PY{l+s+s2}{\PYZdq{}}\PY{p}{)}
\end{Verbatim}
\end{tcolorbox}

    \begin{Verbatim}[commandchars=\\\{\}]
<10111111 | H\_P | 10111111> = 121.0
    \end{Verbatim}

    Remembering from above that the minimum value of \(f_3\) (and of \(f_1\))
is \(6.0\), we see that the solution of QAOA is not good. A main reason
for this is that our parameters \texttt{betas} and \texttt{gammas} were
not optimized. We do this in the next section.

    \hypertarget{qaoa-classical-optimization}{%
\section{QAOA: Classical
Optimization}\label{sec:qaoa-classical-optimization}}

    First, we need to write a function that has as input numerical values
for the QAOA parameters \(\vec{\beta}, \vec{\gamma}\) and computes the
expectation value \(e(\vec{\beta}, \vec{\gamma})\), see \eqref{eq:qaoa2}, as
output. We name this function \texttt{energy\_evaluation} to be
consistent with other Qiskit implementations, most importantly
\href{https://qiskit.org/documentation/stubs/qiskit.algorithms.QAOA.html#qiskit.algorithms.QAOA}{\texttt{qiskit.algorithms.QAOA}}.
The term energy comes from applications of the VQE algorithm and means
the same as expectation. It is \emph{not} connected to the term energy
as used in our electric vehicle use case, where it refers to the energy
the electric cars need to charge.

\begin{tcolorbox}[breakable, size=fbox, boxrule=1pt, pad at break*=1mm,colback=cellbackground, colframe=cellborder]
\prompt{In}{incolor}{22}{\boxspacing}
\begin{Verbatim}[commandchars=\\\{\}]
\PY{k+kn}{from} \PY{n+nn}{typing} \PY{k+kn}{import} \PY{n}{List}

\PY{k}{def} \PY{n+nf}{energy\PYZus{}evaluation}\PY{p}{(}\PY{n}{parameter\PYZus{}values}\PY{p}{:} \PY{n}{List}\PY{p}{[}\PY{n+nb}{float}\PY{p}{]}\PY{p}{)}\PY{p}{:}
    \PY{l+s+sd}{\PYZdq{}\PYZdq{}\PYZdq{}parameter\PYZus{}values is expected to be of }
\PY{l+s+sd}{    the form [beta\PYZus{}0, beta\PYZus{}1, ..., gamma\PYZus{}0, gamma\PYZus{}1, ...]\PYZdq{}\PYZdq{}\PYZdq{}}
    \PY{n}{parameter\PYZus{}bindings} \PY{o}{=} \PY{n+nb}{dict}\PY{p}{(}\PY{n+nb}{zip}\PY{p}{(}\PY{n}{qaoa\PYZus{}ansatz}\PY{o}{.}\PY{n}{parameters}\PY{p}{,} \PY{n}{parameter\PYZus{}values}\PY{p}{)}\PY{p}{)}

    \PY{n}{qaoa\PYZus{}with\PYZus{}parameters} \PY{o}{=} \PY{n}{qaoa\PYZus{}ansatz}\PY{o}{.}\PY{n}{bind\PYZus{}parameters}\PY{p}{(}\PY{n}{parameter\PYZus{}bindings}\PY{p}{)}
    \PY{n}{qaoa\PYZus{}with\PYZus{}parameters\PYZus{}decomposed} \PY{o}{=} \PY{n}{qaoa\PYZus{}with\PYZus{}parameters}\PY{o}{.}\PY{n}{decompose}\PY{p}{(}\PY{n}{reps}\PY{o}{=}\PY{l+m+mi}{3}\PY{p}{)}

    \PY{n}{result} \PY{o}{=} \PY{n}{aer\PYZus{}simulator}\PY{o}{.}\PY{n}{run}\PY{p}{(}\PY{n}{qaoa\PYZus{}with\PYZus{}parameters\PYZus{}decomposed}\PY{p}{)}\PY{o}{.}\PY{n}{result}\PY{p}{(}\PY{p}{)}
    \PY{n}{counts} \PY{o}{=} \PY{n}{result}\PY{o}{.}\PY{n}{get\PYZus{}counts}\PY{p}{(}\PY{p}{)}
    \PY{n}{sqrt\PYZus{}probabilities} \PY{o}{=} \PY{p}{\PYZob{}}\PY{n}{bitstring}\PY{p}{:} \PY{n}{np}\PY{o}{.}\PY{n}{sqrt}\PY{p}{(}\PY{n}{count}\PY{o}{/}\PY{n}{number\PYZus{}of\PYZus{}shots}\PY{p}{)} \PYZbs{}
                          \PY{k}{for} \PY{n}{bitstring}\PY{p}{,} \PY{n}{count} \PY{o+ow}{in} \PY{n}{counts}\PY{o}{.}\PY{n}{items}\PY{p}{(}\PY{p}{)}\PY{p}{\PYZcb{}}
    
    \PY{n}{expectation} \PY{o}{=} \PY{n}{np}\PY{o}{.}\PY{n}{real}\PY{p}{(}\PY{n}{observable}\PY{o}{.}\PY{n}{eval}\PY{p}{(}\PY{n}{sqrt\PYZus{}probabilities}\PY{p}{)}\PY{p}{)} \PY{o}{+} \PY{n}{ising\PYZus{}offset}
    
    \PY{k}{return} \PY{n}{expectation}
\end{Verbatim}
\end{tcolorbox}

    We test our implementation with \texttt{betas} and \texttt{gammas} from Section \ref{sec:qaoa-implementation}.

    \begin{tcolorbox}[breakable, size=fbox, boxrule=1pt, pad at break*=1mm,colback=cellbackground, colframe=cellborder]
\prompt{In}{incolor}{22}{\boxspacing}
\begin{Verbatim}[commandchars=\\\{\}]
\PY{n}{energy\PYZus{}evaluation}\PY{p}{(}\PY{p}{[}\PY{o}{*}\PY{n}{betas}\PY{p}{,} \PY{o}{*}\PY{n}{gammas}\PY{p}{]}\PY{p}{)}
\end{Verbatim}
\end{tcolorbox}

            \begin{tcolorbox}[breakable, size=fbox, boxrule=.5pt, pad at break*=1mm, opacityfill=0]
\prompt{Out}{outcolor}{22}{\boxspacing}
\begin{Verbatim}[commandchars=\\\{\}]
39.09017500000002
\end{Verbatim}
\end{tcolorbox}
        
    We see that we get nearly the same result as above. The small difference
is due to statistical errors stemming from the fact that from a finite
number of shots (see \texttt{number\_shots} in the code) the expectation
value \(\langle \psi | \HP | \psi \rangle\) cannot be computed exactly.
This effect is also known as \textbf{shot noise} \cite{ShotNoiseWeb}.

    Having this function we can use a classical optimizer, as e.g.~provided
in \href{https://qiskit.org/documentation/stubs/qiskit.algorithms.optimizers.html}{\texttt{qiskit.algorithms.optimizers}}.
We will use the COBYLA optimizer and random initial guesses for our
parameters. These can be implemented e.g.~with NumPy's \href{https://numpy.org/doc/stable/reference/random/generator.html#numpy.random.default_rng}{default random generator}.

    \begin{tcolorbox}[breakable, size=fbox, boxrule=1pt, pad at break*=1mm,colback=cellbackground, colframe=cellborder]
\prompt{In}{incolor}{23}{\boxspacing}
\begin{Verbatim}[commandchars=\\\{\}]
\PY{c+c1}{\PYZsh{} Run this cell only once if you want changing random numbers in the cell below}
\PY{n}{random\PYZus{}generator} \PY{o}{=} \PY{n}{np}\PY{o}{.}\PY{n}{random}\PY{o}{.}\PY{n}{default\PYZus{}rng}\PY{p}{(}\PY{l+m+mi}{1234}\PY{p}{)}
\end{Verbatim}
\end{tcolorbox}

    \begin{tcolorbox}[breakable, size=fbox, boxrule=1pt, pad at break*=1mm,colback=cellbackground, colframe=cellborder]
\prompt{In}{incolor}{24}{\boxspacing}
\begin{Verbatim}[commandchars=\\\{\}]
\PY{k+kn}{from} \PY{n+nn}{qiskit}\PY{n+nn}{.}\PY{n+nn}{algorithms}\PY{n+nn}{.}\PY{n+nn}{optimizers} \PY{k+kn}{import} \PY{n}{COBYLA}

\PY{n}{betas\PYZus{}initial\PYZus{}guess} \PY{o}{=} \PY{n}{np}\PY{o}{.}\PY{n}{pi}\PY{o}{*}\PY{n}{random\PYZus{}generator}\PY{o}{.}\PY{n}{random}\PY{p}{(}\PY{n}{qaoa\PYZus{}reps}\PY{p}{)}
\PY{n}{gammas\PYZus{}initial\PYZus{}guess} \PY{o}{=} \PY{l+m+mi}{2}\PY{o}{*}\PY{n}{np}\PY{o}{.}\PY{n}{pi}\PY{o}{*}\PY{n}{random\PYZus{}generator}\PY{o}{.}\PY{n}{random}\PY{p}{(}\PY{n}{qaoa\PYZus{}reps}\PY{p}{)}
\PY{n}{parameter\PYZus{}values\PYZus{}initial\PYZus{}guess} \PY{o}{=} \PY{p}{[}\PY{o}{*}\PY{n}{betas\PYZus{}initial\PYZus{}guess}\PY{p}{,} \PY{o}{*}\PY{n}{gammas\PYZus{}initial\PYZus{}guess}\PY{p}{]}

\PY{n}{cobyla\PYZus{}optimizer} \PY{o}{=} \PY{n}{COBYLA}\PY{p}{(}\PY{p}{)}

\PY{n}{result\PYZus{}optimization} \PY{o}{=} \PY{n}{cobyla\PYZus{}optimizer}\PY{o}{.}\PY{n}{minimize}\PY{p}{(}
    \PY{n}{fun}\PY{o}{=}\PY{n}{energy\PYZus{}evaluation}\PY{p}{,} \PY{n}{x0}\PY{o}{=}\PY{n}{parameter\PYZus{}values\PYZus{}initial\PYZus{}guess}\PY{p}{)}

\PY{n}{parameter\PYZus{}values\PYZus{}optimized} \PY{o}{=} \PY{n}{result\PYZus{}optimization}\PY{o}{.}\PY{n}{x}
\PY{n}{energy\PYZus{}optimized} \PY{o}{=} \PY{n}{result\PYZus{}optimization}\PY{o}{.}\PY{n}{fun}
\PY{c+c1}{\PYZsh{} Number of evalutions of energy\PYZus{}evaluation}
\PY{n}{number\PYZus{}function\PYZus{}evaluations} \PY{o}{=} \PY{n}{result\PYZus{}optimization}\PY{o}{.}\PY{n}{nfev}

\PY{n+nb}{print}\PY{p}{(}\PY{l+s+sa}{f}\PY{l+s+s2}{\PYZdq{}}\PY{l+s+s2}{Optimized parameters: }\PY{l+s+si}{\PYZob{}}\PY{n}{parameter\PYZus{}values\PYZus{}optimized}\PY{l+s+si}{\PYZcb{}}\PY{l+s+s2}{\PYZdq{}}\PY{p}{)}
\PY{n+nb}{print}\PY{p}{(}\PY{l+s+sa}{f}\PY{l+s+s2}{\PYZdq{}}\PY{l+s+s2}{Expectation value: }\PY{l+s+si}{\PYZob{}}\PY{n}{energy\PYZus{}optimized}\PY{l+s+si}{\PYZcb{}}\PY{l+s+s2}{\PYZdq{}}\PY{p}{)}
\PY{n+nb}{print}\PY{p}{(}\PY{l+s+sa}{f}\PY{l+s+s2}{\PYZdq{}}\PY{l+s+s2}{Number function evaluations: }\PY{l+s+si}{\PYZob{}}\PY{n}{number\PYZus{}function\PYZus{}evaluations}\PY{l+s+si}{\PYZcb{}}\PY{l+s+s2}{\PYZdq{}}\PY{p}{)}
\end{Verbatim}
\end{tcolorbox}

\begin{Verbatim}[commandchars=\\\{\}]
Optimized parameters: [3.18220113 1.19369043 5.75551235 1.65220983]
Expectation value: 19.25527500000001
Number function evaluations: 45
\end{Verbatim}

    Let's run the QAOA circuit with the optimized parameters.

    \begin{tcolorbox}[breakable, size=fbox, boxrule=1pt, pad at break*=1mm,colback=cellbackground, colframe=cellborder]
\prompt{In}{incolor}{25}{\boxspacing}
\begin{Verbatim}[commandchars=\\\{\}]
\PY{n}{parameter\PYZus{}bindings} \PY{o}{=} \PY{n+nb}{dict}\PY{p}{(}
    \PY{n+nb}{zip}\PY{p}{(}\PY{n}{qaoa\PYZus{}ansatz}\PY{o}{.}\PY{n}{parameters}\PY{p}{,} \PY{n}{parameter\PYZus{}values\PYZus{}optimized}\PY{p}{)}\PY{p}{)}

\PY{n}{qaoa\PYZus{}with\PYZus{}parameters} \PY{o}{=} \PY{n}{qaoa\PYZus{}ansatz}\PY{o}{.}\PY{n}{bind\PYZus{}parameters}\PY{p}{(}\PY{n}{parameter\PYZus{}bindings}\PY{p}{)}
\PY{n}{qaoa\PYZus{}with\PYZus{}parameters\PYZus{}decomposed} \PY{o}{=} \PY{n}{qaoa\PYZus{}with\PYZus{}parameters}\PY{o}{.}\PY{n}{decompose}\PY{p}{(}\PY{n}{reps}\PY{o}{=}\PY{l+m+mi}{3}\PY{p}{)}

\PY{n}{result} \PY{o}{=} \PY{n}{aer\PYZus{}simulator}\PY{o}{.}\PY{n}{run}\PY{p}{(}\PY{n}{qaoa\PYZus{}with\PYZus{}parameters\PYZus{}decomposed}\PY{p}{)}\PY{o}{.}\PY{n}{result}\PY{p}{(}\PY{p}{)}
\PY{n}{counts} \PY{o}{=} \PY{n}{result}\PY{o}{.}\PY{n}{get\PYZus{}counts}\PY{p}{(}\PY{p}{)}
\end{Verbatim}
\end{tcolorbox}

    \begin{tcolorbox}[breakable, size=fbox, boxrule=1pt, pad at break*=1mm,colback=cellbackground, colframe=cellborder]
\prompt{In}{incolor}{26}{\boxspacing}
\begin{Verbatim}[commandchars=\\\{\}]
\PY{c+c1}{\PYZsh{} Plot the 40 bitstrings with the highest count.}
\PY{n}{plot\PYZus{}histogram}\PY{p}{(}
    \PY{n}{counts}\PY{p}{,}
    \PY{n}{number\PYZus{}to\PYZus{}keep}\PY{o}{=}\PY{l+m+mi}{40}\PY{p}{,}
    \PY{n}{sort}\PY{o}{=}\PY{l+s+s2}{\PYZdq{}}\PY{l+s+s2}{value\PYZus{}desc}\PY{l+s+s2}{\PYZdq{}}\PY{p}{,}
    \PY{n}{bar\PYZus{}labels}\PY{o}{=}\PY{k+kc}{False}\PY{p}{,}
    \PY{n}{figsize}\PY{o}{=}\PY{p}{(}\PY{l+m+mi}{10}\PY{p}{,}\PY{l+m+mi}{5}\PY{p}{)}\PY{p}{)}
\end{Verbatim}
\end{tcolorbox}

\prompt{Out}{outcolor}{26}{}
    
    \begin{center}
    \adjustimage{max size={0.9\linewidth}{0.9\paperheight}}{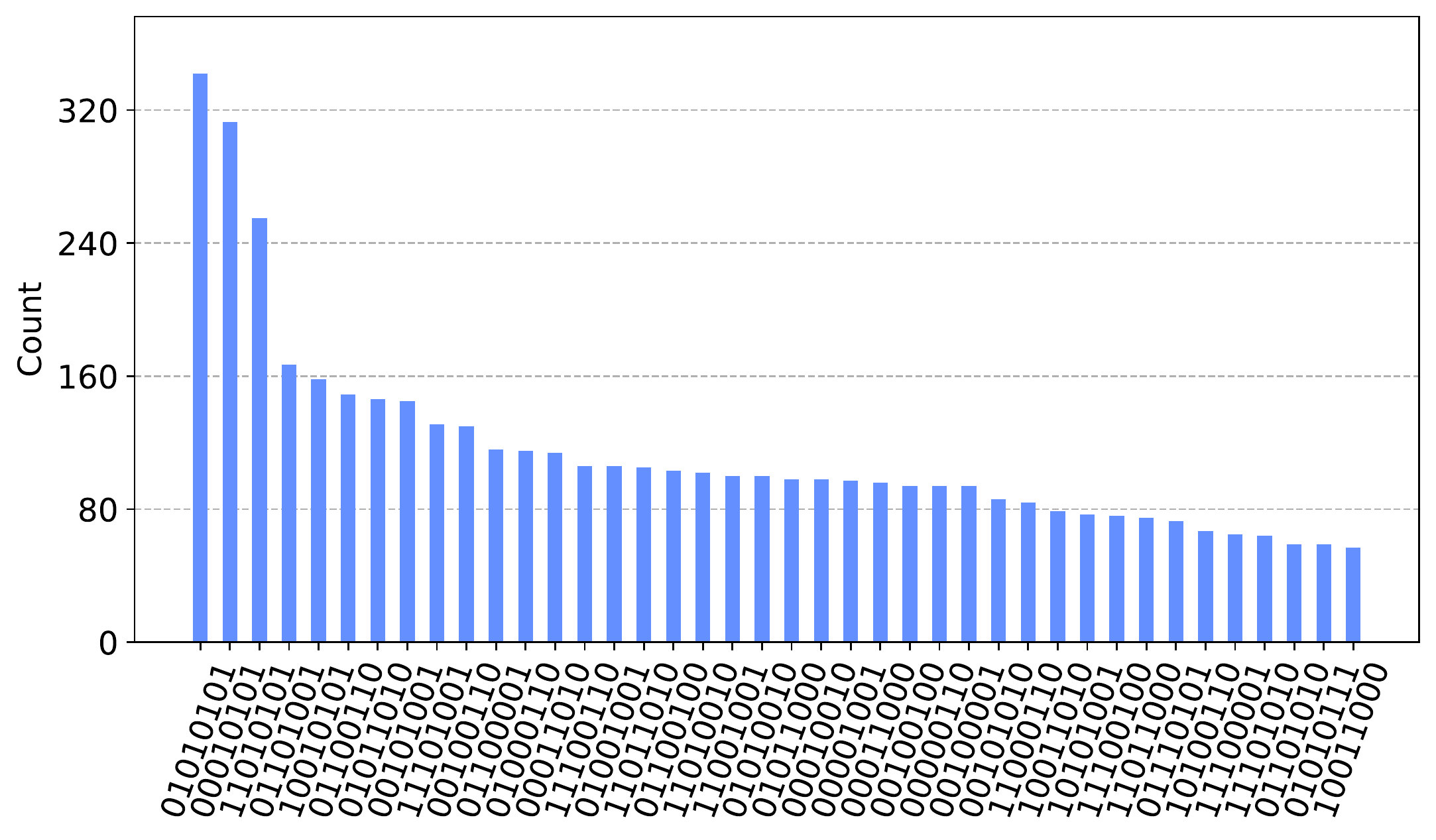}
    \end{center}
    { \hspace*{\fill} \\}

\begin{tcolorbox}[breakable, size=fbox, boxrule=1pt, pad at break*=1mm,colback=cellbackground, colframe=cellborder]
\prompt{In}{incolor}{28}{\boxspacing}
\begin{Verbatim}[commandchars=\\\{\}]
\PY{n}{sqrt\PYZus{}probabilities} \PY{o}{=} \PY{p}{\PYZob{}}\PY{n}{bitstring}\PY{p}{:} \PY{n}{np}\PY{o}{.}\PY{n}{sqrt}\PY{p}{(}\PY{n}{count}\PY{o}{/}\PY{n}{number\PYZus{}of\PYZus{}shots}\PY{p}{)} \PYZbs{}
                      \PY{k}{for} \PY{n}{bitstring}\PY{p}{,} \PY{n}{count} \PY{o+ow}{in} \PY{n}{counts}\PY{o}{.}\PY{n}{items}\PY{p}{(}\PY{p}{)}\PY{p}{\PYZcb{}}

\PY{n}{bitstring\PYZus{}highest\PYZus{}prob} \PY{o}{=} \PY{n+nb}{max}\PY{p}{(}
    \PY{n}{sqrt\PYZus{}probabilities}\PY{p}{,} \PY{n}{key}\PY{o}{=}\PY{k}{lambda} \PY{n}{key}\PY{p}{:} \PY{n}{sqrt\PYZus{}probabilities}\PY{p}{[}\PY{n}{key}\PY{p}{]}\PY{p}{)}
\PY{n+nb}{print}\PY{p}{(}\PY{l+s+sa}{f}\PY{l+s+s2}{\PYZdq{}}\PY{l+s+s2}{bitstring with highest probability: }\PY{l+s+si}{\PYZob{}}\PY{n}{bitstring\PYZus{}highest\PYZus{}prob}\PY{l+s+si}{\PYZcb{}}\PY{l+s+s2}{\PYZdq{}}\PY{p}{)}

\PY{n}{expectation} \PY{o}{=} \PY{n}{np}\PY{o}{.}\PY{n}{real}\PY{p}{(}\PY{n}{observable}\PY{o}{.}\PY{n}{eval}\PY{p}{(}\PY{n}{bitstring\PYZus{}highest\PYZus{}prob}\PY{p}{)}\PY{p}{)} \PYZbs{}
                \PY{o}{+} \PY{n}{ising\PYZus{}offset}
\PY{n+nb}{print}\PY{p}{(}\PY{l+s+sa}{f}\PY{l+s+s2}{\PYZdq{}}\PY{l+s+s2}{\PYZlt{}}\PY{l+s+si}{\PYZob{}}\PY{n}{bitstring\PYZus{}highest\PYZus{}prob}\PY{l+s+si}{\PYZcb{}}\PY{l+s+s2}{ | H\PYZus{}P | }\PY{l+s+si}{\PYZob{}}\PY{n}{bitstring\PYZus{}highest\PYZus{}prob}\PY{l+s+si}{\PYZcb{}}\PY{l+s+s2}{\PYZgt{} = }\PY{l+s+s2}{\PYZdq{}}
        \PY{l+s+sa}{f}\PY{l+s+s2}{\PYZdq{}}\PY{l+s+si}{\PYZob{}}\PY{n}{expectation}\PY{l+s+si}{\PYZcb{}}\PY{l+s+s2}{\PYZdq{}}\PY{p}{)}
\end{Verbatim}
\end{tcolorbox}

    \begin{Verbatim}[commandchars=\\\{\}]
bitstring with highest probability: 01010101
<01010101 | H\_P | 01010101> = 7.6000000000000085
    \end{Verbatim}

    We see that the result is considerably improved compared with the non-optimized
parameters from above. You can also rerun the optimization procedure so
that other inital values are used.

    \hypertarget{optional-advanced-knowedge-gate-synthesis}{%
\section{(Optional) Advanced Knowledge: Gate Synthesis}\label{sec:gate-synthesis}}

    In the remaining part of the notebook we discuss how \eqref{eq:qaoa1} can be
implemented, i.e.~which gates are needed to synthesize the unitary
operations and verify that Qiskit's implementation of QAOA indeed uses
theses gates. This will become technical and you can skip to the next
notebook without problems.

    The calculations below rely on the following identities:

\begin{enumerate}
\def\labelenumi{\arabic{enumi})}
\tightlist
\item For \emph{commuting} matrices \(A\) and \(B\), i.e.~\([A,B] = AB - BA = 0\), we have
\[
\mathrm{e}^{A + B} = \mathrm{e}^A \mathrm{e}^B = \mathrm{e}^B \mathrm{e}^A.
\]
\item For a Pauli-string \(S = P_1 \otimes P_2 \otimes \dots \otimes P_n\),
\(P_j \in \{I, X, Y, Z\}\), we have 
\[
\mathrm{exp}(\i \theta S) = \cos(\theta) I^{\otimes n} + \i \sin(\theta) S \ .
\]
In particular, this means
\[
\mathrm{exp}(\i \theta \sigma_P^{(k)}) = I^{\otimes k} \otimes \mathrm{exp}(\i \theta P) \otimes I^{\otimes (n-k-1)},
\qquad
P \in \{X, Y, Z\}.
\]
\end{enumerate}

    Note that \(\mathrm{e}^A\) and \(\exp(A)\) are the same function and we
use the form that gives the clearest notation.

    \hypertarget{gates-for-uniform-superposition}{%
\subsection{Gates for Uniform
Superposition}\label{gates-for-uniform-superposition}}

    The initial state \(\ket{+}^{\otimes n}\) can easily be obtained by

\[
\ket{+}^{\otimes n}
= H^{\otimes n} \ket{0}^{\otimes n}
= (\underbrace{H \otimes \dots \otimes H}_{n \text{ times}})
|\underbrace{0\dots0}_{n \text{ times}}\rangle,
\qquad
H = \tfrac{1}{\sqrt2} \pmat{1 & 1 \\ 1 & -1} \ .
\] Here, \(H\) is the \href{https://qiskit.org/documentation/stubs/qiskit.circuit.library.HGate.html#qiskit.circuit.library.HGate}{Hadamard gate}.

    \hypertarget{gates-for-mixing-operator}{%
\subsection{Gates for Mixing Operator}\label{gates-for-mixing-operator}}

    For the mixing operator we have
\begin{align*}
\UM(\beta)
&= \mathrm{exp}(- \mathrm{i} \beta \HM)
\\[0.2cm]
&= \mathrm{exp}\bigl(- \mathrm{i} \beta \sum_{k} \sigma_X^{(k)} \bigr)
\\[0.2cm]
&= \prod_{k} \mathrm{exp}\bigl(- \mathrm{i} \beta \sigma_X^{(k)} \bigr)
\\[0.2cm]
&= \bigl(\mathrm{exp}(- \mathrm{i} \beta X) \otimes I \otimes \dots \otimes I\bigr) 
\cdots
\bigl(I \otimes \dots I \otimes \mathrm{exp}(- \mathrm{i} \beta X)\bigr)
%\\
%&= \mathrm{exp}(- \mathrm{i} \beta X) \otimes \dots \otimes \mathrm{exp}(- \mathrm{i} \beta X)
\end{align*}

    Recall that the \href{https://qiskit.org/documentation/stubs/qiskit.circuit.library.RXGate.html}{\(\RX\) gate} is given by

\[
\RX(\theta) = \mathrm{exp}(- \mathrm{i} \tfrac{\theta}{2} X) \ .
\]

This means the \(\RX\) gate applied on qubit \(k\) is

\[
\RX_k(\theta) = I^{\otimes k} \otimes \exp(-\i \tfrac{\theta}{2} X) \otimes I^{n-k-1} \ .
\]

    Thus, the mixing operator can be implemented as

\[
\UM(\beta)
= \RX_0(2 \beta) \RX_1(2 \beta) \cdots \RX_{n-1}(2 \beta) \ .
\]

    Note that since every rotation acts on a different qubit all gates can
be implemented in parallel.

    \hypertarget{gates-for-phase-operator}{%
\subsection{Gates for Phase Operator}\label{gates-for-phase-operator}}

    For the phase operator we have
\begin{align*}
\UP(\gamma)
&= \mathrm{exp}(- \mathrm{i} \gamma \HP)
\\[0.2cm]
&= \mathrm{exp}\Bigl(- \mathrm{i} \gamma 
    \bigl( \sum_k \sum_{l>k} h_{k l} \sigma_Z^{(k)} \sigma_Z^{(l)} 
     + \sum_k h_k^\prime \sigma_Z^{(k)} 
      + h^{\prime \prime}I^{\otimes n} \bigr)
\Bigr)
\\[0.2cm]
&= \mathrm{exp}\Bigl(- \mathrm{i} \gamma \sum_k \sum_{l>k} h_{kl} \sigma_Z^{(k)} \sigma_Z^{(l)} \Bigr) 
\,
\mathrm{exp}\Bigl(- \mathrm{i} \gamma \sum_k h_k^\prime \sigma_Z^{(k)} \Bigr)
\, 
\mathrm{exp}\Bigl(- \mathrm{i} \gamma  h^{\prime \prime}I^{\otimes n} \Bigr) \ .
\end{align*}

    The third factor on the RHS is just a multiplication with a global phase
\(\mathrm{e}^{-\i \gamma h^{\prime \prime}}\). Since this part is not
included in Qiskit's \texttt{ising} (see \texttt{ising\_offset} above)
we ignore it for the remaining section.

    By the same computations as for the mixing operator the second factor
can be written as

\[
\mathrm{exp}\Bigl(- \mathrm{i} \gamma \sum_k h_k^\prime \sigma_Z^{(k)} \Bigr) 
= \RZ_0(2\gamma h_0^\prime) \RZ_1(2\gamma h_1^\prime) \cdots \RZ_{n-1}(2\gamma h_{n-1}^\prime) \ ,
\]

where the \href{https://qiskit.org/documentation/stubs/qiskit.circuit.library.RZGate.html}{\(\RZ\) gate} is given by

\[
\RZ(\theta) = \mathrm{exp}(- \mathrm{i} \tfrac{\theta}{2} Z) \ .
\]

Note that contrary to the mixing operator now the \textbf{rotation angle
also depends on the coefficients \(h_k^\prime\)}.

    For the first factor we have

\begin{align*}
\mathrm{exp}\Bigl(- \mathrm{i} \gamma \sum_k \sum_{l>k} h_{kl} \sigma_Z^{(k)} \sigma_Z^{(l)} \Bigr)
&= \prod_k \prod_{l>k} \mathrm{exp}\bigl(- \mathrm{i} \gamma h_{kl} \sigma_Z^{(k)} \sigma_Z^{(l)} \bigr)
\\[0.2cm]
&= \prod_k \prod_{l>k} \RZZ_{k,l}(2 \gamma h_{kl})
\\[0.2cm]
&= \RZZ_{0,1}(2 \gamma h_{01}) \RZZ_{0,2}(2 \gamma h_{02}) \cdots \RZZ_{n-2,n-1}(2 \gamma h_{n-2,n-1})
\ ,
\end{align*}

    since

\[
\sigma_Z^{(k)} \sigma_Z^{(l)}
= I^{\otimes k} \otimes Z \otimes I^{\otimes(l-k-1)} \otimes Z \otimes I^{\otimes(n-l-1)} \ ,
\]

    and since the \href{https://qiskit.org/documentation/stubs/qiskit.circuit.library.RZZGate.html}{\(\RZZ\) gate} is defined by

\[
\RZZ(\theta) = \mathrm{exp}\bigl(- \mathrm{i} \tfrac{\theta}{2} Z \otimes Z \bigr) \ .
\]

Again note that the \textbf{rotation angle not only depends on
\(\gamma\) but also on the coeffiecients \(h_{kl}\)}.

    \hypertarget{implementation}{%
\subsection{Implementation}}

    Let us verify that indeed we get a quantum circuit with the gates
discussed in the previous three sections. To see this we have to
\texttt{decompose} \texttt{qaoa\_ansatz} a different number of times.
For simplicity we now choose \(p=1\).

    \begin{tcolorbox}[breakable, size=fbox, boxrule=1pt, pad at break*=1mm,colback=cellbackground, colframe=cellborder]
\prompt{In}{incolor}{28}{\boxspacing}
\begin{Verbatim}[commandchars=\\\{\}]
\PY{n}{qaoa\PYZus{}ansatz} \PY{o}{=} \PY{n}{QAOAAnsatz}\PY{p}{(}\PY{n}{cost\PYZus{}operator}\PY{o}{=}\PY{n}{ising}\PY{p}{,} \PY{n}{reps}\PY{o}{=}\PY{l+m+mi}{1}\PY{p}{,} \PY{n}{name}\PY{o}{=}\PY{l+s+s1}{\PYZsq{}}\PY{l+s+s1}{qaoa}\PY{l+s+s1}{\PYZsq{}}\PY{p}{)}
\end{Verbatim}
\end{tcolorbox}

    \hypertarget{decompose-1-time}{%
\subsubsection{Decompose once}\label{decompose-1-time}}

    With this decomposition we see that
\begin{align*}
\ket{\psi_\text{QAOA}} 
&= \UM(\beta_0)\UP(\gamma_0) \ket{+}^{\otimes n}
\\[0.2cm]
&= \mathrm{exp}\Bigl(- \mathrm{i} \beta_0 \sum_{k} \sigma_X^{(k)} \Bigr) \ 
\mathrm{exp}\Bigl(- \mathrm{i} \gamma_0 
    \bigl( \sum_k \sum_{l>k} h_{kl} \sigma_Z^{(k)} \sigma_Z^{(l)} 
     + \sum_k h_k^\prime \sigma_Z^{(k)} \bigr)
\Bigr) \,
H^{\otimes n} \,
\ket{0}^{\otimes n} \ .
\end{align*}

    Recall: Mathematical formulas are read right to left whereas quantum
circuits are read left to right.

    \begin{tcolorbox}[breakable, size=fbox, boxrule=1pt, pad at break*=1mm,colback=cellbackground, colframe=cellborder]
\prompt{In}{incolor}{29}{\boxspacing}
\begin{Verbatim}[commandchars=\\\{\}]
\PY{n}{qaoa\PYZus{}ansatz}\PY{o}{.}\PY{n}{decompose}\PY{p}{(}\PY{n}{reps}\PY{o}{=}\PY{l+m+mi}{1}\PY{p}{)}\PY{o}{.}\PY{n}{draw}\PY{p}{(}\PY{n}{scale}\PY{o}{=}\PY{l+m+mf}{0.5}\PY{p}{,} \PY{n}{fold}\PY{o}{=}\PY{o}{\PYZhy{}}\PY{l+m+mi}{1}\PY{p}{)}
\end{Verbatim}
\end{tcolorbox}

\prompt{Out}{outcolor}{29}{}
    
    \begin{center}
    \adjustimage{max size={0.9\linewidth}{0.9\paperheight}}{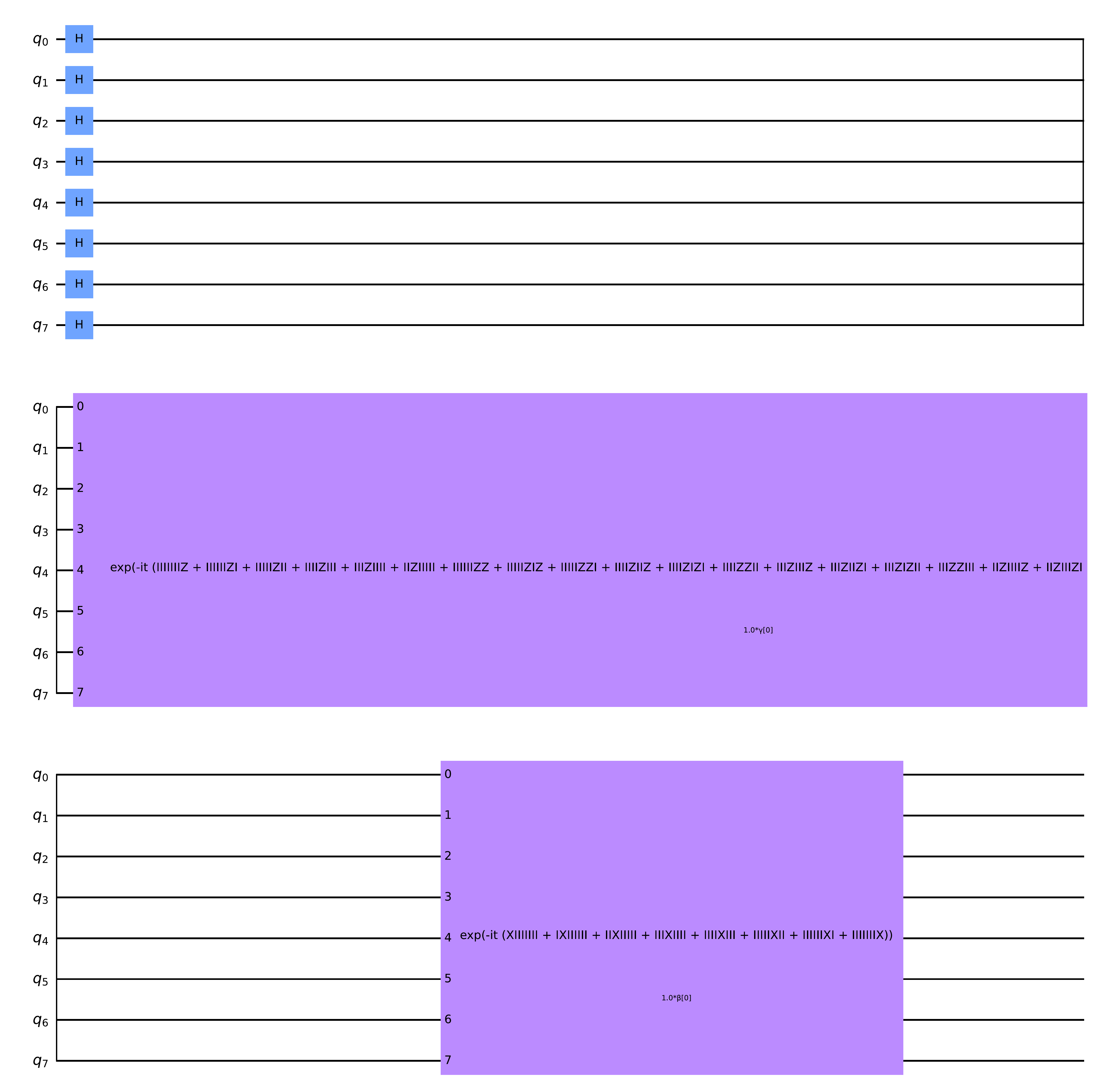}
    \end{center}
    { \hspace*{\fill} \\}

    \hypertarget{decompose-2-times}{%
\subsubsection{Decompose two times}\label{decompose-2-times}}

    Here we see

\[
\UM(\beta_0)
= \prod_{k} \mathrm{exp}\bigl(- \mathrm{i} \beta_0 \sigma_X^{(k)} \bigr) \ ,
\]

and

\[
\UP(\gamma_0)
=
\prod_k \prod_{l>k} \mathrm{exp}\bigl(- \mathrm{i} \gamma_0 h_{kl} \sigma_Z^{(k)} \sigma_Z^{(l)} \bigr)
\prod_{k} \mathrm{exp}\bigl(- \mathrm{i} \gamma_0 h_k^\prime \sigma_Z^{(k)} \bigr) \ .
\]

    \begin{tcolorbox}[breakable, size=fbox, boxrule=1pt, pad at break*=1mm,colback=cellbackground, colframe=cellborder]
\prompt{In}{incolor}{30}{\boxspacing}
\begin{Verbatim}[commandchars=\\\{\}]
\PY{n}{qaoa\PYZus{}ansatz}\PY{o}{.}\PY{n}{decompose}\PY{p}{(}\PY{n}{reps}\PY{o}{=}\PY{l+m+mi}{2}\PY{p}{)}\PY{o}{.}\PY{n}{draw}\PY{p}{(}\PY{n}{scale}\PY{o}{=}\PY{l+m+mf}{0.5}\PY{p}{,} \PY{n}{fold}\PY{o}{=}\PY{o}{\PYZhy{}}\PY{l+m+mi}{1}\PY{p}{)}
\end{Verbatim}
\end{tcolorbox}

\prompt{Out}{outcolor}{30}{}
    
    \begin{center}
    \adjustimage{max size={0.9\linewidth}{0.9\paperheight}}{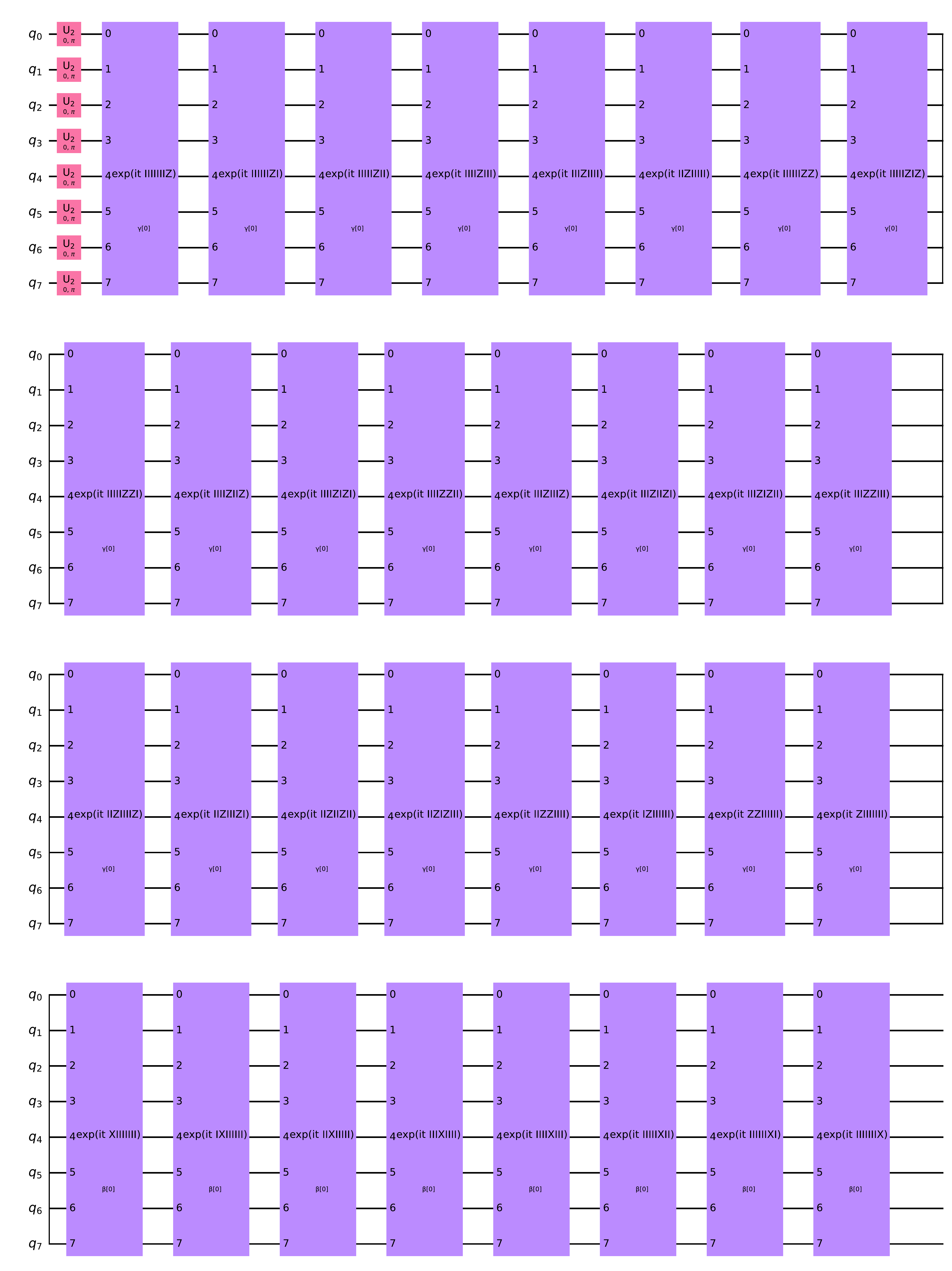}
    \end{center}
    { \hspace*{\fill} \\}

    \hypertarget{decompose-3-times}{%
\subsubsection{Decompose three times}\label{decompose-3-times}}

    With this decomposition we see

\[
\UM(\beta_0)
= \RX_0(2 \beta_0) \RX_1(2 \beta_0) \cdots \RX_{n-1}(2 \beta_0) \ ,
\]

    as well as

    \[
\UP(\gamma_0) 
=
\RZZ_{0,1}(2 \gamma_0 h_{01}) \RZZ_{0,2}(2 \gamma_0 h_{02}) \cdots \RZZ_{n-2,n-1}(2 \gamma_0 h_{n-2,n-1})
\ 
\RZ_0(2\gamma_0 h_0^\prime) 
%\RZ_1(2\gamma h_1^\prime)
\cdots \RZ_{n-1}(2\gamma_0 h_{n-1}^\prime) \ .
\]

    \begin{tcolorbox}[breakable, size=fbox, boxrule=1pt, pad at break*=1mm,colback=cellbackground, colframe=cellborder]
\prompt{In}{incolor}{31}{\boxspacing}
\begin{Verbatim}[commandchars=\\\{\}]
\PY{n}{qaoa\PYZus{}ansatz}\PY{o}{.}\PY{n}{decompose}\PY{p}{(}\PY{n}{reps}\PY{o}{=}\PY{l+m+mi}{3}\PY{p}{)}\PY{o}{.}\PY{n}{draw}\PY{p}{(}\PY{n}{scale}\PY{o}{=}\PY{l+m+mf}{0.5}\PY{p}{,} \PY{n}{fold}\PY{o}{=}\PY{o}{\PYZhy{}}\PY{l+m+mi}{1}\PY{p}{)}
\end{Verbatim}
\end{tcolorbox}

\prompt{Out}{outcolor}{31}{}
    
    \begin{center}
    \adjustimage{max size={0.9\linewidth}{0.9\paperheight}}{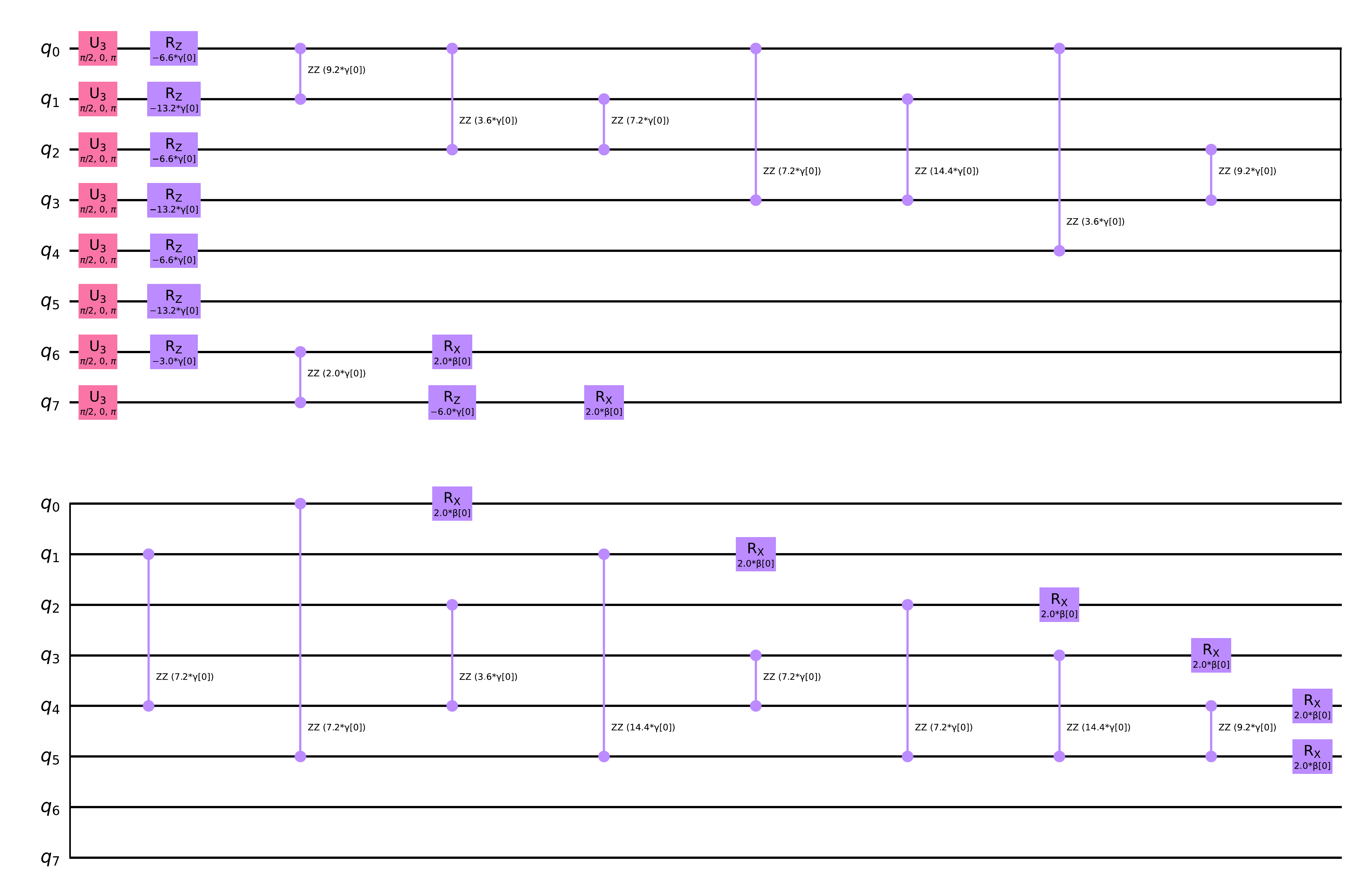}
    \end{center}
    { \hspace*{\fill} \\}

    Compare this also with the coefficient of \texttt{ising} (but recall the
different ordering of Qiskit)

    \begin{tcolorbox}[breakable, size=fbox, boxrule=1pt, pad at break*=1mm,colback=cellbackground, colframe=cellborder]
\prompt{In}{incolor}{32}{\boxspacing}
\begin{Verbatim}[commandchars=\\\{\}]
\PY{n+nb}{print}\PY{p}{(}\PY{n}{ising}\PY{p}{)}
\end{Verbatim}
\end{tcolorbox}

    \begin{Verbatim}[commandchars=\\\{\}]
-3.3 * IIIIIIIZ
- 6.599999999999998 * IIIIIIZI
- 3.3000000000000007 * IIIIIZII
- 6.6 * IIIIZIII
- 3.2999999999999994 * IIIZIIII
- 6.599999999999999 * IIZIIIII
+ 4.6 * IIIIIIZZ
+ 1.8 * IIIIIZIZ
+ 3.6 * IIIIIZZI
+ 3.6 * IIIIZIIZ
+ 7.2 * IIIIZIZI
+ 4.6 * IIIIZZII
+ 1.8 * IIIZIIIZ
+ 3.6 * IIIZIIZI
+ 1.8 * IIIZIZII
+ 3.6 * IIIZZIII
+ 3.6 * IIZIIIIZ
+ 7.2 * IIZIIIZI
+ 3.6 * IIZIIZII
+ 7.2 * IIZIZIII
+ 4.6 * IIZZIIII
- 1.5 * IZIIIIII
+ 1.0 * ZZIIIIII
- 3.0 * ZIIIIIII
    \end{Verbatim}

%% file: notebook_3_latex_source.tex
\hypertarget{introdution}{%
\section{Introdution}\label{sec:notebook-3-introdution}}

    In Notebook \ref{chap:notebook-1} we presented a real-world use case for optimizing charging
schedules for electric cars, reduced it to a proof of concept model, and
transformed it to a QUBO. Then, in Notebook \ref{chap:notebook-2} we presented the quantum
algorithm QAOA and how the associated quantum circuits can be obtained.
However, these circuits cannot directly be executed on a real quantum
computer but they first need to be \textbf{transpiled}. In this notebook
we will implement this process. Moreover, we will present how results
from a quantum computer can be \textbf{post-processed}.

    We will demonstrate these points with the same example as used in
Notebook \ref{chap:notebook-2} enriched with the implementations that where derived there.

    \hypertarget{example-for-this-notebook}{%
\section{Example for this Notebook}\label{sec:notebook-3-example}}

    \begin{tcolorbox}[breakable, size=fbox, boxrule=1pt, pad at break*=1mm,colback=cellbackground, colframe=cellborder]
\prompt{In}{incolor}{1}{\boxspacing}
\begin{Verbatim}[commandchars=\\\{\}]
\PY{k+kn}{from} \PY{n+nn}{codes\PYZus{}notebook\PYZus{}2} \PY{k+kn}{import} \PY{n}{generate\PYZus{}example}
\PY{n}{charging\PYZus{}unit}\PY{p}{,} \PY{n}{car\PYZus{}green}\PY{p}{,} \PY{n}{qcio}\PY{p}{,} \PY{n}{converter}\PY{p}{,} \PY{n}{qubo}\PY{p}{,} \PYZbs{}
    \PY{n}{number\PYZus{}binary\PYZus{}variables}\PY{p}{,} \PY{n}{qubo\PYZus{}minimization\PYZus{}result}\PY{p}{,} \PYZbs{}
    \PY{n}{ising}\PY{p}{,} \PY{n}{ising\PYZus{}offset}\PY{p}{,} \PY{n}{qaoa\PYZus{}reps}\PY{p}{,} \PY{n}{qaoa\PYZus{}circuit} \PY{o}{=} \PY{n}{generate\PYZus{}example}\PY{p}{(}\PY{p}{)}
\end{Verbatim}
\end{tcolorbox}

    \begin{tcolorbox}[breakable, size=fbox, boxrule=1pt, pad at break*=1mm,colback=cellbackground, colframe=cellborder]
\prompt{In}{incolor}{2}{\boxspacing}
\begin{Verbatim}[commandchars=\\\{\}]
\PY{n+nb}{print}\PY{p}{(}\PY{n}{charging\PYZus{}unit}\PY{p}{)}
\PY{n+nb}{print}\PY{p}{(}\PY{n}{car\PYZus{}green}\PY{p}{)}
\end{Verbatim}
\end{tcolorbox}

    \begin{Verbatim}[commandchars=\\\{\}]
Charging unit with
  charging levels: 0, 1, 2, 3
  time slots: 0, 1, 2, 3
  cars to charge: car\_green
Car 'car\_green':
  at charging station at time slots [0, 1, 2]
  requires 4 energy units
    \end{Verbatim}

    \begin{tcolorbox}[breakable, size=fbox, boxrule=1pt, pad at break*=1mm,colback=cellbackground, colframe=cellborder]
\prompt{In}{incolor}{3}{\boxspacing}
\begin{Verbatim}[commandchars=\\\{\}]
\PY{n}{b\PYZus{}min} \PY{o}{=} \PY{n}{qubo\PYZus{}minimization\PYZus{}result}\PY{o}{.}\PY{n}{x}
\PY{n}{f\PYZus{}3\PYZus{}min} \PY{o}{=} \PY{n}{qubo\PYZus{}minimization\PYZus{}result}\PY{o}{.}\PY{n}{fval}
\PY{n+nb}{print}\PY{p}{(}\PY{l+s+s2}{\PYZdq{}}\PY{l+s+s2}{minimum point (binary): b\PYZus{}min = }\PY{l+s+s2}{\PYZdq{}}\PY{p}{,} \PY{n}{b\PYZus{}min}\PY{p}{)}
\PY{n+nb}{print}\PY{p}{(}\PY{l+s+s2}{\PYZdq{}}\PY{l+s+s2}{minimum value: f\PYZus{}3(b\PYZus{}min) = }\PY{l+s+s2}{\PYZdq{}}\PY{p}{,} \PY{n}{f\PYZus{}3\PYZus{}min}\PY{p}{)}
\end{Verbatim}
\end{tcolorbox}

    \begin{Verbatim}[commandchars=\\\{\}]
minimum point (binary): b\_min =  [0. 1. 1. 0. 1. 0. 0. 0.]
minimum value: f\_3(b\_min) =  5.999999999999993
    \end{Verbatim}

    \begin{tcolorbox}[breakable, size=fbox, boxrule=1pt, pad at break*=1mm,colback=cellbackground, colframe=cellborder]
\prompt{In}{incolor}{4}{\boxspacing}
\begin{Verbatim}[commandchars=\\\{\}]
\PY{n}{p\PYZus{}min} \PY{o}{=} \PY{n}{converter}\PY{o}{.}\PY{n}{interpret}\PY{p}{(}\PY{n}{b\PYZus{}min}\PY{p}{)}
\PY{n}{f\PYZus{}1\PYZus{}min} \PY{o}{=} \PY{n}{qcio}\PY{o}{.}\PY{n}{objective}\PY{o}{.}\PY{n}{evaluate}\PY{p}{(}\PY{n}{p\PYZus{}min}\PY{p}{)}
\PY{n}{p\PYZus{}min\PYZus{}feasible} \PY{o}{=} \PY{n}{qcio}\PY{o}{.}\PY{n}{is\PYZus{}feasible}\PY{p}{(}\PY{n}{p\PYZus{}min}\PY{p}{)}

\PY{n+nb}{print}\PY{p}{(}\PY{l+s+s2}{\PYZdq{}}\PY{l+s+s2}{minimum point (integer): p\PYZus{}min = }\PY{l+s+s2}{\PYZdq{}}\PY{p}{,} \PY{n}{p\PYZus{}min}\PY{p}{)}
\PY{n+nb}{print}\PY{p}{(}\PY{l+s+s2}{\PYZdq{}}\PY{l+s+s2}{minimum value: f\PYZus{}1(p\PYZus{}min) = }\PY{l+s+s2}{\PYZdq{}}\PY{p}{,} \PY{n}{f\PYZus{}1\PYZus{}min}\PY{p}{)}
\PY{n+nb}{print}\PY{p}{(}\PY{l+s+s2}{\PYZdq{}}\PY{l+s+s2}{minimum point feasible = }\PY{l+s+s2}{\PYZdq{}}\PY{p}{,} \PY{n}{p\PYZus{}min\PYZus{}feasible}\PY{p}{)}
\end{Verbatim}
\end{tcolorbox}

    \begin{Verbatim}[commandchars=\\\{\}]
minimum point (integer): p\_min =  [2. 1. 1. 0.]
minimum value: f\_1(p\_min) =  6.0
minimum point feasible =  True
    \end{Verbatim}

    \begin{tcolorbox}[breakable, size=fbox, boxrule=1pt, pad at break*=1mm,colback=cellbackground, colframe=cellborder]
\prompt{In}{incolor}{5}{\boxspacing}
\begin{Verbatim}[commandchars=\\\{\}]
\PY{k+kn}{from} \PY{n+nn}{utils} \PY{k+kn}{import} \PY{n}{plot\PYZus{}charging\PYZus{}schedule}
\PY{n}{fig} \PY{o}{=} \PY{n}{plot\PYZus{}charging\PYZus{}schedule}\PY{p}{(}
    \PY{n}{charging\PYZus{}unit}\PY{p}{,} \PY{n}{p\PYZus{}min}\PY{p}{,} \PY{n}{marker\PYZus{}size}\PY{o}{=}\PY{l+m+mi}{30}\PY{p}{)}
\PY{n}{fig}\PY{o}{.}\PY{n}{update\PYZus{}layout}\PY{p}{(}\PY{n}{width}\PY{o}{=}\PY{l+m+mi}{350}\PY{p}{,} \PY{n}{height}\PY{o}{=}\PY{l+m+mi}{300}\PY{p}{)}
\PY{n}{fig}\PY{o}{.}\PY{n}{show}\PY{p}{(}\PY{p}{)}
\end{Verbatim}
\end{tcolorbox}

    \begin{tcolorbox}[breakable, size=fbox, boxrule=1pt, pad at break*=1mm,colback=cellbackground, colframe=cellborder]
\prompt{In}{incolor}{6}{\boxspacing}
\begin{Verbatim}[commandchars=\\\{\}]
\PY{n+nb}{print}\PY{p}{(}\PY{n}{ising}\PY{p}{)}
\end{Verbatim}
\end{tcolorbox}

    \begin{Verbatim}[commandchars=\\\{\}]
-3.3 * IIIIIIIZ
- 6.599999999999998 * IIIIIIZI
- 3.3000000000000007 * IIIIIZII
- 6.6 * IIIIZIII
- 3.2999999999999994 * IIIZIIII
- 6.599999999999999 * IIZIIIII
+ 4.6 * IIIIIIZZ
+ 1.8 * IIIIIZIZ
+ 3.6 * IIIIIZZI
+ 3.6 * IIIIZIIZ
+ 7.2 * IIIIZIZI
+ 4.6 * IIIIZZII
+ 1.8 * IIIZIIIZ
+ 3.6 * IIIZIIZI
+ 1.8 * IIIZIZII
+ 3.6 * IIIZZIII
+ 3.6 * IIZIIIIZ
+ 7.2 * IIZIIIZI
+ 3.6 * IIZIIZII
+ 7.2 * IIZIZIII
+ 4.6 * IIZZIIII
- 1.5 * IZIIIIII
+ 1.0 * ZZIIIIII
- 3.0 * ZIIIIIII
    \end{Verbatim}

    \hypertarget{access-to-ibmq}{%
\section{Access to IBMQ}\label{sec:access-to-ibmq}}

    We start this notebook with the access to IBMQ systems. Note that this
can (and probably will) change in the future and also depends on your
membership to an IBMQ Hub. We therefore cannot guarantee that the
following code cells work for everybody. A starting point what things
need to be adapted can be found
\href{https://quantum-computing.ibm.com/composer/docs/iqx/manage/account/ibmq}{here}
or ask your local IBMQ Hub admin.

    In order to access IBMQ systems one needs a personal token. For a more
comfortable access we can save this token in a binary \texttt{pickle}
file with the provided function \texttt{save\_token}. After having saved
your token you can load it with \texttt{load\_token}.

    If you \textbf{don't have access to real backends} you can skip the
following four cells and use a \textbf{fake backend} as described next.

\begin{center}\rule{0.9\linewidth}{0.5pt}\end{center}

    \begin{tcolorbox}[breakable, size=fbox, boxrule=1pt, pad at break*=1mm,colback=cellbackground, colframe=cellborder]
\prompt{In}{incolor}{ }{\boxspacing}
\begin{Verbatim}[commandchars=\\\{\}]
\PY{k+kn}{from} \PY{n+nn}{utils} \PY{k+kn}{import} \PY{n}{save\PYZus{}token}\PY{p}{,} \PY{n}{load\PYZus{}token}

\PY{c+c1}{\PYZsh{} Note: For different URLs you might need different tokens.}
\PY{c+c1}{\PYZsh{} Note: You only have to do this once per token.}
\PY{c+c1}{\PYZsh{} save\PYZus{}token(}
\PY{c+c1}{\PYZsh{}     token=\PYZdq{}\PYZus{}your\PYZus{}token\PYZus{}\PYZdq{},}
\PY{c+c1}{\PYZsh{}     file\PYZus{}name=\PYZdq{}\PYZus{}filename\PYZus{}for\PYZus{}your\PYZus{}saved\PYZus{}token\PYZus{}\PYZdq{})}

\PY{c+c1}{\PYZsh{} Use the following line if you have saved your token:}
\PY{c+c1}{\PYZsh{} token = load\PYZus{}token(\PYZdq{}\PYZus{}filename\PYZus{}for\PYZus{}your\PYZus{}saved\PYZus{}token\PYZus{}\PYZdq{}) }
\PY{c+c1}{\PYZsh{} Otherwise, put your token in the following line:}
\PY{n}{token} \PY{o}{=} \PY{l+s+s2}{\PYZdq{}}\PY{l+s+s2}{\PYZus{}your\PYZus{}token\PYZus{}}\PY{l+s+s2}{\PYZdq{}} 
\end{Verbatim}
\end{tcolorbox}

    \begin{tcolorbox}[breakable, size=fbox, boxrule=1pt, pad at break*=1mm,colback=cellbackground, colframe=cellborder]
\prompt{In}{incolor}{ }{\boxspacing}
\begin{Verbatim}[commandchars=\\\{\}]
\PY{k+kn}{from} \PY{n+nn}{qiskit} \PY{k+kn}{import} \PY{n}{IBMQ}
\PY{k+kn}{from} \PY{n+nn}{qiskit}\PY{n+nn}{.}\PY{n+nn}{providers}\PY{n+nn}{.}\PY{n+nn}{ibmq}\PY{n+nn}{.}\PY{n+nn}{exceptions} \PY{k+kn}{import} \PY{n}{IBMQAccountError}\PY{p}{,} \PYZbs{}
    \PY{n}{IBMQProviderError}
\PY{k+kn}{from} \PY{n+nn}{qiskit}\PY{n+nn}{.}\PY{n+nn}{providers}\PY{n+nn}{.}\PY{n+nn}{ibmq}\PY{n+nn}{.}\PY{n+nn}{api}\PY{n+nn}{.}\PY{n+nn}{exceptions} \PY{k+kn}{import} \PY{n}{RequestsApiError}

\PY{c+c1}{\PYZsh{} The following URL belongs to the Fraunhofer\PYZhy{}DE hub:}
\PY{n}{api\PYZus{}url} \PY{o}{=} \PY{l+s+s2}{\PYZdq{}}\PY{l+s+s2}{https://auth.de.quantum\PYZhy{}computing.ibm.com/api}\PY{l+s+s2}{\PYZdq{}}
\PY{c+c1}{\PYZsh{} The following URL belongs to the US hub:}
\PY{c+c1}{\PYZsh{} api\PYZus{}url = \PYZdq{}https://auth.quantum\PYZhy{}computing.ibm.com/api\PYZdq{}}

\PY{k}{try}\PY{p}{:}
    \PY{n}{IBMQ}\PY{o}{.}\PY{n}{enable\PYZus{}account}\PY{p}{(}\PY{n}{token}\PY{p}{,} \PY{n}{api\PYZus{}url}\PY{p}{)}
\PY{k}{except} \PY{n}{IBMQAccountError} \PY{k}{as} \PY{n}{e}\PY{p}{:}
    \PY{k}{if} \PY{o+ow}{not} \PY{p}{(}\PY{n}{e}\PY{o}{.}\PY{n}{args}\PY{p}{[}\PY{l+m+mi}{0}\PY{p}{]} \PY{o}{==} \PY{l+s+s2}{\PYZdq{}}\PY{l+s+s2}{An IBM Quantum Experience account }\PY{l+s+s2}{\PYZdq{}}
                         \PY{l+s+s2}{\PYZdq{}}\PY{l+s+s2}{is already in use for the session.}\PY{l+s+s2}{\PYZdq{}}\PY{p}{)}\PY{p}{:}
        \PY{k}{raise} \PY{n}{IBMQAccountError}
\PY{k}{except} \PY{n}{RequestsApiError} \PY{k}{as} \PY{n}{e}\PY{p}{:}
    \PY{n+nb}{print}\PY{p}{(}\PY{l+s+sa}{f}\PY{l+s+s2}{\PYZdq{}}\PY{l+s+s2}{Error: }\PY{l+s+si}{\PYZob{}}\PY{n}{e}\PY{o}{.}\PY{n}{message}\PY{l+s+si}{\PYZcb{}}\PY{l+s+se}{\PYZbs{}n}\PY{l+s+s2}{\PYZdq{}}
            \PY{l+s+s2}{\PYZdq{}}\PY{l+s+s2}{\PYZhy{}\PYZgt{} Check if the URL and your token are correct.}\PY{l+s+s2}{\PYZdq{}}\PY{p}{)}
\end{Verbatim}
\end{tcolorbox}

    \begin{tcolorbox}[breakable, size=fbox, boxrule=1pt, pad at break*=1mm,colback=cellbackground, colframe=cellborder]
\prompt{In}{incolor}{ }{\boxspacing}
\begin{Verbatim}[commandchars=\\\{\}]
\PY{k+kn}{from} \PY{n+nn}{qiskit}\PY{n+nn}{.}\PY{n+nn}{providers}\PY{n+nn}{.}\PY{n+nn}{ibmq}\PY{n+nn}{.}\PY{n+nn}{exceptions} \PY{k+kn}{import} \PY{n}{IBMQProviderError}

\PY{k}{try}\PY{p}{:}
\PY{c+c1}{\PYZsh{} If you belong to the Fraunhofer Hub your parameters should}
\PY{c+c1}{\PYZsh{} look likes this EXCEPT you need to adapat project=\PYZsq{}...\PYZsq{}:}
    \PY{n}{provider} \PY{o}{=} \PY{n}{IBMQ}\PY{o}{.}\PY{n}{get\PYZus{}provider}\PY{p}{(}
        \PY{n}{hub}\PY{o}{=}\PY{l+s+s1}{\PYZsq{}}\PY{l+s+s1}{fraunhofer\PYZhy{}de}\PY{l+s+s1}{\PYZsq{}}\PY{p}{,} 
        \PY{n}{group}\PY{o}{=}\PY{l+s+s1}{\PYZsq{}}\PY{l+s+s1}{fhg\PYZhy{}all}\PY{l+s+s1}{\PYZsq{}}\PY{p}{,}
        \PY{n}{project}\PY{o}{=}\PY{l+s+s1}{\PYZsq{}}\PY{l+s+s1}{fiao01}\PY{l+s+s1}{\PYZsq{}}\PY{p}{)}
\PY{c+c1}{\PYZsh{} If you use the open access quantum systems your parameters}
\PY{c+c1}{\PYZsh{} should look likes this:}
\PY{c+c1}{\PYZsh{}     provider = IBMQ.get\PYZus{}provider(}
\PY{c+c1}{\PYZsh{}         group=\PYZsq{}open\PYZsq{}, }
\PY{c+c1}{\PYZsh{}         project=\PYZsq{}main\PYZsq{})}
\PY{k}{except} \PY{n}{IBMQProviderError} \PY{k}{as} \PY{n}{e}\PY{p}{:}
    \PY{n+nb}{print}\PY{p}{(}\PY{l+s+sa}{f}\PY{l+s+s2}{\PYZdq{}}\PY{l+s+s2}{Error: }\PY{l+s+si}{\PYZob{}}\PY{n}{e}\PY{o}{.}\PY{n}{message}\PY{l+s+si}{\PYZcb{}}\PY{l+s+se}{\PYZbs{}n}\PY{l+s+s2}{\PYZdq{}}
          \PY{l+s+s2}{\PYZdq{}}\PY{l+s+s2}{\PYZhy{}\PYZgt{} Check your parameters in IBMQ.get\PYZus{}provider}\PY{l+s+se}{\PYZbs{}n}\PY{l+s+s2}{\PYZdq{}}
          \PY{l+s+s2}{\PYZdq{}}\PY{l+s+s2}{\PYZhy{}\PYZgt{} If you don}\PY{l+s+s2}{\PYZsq{}}\PY{l+s+s2}{t have access to IBMQ quantum systems }\PY{l+s+s2}{\PYZdq{}}
          \PY{l+s+s2}{\PYZdq{}}\PY{l+s+s2}{use the fake backend below.}\PY{l+s+s2}{\PYZdq{}}\PY{p}{)}
\end{Verbatim}
\end{tcolorbox}

    \begin{tcolorbox}[breakable, size=fbox, boxrule=1pt, pad at break*=1mm,colback=cellbackground, colframe=cellborder]
\prompt{In}{incolor}{ }{\boxspacing}
\begin{Verbatim}[commandchars=\\\{\}]
\PY{k+kn}{from} \PY{n+nn}{qiskit}\PY{n+nn}{.}\PY{n+nn}{providers}\PY{n+nn}{.}\PY{n+nn}{exceptions} \PY{k+kn}{import} \PY{n}{QiskitBackendNotFoundError}

\PY{c+c1}{\PYZsh{} Depending on your access to real backends choose an appropriate backend here.}
\PY{k}{try}\PY{p}{:}
    \PY{n}{real\PYZus{}backend} \PY{o}{=} \PY{n}{provider}\PY{o}{.}\PY{n}{get\PYZus{}backend}\PY{p}{(}\PY{l+s+s2}{\PYZdq{}}\PY{l+s+s2}{ibmq\PYZus{}ehningen}\PY{l+s+s2}{\PYZdq{}}\PY{p}{)}
\PY{k}{except} \PY{n+ne}{NameError}\PY{p}{:}
    \PY{n+nb}{print}\PY{p}{(}\PY{l+s+s2}{\PYZdq{}}\PY{l+s+s2}{Error: You first have to instantiate a provider object }\PY{l+s+s2}{\PYZdq{}}
          \PY{l+s+s2}{\PYZdq{}}\PY{l+s+s2}{with the code in the cell above.}\PY{l+s+s2}{\PYZdq{}}\PY{p}{)}
\PY{k}{except} \PY{n}{QiskitBackendNotFoundError} \PY{k}{as} \PY{n}{e}\PY{p}{:}
    \PY{n}{list\PYZus{}backends} \PY{o}{=} \PY{p}{[}\PY{n}{backend}\PY{o}{.}\PY{n}{name}\PY{p}{(}\PY{p}{)} \PY{k}{for} \PY{n}{backend} \PY{o+ow}{in} \PY{n}{provider}\PY{o}{.}\PY{n}{backends}\PY{p}{(}\PY{p}{)}\PY{p}{]}
    \PY{n+nb}{print}\PY{p}{(}\PY{l+s+sa}{f}\PY{l+s+s2}{\PYZdq{}}\PY{l+s+s2}{Error: backend not found. You only have access to: }\PY{l+s+si}{\PYZob{}}\PY{n}{list\PYZus{}backends}\PY{l+s+si}{\PYZcb{}}\PY{l+s+s2}{\PYZdq{}}\PY{p}{)}
\end{Verbatim}
\end{tcolorbox}

\begin{center}\rule{0.9\linewidth}{0.5pt}\end{center}

    Fake backends are provided among others in
\href{https://qiskit.org/documentation/apidoc/providers_fake_provider.html}{\texttt{qiskit.providers.fake\_provider}}.

    \begin{tcolorbox}[breakable, size=fbox, boxrule=1pt, pad at break*=1mm,colback=cellbackground, colframe=cellborder]
\prompt{In}{incolor}{7}{\boxspacing}
\begin{Verbatim}[commandchars=\\\{\}]
\PY{k+kn}{from} \PY{n+nn}{qiskit}\PY{n+nn}{.}\PY{n+nn}{providers}\PY{n+nn}{.}\PY{n+nn}{fake\PYZus{}provider} \PY{k+kn}{import} \PY{n}{FakeKolkata}
\PY{n}{fake\PYZus{}kolkata} \PY{o}{=} \PY{n}{FakeKolkata}\PY{p}{(}\PY{p}{)}
\end{Verbatim}
\end{tcolorbox}

    In the next cell we decide which backend we want to use for the remaining notebook.

    \begin{tcolorbox}[breakable, size=fbox, boxrule=1pt, pad at break*=1mm,colback=cellbackground, colframe=cellborder]
\prompt{In}{incolor}{8}{\boxspacing}
\begin{Verbatim}[commandchars=\\\{\}]
\PY{c+c1}{\PYZsh{} backend = real\PYZus{}backend}
\PY{n}{backend} \PY{o}{=} \PY{n}{fake\PYZus{}kolkata}

\PY{n}{number\PYZus{}shots} \PY{o}{=} \PY{l+m+mi}{8000}
\PY{n}{backend}\PY{o}{.}\PY{n}{set\PYZus{}options}\PY{p}{(}\PY{n}{shots}\PY{o}{=}\PY{n}{number\PYZus{}shots}\PY{p}{)}
\end{Verbatim}
\end{tcolorbox}

    \hypertarget{transpilation-pipeline}{%
\section{Transpilation Pipeline}\label{sec:transpilation-pipeline}}

    It is important to know that \textbf{current quantum computers} (still)
suffer from
\begin{itemize}
    \tightlist
    \item a limited set of gates that they can execute (see basis
    gates below),
    \item noisy qubits and erroneous gates (see x, sx and cx errors
    below),
    \item limited connectivity between individual qubits (see coupling
    map below),
    \item measurement errors (see error map below), and
    \item many other shortcomings.
\end{itemize}

Among others the task of a \textbf{transpilation} is to address these
issues. Before we build a transpilation pipeline let us shortly visualize
these problems.

    \hypertarget{analysis-of-quantum-backend}{%
\subsection{Analysis of Quantum
Backend}\label{sec:analysis-of-quantum-backend}}

    In the following code cells we access and visualize different properties
of our (fake) quantum backend. If you don't have access to a real
quantum computer you can look at the plots for the system ibmq\_ehningen
in Appendix \ref{app:properties-ehningen}.

    \begin{tcolorbox}[breakable, size=fbox, boxrule=1pt, pad at break*=1mm,colback=cellbackground, colframe=cellborder]
\prompt{In}{incolor}{9}{\boxspacing}
\begin{Verbatim}[commandchars=\\\{\}]
\PY{n+nb}{print}\PY{p}{(}\PY{l+s+sa}{f}\PY{l+s+s2}{\PYZdq{}}\PY{l+s+s2}{Basis gates of }\PY{l+s+si}{\PYZob{}}\PY{n}{backend}\PY{o}{.}\PY{n}{name}\PY{p}{(}\PY{p}{)}\PY{l+s+si}{\PYZcb{}}\PY{l+s+s2}{: }\PY{l+s+s2}{\PYZdq{}}
      \PY{l+s+sa}{f}\PY{l+s+s2}{\PYZdq{}}\PY{l+s+si}{\PYZob{}}\PY{n}{backend}\PY{o}{.}\PY{n}{configuration}\PY{p}{(}\PY{p}{)}\PY{o}{.}\PY{n}{basis\PYZus{}gates}\PY{l+s+si}{\PYZcb{}}\PY{l+s+s2}{\PYZdq{}}\PY{p}{)}
\end{Verbatim}
\end{tcolorbox}

    \begin{Verbatim}[commandchars=\\\{\}]
Basis gates of fake\_kolkata: ['id', 'rz', 'sx', 'x', 'cx', 'reset']
    \end{Verbatim}

    \begin{tcolorbox}[breakable, size=fbox, boxrule=1pt, pad at break*=1mm,colback=cellbackground, colframe=cellborder]
\prompt{In}{incolor}{10}{\boxspacing}
\begin{Verbatim}[commandchars=\\\{\}]
\PY{k+kn}{from} \PY{n+nn}{qiskit}\PY{n+nn}{.}\PY{n+nn}{visualization} \PY{k+kn}{import} \PY{n}{plot\PYZus{}gate\PYZus{}map}

\PY{n+nb}{print}\PY{p}{(}\PY{l+s+s2}{\PYZdq{}}\PY{l+s+s2}{Coupling map of}\PY{l+s+s2}{\PYZdq{}}\PY{p}{,} \PY{n}{backend}\PY{o}{.}\PY{n}{name}\PY{p}{(}\PY{p}{)}\PY{p}{)}
\PY{n}{plot\PYZus{}gate\PYZus{}map}\PY{p}{(}\PY{n}{backend}\PY{p}{)}
\end{Verbatim}
\end{tcolorbox}

    \begin{Verbatim}[commandchars=\\\{\}]
Coupling map of fake\_kolkata
    \end{Verbatim}

\prompt{Out}{outcolor}{10}{}
    
    \begin{center}
    \adjustimage{max size={0.9\linewidth}{0.9\paperheight}}{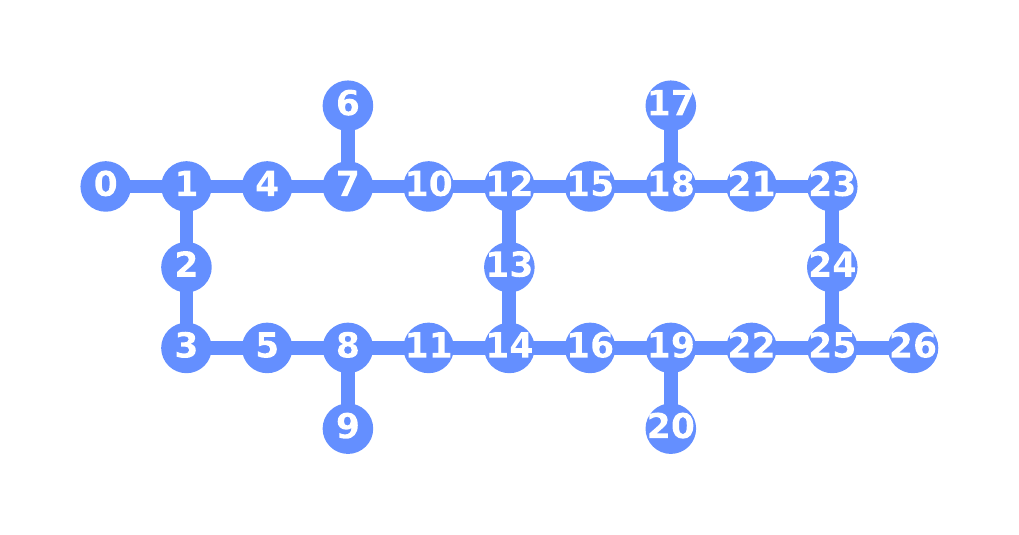}
    \end{center}
    { \hspace*{\fill} \\}

    \begin{tcolorbox}[breakable, size=fbox, boxrule=1pt, pad at break*=1mm,colback=cellbackground, colframe=cellborder]
\prompt{In}{incolor}{11}{\boxspacing}
\begin{Verbatim}[commandchars=\\\{\}]
\PY{n}{backend}\PY{o}{.}\PY{n}{configuration}\PY{p}{(}\PY{p}{)}\PY{o}{.}\PY{n}{coupling\PYZus{}map}
\end{Verbatim}
\end{tcolorbox}

            \begin{tcolorbox}[breakable, size=fbox, boxrule=.5pt, pad at break*=1mm, opacityfill=0]
\prompt{Out}{outcolor}{11}{\boxspacing}
\begin{Verbatim}[commandchars=\\\{\}]
[[0, 1],
 [1, 0],
 [1, 2],
 [1, 4],
 [2, 1],
 [2, 3],
 [3, 2],
 [3, 5],
 [4, 1],
 ...
 [24, 25],
 [25, 22],
 [25, 24],
 [25, 26],
 [26, 25]]
\end{Verbatim}
\end{tcolorbox}
        
    \begin{tcolorbox}[breakable, size=fbox, boxrule=1pt, pad at break*=1mm,colback=cellbackground, colframe=cellborder]
\prompt{In}{incolor}{12}{\boxspacing}
\begin{Verbatim}[commandchars=\\\{\}]
\PY{k+kn}{from} \PY{n+nn}{qiskit}\PY{n+nn}{.}\PY{n+nn}{visualization} \PY{k+kn}{import} \PY{n}{plot\PYZus{}error\PYZus{}map}

\PY{n}{plot\PYZus{}error\PYZus{}map}\PY{p}{(}\PY{n}{backend}\PY{p}{,} \PY{n}{figsize}\PY{o}{=}\PY{p}{(}\PY{l+m+mi}{10}\PY{p}{,}\PY{l+m+mi}{6}\PY{p}{)}\PY{p}{)}
\end{Verbatim}
\end{tcolorbox}

\prompt{Out}{outcolor}{12}{}
    
    \begin{center}
    \adjustimage{max size={0.9\linewidth}{0.9\paperheight}}{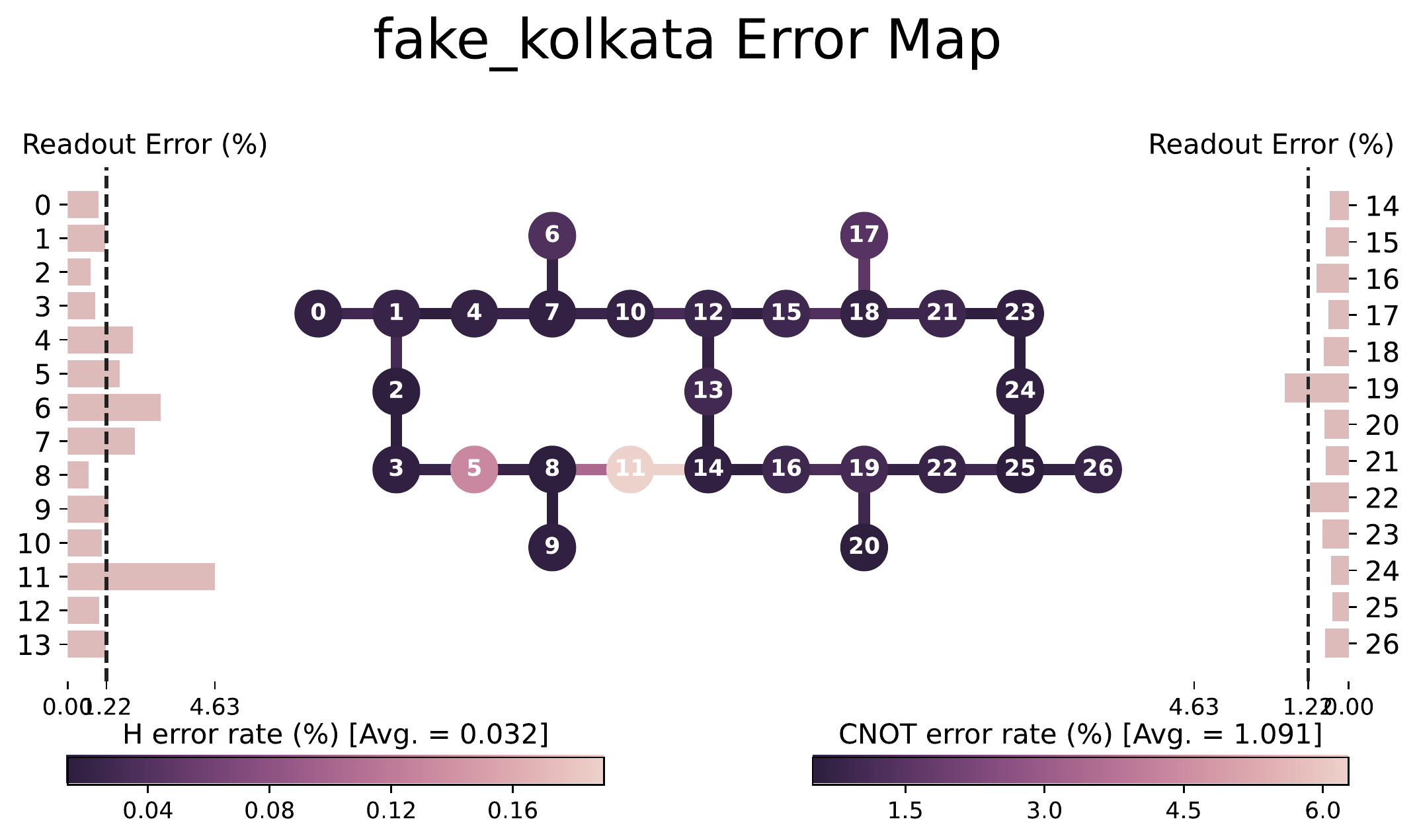}
    \end{center}
    { \hspace*{\fill} \\}

    \begin{tcolorbox}[breakable, size=fbox, boxrule=1pt, pad at break*=1mm,colback=cellbackground, colframe=cellborder]
\prompt{In}{incolor}{13}{\boxspacing}
\begin{Verbatim}[commandchars=\\\{\}]
\PY{n}{x\PYZus{}errors} \PY{o}{=} \PY{p}{[}\PY{p}{]}
\PY{n}{sx\PYZus{}errors} \PY{o}{=} \PY{p}{[}\PY{p}{]}
\PY{n}{cx\PYZus{}errors} \PY{o}{=} \PY{p}{[}\PY{p}{]}

\PY{n}{backend\PYZus{}properties} \PY{o}{=} \PY{n}{backend}\PY{o}{.}\PY{n}{properties}\PY{p}{(}\PY{p}{)}

\PY{k}{for} \PY{n}{qubit} \PY{o+ow}{in} \PY{n+nb}{range}\PY{p}{(}\PY{n}{backend}\PY{o}{.}\PY{n}{configuration}\PY{p}{(}\PY{p}{)}\PY{o}{.}\PY{n}{n\PYZus{}qubits}\PY{p}{)}\PY{p}{:}
    \PY{n}{x\PYZus{}errors}\PY{o}{.}\PY{n}{append}\PY{p}{(}\PY{n}{backend\PYZus{}properties}\PY{o}{.}\PY{n}{gate\PYZus{}error}\PY{p}{(}\PY{l+s+s2}{\PYZdq{}}\PY{l+s+s2}{x}\PY{l+s+s2}{\PYZdq{}}\PY{p}{,} \PY{n}{qubit}\PY{p}{)}\PY{p}{)}
    \PY{n}{sx\PYZus{}errors}\PY{o}{.}\PY{n}{append}\PY{p}{(}\PY{n}{backend\PYZus{}properties}\PY{o}{.}\PY{n}{gate\PYZus{}error}\PY{p}{(}\PY{l+s+s2}{\PYZdq{}}\PY{l+s+s2}{sx}\PY{l+s+s2}{\PYZdq{}}\PY{p}{,} \PY{n}{qubit}\PY{p}{)}\PY{p}{)}
    
\PY{k}{for} \PY{n}{qubit0}\PY{p}{,} \PY{n}{qubit1} \PY{o+ow}{in} \PY{n}{backend}\PY{o}{.}\PY{n}{configuration}\PY{p}{(}\PY{p}{)}\PY{o}{.}\PY{n}{coupling\PYZus{}map}\PY{p}{:}
    \PY{n}{cx\PYZus{}errors}\PY{o}{.}\PY{n}{append}\PY{p}{(}\PY{n}{backend\PYZus{}properties}\PY{o}{.}\PY{n}{gate\PYZus{}error}\PY{p}{(}\PY{l+s+s2}{\PYZdq{}}\PY{l+s+s2}{cx}\PY{l+s+s2}{\PYZdq{}}\PY{p}{,} \PY{p}{[}\PY{n}{qubit0}\PY{p}{,} \PY{n}{qubit1}\PY{p}{]}\PY{p}{)}\PY{p}{)}
\end{Verbatim}
\end{tcolorbox}

\begin{tcolorbox}[breakable, size=fbox, boxrule=1pt, pad at break*=1mm,colback=cellbackground, colframe=cellborder]
\prompt{In}{incolor}{14}{\boxspacing}
\begin{Verbatim}[commandchars=\\\{\}]
\PY{k+kn}{import} \PY{n+nn}{numpy} \PY{k}{as} \PY{n+nn}{np}
\PY{k+kn}{import} \PY{n+nn}{plotly}\PY{n+nn}{.}\PY{n+nn}{graph\PYZus{}objects} \PY{k}{as} \PY{n+nn}{go}

\PY{n}{fig} \PY{o}{=} \PY{n}{go}\PY{o}{.}\PY{n}{Figure}\PY{p}{(}\PY{p}{)}
\PY{n}{fig}\PY{o}{.}\PY{n}{add\PYZus{}trace}\PY{p}{(}\PY{n}{go}\PY{o}{.}\PY{n}{Scatter}\PY{p}{(}
    \PY{n}{x}\PY{o}{=}\PY{n}{np}\PY{o}{.}\PY{n}{arange}\PY{p}{(}\PY{n}{backend}\PY{o}{.}\PY{n}{configuration}\PY{p}{(}\PY{p}{)}\PY{o}{.}\PY{n}{n\PYZus{}qubits}\PY{p}{)}\PY{p}{,}
    \PY{n}{y}\PY{o}{=}\PY{n}{x\PYZus{}errors}\PY{p}{,}
    \PY{n}{mode}\PY{o}{=}\PY{l+s+s2}{\PYZdq{}}\PY{l+s+s2}{lines+markers}\PY{l+s+s2}{\PYZdq{}}\PY{p}{,}
    \PY{n}{name}\PY{o}{=}\PY{l+s+s2}{\PYZdq{}}\PY{l+s+s2}{x error}\PY{l+s+s2}{\PYZdq{}}\PY{p}{,}
    \PY{n}{marker\PYZus{}size}\PY{o}{=}\PY{l+m+mi}{10}
\PY{p}{)}\PY{p}{)}
\PY{n}{fig}\PY{o}{.}\PY{n}{add\PYZus{}trace}\PY{p}{(}\PY{n}{go}\PY{o}{.}\PY{n}{Scatter}\PY{p}{(}
    \PY{n}{x}\PY{o}{=}\PY{n}{np}\PY{o}{.}\PY{n}{arange}\PY{p}{(}\PY{n}{backend}\PY{o}{.}\PY{n}{configuration}\PY{p}{(}\PY{p}{)}\PY{o}{.}\PY{n}{n\PYZus{}qubits}\PY{p}{)}\PY{p}{,}
    \PY{n}{y}\PY{o}{=}\PY{n}{sx\PYZus{}errors}\PY{p}{,}
    \PY{n}{mode}\PY{o}{=}\PY{l+s+s2}{\PYZdq{}}\PY{l+s+s2}{lines+markers}\PY{l+s+s2}{\PYZdq{}}\PY{p}{,}
    \PY{n}{name}\PY{o}{=}\PY{l+s+s2}{\PYZdq{}}\PY{l+s+s2}{sx error}\PY{l+s+s2}{\PYZdq{}}\PY{p}{,}
    \PY{n}{marker\PYZus{}symbol}\PY{o}{=}\PY{l+s+s2}{\PYZdq{}}\PY{l+s+s2}{x}\PY{l+s+s2}{\PYZdq{}}
\PY{p}{)}\PY{p}{)}
\PY{n}{fig}\PY{o}{.}\PY{n}{update\PYZus{}xaxes}\PY{p}{(}
    \PY{n}{title}\PY{o}{=}\PY{l+s+s2}{\PYZdq{}}\PY{l+s+s2}{qubit}\PY{l+s+s2}{\PYZdq{}}\PY{p}{)}
\PY{n}{fig}\PY{o}{.}\PY{n}{update\PYZus{}yaxes}\PY{p}{(}
    \PY{n}{title}\PY{o}{=}\PY{l+s+s2}{\PYZdq{}}\PY{l+s+s2}{error}\PY{l+s+s2}{\PYZdq{}}\PY{p}{)}
\end{Verbatim}
\end{tcolorbox}

\includegraphics{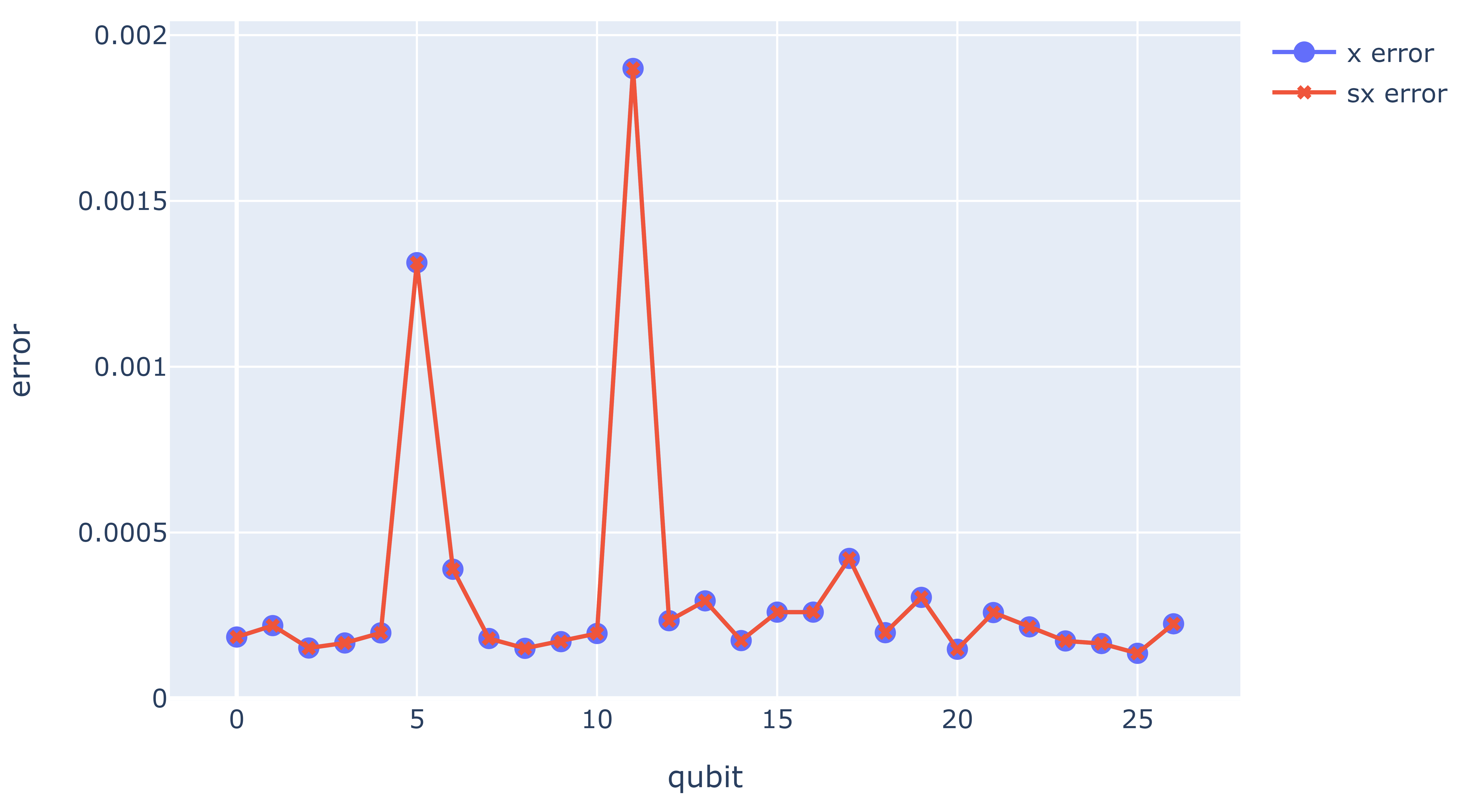}
    
    \begin{tcolorbox}[breakable, size=fbox, boxrule=1pt, pad at break*=1mm,colback=cellbackground, colframe=cellborder]
\prompt{In}{incolor}{15}{\boxspacing}
\begin{Verbatim}[commandchars=\\\{\}]
\PY{n}{fig} \PY{o}{=} \PY{n}{go}\PY{o}{.}\PY{n}{Figure}\PY{p}{(}\PY{p}{)}
\PY{n}{fig}\PY{o}{.}\PY{n}{add\PYZus{}trace}\PY{p}{(}\PY{n}{go}\PY{o}{.}\PY{n}{Scatter}\PY{p}{(}
    \PY{n}{x}\PY{o}{=}\PY{n+nb}{list}\PY{p}{(}\PY{n+nb}{map}\PY{p}{(}\PY{n+nb}{str}\PY{p}{,} \PY{n}{backend}\PY{o}{.}\PY{n}{configuration}\PY{p}{(}\PY{p}{)}\PY{o}{.}\PY{n}{coupling\PYZus{}map}\PY{p}{)}\PY{p}{)}\PY{p}{,}
    \PY{n}{y}\PY{o}{=}\PY{n}{cx\PYZus{}errors}\PY{p}{,}
    \PY{n}{mode}\PY{o}{=}\PY{l+s+s2}{\PYZdq{}}\PY{l+s+s2}{lines+markers}\PY{l+s+s2}{\PYZdq{}}\PY{p}{,}
    \PY{n}{name}\PY{o}{=}\PY{l+s+s2}{\PYZdq{}}\PY{l+s+s2}{CNOT error}\PY{l+s+s2}{\PYZdq{}}
\PY{p}{)}\PY{p}{)}
\PY{n}{fig}\PY{o}{.}\PY{n}{update\PYZus{}xaxes}\PY{p}{(}
    \PY{n}{title}\PY{o}{=}\PY{l+s+s2}{\PYZdq{}}\PY{l+s+s2}{[qubit0, qubit1]}\PY{l+s+s2}{\PYZdq{}}\PY{p}{)}
\PY{n}{fig}\PY{o}{.}\PY{n}{update\PYZus{}yaxes}\PY{p}{(}
    \PY{n}{title}\PY{o}{=}\PY{l+s+s2}{\PYZdq{}}\PY{l+s+s2}{CNOT error}\PY{l+s+s2}{\PYZdq{}}\PY{p}{)}
\end{Verbatim}
\end{tcolorbox}

\includegraphics{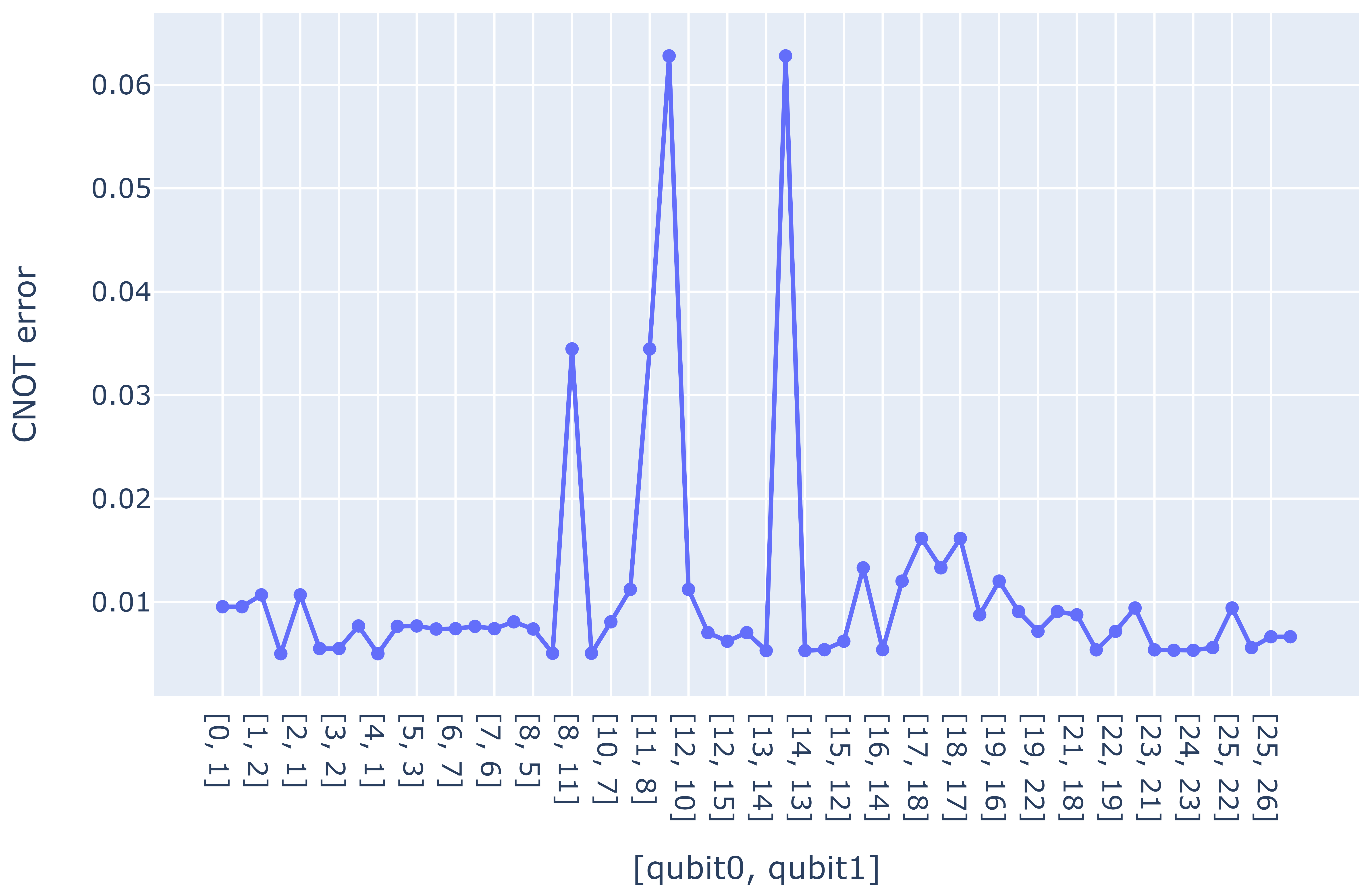}

    \hypertarget{standard-transpilation}{%
\subsection{Standard Transpilation}\label{sec:standard-transpilation}}

    In the following, we demonstrate how to build a transpilation pipeline based on the QAOA
example from Notebook \ref{chap:notebook-2}. Since the transpilation process in Qiskit is
changed and improved frequently we only briefly sketch how a good
transpilation pipeline looks like today (end of 2022). For future use \cite{QiskitTranspilerWeb,QuantumEnablementWeb} could provide a state of the art pipeline.

    Let's begin by drawing the QAOA circuit.

    \begin{tcolorbox}[breakable, size=fbox, boxrule=1pt, pad at break*=1mm,colback=cellbackground, colframe=cellborder]
\prompt{In}{incolor}{16}{\boxspacing}
\begin{Verbatim}[commandchars=\\\{\}]
\PY{n}{qaoa\PYZus{}circuit}\PY{o}{.}\PY{n}{decompose}\PY{p}{(}\PY{n}{reps}\PY{o}{=}\PY{l+m+mi}{3}\PY{p}{)}\PY{o}{.}\PY{n}{draw}\PY{p}{(}\PY{n}{scale}\PY{o}{=}\PY{l+m+mi}{1}\PY{p}{)}
\end{Verbatim}
\end{tcolorbox}

\prompt{Out}{outcolor}{16}{}
    
    \begin{center}
    \adjustimage{max size={0.9\linewidth}{0.9\paperheight}}{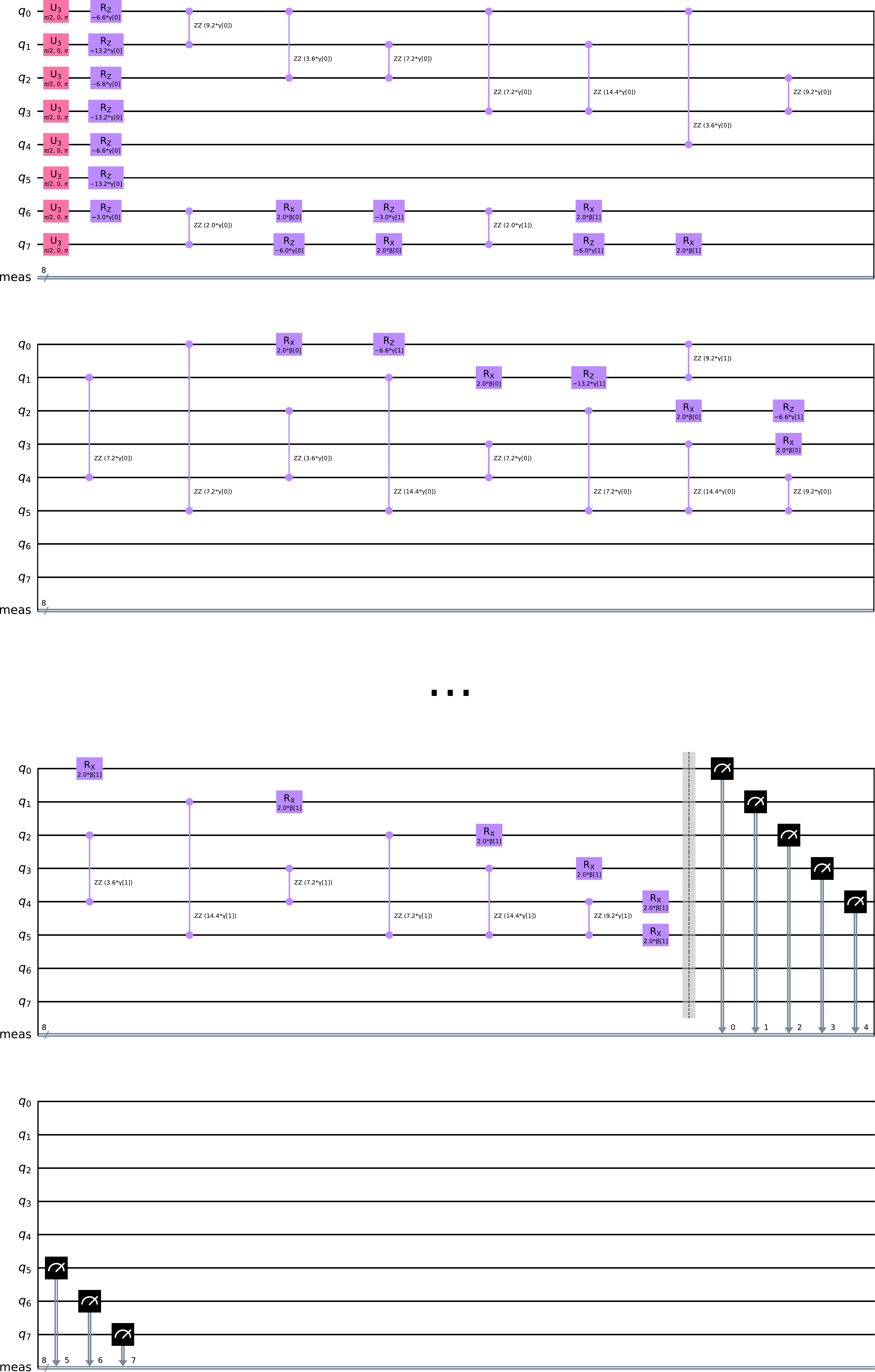}
    \end{center}
    { \hspace*{\fill} \\}

    Now, we want to use \textbf{Qiskit's transpiler} to transpile this
circuit. It is important to know that the transpiler has
\textbf{stochastic components} which can be varied to obtain a good
transpilation of a given quantum circuit. The stochastic components can
be controlled with a \textbf{seed}. If we provide the transpiler with a
list of seeds we will get a list of differently transpiled circuits.

    \begin{tcolorbox}[breakable, size=fbox, boxrule=1pt, pad at break*=1mm,colback=cellbackground, colframe=cellborder]
\prompt{In}{incolor}{17}{\boxspacing}
\begin{Verbatim}[commandchars=\\\{\}]
\PY{k+kn}{from} \PY{n+nn}{qiskit}\PY{n+nn}{.}\PY{n+nn}{compiler} \PY{k+kn}{import} \PY{n}{transpile}

\PY{n}{seeds\PYZus{}for\PYZus{}transpiler} \PY{o}{=} \PY{p}{[}\PY{n}{k} \PY{k}{for} \PY{n}{k} \PY{o+ow}{in} \PY{n+nb}{range}\PY{p}{(}\PY{l+m+mi}{40}\PY{p}{)}\PY{p}{]}
\PY{n}{number\PYZus{}seeds\PYZus{}for\PYZus{}transpiler} \PY{o}{=} \PY{n+nb}{len}\PY{p}{(}\PY{n}{seeds\PYZus{}for\PYZus{}transpiler}\PY{p}{)}

\PY{n}{qaoa\PYZus{}circuit\PYZus{}transpilations} \PY{o}{=} \PY{n}{transpile}\PY{p}{(}
    \PY{p}{[}\PY{n}{qaoa\PYZus{}circuit}\PY{p}{]}\PY{o}{*}\PY{n}{number\PYZus{}seeds\PYZus{}for\PYZus{}transpiler}\PY{p}{,}
    \PY{n}{backend}\PY{o}{=}\PY{n}{backend}\PY{p}{,}
    \PY{n}{optimization\PYZus{}level}\PY{o}{=}\PY{l+m+mi}{3}\PY{p}{,}
    \PY{n}{seed\PYZus{}transpiler}\PY{o}{=}\PY{n}{seeds\PYZus{}for\PYZus{}transpiler}\PY{p}{)}

\PY{n+nb}{print}\PY{p}{(}\PY{l+s+sa}{f}\PY{l+s+s2}{\PYZdq{}}\PY{l+s+s2}{We have }\PY{l+s+si}{\PYZob{}}\PY{n+nb}{len}\PY{p}{(}\PY{n}{qaoa\PYZus{}circuit\PYZus{}transpilations}\PY{p}{)}\PY{l+s+si}{\PYZcb{}}\PY{l+s+s2}{ }\PY{l+s+s2}{\PYZdq{}}
      \PY{l+s+s2}{\PYZdq{}}\PY{l+s+s2}{different transpilations of the QAOA circuit.}\PY{l+s+s2}{\PYZdq{}}\PY{p}{)}
\end{Verbatim}
\end{tcolorbox}

    \begin{Verbatim}[commandchars=\\\{\}]
We have 40 different transpilations of the QAOA circuit.
    \end{Verbatim}

    \begin{tcolorbox}[breakable, size=fbox, boxrule=1pt, pad at break*=1mm,colback=cellbackground, colframe=cellborder]
\prompt{In}{incolor}{18}{\boxspacing}
\begin{Verbatim}[commandchars=\\\{\}]
\PY{c+c1}{\PYZsh{} Put here a number between 0 and number\PYZus{}seeds\PYZus{}for\PYZus{}transpiler}
\PY{n}{number} \PY{o}{=} \PY{l+m+mi}{6} 

\PY{n}{qaoa\PYZus{}circuit\PYZus{}transpilations}\PY{p}{[}\PY{n}{number}\PY{p}{]}\PY{o}{.}\PY{n}{draw}\PY{p}{(}\PY{p}{)}
\end{Verbatim}
\end{tcolorbox}

\prompt{Out}{outcolor}{18}{}
    
    \begin{center}
    \adjustimage{max size={0.9\linewidth}{0.9\paperheight}}{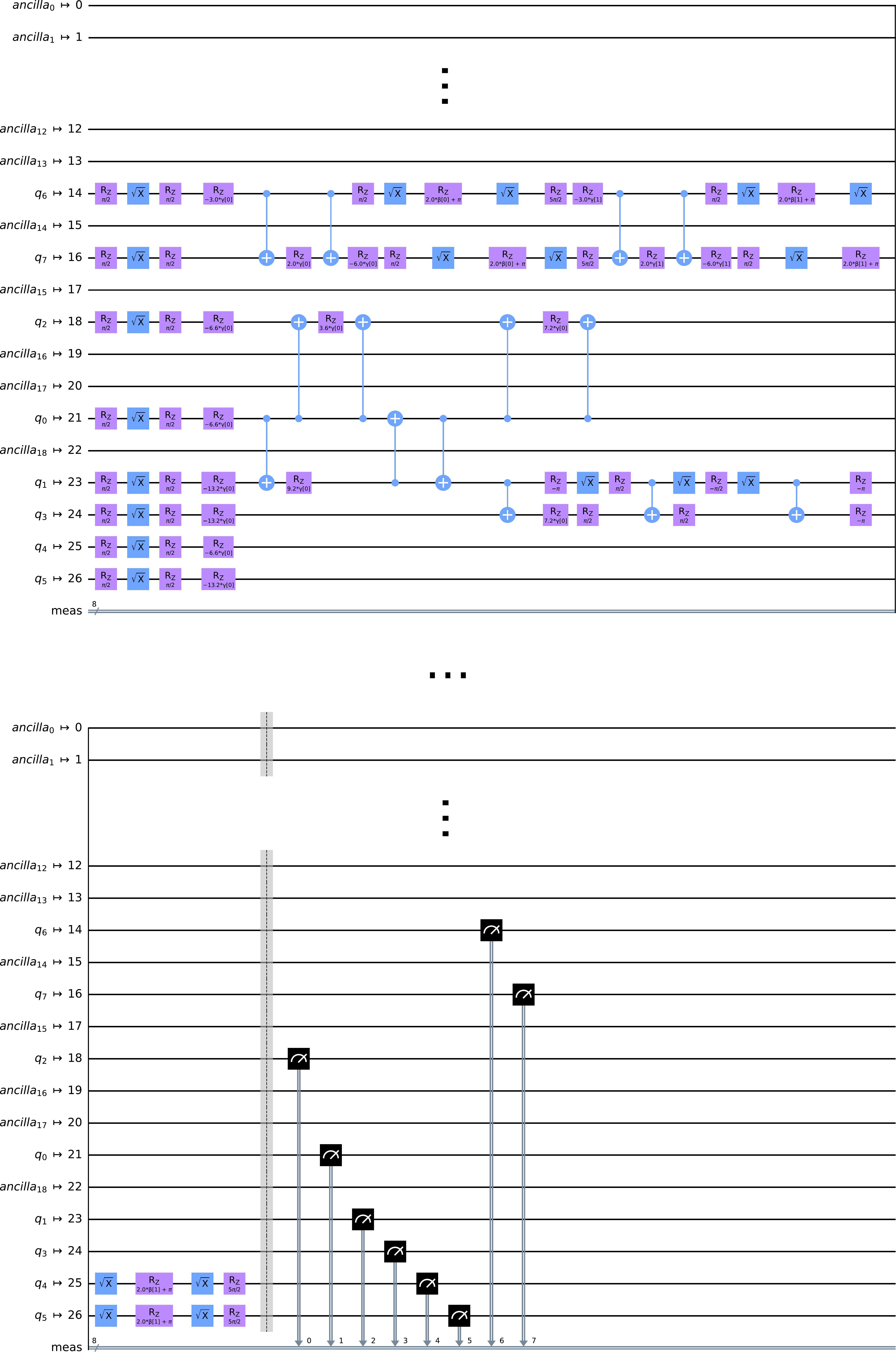}
    \end{center}
    { \hspace*{\fill} \\}

    Clearly, now the question arises which circuit is the best. There are many
metrics one can consider in order to answer this question. Probably the
most simple one is to count the \(\CNOT\) gates and take the circuit with
the smallest count. The background for this metric is that the \(\CNOT\)
gate is the gate with highest error rate (see above).

    \begin{tcolorbox}[breakable, size=fbox, boxrule=1pt, pad at break*=1mm,colback=cellbackground, colframe=cellborder]
\prompt{In}{incolor}{19}{\boxspacing}
\begin{Verbatim}[commandchars=\\\{\}]
\PY{n}{number\PYZus{}cnots} \PY{o}{=} \PY{n}{np}\PY{o}{.}\PY{n}{array}\PY{p}{(}\PY{p}{[}\PY{n}{circuit}\PY{o}{.}\PY{n}{count\PYZus{}ops}\PY{p}{(}\PY{p}{)}\PY{p}{[}\PY{l+s+s2}{\PYZdq{}}\PY{l+s+s2}{cx}\PY{l+s+s2}{\PYZdq{}}\PY{p}{]}
                         \PY{k}{for} \PY{n}{circuit} \PY{o+ow}{in} \PY{n}{qaoa\PYZus{}circuit\PYZus{}transpilations}\PY{p}{]}\PY{p}{)}
\PY{n}{index\PYZus{}circuit\PYZus{}with\PYZus{}least\PYZus{}cnots} \PY{o}{=} \PY{n}{np}\PY{o}{.}\PY{n}{argmin}\PY{p}{(}\PY{n}{number\PYZus{}cnots}\PY{p}{)}
\PY{n+nb}{print}\PY{p}{(}\PY{l+s+sa}{f}\PY{l+s+s2}{\PYZdq{}}\PY{l+s+s2}{The circuit with index }\PY{l+s+si}{\PYZob{}}\PY{n}{index\PYZus{}circuit\PYZus{}with\PYZus{}least\PYZus{}cnots}\PY{l+s+si}{\PYZcb{}}\PY{l+s+s2}{ }\PY{l+s+s2}{\PYZdq{}}
      \PY{l+s+s2}{\PYZdq{}}\PY{l+s+s2}{has the least CNOTs.}\PY{l+s+s2}{\PYZdq{}}\PY{p}{)}
\end{Verbatim}
\end{tcolorbox}

    \begin{Verbatim}[commandchars=\\\{\}]
The circuit with index 12 has the least CNOTs.
    \end{Verbatim}

    \begin{tcolorbox}[breakable, size=fbox, boxrule=1pt, pad at break*=1mm,colback=cellbackground, colframe=cellborder]
\prompt{In}{incolor}{20}{\boxspacing}
\begin{Verbatim}[commandchars=\\\{\}]
\PY{n}{qaoa\PYZus{}circuit\PYZus{}best} \PY{o}{=} \PYZbs{}
    \PY{n}{qaoa\PYZus{}circuit\PYZus{}transpilations}\PY{p}{[}\PY{n}{index\PYZus{}circuit\PYZus{}with\PYZus{}least\PYZus{}cnots}\PY{p}{]}
\end{Verbatim}
\end{tcolorbox}

    Now that we have our best transpiled circuit we can apply some
techniques to mitigate errors.

    \hypertarget{lowering-decoherence-and-dephasing-via-dynamical-decoupling}{%
\subsection{Lowering Decoherence and Dephasing via Dynamical
Decoupling}\label{sec:dynamical-decoupling}}

    Dynamical decoupling is a technique to lower decoherence and dephasing.
In its simplest form we apply a $\X$-$\X$-gate sequence. For further details see \cite{QuantumEnablementWeb2}.

    \begin{tcolorbox}[breakable, size=fbox, boxrule=1pt, pad at break*=1mm,colback=cellbackground, colframe=cellborder]
\prompt{In}{incolor}{21}{\boxspacing}
\begin{Verbatim}[commandchars=\\\{\}]
\PY{k+kn}{from} \PY{n+nn}{qiskit}\PY{n+nn}{.}\PY{n+nn}{circuit}\PY{n+nn}{.}\PY{n+nn}{library} \PY{k+kn}{import} \PY{n}{XGate}
\PY{k+kn}{from} \PY{n+nn}{qiskit}\PY{n+nn}{.}\PY{n+nn}{transpiler} \PY{k+kn}{import} \PY{n}{PassManager}\PY{p}{,} \PY{n}{InstructionDurations}
\PY{k+kn}{from} \PY{n+nn}{qiskit}\PY{n+nn}{.}\PY{n+nn}{transpiler}\PY{n+nn}{.}\PY{n+nn}{passes} \PY{k+kn}{import} \PY{n}{PadDynamicalDecoupling}\PY{p}{,} \PY{n}{ALAPScheduleAnalysis}

\PY{n}{instruction\PYZus{}durations} \PY{o}{=} \PY{n}{InstructionDurations}\PY{o}{.}\PY{n}{from\PYZus{}backend}\PY{p}{(}\PY{n}{backend}\PY{p}{)}
\PY{n}{dynamical\PYZus{}decoupling\PYZus{}sequence} \PY{o}{=} \PY{p}{[}\PY{n}{XGate}\PY{p}{(}\PY{p}{)}\PY{p}{,} \PY{n}{XGate}\PY{p}{(}\PY{p}{)}\PY{p}{]}
\PY{n}{pulse\PYZus{}alignment} \PY{o}{=} \PY{n}{backend}\PY{o}{.}\PY{n}{configuration}\PY{p}{(}\PY{p}{)}\PY{o}{.}\PY{n}{timing\PYZus{}constraints}\PY{p}{[}\PY{l+s+s2}{\PYZdq{}}\PY{l+s+s2}{pulse\PYZus{}alignment}\PY{l+s+s2}{\PYZdq{}}\PY{p}{]}
\PY{n}{pass\PYZus{}manager} \PY{o}{=} \PY{n}{PassManager}\PY{p}{(}\PY{p}{[}
    \PY{n}{ALAPScheduleAnalysis}\PY{p}{(}\PY{n}{instruction\PYZus{}durations}\PY{p}{)}\PY{p}{,}
    \PY{n}{PadDynamicalDecoupling}\PY{p}{(}
        \PY{n}{instruction\PYZus{}durations}\PY{p}{,}
        \PY{n}{dynamical\PYZus{}decoupling\PYZus{}sequence}\PY{p}{,}
        \PY{n}{pulse\PYZus{}alignment}\PY{o}{=}\PY{n}{pulse\PYZus{}alignment}
    \PY{p}{)}
\PY{p}{]}\PY{p}{)}

\PY{n}{qaoa\PYZus{}circuit\PYZus{}best} \PY{o}{=} \PY{n}{pass\PYZus{}manager}\PY{o}{.}\PY{n}{run}\PY{p}{(}\PY{n}{qaoa\PYZus{}circuit\PYZus{}best}\PY{p}{)}
\end{Verbatim}
\end{tcolorbox}

    \begin{tcolorbox}[breakable, size=fbox, boxrule=1pt, pad at break*=1mm,colback=cellbackground, colframe=cellborder]
\prompt{In}{incolor}{22}{\boxspacing}
\begin{Verbatim}[commandchars=\\\{\}]
\PY{n}{qaoa\PYZus{}circuit\PYZus{}best}\PY{o}{.}\PY{n}{draw}\PY{p}{(}\PY{p}{)}
\end{Verbatim}
\end{tcolorbox}

\prompt{Out}{outcolor}{22}{}
    
    \begin{center}
    \adjustimage{max size={0.9\linewidth}{0.9\paperheight}}{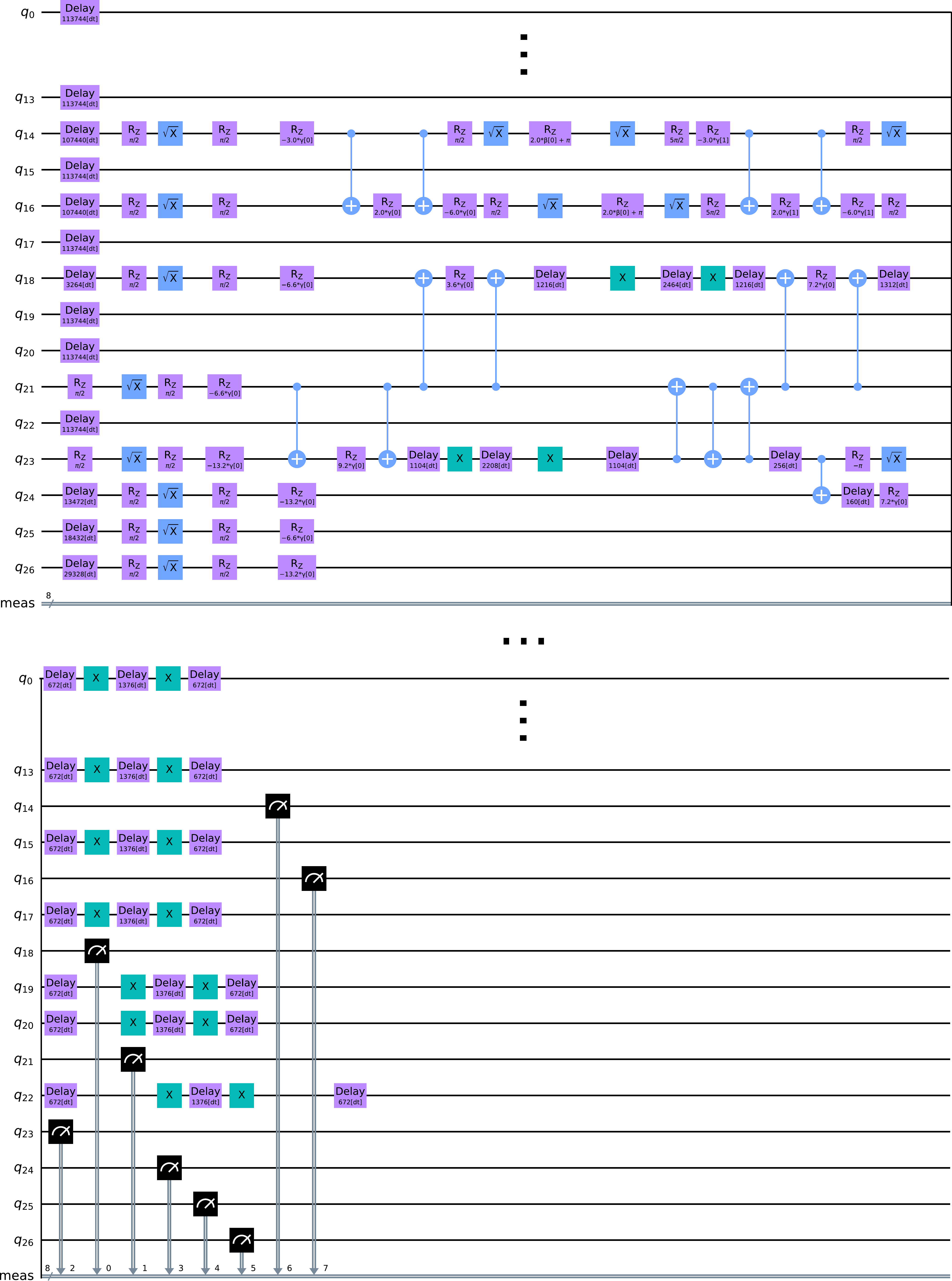}
    \end{center}
    { \hspace*{\fill} \\}

    \hypertarget{measurement-error-mititgation}{%
\subsection{Measurement Error
Mititgation}\label{sec:measurement-error-mititgation}}

    Measurement errors \cite{ShotNoiseWeb} can e.g.~be mitigated with the M3 package \cite{MthreeWeb,QuantumEnablementWeb3}.

    \begin{tcolorbox}[breakable, size=fbox, boxrule=1pt, pad at break*=1mm,colback=cellbackground, colframe=cellborder]
\prompt{In}{incolor}{23}{\boxspacing}
\begin{Verbatim}[commandchars=\\\{\}]
\PY{k+kn}{import} \PY{n+nn}{mthree}
\PY{n}{mit} \PY{o}{=} \PY{n}{mthree}\PY{o}{.}\PY{n}{M3Mitigation}\PY{p}{(}\PY{n}{backend}\PY{p}{)}
\PY{n}{mit\PYZus{}final\PYZus{}measurement\PYZus{}mapping} \PY{o}{=} \PYZbs{}
    \PY{n}{mthree}\PY{o}{.}\PY{n}{utils}\PY{o}{.}\PY{n}{final\PYZus{}measurement\PYZus{}mapping}\PY{p}{(}\PY{n}{qaoa\PYZus{}circuit\PYZus{}best}\PY{p}{)}
\end{Verbatim}
\end{tcolorbox}

    \hypertarget{run-transpiled-circuit-on-backend}{%
\subsection{Run Transpiled Circuit on
Backend}\label{sec:run-transpiled-circuit-on-backend}}

    Now let's run our circuit on the backend.

    \begin{tcolorbox}[breakable, size=fbox, boxrule=1pt, pad at break*=1mm,colback=cellbackground, colframe=cellborder]
\prompt{In}{incolor}{24}{\boxspacing}
\begin{Verbatim}[commandchars=\\\{\}]
\PY{c+c1}{\PYZsh{} The below provided values for the QAOA parameters }
\PY{c+c1}{\PYZsh{} stem from an optimization as discussed in Notebook 2}
\PY{n}{betas} \PY{o}{=} \PY{p}{[}\PY{l+m+mf}{3.99890724}\PY{p}{,} \PY{l+m+mf}{2.72012026}\PY{p}{]}
\PY{n}{gammas} \PY{o}{=} \PY{p}{[}\PY{l+m+mf}{6.11303759}\PY{p}{,} \PY{l+m+mf}{1.75840967}\PY{p}{]}
\PY{n}{qaoa\PYZus{}parameter\PYZus{}values} \PY{o}{=} \PY{p}{[}\PY{o}{*}\PY{n}{betas}\PY{p}{,} \PY{o}{*}\PY{n}{gammas}\PY{p}{]}

\PY{n}{parameter\PYZus{}bindings} \PY{o}{=} \PYZbs{}
    \PY{n+nb}{dict}\PY{p}{(}\PY{n+nb}{zip}\PY{p}{(}\PY{n}{qaoa\PYZus{}circuit\PYZus{}best}\PY{o}{.}\PY{n}{parameters}\PY{p}{,} \PY{n}{qaoa\PYZus{}parameter\PYZus{}values}\PY{p}{)}\PY{p}{)}
\PY{n}{qaoa\PYZus{}circuit\PYZus{}best\PYZus{}with\PYZus{}parameters} \PY{o}{=} \PYZbs{}
    \PY{n}{qaoa\PYZus{}circuit\PYZus{}best}\PY{o}{.}\PY{n}{bind\PYZus{}parameters}\PY{p}{(}\PY{n}{parameter\PYZus{}bindings}\PY{p}{)}
\end{Verbatim}
\end{tcolorbox}

Note: If you have a real backend you will get in the \textbf{job queue}
when executing the next cells. Instead you can also \textbf{proceed to Section \ref{sec:postprocessing}}
and work with the provided data.

    \begin{tcolorbox}[breakable, size=fbox, boxrule=1pt, pad at break*=1mm,colback=cellbackground, colframe=cellborder]
\prompt{In}{incolor}{25}{\boxspacing}
\begin{Verbatim}[commandchars=\\\{\}]
\PY{c+c1}{\PYZsh{} Run job to get calibration data for measurement error mitigation. }
\PY{c+c1}{\PYZsh{} Ignore a possible warning.}
\PY{n}{mit}\PY{o}{.}\PY{n}{cals\PYZus{}from\PYZus{}system}\PY{p}{(}\PY{n}{mit\PYZus{}final\PYZus{}measurement\PYZus{}mapping}\PY{p}{,} \PY{n}{async\PYZus{}cal}\PY{o}{=}\PY{k+kc}{True}\PY{p}{)}
\end{Verbatim}
\end{tcolorbox}

    \begin{tcolorbox}[breakable, size=fbox, boxrule=1pt, pad at break*=1mm,colback=cellbackground, colframe=cellborder]
\prompt{In}{incolor}{26}{\boxspacing}
\begin{Verbatim}[commandchars=\\\{\}]
\PY{c+c1}{\PYZsh{} Run transpiled QAOA circuit}
\PY{n}{job} \PY{o}{=} \PY{n}{backend}\PY{o}{.}\PY{n}{run}\PY{p}{(}\PY{n}{qaoa\PYZus{}circuit\PYZus{}best\PYZus{}with\PYZus{}parameters}\PY{p}{,} \PY{n}{job\PYZus{}name}\PY{o}{=}\PY{l+s+s2}{\PYZdq{}}\PY{l+s+s2}{qaoa\PYZus{}test\PYZus{}job}\PY{l+s+s2}{\PYZdq{}}\PY{p}{)}
\PY{n+nb}{print}\PY{p}{(}\PY{l+s+s2}{\PYZdq{}}\PY{l+s+s2}{job is sent to backend}\PY{l+s+s2}{\PYZdq{}}\PY{p}{)}
\PY{n+nb}{print}\PY{p}{(}\PY{l+s+s2}{\PYZdq{}}\PY{l+s+s2}{job\PYZus{}id: }\PY{l+s+s2}{\PYZdq{}}\PY{p}{,} \PY{n}{job}\PY{o}{.}\PY{n}{job\PYZus{}id}\PY{p}{(}\PY{p}{)}\PY{p}{)}
\end{Verbatim}
\end{tcolorbox}

    \begin{Verbatim}[commandchars=\\\{\}]
job is sent to backend
job\_id:  4cb2e677-b52a-4d20-a854-2ffac1905101
    \end{Verbatim}

    \begin{tcolorbox}[breakable, size=fbox, boxrule=1pt, pad at break*=1mm,colback=cellbackground, colframe=cellborder]
\prompt{In}{incolor}{27}{\boxspacing}
\begin{Verbatim}[commandchars=\\\{\}]
\PY{k+kn}{from} \PY{n+nn}{qiskit\PYZus{}aer}\PY{n+nn}{.}\PY{n+nn}{jobs} \PY{k+kn}{import} \PY{n}{AerJob}

\PY{c+c1}{\PYZsh{} Write info about job into job\PYZus{}list.txt (if using a real backend)}
\PY{k}{if} \PY{o+ow}{not} \PY{n+nb}{isinstance}\PY{p}{(}\PY{n}{job}\PY{p}{,} \PY{n}{AerJob}\PY{p}{)}\PY{p}{:}
    \PY{k}{with} \PY{n+nb}{open}\PY{p}{(}\PY{l+s+s2}{\PYZdq{}}\PY{l+s+s2}{job\PYZus{}list.txt}\PY{l+s+s2}{\PYZdq{}}\PY{p}{,} \PY{l+s+s2}{\PYZdq{}}\PY{l+s+s2}{at}\PY{l+s+s2}{\PYZdq{}}\PY{p}{)} \PY{k}{as} \PY{n}{f}\PY{p}{:}
        \PY{n}{f}\PY{o}{.}\PY{n}{write}\PY{p}{(}\PY{l+s+sa}{f}\PY{l+s+s2}{\PYZdq{}}\PY{l+s+s2}{creation date: }\PY{l+s+si}{\PYZob{}}\PY{n}{job}\PY{o}{.}\PY{n}{creation\PYZus{}date}\PY{p}{(}\PY{p}{)}\PY{l+s+si}{\PYZcb{}}\PY{l+s+s2}{, }\PY{l+s+s2}{\PYZdq{}}
            \PY{o}{+} \PY{l+s+sa}{f}\PY{l+s+s2}{\PYZdq{}}\PY{l+s+s2}{backend: }\PY{l+s+si}{\PYZob{}}\PY{n}{backend}\PY{o}{.}\PY{n}{name}\PY{p}{(}\PY{p}{)}\PY{l+s+si}{\PYZcb{}}\PY{l+s+s2}{, }\PY{l+s+s2}{\PYZdq{}}
            \PY{o}{+} \PY{l+s+sa}{f}\PY{l+s+s2}{\PYZdq{}}\PY{l+s+s2}{job name: }\PY{l+s+si}{\PYZob{}}\PY{n}{job}\PY{o}{.}\PY{n}{name}\PY{p}{(}\PY{p}{)}\PY{l+s+si}{\PYZcb{}}\PY{l+s+s2}{, }\PY{l+s+s2}{\PYZdq{}}
            \PY{o}{+} \PY{l+s+sa}{f}\PY{l+s+s2}{\PYZdq{}}\PY{l+s+s2}{job id: }\PY{l+s+si}{\PYZob{}}\PY{n}{job}\PY{o}{.}\PY{n}{job\PYZus{}id}\PY{p}{(}\PY{p}{)}\PY{l+s+si}{\PYZcb{}}\PY{l+s+se}{\PYZbs{}n}\PY{l+s+s2}{\PYZdq{}}\PY{p}{)}
\end{Verbatim}
\end{tcolorbox}

    \begin{tcolorbox}[breakable, size=fbox, boxrule=1pt, pad at break*=1mm,colback=cellbackground, colframe=cellborder]
\prompt{In}{incolor}{28}{\boxspacing}
\begin{Verbatim}[commandchars=\\\{\}]
\PY{c+c1}{\PYZsh{} Get measurement error mitigation matrices.}
\PY{n}{mit\PYZus{}matrices} \PY{o}{=} \PY{n}{mit}\PY{o}{.}\PY{n}{cals\PYZus{}to\PYZus{}matrices}\PY{p}{(}\PY{p}{)}
\end{Verbatim}
\end{tcolorbox}

    \begin{tcolorbox}[breakable, size=fbox, boxrule=1pt, pad at break*=1mm,colback=cellbackground, colframe=cellborder]
\prompt{In}{incolor}{29}{\boxspacing}
\begin{Verbatim}[commandchars=\\\{\}]
\PY{c+c1}{\PYZsh{} Wait for jobs to finish.}
\PY{n+nb}{print}\PY{p}{(}\PY{l+s+s2}{\PYZdq{}}\PY{l+s+s2}{Status qaoa job:}\PY{l+s+s2}{\PYZdq{}}\PY{p}{,} \PY{n}{job}\PY{o}{.}\PY{n}{status}\PY{p}{(}\PY{p}{)}\PY{p}{)}
\end{Verbatim}
\end{tcolorbox}

    \begin{Verbatim}[commandchars=\\\{\}]
Status qaoa job: JobStatus.DONE
    \end{Verbatim}

    \hypertarget{save-results}{%
\subsubsection{Save results}\label{sec:save-results}}

    \begin{tcolorbox}[breakable, size=fbox, boxrule=1pt, pad at break*=1mm,colback=cellbackground, colframe=cellborder]
\prompt{In}{incolor}{30}{\boxspacing}
\begin{Verbatim}[commandchars=\\\{\}]
\PY{k+kn}{from} \PY{n+nn}{pathlib} \PY{k+kn}{import} \PY{n}{Path}
\PY{k+kn}{import} \PY{n+nn}{pickle}
\PY{k+kn}{from} \PY{n+nn}{utils} \PY{k+kn}{import} \PY{n}{convert\PYZus{}to\PYZus{}date\PYZus{}and\PYZus{}time\PYZus{}string}

\PY{k}{assert} \PY{n}{job}\PY{o}{.}\PY{n}{done}\PY{p}{(}\PY{p}{)}\PY{p}{,} \PY{l+s+s2}{\PYZdq{}}\PY{l+s+s2}{Job is not finished.}\PY{l+s+s2}{\PYZdq{}}

\PY{n}{result} \PY{o}{=} \PY{n}{job}\PY{o}{.}\PY{n}{result}\PY{p}{(}\PY{p}{)}

\PY{n}{experiment\PYZus{}data} \PY{o}{=} \PY{p}{\PYZob{}}
    \PY{l+s+s2}{\PYZdq{}}\PY{l+s+s2}{result}\PY{l+s+s2}{\PYZdq{}}\PY{p}{:} \PY{n}{result}\PY{p}{,}
    \PY{l+s+s2}{\PYZdq{}}\PY{l+s+s2}{qaoa\PYZus{}circuit\PYZus{}best\PYZus{}with\PYZus{}parameters}\PY{l+s+s2}{\PYZdq{}}\PY{p}{:} \PY{n}{qaoa\PYZus{}circuit\PYZus{}best\PYZus{}with\PYZus{}parameters}\PY{p}{,}
    \PY{l+s+s2}{\PYZdq{}}\PY{l+s+s2}{backend\PYZus{}properties}\PY{l+s+s2}{\PYZdq{}}\PY{p}{:} \PY{n}{backend}\PY{o}{.}\PY{n}{properties}\PY{p}{(}\PY{p}{)}\PY{p}{,}
    \PY{l+s+s2}{\PYZdq{}}\PY{l+s+s2}{mit\PYZus{}matrices}\PY{l+s+s2}{\PYZdq{}}\PY{p}{:} \PY{n}{mit\PYZus{}matrices}\PY{p}{,}
    \PY{l+s+s2}{\PYZdq{}}\PY{l+s+s2}{mit\PYZus{}final\PYZus{}measurement\PYZus{}mapping}\PY{l+s+s2}{\PYZdq{}}\PY{p}{:} \PY{n}{mit\PYZus{}final\PYZus{}measurement\PYZus{}mapping}\PY{p}{,}
    \PY{c+c1}{\PYZsh{} and all other meta data you need}
\PY{p}{\PYZcb{}}

\PY{c+c1}{\PYZsh{} Add more data in the case of real backend.}
\PY{k}{if} \PY{o+ow}{not} \PY{n+nb}{isinstance}\PY{p}{(}\PY{n}{job}\PY{p}{,} \PY{n}{AerJob}\PY{p}{)}\PY{p}{:}
    \PY{n}{experiment\PYZus{}data}\PY{p}{[}\PY{l+s+s2}{\PYZdq{}}\PY{l+s+s2}{job\PYZus{}properties}\PY{l+s+s2}{\PYZdq{}}\PY{p}{]} \PY{o}{=} \PY{n}{job}\PY{o}{.}\PY{n}{properties}\PY{p}{(}\PY{p}{)}
    \PY{c+c1}{\PYZsh{} and all other meta data you need}

\PY{n}{result\PYZus{}time\PYZus{}stamp} \PY{o}{=} \PY{n}{convert\PYZus{}to\PYZus{}date\PYZus{}and\PYZus{}time\PYZus{}string}\PY{p}{(}\PY{n}{result}\PY{o}{.}\PY{n}{date}\PY{p}{)}
\PY{n}{experiment\PYZus{}data\PYZus{}file\PYZus{}path} \PY{o}{=} \PY{n}{Path}\PY{p}{(}
    \PY{l+s+sa}{f}\PY{l+s+s2}{\PYZdq{}}\PY{l+s+si}{\PYZob{}}\PY{n}{result\PYZus{}time\PYZus{}stamp}\PY{l+s+si}{\PYZcb{}}\PY{l+s+s2}{\PYZus{}}\PY{l+s+s2}{\PYZdq{}}
    \PY{l+s+sa}{f}\PY{l+s+s2}{\PYZdq{}}\PY{l+s+si}{\PYZob{}}\PY{n}{backend}\PY{o}{.}\PY{n}{name}\PY{p}{(}\PY{p}{)}\PY{l+s+si}{\PYZcb{}}\PY{l+s+s2}{\PYZus{}}\PY{l+s+s2}{\PYZdq{}}
    \PY{l+s+s2}{\PYZdq{}}\PY{l+s+s2}{qaoa\PYZus{}experiment\PYZus{}data}\PY{l+s+s2}{\PYZdq{}}\PY{p}{)}\PY{o}{.}\PY{n}{with\PYZus{}suffix}\PY{p}{(}\PY{l+s+s2}{\PYZdq{}}\PY{l+s+s2}{.pickle}\PY{l+s+s2}{\PYZdq{}}\PY{p}{)}
\PY{k}{with} \PY{n+nb}{open}\PY{p}{(}\PY{n}{experiment\PYZus{}data\PYZus{}file\PYZus{}path}\PY{p}{,} \PY{l+s+s2}{\PYZdq{}}\PY{l+s+s2}{wb}\PY{l+s+s2}{\PYZdq{}}\PY{p}{)} \PY{k}{as} \PY{n}{f}\PY{p}{:}
    \PY{n}{pickle}\PY{o}{.}\PY{n}{dump}\PY{p}{(}\PY{n}{experiment\PYZus{}data}\PY{p}{,} \PY{n}{f}\PY{p}{)}
\end{Verbatim}
\end{tcolorbox}

    \hypertarget{postprocessing}{%
\subsection{Postprocessing}\label{sec:postprocessing}}

    In the rest of the notebook we are concerned with the processing of our
result from the backend for which we will use the big data tool \href{https://pandas.pydata.org/docs/index.html}{\textbf{pandas}}.
But first let us load our saved results.

    \begin{tcolorbox}[breakable, size=fbox, boxrule=1pt, pad at break*=1mm,colback=cellbackground, colframe=cellborder]
\prompt{In}{incolor}{31}{\boxspacing}
\begin{Verbatim}[commandchars=\\\{\}]
\PY{k+kn}{from} \PY{n+nn}{pathlib} \PY{k+kn}{import} \PY{n}{Path}
\PY{k+kn}{import} \PY{n+nn}{pickle}

\PY{c+c1}{\PYZsh{} Load your data ... }
\PY{c+c1}{\PYZsh{} experiment\PYZus{}data\PYZus{}file\PYZus{}path = \PYZbs{}}
\PY{c+c1}{\PYZsh{}     Path(\PYZdq{}path\PYZus{}to\PYZus{}your\PYZus{}experiment\PYZus{}data\PYZdq{}).with\PYZus{}suffix(\PYZdq{}.pickle\PYZdq{})}
\PY{c+c1}{\PYZsh{}}
\PY{c+c1}{\PYZsh{} or use the provided data from ibmq\PYZus{}ehningen ...}
\PY{c+c1}{\PYZsh{} experiment\PYZus{}data\PYZus{}file\PYZus{}path = \PYZbs{}}
\PY{c+c1}{\PYZsh{}     Path(\PYZdq{}2022\PYZus{}09\PYZus{}22\PYZhy{}14h47m\PYZus{}ibmq\PYZus{}ehningen\PYZus{}qaoa\PYZus{}experiment\PYZus{}data\PYZdq{}).with\PYZus{}suffix(\PYZdq{}.pickle\PYZdq{})}
\PY{n}{experiment\PYZus{}data\PYZus{}file\PYZus{}path} \PY{o}{=} \PYZbs{}
    \PY{n}{Path}\PY{p}{(}\PY{l+s+s2}{\PYZdq{}}\PY{l+s+s2}{2022\PYZus{}09\PYZus{}23\PYZhy{}08h42m\PYZus{}ibmq\PYZus{}ehningen\PYZus{}qaoa\PYZus{}experiment\PYZus{}data}\PY{l+s+s2}{\PYZdq{}}\PY{p}{)}\PY{o}{.}\PY{n}{with\PYZus{}suffix}\PY{p}{(}\PY{l+s+s2}{\PYZdq{}}\PY{l+s+s2}{.pickle}\PY{l+s+s2}{\PYZdq{}}\PY{p}{)}
\PY{c+c1}{\PYZsh{}}
\PY{c+c1}{\PYZsh{} or from fake\PYZus{}kolakata}
\PY{c+c1}{\PYZsh{} experiment\PYZus{}data\PYZus{}file\PYZus{}path = \PYZbs{}}
\PY{c+c1}{\PYZsh{}     Path(\PYZdq{}2022\PYZus{}12\PYZus{}06\PYZhy{}15h31m\PYZus{}fake\PYZus{}kolkata\PYZus{}qaoa\PYZus{}experiment\PYZus{}data\PYZdq{}).with\PYZus{}suffix(\PYZdq{}.pickle\PYZdq{})}

\PY{k}{with} \PY{n+nb}{open}\PY{p}{(}\PY{n}{experiment\PYZus{}data\PYZus{}file\PYZus{}path}\PY{p}{,} \PY{l+s+s2}{\PYZdq{}}\PY{l+s+s2}{rb}\PY{l+s+s2}{\PYZdq{}}\PY{p}{)} \PY{k}{as} \PY{n}{f}\PY{p}{:}
    \PY{n}{experiment\PYZus{}data} \PY{o}{=} \PY{n}{pickle}\PY{o}{.}\PY{n}{load}\PY{p}{(}\PY{n}{f}\PY{p}{)}
    
\PY{n}{result} \PY{o}{=} \PY{n}{experiment\PYZus{}data}\PY{p}{[}\PY{l+s+s2}{\PYZdq{}}\PY{l+s+s2}{result}\PY{l+s+s2}{\PYZdq{}}\PY{p}{]}
\PY{n}{mit\PYZus{}matrices} \PY{o}{=} \PY{n}{experiment\PYZus{}data}\PY{p}{[}\PY{l+s+s2}{\PYZdq{}}\PY{l+s+s2}{mit\PYZus{}matrices}\PY{l+s+s2}{\PYZdq{}}\PY{p}{]}
\PY{n}{mit\PYZus{}final\PYZus{}measurement\PYZus{}mapping} \PY{o}{=} \PY{n}{experiment\PYZus{}data}\PY{p}{[}\PY{l+s+s2}{\PYZdq{}}\PY{l+s+s2}{mit\PYZus{}final\PYZus{}measurement\PYZus{}mapping}\PY{l+s+s2}{\PYZdq{}}\PY{p}{]}
\end{Verbatim}
\end{tcolorbox}

    Let us retrieve the counts from our \texttt{result} object and put it in
a \texttt{DataFrame} together with the respective bitstring. Then, let
us add a column with the probabilities, which we can compute from the
counts.

    \begin{tcolorbox}[breakable, size=fbox, boxrule=1pt, pad at break*=1mm,colback=cellbackground, colframe=cellborder]
\prompt{In}{incolor}{32}{\boxspacing}
\begin{Verbatim}[commandchars=\\\{\}]
\PY{k+kn}{import} \PY{n+nn}{pandas} \PY{k}{as} \PY{n+nn}{pd}

\PY{n}{experiment\PYZus{}counts} \PY{o}{=} \PY{n}{result}\PY{o}{.}\PY{n}{get\PYZus{}counts}\PY{p}{(}\PY{p}{)}
\PY{n}{number\PYZus{}shots} \PY{o}{=} \PY{n+nb}{sum}\PY{p}{(}\PY{n}{experiment\PYZus{}counts}\PY{o}{.}\PY{n}{values}\PY{p}{(}\PY{p}{)}\PY{p}{)}

\PY{c+c1}{\PYZsh{} Create the DataFrame from experiment\PYZus{}counts.}
\PY{n}{experiment\PYZus{}df} \PY{o}{=} \PY{n}{pd}\PY{o}{.}\PY{n}{DataFrame}\PY{o}{.}\PY{n}{from\PYZus{}dict}\PY{p}{(}\PY{n}{data}\PY{o}{=}\PY{p}{\PYZob{}}
    \PY{l+s+s2}{\PYZdq{}}\PY{l+s+s2}{bit\PYZus{}string}\PY{l+s+s2}{\PYZdq{}}\PY{p}{:} \PY{n}{experiment\PYZus{}counts}\PY{o}{.}\PY{n}{keys}\PY{p}{(}\PY{p}{)}\PY{p}{,}
    \PY{l+s+s2}{\PYZdq{}}\PY{l+s+s2}{count}\PY{l+s+s2}{\PYZdq{}}\PY{p}{:} \PY{n}{experiment\PYZus{}counts}\PY{o}{.}\PY{n}{values}\PY{p}{(}\PY{p}{)}\PY{p}{\PYZcb{}}\PY{p}{)}

\PY{c+c1}{\PYZsh{} Add column probability.}
\PY{n}{experiment\PYZus{}df}\PY{p}{[}\PY{l+s+s2}{\PYZdq{}}\PY{l+s+s2}{probability}\PY{l+s+s2}{\PYZdq{}}\PY{p}{]} \PY{o}{=} \PYZbs{}
    \PY{n}{experiment\PYZus{}df}\PY{p}{[}\PY{l+s+s2}{\PYZdq{}}\PY{l+s+s2}{count}\PY{l+s+s2}{\PYZdq{}}\PY{p}{]}\PY{o}{/}\PY{n}{number\PYZus{}shots}

\PY{n}{experiment\PYZus{}df}
\end{Verbatim}
\end{tcolorbox}

            \begin{tcolorbox}[breakable, size=fbox, boxrule=.5pt, pad at break*=1mm, opacityfill=0]
\prompt{Out}{outcolor}{32}{\boxspacing}
\begin{Verbatim}[commandchars=\\\{\}]
    bit\_string  count  probability
0     00000000     49     0.006125
1     00000001     58     0.007250
2     00010000    112     0.014000
3     00010001    167     0.020875
4     00010010    130     0.016250
..         {\ldots}    {\ldots}          {\ldots}
249   11111011      6     0.000750
250   11111100      6     0.000750
251   11111101      7     0.000875
252   11111110     12     0.001500
253   11111111      4     0.000500

[254 rows x 3 columns]
\end{Verbatim}
\end{tcolorbox}
        
    Next, we create a \texttt{DataFrame} with the probabilities stemming from
applying the measurement error mitigation to the counts from our
experiment.

    \begin{tcolorbox}[breakable, size=fbox, boxrule=1pt, pad at break*=1mm,colback=cellbackground, colframe=cellborder]
\prompt{In}{incolor}{33}{\boxspacing}
\begin{Verbatim}[commandchars=\\\{\}]
\PY{k+kn}{import} \PY{n+nn}{mthree}

\PY{n}{mit} \PY{o}{=} \PY{n}{mthree}\PY{o}{.}\PY{n}{M3Mitigation}\PY{p}{(}\PY{p}{)}
\PY{n}{mit}\PY{o}{.}\PY{n}{cals\PYZus{}from\PYZus{}matrices}\PY{p}{(}\PY{n}{mit\PYZus{}matrices}\PY{p}{)}

\PY{n}{experiment\PYZus{}quasi\PYZus{}distribution} \PY{o}{=} \PYZbs{}
    \PY{n}{mit}\PY{o}{.}\PY{n}{apply\PYZus{}correction}\PY{p}{(}\PY{n}{experiment\PYZus{}counts}\PY{p}{,} \PY{n}{mit\PYZus{}final\PYZus{}measurement\PYZus{}mapping}\PY{p}{)}
\PY{n}{experiment\PYZus{}mitigated\PYZus{}probabilities} \PY{o}{=} \PYZbs{}
    \PY{n}{experiment\PYZus{}quasi\PYZus{}distribution}\PY{o}{.}\PY{n}{nearest\PYZus{}probability\PYZus{}distribution}\PY{p}{(}\PY{p}{)}

\PY{n}{experiment\PYZus{}mitigated\PYZus{}df} \PY{o}{=} \PY{n}{pd}\PY{o}{.}\PY{n}{DataFrame}\PY{o}{.}\PY{n}{from\PYZus{}dict}\PY{p}{(}\PY{n}{data}\PY{o}{=}\PY{p}{\PYZob{}}
    \PY{l+s+s2}{\PYZdq{}}\PY{l+s+s2}{bit\PYZus{}string}\PY{l+s+s2}{\PYZdq{}}\PY{p}{:} \PY{n}{experiment\PYZus{}mitigated\PYZus{}probabilities}\PY{o}{.}\PY{n}{keys}\PY{p}{(}\PY{p}{)}\PY{p}{,}
    \PY{l+s+s2}{\PYZdq{}}\PY{l+s+s2}{probability\PYZus{}mit}\PY{l+s+s2}{\PYZdq{}}\PY{p}{:} \PY{n}{experiment\PYZus{}mitigated\PYZus{}probabilities}\PY{o}{.}\PY{n}{values}\PY{p}{(}\PY{p}{)}\PY{p}{\PYZcb{}}\PY{p}{)}

\PY{n}{experiment\PYZus{}mitigated\PYZus{}df}
\end{Verbatim}
\end{tcolorbox}

            \begin{tcolorbox}[breakable, size=fbox, boxrule=.5pt, pad at break*=1mm, opacityfill=0]
\prompt{Out}{outcolor}{33}{\boxspacing}
\begin{Verbatim}[commandchars=\\\{\}]
    bit\_string  probability\_mit
0     01100001         0.000017
1     10000000         0.000019
2     10000101         0.000022
3     01001010         0.000024
4     01100100         0.000030
..         {\ldots}              {\ldots}
246   00011000         0.025836
247   00110110         0.026571
248   00010110         0.029771
249   00011110         0.034682
250   00111110         0.034965

[251 rows x 2 columns]
\end{Verbatim}
\end{tcolorbox}
        
    Now, let's merge our two \texttt{DataFrame}s and visualize the result.

    \begin{tcolorbox}[breakable, size=fbox, boxrule=1pt, pad at break*=1mm,colback=cellbackground, colframe=cellborder]
\prompt{In}{incolor}{34}{\boxspacing}
\begin{Verbatim}[commandchars=\\\{\}]
\PY{n}{experiment\PYZus{}df} \PY{o}{=} \PY{n}{experiment\PYZus{}df}\PY{o}{.}\PY{n}{merge}\PY{p}{(}
    \PY{n}{experiment\PYZus{}mitigated\PYZus{}df}\PY{p}{,} \PY{n}{how}\PY{o}{=}\PY{l+s+s2}{\PYZdq{}}\PY{l+s+s2}{outer}\PY{l+s+s2}{\PYZdq{}}\PY{p}{,} \PY{n}{on}\PY{o}{=}\PY{l+s+s2}{\PYZdq{}}\PY{l+s+s2}{bit\PYZus{}string}\PY{l+s+s2}{\PYZdq{}}\PY{p}{)}
\PY{n}{experiment\PYZus{}df}\PY{o}{.}\PY{n}{fillna}\PY{p}{(}\PY{l+m+mf}{0.0}\PY{p}{,} \PY{n}{inplace}\PY{o}{=}\PY{k+kc}{True}\PY{p}{)}

\PY{n}{experiment\PYZus{}df}
\end{Verbatim}
\end{tcolorbox}

            \begin{tcolorbox}[breakable, size=fbox, boxrule=.5pt, pad at break*=1mm, opacityfill=0]
\prompt{Out}{outcolor}{34}{\boxspacing}
\begin{Verbatim}[commandchars=\\\{\}]
    bit\_string  count  probability  probability\_mit
0     00000000     49     0.006125         0.005830
1     00000001     58     0.007250         0.007232
2     00010000    112     0.014000         0.013703
3     00010001    167     0.020875         0.022115
4     00010010    130     0.016250         0.016376
..         {\ldots}    {\ldots}          {\ldots}              {\ldots}
249   11111011      6     0.000750         0.000755
250   11111100      6     0.000750         0.000704
251   11111101      7     0.000875         0.000893
252   11111110     12     0.001500         0.001518
253   11111111      4     0.000500         0.000455

[254 rows x 4 columns]
\end{Verbatim}
\end{tcolorbox}
        
    \begin{tcolorbox}[breakable, size=fbox, boxrule=1pt, pad at break*=1mm,colback=cellbackground, colframe=cellborder]
\prompt{In}{incolor}{35}{\boxspacing}
\begin{Verbatim}[commandchars=\\\{\}]
\PY{c+c1}{\PYZsh{} Plot first 40 bit\PYZus{}strings and probability}
\PY{n}{experiment\PYZus{}df}\PY{o}{.}\PY{n}{iloc}\PY{p}{[}\PY{l+m+mi}{0}\PY{p}{:}\PY{l+m+mi}{40}\PY{p}{,}\PY{p}{:}\PY{p}{]}\PY{o}{.}\PY{n}{plot}\PY{o}{.}\PY{n}{bar}\PY{p}{(}
    \PY{n}{x}\PY{o}{=}\PY{l+s+s2}{\PYZdq{}}\PY{l+s+s2}{bit\PYZus{}string}\PY{l+s+s2}{\PYZdq{}}\PY{p}{,} \PY{n}{y}\PY{o}{=}\PY{p}{[}\PY{l+s+s2}{\PYZdq{}}\PY{l+s+s2}{probability}\PY{l+s+s2}{\PYZdq{}}\PY{p}{,} \PY{l+s+s2}{\PYZdq{}}\PY{l+s+s2}{probability\PYZus{}mit}\PY{l+s+s2}{\PYZdq{}}\PY{p}{]}\PY{p}{)}
\end{Verbatim}
\end{tcolorbox}

            \begin{tcolorbox}[breakable, size=fbox, boxrule=.5pt, pad at break*=1mm, opacityfill=0]
\prompt{Out}{outcolor}{35}{\boxspacing}
\begin{Verbatim}[commandchars=\\\{\}]
<AxesSubplot:xlabel='bit\_string'>
\end{Verbatim}
\end{tcolorbox}
        
    \begin{center}
    \adjustimage{max size={0.9\linewidth}{0.9\paperheight}}{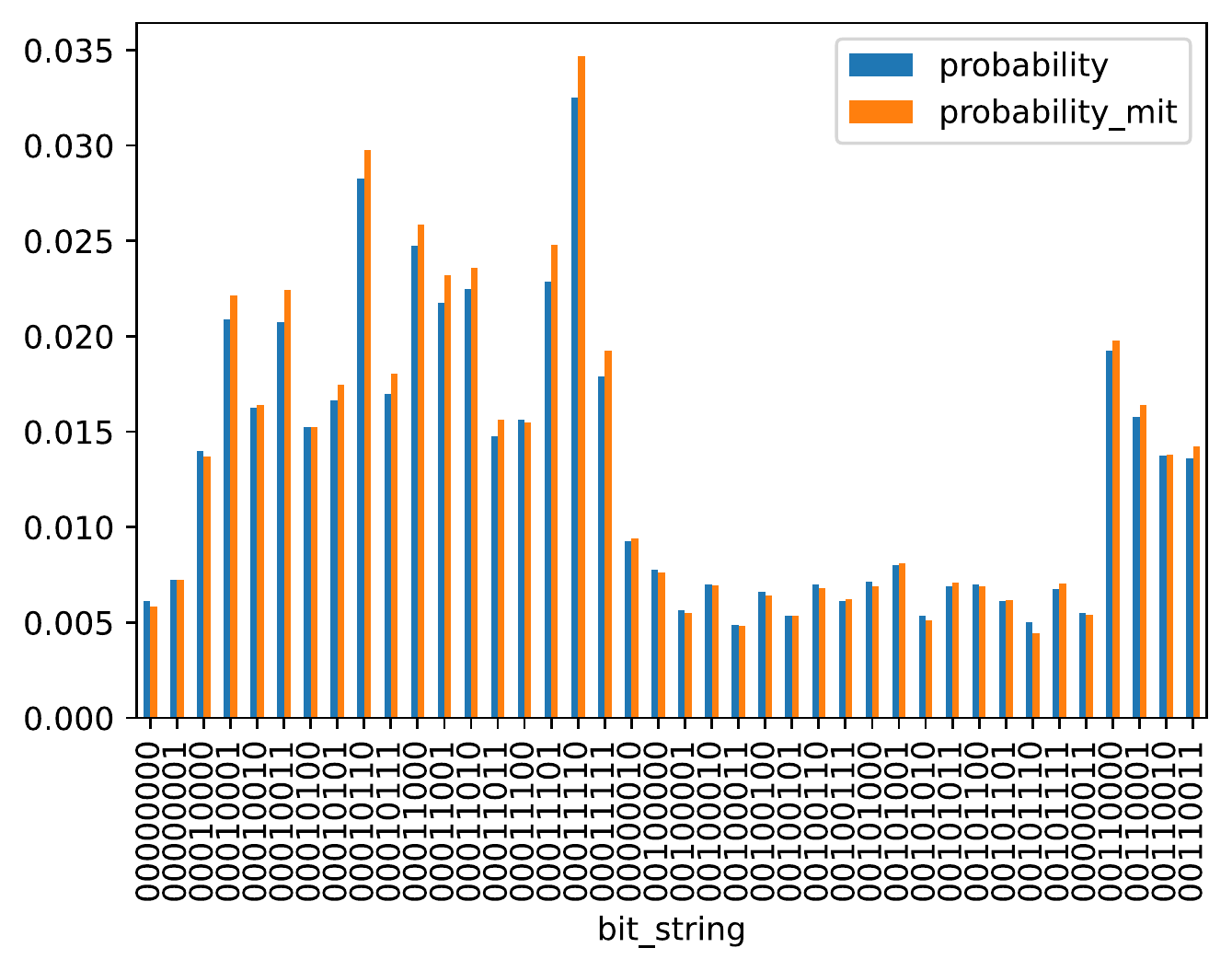}
    \end{center}
    { \hspace*{\fill} \\}
    
    For the remaining section let's focus on the 20 data with the highest
probability. We first do some preparations, namely adding columns with
\texttt{bit\_array} and \texttt{integer\_array}, so that we can
apply \texttt{qubo.objective.evaluate} and \texttt{qcio.is\_feasible} later
(see Notebook \ref{chap:notebook-1}, also note the different ordering conventions).

    \begin{tcolorbox}[breakable, size=fbox, boxrule=1pt, pad at break*=1mm,colback=cellbackground, colframe=cellborder]
\prompt{In}{incolor}{36}{\boxspacing}
\begin{Verbatim}[commandchars=\\\{\}]
\PY{k+kn}{import} \PY{n+nn}{numpy} \PY{k}{as} \PY{n+nn}{np}

\PY{n}{experiment\PYZus{}df\PYZus{}top20} \PY{o}{=} \PYZbs{}
    \PY{n}{experiment\PYZus{}df}\PY{o}{.}\PY{n}{sort\PYZus{}values}\PY{p}{(}\PY{l+s+s2}{\PYZdq{}}\PY{l+s+s2}{probability}\PY{l+s+s2}{\PYZdq{}}\PY{p}{,} \PY{n}{ascending}\PY{o}{=}\PY{k+kc}{False}\PY{p}{)}\PY{o}{.}\PY{n}{iloc}\PY{p}{[}\PY{l+m+mi}{0}\PY{p}{:}\PY{l+m+mi}{20}\PY{p}{,} \PY{p}{:}\PY{p}{]}

\PY{n}{experiment\PYZus{}df\PYZus{}top20}\PY{p}{[}\PY{l+s+s2}{\PYZdq{}}\PY{l+s+s2}{bit\PYZus{}array}\PY{l+s+s2}{\PYZdq{}}\PY{p}{]} \PY{o}{=} \PYZbs{}
    \PY{n}{experiment\PYZus{}df\PYZus{}top20}\PY{p}{[}\PY{l+s+s2}{\PYZdq{}}\PY{l+s+s2}{bit\PYZus{}string}\PY{l+s+s2}{\PYZdq{}}\PY{p}{]}\PY{o}{.}\PY{n}{apply}\PY{p}{(}
        \PY{k}{lambda} \PY{n}{bitstring}\PY{p}{:} \PY{n}{np}\PY{o}{.}\PY{n}{fromiter}\PY{p}{(}\PY{n}{bitstring}\PY{p}{,} \PY{n}{dtype}\PY{o}{=}\PY{n+nb}{int}\PY{p}{)}\PY{p}{[}\PY{p}{:}\PY{p}{:}\PY{o}{\PYZhy{}}\PY{l+m+mi}{1}\PY{p}{]}\PY{p}{)}

\PY{n}{experiment\PYZus{}df\PYZus{}top20}\PY{p}{[}\PY{l+s+s2}{\PYZdq{}}\PY{l+s+s2}{integer\PYZus{}array}\PY{l+s+s2}{\PYZdq{}}\PY{p}{]} \PY{o}{=} \PYZbs{}
    \PY{n}{experiment\PYZus{}df\PYZus{}top20}\PY{p}{[}\PY{l+s+s2}{\PYZdq{}}\PY{l+s+s2}{bit\PYZus{}array}\PY{l+s+s2}{\PYZdq{}}\PY{p}{]}\PY{o}{.}\PY{n}{apply}\PY{p}{(}\PY{n}{converter}\PY{o}{.}\PY{n}{interpret}\PY{p}{)}

\PY{c+c1}{\PYZsh{} Change order of columns.}
\PY{n}{experiment\PYZus{}df\PYZus{}top20} \PY{o}{=} \PY{n}{experiment\PYZus{}df\PYZus{}top20}\PY{p}{[}
    \PY{p}{[}\PY{l+s+s2}{\PYZdq{}}\PY{l+s+s2}{bit\PYZus{}string}\PY{l+s+s2}{\PYZdq{}}\PY{p}{,} \PY{l+s+s2}{\PYZdq{}}\PY{l+s+s2}{bit\PYZus{}array}\PY{l+s+s2}{\PYZdq{}}\PY{p}{,}\PY{l+s+s2}{\PYZdq{}}\PY{l+s+s2}{integer\PYZus{}array}\PY{l+s+s2}{\PYZdq{}}\PY{p}{,}
     \PY{l+s+s2}{\PYZdq{}}\PY{l+s+s2}{count}\PY{l+s+s2}{\PYZdq{}}\PY{p}{,} \PY{l+s+s2}{\PYZdq{}}\PY{l+s+s2}{probability}\PY{l+s+s2}{\PYZdq{}}\PY{p}{,} \PY{l+s+s2}{\PYZdq{}}\PY{l+s+s2}{probability\PYZus{}mit}\PY{l+s+s2}{\PYZdq{}}\PY{p}{]}\PY{p}{]}

\PY{n}{experiment\PYZus{}df\PYZus{}top20}\PY{o}{.}\PY{n}{reset\PYZus{}index}\PY{p}{(}\PY{n}{inplace}\PY{o}{=}\PY{k+kc}{True}\PY{p}{,} \PY{n}{drop}\PY{o}{=}\PY{k+kc}{True}\PY{p}{)}

\PY{n}{experiment\PYZus{}df\PYZus{}top20}
\end{Verbatim}
\end{tcolorbox}

            \begin{tcolorbox}[breakable, size=fbox, boxrule=.5pt, pad at break*=1mm, opacityfill=0]
\prompt{Out}{outcolor}{36}{\boxspacing}
\begin{Verbatim}[commandchars=\\\{\}]
   bit\_string                 bit\_array         integer\_array  count  \textbackslash{}
0    00111110  [0, 1, 1, 1, 1, 1, 0, 0]  [2.0, 3.0, 3.0, 0.0]    261
1    00011110  [0, 1, 1, 1, 1, 0, 0, 0]  [2.0, 3.0, 1.0, 0.0]    260
2    00010110  [0, 1, 1, 0, 1, 0, 0, 0]  [2.0, 1.0, 1.0, 0.0]    226
3    00110110  [0, 1, 1, 0, 1, 1, 0, 0]  [2.0, 1.0, 3.0, 0.0]    204
4    00011000  [0, 0, 0, 1, 1, 0, 0, 0]  [0.0, 2.0, 1.0, 0.0]    198
5    00011101  [1, 0, 1, 1, 1, 0, 0, 0]  [1.0, 3.0, 1.0, 0.0]    183
6    00011010  [0, 1, 0, 1, 1, 0, 0, 0]  [2.0, 2.0, 1.0, 0.0]    180
7    00011001  [1, 0, 0, 1, 1, 0, 0, 0]  [1.0, 2.0, 1.0, 0.0]    174
8    00010001  [1, 0, 0, 0, 1, 0, 0, 0]  [1.0, 0.0, 1.0, 0.0]    167
9    00010011  [1, 1, 0, 0, 1, 0, 0, 0]  [3.0, 0.0, 1.0, 0.0]    166
10   00110100  [0, 0, 1, 0, 1, 1, 0, 0]  [0.0, 1.0, 3.0, 0.0]    159
11   00110000  [0, 0, 0, 0, 1, 1, 0, 0]  [0.0, 0.0, 3.0, 0.0]    154
12   00111001  [1, 0, 0, 1, 1, 1, 0, 0]  [1.0, 2.0, 3.0, 0.0]    152
13   00111010  [0, 1, 0, 1, 1, 1, 0, 0]  [2.0, 2.0, 3.0, 0.0]    151
14   00011111  [1, 1, 1, 1, 1, 0, 0, 0]  [3.0, 3.0, 1.0, 0.0]    143
15   00111000  [0, 0, 0, 1, 1, 1, 0, 0]  [0.0, 2.0, 3.0, 0.0]    140
16   00010111  [1, 1, 1, 0, 1, 0, 0, 0]  [3.0, 1.0, 1.0, 0.0]    136
17   00111100  [0, 0, 1, 1, 1, 1, 0, 0]  [0.0, 3.0, 3.0, 0.0]    134
18   00110111  [1, 1, 1, 0, 1, 1, 0, 0]  [3.0, 1.0, 3.0, 0.0]    133
19   00010101  [1, 0, 1, 0, 1, 0, 0, 0]  [1.0, 1.0, 1.0, 0.0]    133

    probability  probability\_mit
0      0.032625         0.034965
1      0.032500         0.034682
2      0.028250         0.029771
3      0.025500         0.026571
4      0.024750         0.025836
5      0.022875         0.024802
6      0.022500         0.023610
7      0.021750         0.023181
8      0.020875         0.022115
9      0.020750         0.022423
10     0.019875         0.020464
11     0.019250         0.019782
12     0.019000         0.020227
13     0.018875         0.019463
14     0.017875         0.019263
15     0.017500         0.017681
16     0.017000         0.018043
17     0.016750         0.017082
18     0.016625         0.017772
19     0.016625         0.017476
\end{Verbatim}
\end{tcolorbox}
        
    \begin{tcolorbox}[breakable, size=fbox, boxrule=1pt, pad at break*=1mm,colback=cellbackground, colframe=cellborder]
\prompt{In}{incolor}{37}{\boxspacing}
\begin{Verbatim}[commandchars=\\\{\}]
\PY{n}{experiment\PYZus{}df\PYZus{}top20}\PY{o}{.}\PY{n}{plot}\PY{o}{.}\PY{n}{bar}\PY{p}{(}
    \PY{n}{x}\PY{o}{=}\PY{l+s+s2}{\PYZdq{}}\PY{l+s+s2}{bit\PYZus{}string}\PY{l+s+s2}{\PYZdq{}}\PY{p}{,} \PY{n}{y}\PY{o}{=}\PY{p}{[}\PY{l+s+s2}{\PYZdq{}}\PY{l+s+s2}{probability}\PY{l+s+s2}{\PYZdq{}}\PY{p}{,} \PY{l+s+s2}{\PYZdq{}}\PY{l+s+s2}{probability\PYZus{}mit}\PY{l+s+s2}{\PYZdq{}}\PY{p}{]}\PY{p}{)}
\end{Verbatim}
\end{tcolorbox}

            \begin{tcolorbox}[breakable, size=fbox, boxrule=.5pt, pad at break*=1mm, opacityfill=0]
\prompt{Out}{outcolor}{37}{\boxspacing}
\begin{Verbatim}[commandchars=\\\{\}]
<AxesSubplot:xlabel='bit\_string'>
\end{Verbatim}
\end{tcolorbox}
        
    \begin{center}
    \adjustimage{max size={0.9\linewidth}{0.9\paperheight}}{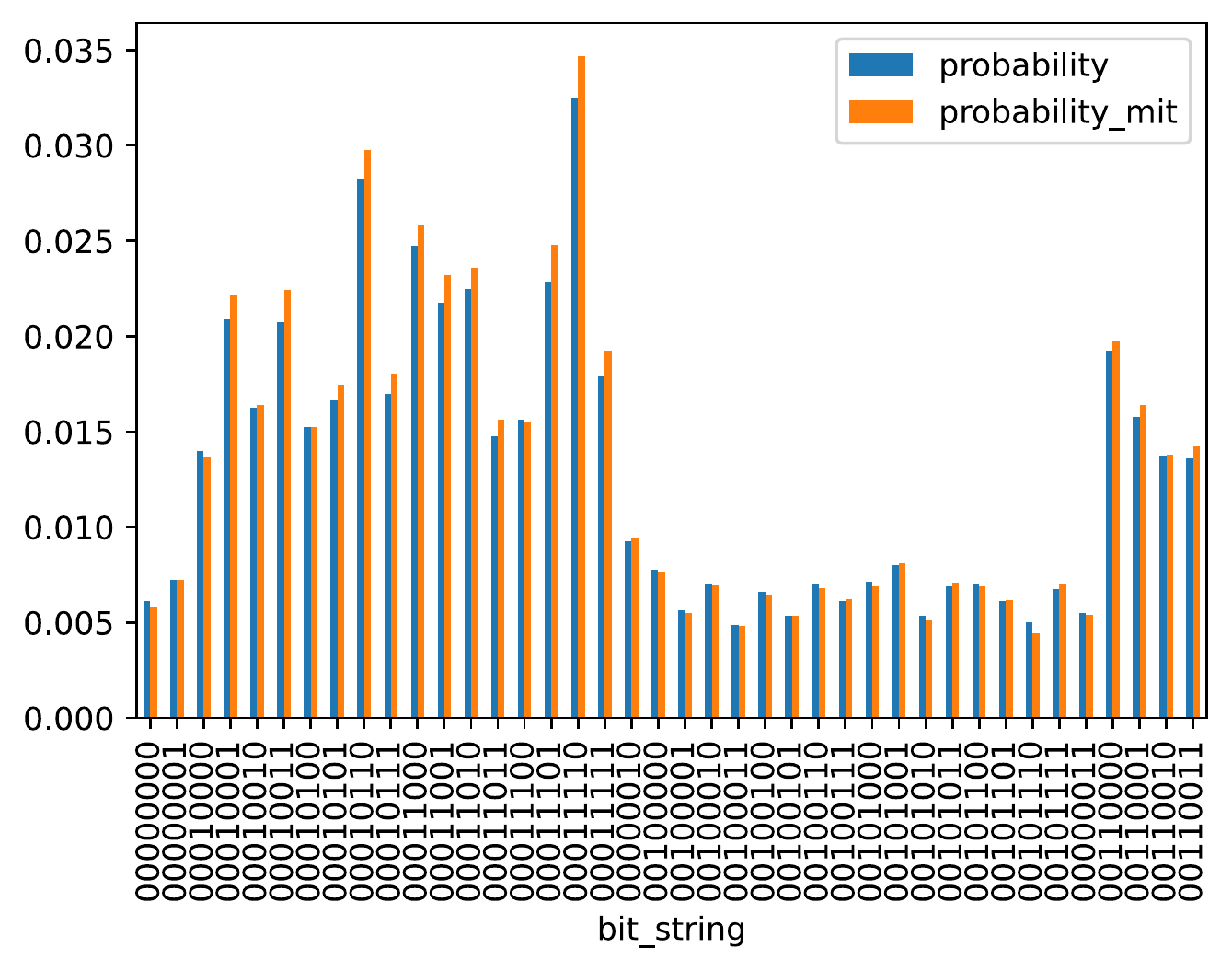}
    \end{center}
    { \hspace*{\fill} \\}
    
    Since we only have 20 rows in our dataframe it is not too costly to
apply \texttt{qubo.objective.evaluate} and \texttt{qcio.is\_feasible}.
This gives us a good overview what could be a good solution candidate (i.e.~it is feasible and has low cost).

    \begin{tcolorbox}[breakable, size=fbox, boxrule=1pt, pad at break*=1mm,colback=cellbackground, colframe=cellborder]
\prompt{In}{incolor}{38}{\boxspacing}
\begin{Verbatim}[commandchars=\\\{\}]
\PY{n}{experiment\PYZus{}df\PYZus{}top20}\PY{p}{[}\PY{l+s+s2}{\PYZdq{}}\PY{l+s+s2}{cost}\PY{l+s+s2}{\PYZdq{}}\PY{p}{]} \PY{o}{=} \PYZbs{}
    \PY{n}{experiment\PYZus{}df\PYZus{}top20}\PY{p}{[}\PY{l+s+s2}{\PYZdq{}}\PY{l+s+s2}{bit\PYZus{}array}\PY{l+s+s2}{\PYZdq{}}\PY{p}{]}\PY{o}{.}\PY{n}{apply}\PY{p}{(}\PY{n}{qubo}\PY{o}{.}\PY{n}{objective}\PY{o}{.}\PY{n}{evaluate}\PY{p}{)}

\PY{n}{experiment\PYZus{}df\PYZus{}top20}\PY{p}{[}\PY{l+s+s2}{\PYZdq{}}\PY{l+s+s2}{is\PYZus{}feasible}\PY{l+s+s2}{\PYZdq{}}\PY{p}{]} \PY{o}{=} \PYZbs{}
    \PY{n}{experiment\PYZus{}df\PYZus{}top20}\PY{p}{[}\PY{l+s+s2}{\PYZdq{}}\PY{l+s+s2}{integer\PYZus{}array}\PY{l+s+s2}{\PYZdq{}}\PY{p}{]}\PY{o}{.}\PY{n}{apply}\PY{p}{(}\PY{n}{qcio}\PY{o}{.}\PY{n}{is\PYZus{}feasible}\PY{p}{)}

\PY{n}{experiment\PYZus{}df\PYZus{}top20}
\end{Verbatim}
\end{tcolorbox}

            \begin{tcolorbox}[breakable, size=fbox, boxrule=.5pt, pad at break*=1mm, opacityfill=0]
\prompt{Out}{outcolor}{38}{\boxspacing}
\begin{Verbatim}[commandchars=\\\{\}]
   bit\_string                 bit\_array         integer\_array  count  \textbackslash{}
0    00111110  [0, 1, 1, 1, 1, 1, 0, 0]  [2.0, 3.0, 3.0, 0.0]    261
1    00011110  [0, 1, 1, 1, 1, 0, 0, 0]  [2.0, 3.0, 1.0, 0.0]    260
2    00010110  [0, 1, 1, 0, 1, 0, 0, 0]  [2.0, 1.0, 1.0, 0.0]    226
3    00110110  [0, 1, 1, 0, 1, 1, 0, 0]  [2.0, 1.0, 3.0, 0.0]    204
4    00011000  [0, 0, 0, 1, 1, 0, 0, 0]  [0.0, 2.0, 1.0, 0.0]    198
5    00011101  [1, 0, 1, 1, 1, 0, 0, 0]  [1.0, 3.0, 1.0, 0.0]    183
6    00011010  [0, 1, 0, 1, 1, 0, 0, 0]  [2.0, 2.0, 1.0, 0.0]    180
7    00011001  [1, 0, 0, 1, 1, 0, 0, 0]  [1.0, 2.0, 1.0, 0.0]    174
8    00010001  [1, 0, 0, 0, 1, 0, 0, 0]  [1.0, 0.0, 1.0, 0.0]    167
9    00010011  [1, 1, 0, 0, 1, 0, 0, 0]  [3.0, 0.0, 1.0, 0.0]    166
10   00110100  [0, 0, 1, 0, 1, 1, 0, 0]  [0.0, 1.0, 3.0, 0.0]    159
11   00110000  [0, 0, 0, 0, 1, 1, 0, 0]  [0.0, 0.0, 3.0, 0.0]    154
12   00111001  [1, 0, 0, 1, 1, 1, 0, 0]  [1.0, 2.0, 3.0, 0.0]    152
13   00111010  [0, 1, 0, 1, 1, 1, 0, 0]  [2.0, 2.0, 3.0, 0.0]    151
14   00011111  [1, 1, 1, 1, 1, 0, 0, 0]  [3.0, 3.0, 1.0, 0.0]    143
15   00111000  [0, 0, 0, 1, 1, 1, 0, 0]  [0.0, 2.0, 3.0, 0.0]    140
16   00010111  [1, 1, 1, 0, 1, 0, 0, 0]  [3.0, 1.0, 1.0, 0.0]    136
17   00111100  [0, 0, 1, 1, 1, 1, 0, 0]  [0.0, 3.0, 3.0, 0.0]    134
18   00110111  [1, 1, 1, 0, 1, 1, 0, 0]  [3.0, 1.0, 3.0, 0.0]    133
19   00010101  [1, 0, 1, 0, 1, 0, 0, 0]  [1.0, 1.0, 1.0, 0.0]    133

    probability  probability\_mit  cost  is\_feasible
0      0.032625         0.034965  79.6        False
1      0.032500         0.034682  28.4        False
2      0.028250         0.029771   6.0         True
3      0.025500         0.026571  28.4        False
4      0.024750         0.025836   8.6        False
5      0.022875         0.024802  14.6        False
6      0.022500         0.023610  12.6        False
7      0.021750         0.023181   6.0         True
8      0.020875         0.022115  16.4        False
9      0.020750         0.022423  10.0         True
10     0.019875         0.020464  10.0         True
11     0.019250         0.019782  12.6        False
12     0.019000         0.020227  28.4        False
13     0.018875         0.019463  49.4        False
14     0.017875         0.019263  51.4        False
15     0.017500         0.017681  16.6        False
16     0.017000         0.018043  14.6        False
17     0.016750         0.017082  32.4        False
18     0.016625         0.017772  51.4        False
19     0.016625         0.017476   6.6        False
\end{Verbatim}
\end{tcolorbox}
        
    Let's plot two solutions from the dataframe above.

    \begin{tcolorbox}[breakable, size=fbox, boxrule=1pt, pad at break*=1mm,colback=cellbackground, colframe=cellborder]
\prompt{In}{incolor}{39}{\boxspacing}
\begin{Verbatim}[commandchars=\\\{\}]
\PY{n}{row\PYZus{}number\PYZus{}1} \PY{o}{=} \PY{l+m+mi}{2} \PY{c+c1}{\PYZsh{} choose a number between 0 and 19}
\PY{n}{row\PYZus{}number\PYZus{}2} \PY{o}{=} \PY{l+m+mi}{6} \PY{c+c1}{\PYZsh{} choose a number between 0 and 19}
\end{Verbatim}
\end{tcolorbox}

    \begin{tcolorbox}[breakable, size=fbox, boxrule=1pt, pad at break*=1mm,colback=cellbackground, colframe=cellborder]
\prompt{In}{incolor}{40}{\boxspacing}
\begin{Verbatim}[commandchars=\\\{\}]
\PY{n}{fig} \PY{o}{=} \PY{n}{plot\PYZus{}charging\PYZus{}schedule}\PY{p}{(}
    \PY{n}{charging\PYZus{}unit}\PY{p}{,}
    \PY{n}{experiment\PYZus{}df\PYZus{}top20}\PY{p}{[}\PY{l+s+s2}{\PYZdq{}}\PY{l+s+s2}{integer\PYZus{}array}\PY{l+s+s2}{\PYZdq{}}\PY{p}{]}\PY{o}{.}\PY{n}{iloc}\PY{p}{[}\PY{n}{row\PYZus{}number\PYZus{}1}\PY{p}{]}\PY{p}{,}
    \PY{n}{marker\PYZus{}size}\PY{o}{=}\PY{l+m+mi}{20}\PY{p}{)}
\PY{n}{fig}\PY{o}{.}\PY{n}{update\PYZus{}layout}\PY{p}{(}\PY{n}{width}\PY{o}{=}\PY{l+m+mi}{400}\PY{p}{,} \PY{n}{height}\PY{o}{=}\PY{l+m+mi}{300}\PY{p}{)}
\PY{n}{fig}\PY{o}{.}\PY{n}{show}\PY{p}{(}\PY{p}{)}
\end{Verbatim}
\end{tcolorbox}
\includegraphics[width=7cm]{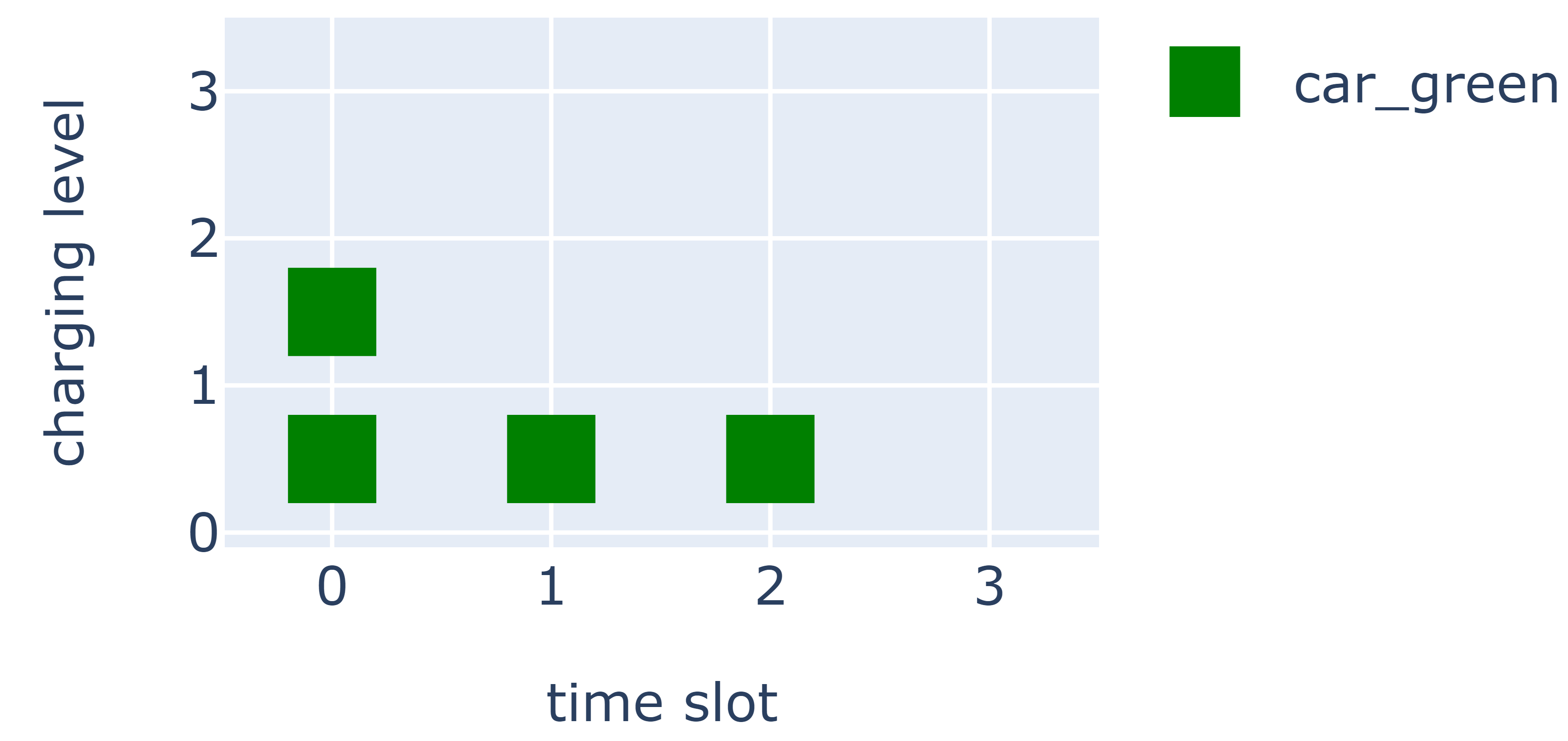}

    \begin{tcolorbox}[breakable, size=fbox, boxrule=1pt, pad at break*=1mm,colback=cellbackground, colframe=cellborder]
\prompt{In}{incolor}{41}{\boxspacing}
\begin{Verbatim}[commandchars=\\\{\}]
\PY{n}{fig} \PY{o}{=} \PY{n}{plot\PYZus{}charging\PYZus{}schedule}\PY{p}{(}
    \PY{n}{charging\PYZus{}unit}\PY{p}{,}
    \PY{n}{experiment\PYZus{}df\PYZus{}top20}\PY{p}{[}\PY{l+s+s2}{\PYZdq{}}\PY{l+s+s2}{integer\PYZus{}array}\PY{l+s+s2}{\PYZdq{}}\PY{p}{]}\PY{o}{.}\PY{n}{iloc}\PY{p}{[}\PY{n}{row\PYZus{}number\PYZus{}2}\PY{p}{]}\PY{p}{,}
    \PY{n}{marker\PYZus{}size}\PY{o}{=}\PY{l+m+mi}{20}\PY{p}{)}
\PY{n}{fig}\PY{o}{.}\PY{n}{update\PYZus{}layout}\PY{p}{(}\PY{n}{width}\PY{o}{=}\PY{l+m+mi}{400}\PY{p}{,} \PY{n}{height}\PY{o}{=}\PY{l+m+mi}{300}\PY{p}{)}
\PY{n}{fig}\PY{o}{.}\PY{n}{show}\PY{p}{(}\PY{p}{)}
\end{Verbatim}
\end{tcolorbox}
\includegraphics[width=7cm]{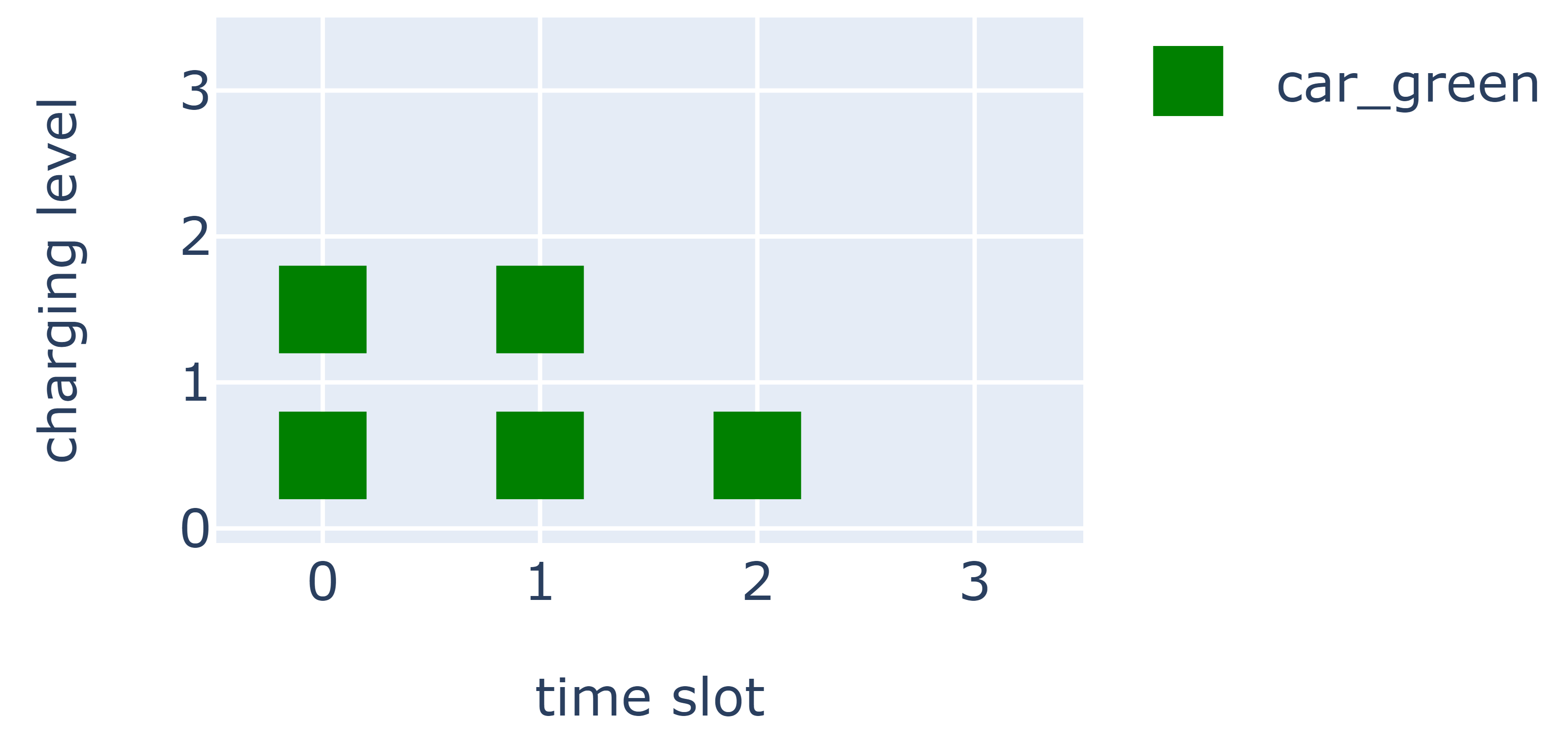}   
    
    \hypertarget{compare-with-exact-simulation}{%
\subsection{Compare with Exact Simulation}\label{sec:compare-with-exact-simulation}}

    In order to judge how good QAOA was executed on the real quantum
computer we can compare it with the result of an exact simulation (a
statevector simulation without shot noise).

    \begin{tcolorbox}[breakable, size=fbox, boxrule=1pt, pad at break*=1mm,colback=cellbackground, colframe=cellborder]
\prompt{In}{incolor}{42}{\boxspacing}
\begin{Verbatim}[commandchars=\\\{\}]
\PY{k+kn}{from} \PY{n+nn}{qiskit}\PY{n+nn}{.}\PY{n+nn}{providers}\PY{n+nn}{.}\PY{n+nn}{aer} \PY{k+kn}{import} \PY{n}{AerSimulator}
\PY{k+kn}{from} \PY{n+nn}{qiskit}\PY{n+nn}{.}\PY{n+nn}{compiler} \PY{k+kn}{import} \PY{n}{transpile}

\PY{n}{aer\PYZus{}simulator} \PY{o}{=} \PY{n}{AerSimulator}\PY{p}{(}\PY{n}{method}\PY{o}{=}\PY{l+s+s2}{\PYZdq{}}\PY{l+s+s2}{statevector}\PY{l+s+s2}{\PYZdq{}}\PY{p}{)}

\PY{n}{qaoa\PYZus{}circuit\PYZus{}for\PYZus{}exact\PYZus{}simulation} \PY{o}{=} \PY{n}{qaoa\PYZus{}circuit}\PY{o}{.}\PY{n}{copy}\PY{p}{(}\PY{p}{)}
\PY{n}{qaoa\PYZus{}circuit\PYZus{}for\PYZus{}exact\PYZus{}simulation}\PY{o}{.}\PY{n}{remove\PYZus{}final\PYZus{}measurements}\PY{p}{(}\PY{p}{)}
\PY{n}{qaoa\PYZus{}circuit\PYZus{}for\PYZus{}exact\PYZus{}simulation}\PY{o}{.}\PY{n}{save\PYZus{}state}\PY{p}{(}\PY{p}{)}

\PY{n}{qaoa\PYZus{}circuit\PYZus{}for\PYZus{}exact\PYZus{}simulation} \PY{o}{=} \PY{n}{transpile}\PY{p}{(}
    \PY{n}{qaoa\PYZus{}circuit\PYZus{}for\PYZus{}exact\PYZus{}simulation}\PY{p}{,}
    \PY{n}{basis\PYZus{}gates}\PY{o}{=}\PY{n}{aer\PYZus{}simulator}\PY{o}{.}\PY{n}{configuration}\PY{p}{(}\PY{p}{)}\PY{o}{.}\PY{n}{basis\PYZus{}gates}\PY{p}{)}

\PY{n}{parameter\PYZus{}bindings} \PY{o}{=} \PY{n+nb}{dict}\PY{p}{(}
    \PY{n+nb}{zip}\PY{p}{(}\PY{n}{qaoa\PYZus{}circuit\PYZus{}for\PYZus{}exact\PYZus{}simulation}\PY{o}{.}\PY{n}{parameters}\PY{p}{,} \PY{n}{qaoa\PYZus{}parameter\PYZus{}values}\PY{p}{)}\PY{p}{)}
\PY{n}{qaoa\PYZus{}circuit\PYZus{}for\PYZus{}exact\PYZus{}simulation} \PY{o}{=} \PYZbs{}
    \PY{n}{qaoa\PYZus{}circuit\PYZus{}for\PYZus{}exact\PYZus{}simulation}\PY{o}{.}\PY{n}{bind\PYZus{}parameters}\PY{p}{(}\PY{n}{parameter\PYZus{}bindings}\PY{p}{)}

\PY{n}{job} \PY{o}{=} \PY{n}{aer\PYZus{}simulator}\PY{o}{.}\PY{n}{run}\PY{p}{(}\PY{n}{qaoa\PYZus{}circuit\PYZus{}for\PYZus{}exact\PYZus{}simulation}\PY{p}{)}
\PY{n}{result\PYZus{}exact\PYZus{}simulation} \PY{o}{=} \PY{n}{job}\PY{o}{.}\PY{n}{result}\PY{p}{(}\PY{p}{)}
\end{Verbatim}
\end{tcolorbox}

Such a simulation has a \href{https://qiskit.org/documentation/stubs/qiskit.quantum_info.Statevector.html#qiskit.quantum_info.Statevector}{\texttt{Statevetor}} object as result.

    \begin{tcolorbox}[breakable, size=fbox, boxrule=1pt, pad at break*=1mm,colback=cellbackground, colframe=cellborder]
\prompt{In}{incolor}{43}{\boxspacing}
\begin{Verbatim}[commandchars=\\\{\}]
\PY{n}{statevector\PYZus{}exact\PYZus{}simulation} \PY{o}{=} \PY{n}{result\PYZus{}exact\PYZus{}simulation}\PY{o}{.}\PY{n}{get\PYZus{}statevector}\PY{p}{(}\PY{p}{)}
\PY{n+nb}{print}\PY{p}{(}\PY{n}{statevector\PYZus{}exact\PYZus{}simulation}\PY{p}{)}
\end{Verbatim}
\end{tcolorbox}

    \begin{Verbatim}[commandchars=\\\{\}]
Statevector([-2.19237435e-03+4.73527294e-05j,
              9.36596020e-03+4.63441187e-03j,
              2.20883527e-06+5.24193322e-04j,
              8.90791252e-03+6.83746374e-03j,
              9.36596020e-03+4.63441187e-03j,
              2.09261296e-02+1.38949582e-01j,
              5.14211808e-06-9.66312497e-04j,
             -4.59808754e-03+3.90829328e-02j,
              2.20883527e-06+5.24193322e-04j,
              ...
              8.24173563e-04-8.22140766e-04j,
             -5.44605429e-04-1.92611140e-03j,
              3.05595401e-03-2.21578751e-03j,
              8.24173563e-04-8.22140766e-04j,
              1.98750175e-03-1.13018901e-04j],
            dims=(2, 2, 2, 2, 2, 2, 2, 2))
    \end{Verbatim}

    Let's transform this into a  \href{https://qiskit.org/documentation/stubs/qiskit.opflow.state_fns.DictStateFn.html#qiskit.opflow.state_fns.DictStateFn}{\texttt{DictStateFn}} to be compatible with the
data type of the result from the real backend.

    \begin{tcolorbox}[breakable, size=fbox, boxrule=1pt, pad at break*=1mm,colback=cellbackground, colframe=cellborder]
\prompt{In}{incolor}{44}{\boxspacing}
\begin{Verbatim}[commandchars=\\\{\}]
\PY{k+kn}{from} \PY{n+nn}{qiskit}\PY{n+nn}{.}\PY{n+nn}{opflow} \PY{k+kn}{import} \PY{n}{VectorStateFn}

\PY{n}{dict\PYZus{}state\PYZus{}fn\PYZus{}exact\PYZus{}simulation} \PY{o}{=} \PYZbs{}
    \PY{n}{VectorStateFn}\PY{p}{(}\PY{n}{statevector\PYZus{}exact\PYZus{}simulation}\PY{p}{)}\PY{o}{.}\PY{n}{to\PYZus{}dict\PYZus{}fn}\PY{p}{(}\PY{p}{)}
\end{Verbatim}
\end{tcolorbox}

    \begin{tcolorbox}[breakable, size=fbox, boxrule=1pt, pad at break*=1mm,colback=cellbackground, colframe=cellborder]
\prompt{In}{incolor}{45}{\boxspacing}
\begin{Verbatim}[commandchars=\\\{\}]
\PY{n}{exact\PYZus{}simulation\PYZus{}df} \PY{o}{=} \PY{n}{pd}\PY{o}{.}\PY{n}{DataFrame}\PY{p}{(}\PY{n}{data}\PY{o}{=}\PY{p}{\PYZob{}}
    \PY{l+s+s2}{\PYZdq{}}\PY{l+s+s2}{bit\PYZus{}string}\PY{l+s+s2}{\PYZdq{}}\PY{p}{:} \PY{n}{dict\PYZus{}state\PYZus{}fn\PYZus{}exact\PYZus{}simulation}\PY{o}{.}\PY{n}{primitive}\PY{o}{.}\PY{n}{keys}\PY{p}{(}\PY{p}{)}\PY{p}{,}
    \PY{l+s+s2}{\PYZdq{}}\PY{l+s+s2}{amplitude}\PY{l+s+s2}{\PYZdq{}}\PY{p}{:} \PY{n}{dict\PYZus{}state\PYZus{}fn\PYZus{}exact\PYZus{}simulation}\PY{o}{.}\PY{n}{primitive}\PY{o}{.}\PY{n}{values}\PY{p}{(}\PY{p}{)}
\PY{p}{\PYZcb{}}\PY{p}{)}
\PY{n}{exact\PYZus{}simulation\PYZus{}df}\PY{p}{[}\PY{l+s+s2}{\PYZdq{}}\PY{l+s+s2}{probability}\PY{l+s+s2}{\PYZdq{}}\PY{p}{]} \PY{o}{=} \PYZbs{}
    \PY{n}{exact\PYZus{}simulation\PYZus{}df}\PY{p}{[}\PY{l+s+s2}{\PYZdq{}}\PY{l+s+s2}{amplitude}\PY{l+s+s2}{\PYZdq{}}\PY{p}{]}\PY{o}{.}\PY{n}{abs}\PY{p}{(}\PY{p}{)}\PY{o}{*}\PY{o}{*}\PY{l+m+mi}{2}
\end{Verbatim}
\end{tcolorbox}

    \begin{tcolorbox}[breakable, size=fbox, boxrule=1pt, pad at break*=1mm,colback=cellbackground, colframe=cellborder]
\prompt{In}{incolor}{46}{\boxspacing}
\begin{Verbatim}[commandchars=\\\{\}]
\PY{n}{exact\PYZus{}simulation\PYZus{}df}
\end{Verbatim}
\end{tcolorbox}

            \begin{tcolorbox}[breakable, size=fbox, boxrule=.5pt, pad at break*=1mm, opacityfill=0]
\prompt{Out}{outcolor}{46}{\boxspacing}
\begin{Verbatim}[commandchars=\\\{\}]
    bit\_string           amplitude   probability
0     00000000 -0.002192+0.000047j  4.808748e-06
1     00000001  0.009366+0.004634j  1.091990e-04
2     00000010  0.000002+0.000524j  2.747835e-07
3     00000011  0.008908+0.006837j  1.261018e-04
4     00000100  0.009366+0.004634j  1.091990e-04
..         {\ldots}                 {\ldots}           {\ldots}
251   11111011  0.000824-0.000822j  1.355178e-06
252   11111100 -0.000545-0.001926j  4.006500e-06
253   11111101  0.003056-0.002216j  1.424857e-05
254   11111110  0.000824-0.000822j  1.355178e-06
255   11111111  0.001988-0.000113j  3.962936e-06

[256 rows x 3 columns]
\end{Verbatim}
\end{tcolorbox}
        
    Now, we can merge the dataframes from the real backend and from the
simulation and plot the results.

    \begin{tcolorbox}[breakable, size=fbox, boxrule=1pt, pad at break*=1mm,colback=cellbackground, colframe=cellborder]
\prompt{In}{incolor}{47}{\boxspacing}
\begin{Verbatim}[commandchars=\\\{\}]
\PY{n}{experiment\PYZus{}df} \PY{o}{=} \PY{n}{experiment\PYZus{}df}\PY{o}{.}\PY{n}{merge}\PY{p}{(}
    \PY{n}{exact\PYZus{}simulation\PYZus{}df}\PY{p}{[}\PY{p}{[}\PY{l+s+s2}{\PYZdq{}}\PY{l+s+s2}{bit\PYZus{}string}\PY{l+s+s2}{\PYZdq{}}\PY{p}{,} \PY{l+s+s2}{\PYZdq{}}\PY{l+s+s2}{probability}\PY{l+s+s2}{\PYZdq{}}\PY{p}{]}\PY{p}{]}\PY{p}{,}
    \PY{n}{how}\PY{o}{=}\PY{l+s+s2}{\PYZdq{}}\PY{l+s+s2}{outer}\PY{l+s+s2}{\PYZdq{}}\PY{p}{,}
    \PY{n}{on}\PY{o}{=}\PY{l+s+s2}{\PYZdq{}}\PY{l+s+s2}{bit\PYZus{}string}\PY{l+s+s2}{\PYZdq{}}\PY{p}{,}
    \PY{n}{suffixes}\PY{o}{=}\PY{p}{[}\PY{l+s+s2}{\PYZdq{}}\PY{l+s+s2}{\PYZdq{}}\PY{p}{,} \PY{l+s+s2}{\PYZdq{}}\PY{l+s+s2}{\PYZus{}exact}\PY{l+s+s2}{\PYZdq{}}\PY{p}{]}\PY{p}{)}
\PY{n}{experiment\PYZus{}df}\PY{o}{.}\PY{n}{fillna}\PY{p}{(}\PY{l+m+mf}{0.0}\PY{p}{,} \PY{n}{inplace}\PY{o}{=}\PY{k+kc}{True}\PY{p}{)}
\end{Verbatim}
\end{tcolorbox}

    \begin{tcolorbox}[breakable, size=fbox, boxrule=1pt, pad at break*=1mm,colback=cellbackground, colframe=cellborder]
\prompt{In}{incolor}{48}{\boxspacing}
\begin{Verbatim}[commandchars=\\\{\}]
\PY{n}{experiment\PYZus{}df}
\end{Verbatim}
\end{tcolorbox}

            \begin{tcolorbox}[breakable, size=fbox, boxrule=.5pt, pad at break*=1mm, opacityfill=0]
\prompt{Out}{outcolor}{48}{\boxspacing}
\begin{Verbatim}[commandchars=\\\{\}]
    bit\_string  count  probability  probability\_mit  probability\_exact
0     00000000   49.0     0.006125         0.005830       4.808748e-06
1     00000001   58.0     0.007250         0.007232       1.091990e-04
2     00010000  112.0     0.014000         0.013703       1.091990e-04
3     00010001  167.0     0.020875         0.022115       1.974489e-02
4     00010010  130.0     0.016250         0.016376       9.337863e-07
..         {\ldots}    {\ldots}          {\ldots}              {\ldots}                {\ldots}
251   11111101    7.0     0.000875         0.000893       1.424857e-05
252   11111110   12.0     0.001500         0.001518       1.355178e-06
253   11111111    4.0     0.000500         0.000455       3.962936e-06
254   01101000    0.0     0.000000         0.000000       7.846307e-08
255   11101101    0.0     0.000000         0.000000       9.549872e-06

[256 rows x 5 columns]
\end{Verbatim}
\end{tcolorbox}
        
We conclude this notebook with two plots comparing the probabilities of the results stemming from the real backend 
(with and without measurement error mitigation) and from the exact simulation.

    \begin{tcolorbox}[breakable, size=fbox, boxrule=1pt, pad at break*=1mm,colback=cellbackground, colframe=cellborder]
\prompt{In}{incolor}{49}{\boxspacing}
\begin{Verbatim}[commandchars=\\\{\}]
\PY{n}{experiment\PYZus{}df}\PY{o}{.}\PY{n}{sort\PYZus{}values}\PY{p}{(}
    \PY{l+s+s2}{\PYZdq{}}\PY{l+s+s2}{probability\PYZus{}exact}\PY{l+s+s2}{\PYZdq{}}\PY{p}{,} \PY{n}{ascending}\PY{o}{=}\PY{k+kc}{False}\PY{p}{)}\PY{o}{.}\PY{n}{iloc}\PY{p}{[}\PY{l+m+mi}{0}\PY{p}{:}\PY{l+m+mi}{40}\PY{p}{,} \PY{p}{:}\PY{p}{]}\PY{o}{.}\PY{n}{plot}\PY{o}{.}\PY{n}{bar}\PY{p}{(}
        \PY{n}{x}\PY{o}{=}\PY{l+s+s2}{\PYZdq{}}\PY{l+s+s2}{bit\PYZus{}string}\PY{l+s+s2}{\PYZdq{}}\PY{p}{,} 
        \PY{n}{y}\PY{o}{=}\PY{p}{[}\PY{l+s+s2}{\PYZdq{}}\PY{l+s+s2}{probability}\PY{l+s+s2}{\PYZdq{}}\PY{p}{,} \PY{l+s+s2}{\PYZdq{}}\PY{l+s+s2}{probability\PYZus{}mit}\PY{l+s+s2}{\PYZdq{}}\PY{p}{,} \PY{l+s+s2}{\PYZdq{}}\PY{l+s+s2}{probability\PYZus{}exact}\PY{l+s+s2}{\PYZdq{}}\PY{p}{]}\PY{p}{)}
\end{Verbatim}
\end{tcolorbox}

            \begin{tcolorbox}[breakable, size=fbox, boxrule=.5pt, pad at break*=1mm, opacityfill=0]
\prompt{Out}{outcolor}{49}{\boxspacing}
\begin{Verbatim}[commandchars=\\\{\}]
<AxesSubplot:xlabel='bit\_string'>
\end{Verbatim}
\end{tcolorbox}
        
    \begin{center}
    \adjustimage{max size={0.9\linewidth}{0.9\paperheight}}{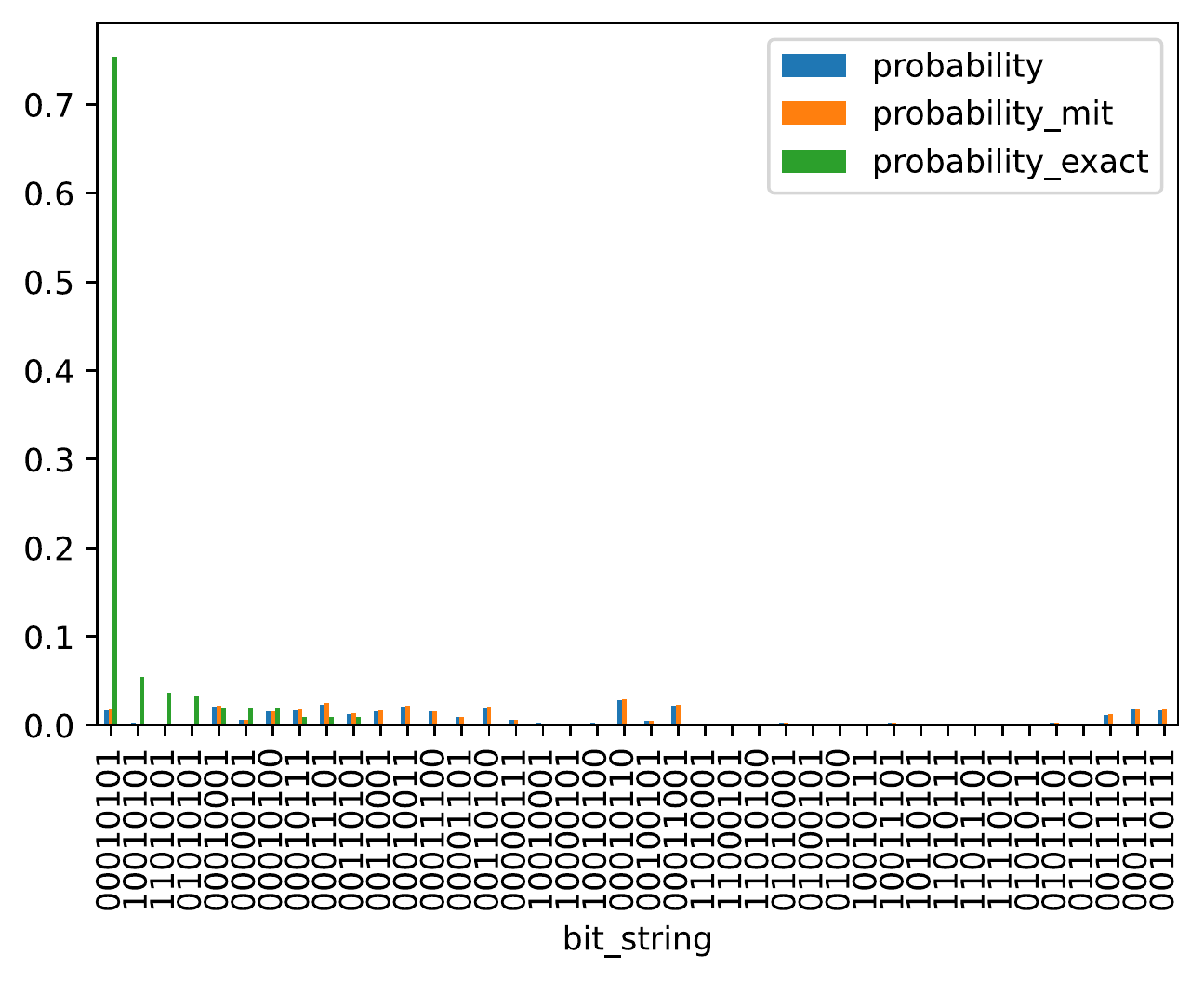}
    \end{center}
    { \hspace*{\fill} \\}
    
    \begin{tcolorbox}[breakable, size=fbox, boxrule=1pt, pad at break*=1mm,colback=cellbackground, colframe=cellborder]
\prompt{In}{incolor}{50}{\boxspacing}
\begin{Verbatim}[commandchars=\\\{\}]
\PY{n}{experiment\PYZus{}df}\PY{o}{.}\PY{n}{sort\PYZus{}values}\PY{p}{(}
    \PY{l+s+s2}{\PYZdq{}}\PY{l+s+s2}{probability}\PY{l+s+s2}{\PYZdq{}}\PY{p}{,} \PY{n}{ascending}\PY{o}{=}\PY{k+kc}{False}\PY{p}{)}\PY{o}{.}\PY{n}{iloc}\PY{p}{[}\PY{l+m+mi}{0}\PY{p}{:}\PY{l+m+mi}{40}\PY{p}{,} \PY{p}{:}\PY{p}{]}\PY{o}{.}\PY{n}{plot}\PY{o}{.}\PY{n}{bar}\PY{p}{(}
        \PY{n}{x}\PY{o}{=}\PY{l+s+s2}{\PYZdq{}}\PY{l+s+s2}{bit\PYZus{}string}\PY{l+s+s2}{\PYZdq{}}\PY{p}{,} 
        \PY{n}{y}\PY{o}{=}\PY{p}{[}\PY{l+s+s2}{\PYZdq{}}\PY{l+s+s2}{probability}\PY{l+s+s2}{\PYZdq{}}\PY{p}{,} \PY{l+s+s2}{\PYZdq{}}\PY{l+s+s2}{probability\PYZus{}mit}\PY{l+s+s2}{\PYZdq{}}\PY{p}{,} \PY{l+s+s2}{\PYZdq{}}\PY{l+s+s2}{probability\PYZus{}exact}\PY{l+s+s2}{\PYZdq{}}\PY{p}{]}\PY{p}{)}
\end{Verbatim}
\end{tcolorbox}

            \begin{tcolorbox}[breakable, size=fbox, boxrule=.5pt, pad at break*=1mm, opacityfill=0]
\prompt{Out}{outcolor}{50}{\boxspacing}
\begin{Verbatim}[commandchars=\\\{\}]
<AxesSubplot:xlabel='bit\_string'>
\end{Verbatim}
\end{tcolorbox}
        
    \begin{center}
    \adjustimage{max size={0.9\linewidth}{0.9\paperheight}}{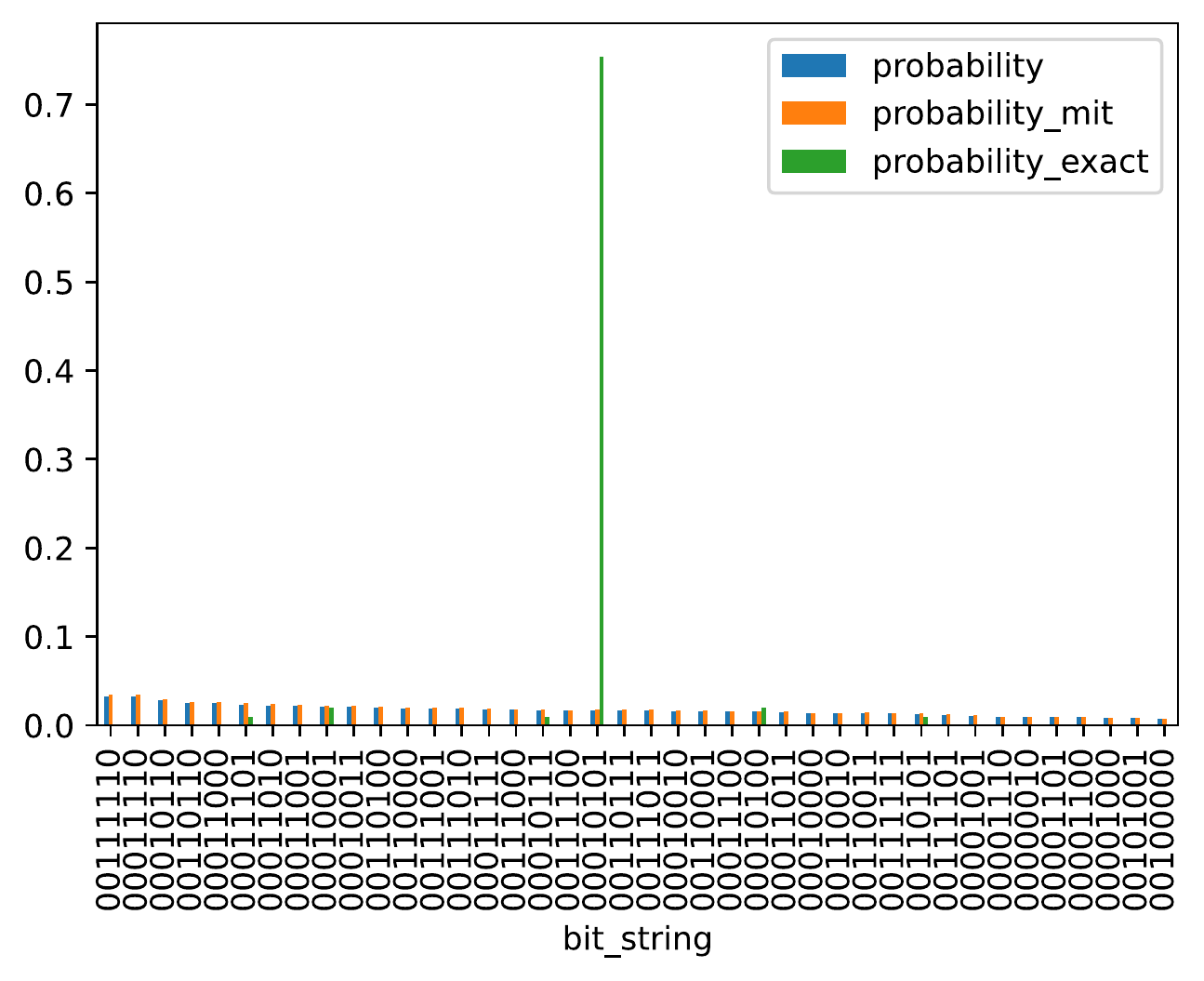}
    \end{center}
    { \hspace*{\fill} \\}

%% file: notebook_4_latex_source.tex
\hypertarget{introduction}{%
\section{Introduction}\label{sec:notebook-4-introduction}}

    In Notebook \ref{chap:notebook-1} we presented a real-world use case for optimizing charging
schedules for electric cars, reduced it to a proof of concept model, and
transformed it to a QUBO. Then, in Notebook \ref{chap:notebook-2} we presented the quantum
algorithm QAOA and how the associated quantum circuits can be obtained.
Subsequently, in Notebook \ref{chap:notebook-3} we explained how these quantum circuits can
be transpiled and run on real quantum computers, and how the results of
such experiments can be postprocessed. Building on this pipeline one can
build a \textbf{series of experiments} in order to \textbf{study how
good quantum computing} (with all its \textbf{limitations} in the
current \textbf{NISQ era}) can be employed \textbf{for our charging
schedule optimization use case}. It is exactly a series of such
experiments that we will present in this notebook.

    \emph{Note: Our main aim with this notebook is to provide and discuss
results from quantum computing experiments. In order to keep this
notebook at a reasonable scope we thus don't provide the detailed codes
but only give the most important parts of them. Together with the
knowledge of the previous notebooks the reader should be able to develop
codes by him/herself (if this is desired)}

    We begin with introducing the two example series on which all our
experiments will be based.

    \hypertarget{examples-for-this-notebook}{%
\section{Examples for this Notebook}\label{sec:examples-for-this-notebook}}

    For all examples in this notebook we consider \textbf{1 charging
station} with \textbf{4 charging levels} and \textbf{4 available time
slots}.

    \hypertarget{example-series-1}{%
\subsection{Example Series 1}\label{sec:example-series-1}}

As a first series of examples (denoted by example1pX, X=0, 1, 2 or 3) we
assume that \textbf{1 car} is at the charging station and needs to
\textbf{charge 4 energy units}. The \textbf{examples differ} in the
\textbf{duration} that the car is \textbf{at the charging station},
namely:

\begin{itemize}
    \tightlist
    \item example1p0 \(\to\) car is at charging station at time slot 0.
    \item example1p1 \(\to\) car is at charging station at time slots 0, 1.
    \item example1p2 \(\to\) car is at charging station at time slots 0, 1, 2.
    \item example1p3 \(\to\) car is at charging station at time slots 0, 1, 2, 3.
\end{itemize}

\begin{figure}
    \begin{center}
        \includegraphics[width=7cm]{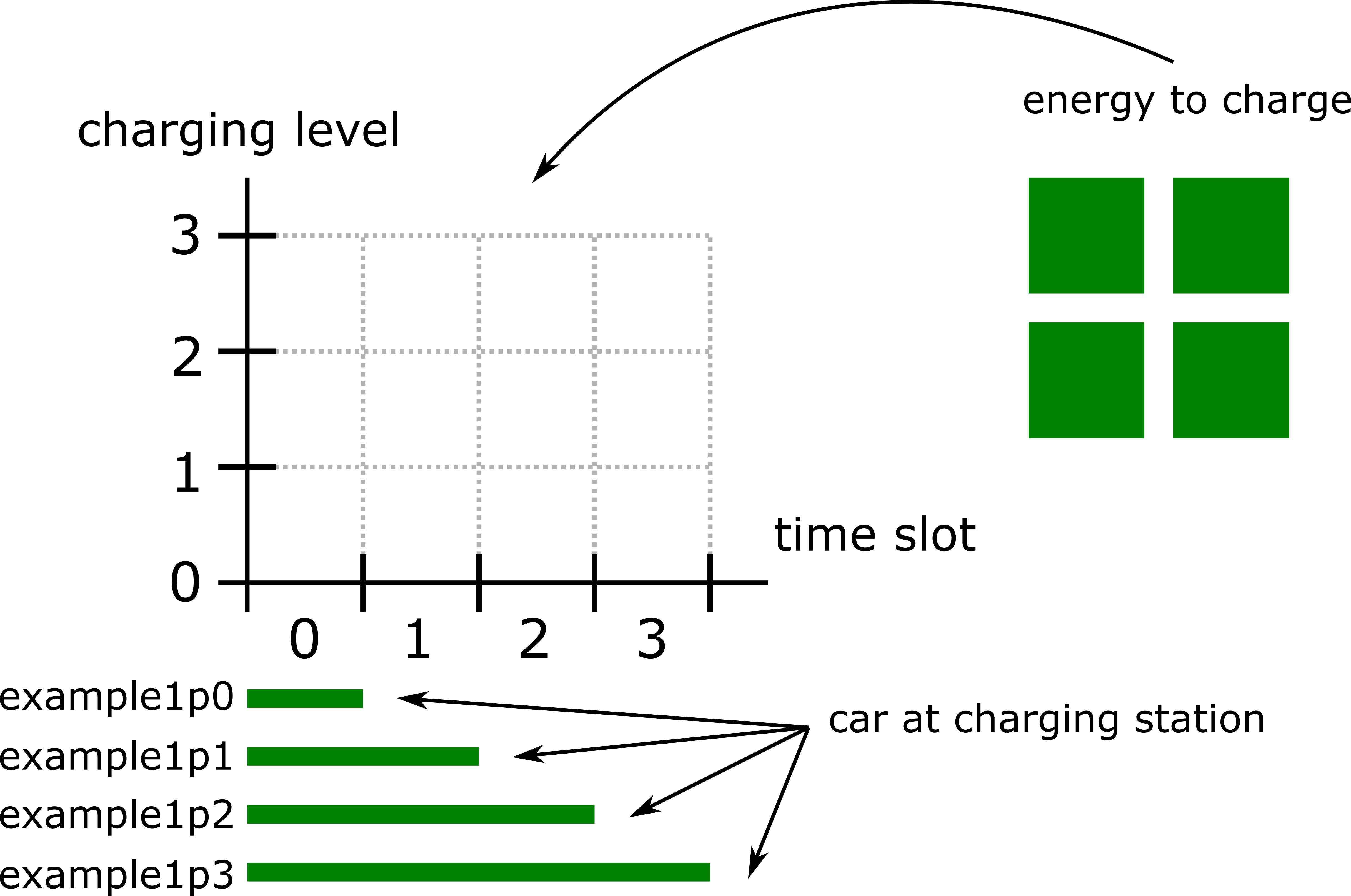}
        \caption{Visualization of Example Series 1.}
        \label{fig:notebook-4-example-series-1}
    \end{center}
\end{figure}

    \hypertarget{example-series-2}{%
\subsection{Example Series 2}\label{sec:example-series-2}}

    Our second series of examples (denoted by example2pX, X=0, 1, 2, 3 or 4)
considers the situation where \textbf{2 cars} are at the charging
station and \textbf{both} need to \textbf{charge 4 energy units}. Again,
the \textbf{examples differ} in the \textbf{duration} that the cars are
\textbf{at the charging station}:

\begin{itemize}
    \tightlist
    \item example2p0 \(\to\) time slots green car: 0, time slots orange car: 1.
    \item example2p1 \(\to\) time slots green car: 0, 1, time slots orange car: 1, 2.
    \item example2p2 \(\to\) time slots green car: 0, 1, 2, time slots orange car: 1, 2, 3.
    \item example2p3 \(\to\) time slots green car: 0, 1, 2, 3, time slots orange car: 1, 2, 3.
    \item example2p4 \(\to\) time slots green car: 0, 1, 2, 3, time slots orange car: 0, 1, 2, 3.
\end{itemize}

\begin{figure}
    \begin{center}
        \includegraphics[width=7cm]{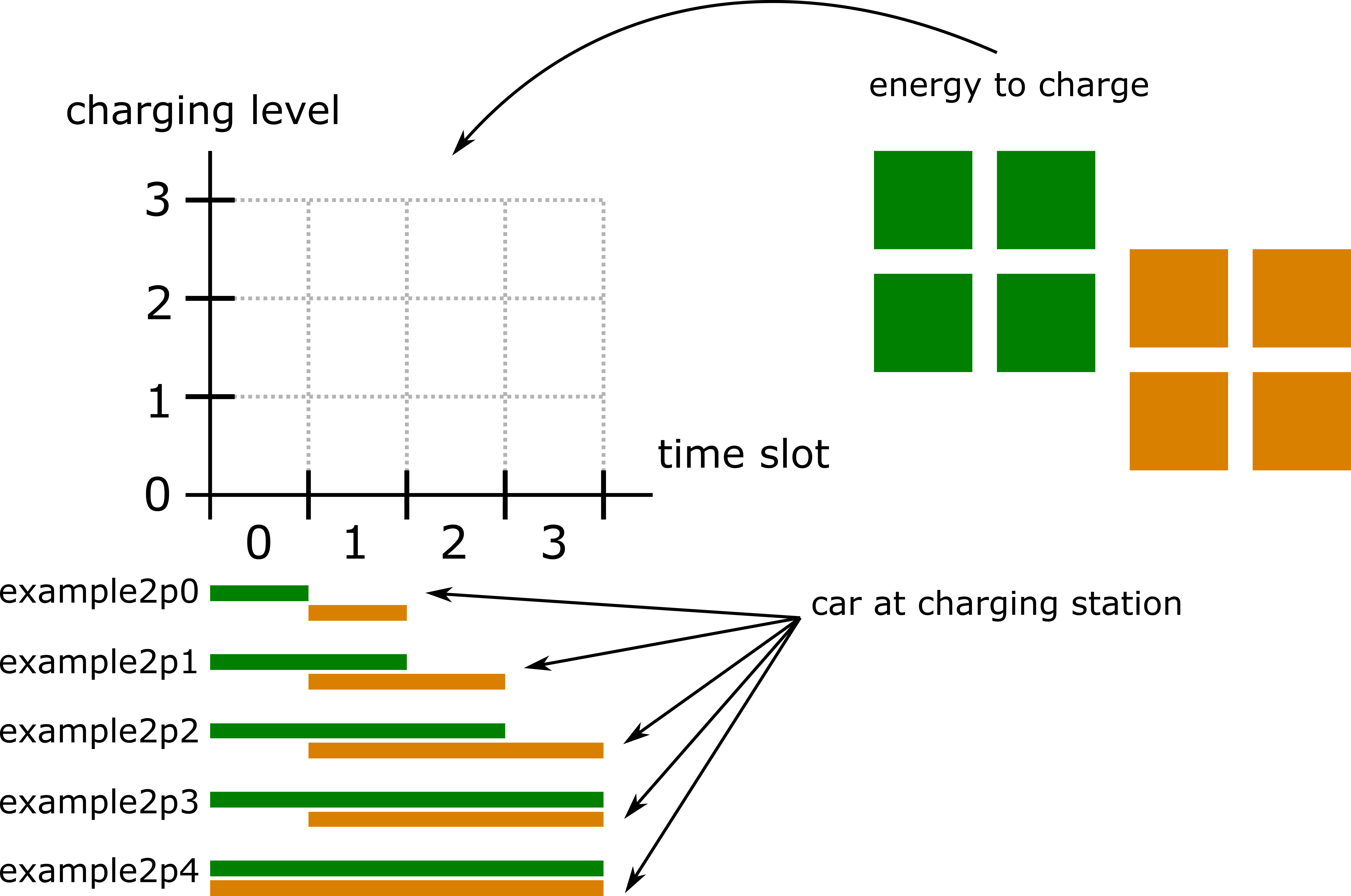}
        \caption{Visualization of Example Series 2.}
        \label{fig:notebook-4-example-series-2}
    \end{center}
\end{figure}

    Our first experiments are concerned with the classical optimization part
of QAOA.

    \hypertarget{classical-optimization}{%
\section{Classical Optimization}\label{sec:notebook-4-classical-optimization}}

    Recall from Notebook \ref{chap:notebook-2} that we want to find parameters
\(\vec{\beta}\) and \(\vec{\gamma}\) such that the expectation value
\(e\), given by

\[
e(\vec{\beta}, \vec{\gamma})
=
\langle \psi_\mathrm{QAOA}(\vec{\beta}, \vec{\gamma}) | \HP |\psi_\mathrm{QAOA}(\vec{\beta}, \vec{\gamma}) \rangle \ ,
\]

is minimized. This means we search for \(\vec{\beta}^\ast\) and
\(\vec{\gamma}^\ast\) that satisfy

\[
(\vec{\beta}^\ast, \vec{\gamma}^\ast) 
= \underset{\vec{\beta}, \vec{\gamma}}{\mathrm{argmin}} \; e(\vec{\beta}, \vec{\gamma}) \ .
\]

    \hypertarget{optimization-landscape}{%
\subsection{Optimization Landscape}\label{optimization-landscape}}

    For \(p=1\) we only have \textbf{two parameters} \(\beta_0\) and
\(\gamma_0\) so that the \textbf{expectation value}
\(e(\beta_0, \gamma_0)\) can be \textbf{visualized as a heatmap}. We
have done this for the following setting:

\begin{itemize}
    \tightlist
    \item We visualize the optimization landscape for parameters in the domain
    \([0, \pi] \times [0, 2 \pi]\). For this purpose we use a fine,
    equidistant discretization of \([0, \pi] \times [0, 2 \pi]\) with
    \(100 \times 200\) gridpoints \((\beta_{0,j}, \gamma_{0,k})\),
    \(j=0, \dots, 99\), \(k=0, \dots, 199\). 
    \item For every grid point \((\beta_{0,j}, \gamma_{0,k})\) we compute the expectation value
    \(e(\beta_{0,j}, \gamma_{0,k})\). This gives a \(100 \times 200\) matrix
    \(E = (e(\beta_{0,j}, \gamma_{0,k}))_{j,k}\).
    \item For the transformation to a QUBO we use as value for the penalty \(\varrho\) the minimum value
    such that the solution of \eqref{eq:qubo} is a solution of \eqref{eq:qcio}.
    \item We use an exact state vector simulation to compute the expectation value.
\end{itemize}

    In the following figures we see the results, where the
\textbf{expectation value} \(e(\beta_{0,j}, \gamma_{0,k})\) is
\textbf{visualized} by the \textbf{color} of the point
\((\beta_{0,j}, \gamma_{0,k})\) (see the colorscale in the plots).

\begin{figure}
    \begin{center}
        \includegraphics{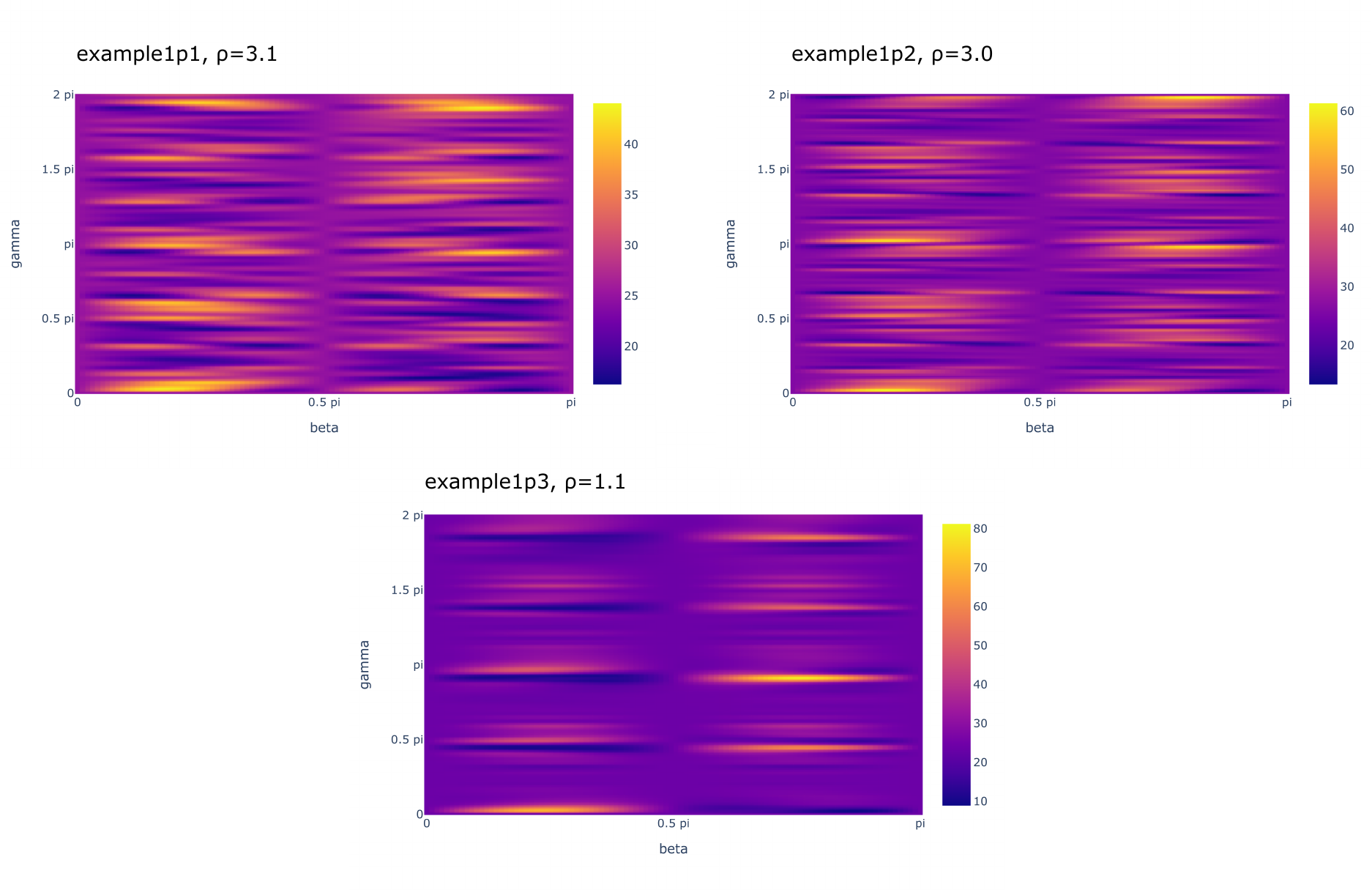}
        \caption{Optimization landscapes for Example Series 1.}
        \label{fig:notebook-4-optimization-landscape-1}
    \end{center}
\end{figure}

\begin{figure}
    \begin{center}
        \includegraphics{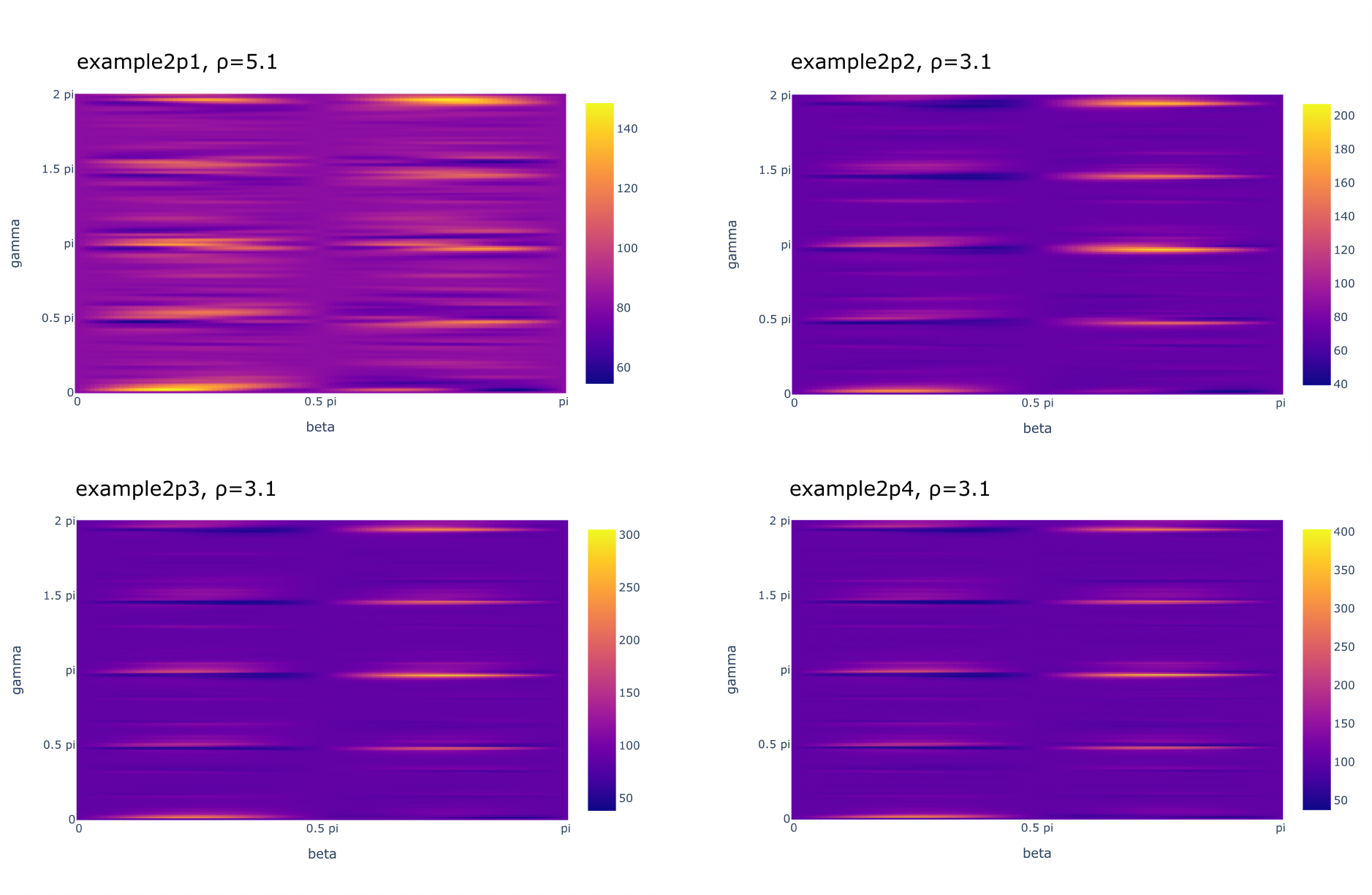}
        \caption{Optimization landscapes for Example Series 2.}
        \label{fig:notebook-4-optimization-landscape-2}
    \end{center}
\end{figure}

    We see that the optimization landscapes are fairly complicated with lots
of \textbf{local extrema}. This indicates that it is a \textbf{hard
task} for classical optimizers to find the \textbf{global minimum}
(depending on the initial guess (=starting point) for a local optimizer
it will get stuck in a local minimum). Moreover, we already can see that
for the bigger examples gradients vanish, see further the so-called
\textbf{barren plateaus} phenomenon \cite{CerEtAl21,McCEtAl18}.

    In the next section we report results from a classical optimizer and
will indeed experience that finding good parameters \(\vec{\beta}\) and
\(\vec{\gamma}\) is a difficult task.

\begin{center}\rule{0.5\linewidth}{0.5pt}\end{center}
Code snippet:

\begin{tcolorbox}[breakable, size=fbox, boxrule=1pt, pad at break*=1mm,colback=cellbackground, colframe=cellborder]
\begin{Verbatim}[commandchars=\\\{\}]
\PY{k+kn}{from} \PY{n+nn}{joblib} \PY{k+kn}{import} \PY{n}{Parallel}\PY{p}{,} \PY{n}{delayed}
\PY{k+kn}{import} \PY{n+nn}{plotly}\PY{n+nn}{.}\PY{n+nn}{graph\PYZus{}objects} \PY{k}{as} \PY{n+nn}{go}

\PY{n}{beta\PYZus{}mesh} \PY{o}{=} \PY{n}{np}\PY{o}{.}\PY{n}{linspace}\PY{p}{(}\PY{l+m+mi}{0}\PY{p}{,} \PY{n}{np}\PY{o}{.}\PY{n}{pi}\PY{p}{,} \PY{n}{beta\PYZus{}mesh\PYZus{}grid\PYZus{}points}\PY{p}{)}
\PY{n}{gamma\PYZus{}mesh} \PY{o}{=} \PY{n}{np}\PY{o}{.}\PY{n}{linspace}\PY{p}{(}\PY{l+m+mi}{0}\PY{p}{,} \PY{l+m+mi}{2}\PY{o}{*}\PY{n}{np}\PY{o}{.}\PY{n}{pi}\PY{p}{,} \PY{n}{gamma\PYZus{}mesh\PYZus{}grid\PYZus{}points}\PY{p}{)}
\PY{n}{energies} \PY{o}{=} \PY{n}{Parallel}\PY{p}{(}\PY{n}{n\PYZus{}jobs}\PY{o}{=}\PY{o}{\PYZhy{}}\PY{l+m+mi}{1}\PY{p}{)}\PY{p}{(}
    \PY{n}{delayed}\PY{p}{(}
        \PY{n}{energy\PYZus{}evaluation} \PY{c+c1}{\PYZsh{} this was defined in Notebook 2}
    \PY{p}{)}\PY{p}{(}
        \PY{p}{[}\PY{n}{beta}\PY{p}{,} \PY{n}{gamma}\PY{p}{]}
    \PY{p}{)} \PY{k}{for} \PY{n}{beta} \PY{o+ow}{in} \PY{n}{beta\PYZus{}mesh} \PY{k}{for} \PY{n}{gamma} \PY{o+ow}{in} \PY{n}{gamma\PYZus{}mesh} 
\PY{p}{)}
\PY{n}{energy\PYZus{}matrix} \PY{o}{=} \PY{n}{np}\PY{o}{.}\PY{n}{reshape}\PY{p}{(}
    \PY{n}{energies}\PY{p}{,} 
    \PY{p}{(}\PY{n}{gamma\PYZus{}mesh}\PY{o}{.}\PY{n}{size}\PY{p}{,} \PY{n}{beta\PYZus{}mesh}\PY{o}{.}\PY{n}{size}\PY{p}{)}\PY{p}{,}
    \PY{n}{order}\PY{o}{=}\PY{l+s+s1}{\PYZsq{}}\PY{l+s+s1}{F}\PY{l+s+s1}{\PYZsq{}}\PY{p}{)}

\PY{n}{fig} \PY{o}{=} \PY{n}{go}\PY{o}{.}\PY{n}{Figure}\PY{p}{(}
    \PY{n}{data} \PY{o}{=}
        \PY{n}{go}\PY{o}{.}\PY{n}{Heatmap}\PY{p}{(}
            \PY{n}{z}\PY{o}{=}\PY{n}{np}\PY{o}{.}\PY{n}{real}\PY{p}{(}\PY{n}{energy\PYZus{}matrix}\PY{p}{)}\PY{p}{,}
            \PY{n}{x}\PY{o}{=}\PY{n}{beta\PYZus{}mesh}\PY{p}{,}
            \PY{n}{y}\PY{o}{=}\PY{n}{gamma\PYZus{}mesh}\PY{p}{,}
\PY{p}{)}\PY{p}{)}
\end{Verbatim}
\end{tcolorbox}

\begin{center}\rule{0.5\linewidth}{0.5pt}\end{center}

    \hypertarget{results-with-optimizer-cobyla}{%
\subsection{Results with Optimizer
COBYLA}\label{sec:results-cobyla}}

For now following results we used the \textbf{optimizer \href{https://qiskit.org/documentation/stubs/qiskit.algorithms.optimizers.COBYLA.html}{COBYLA}}
to minimize \(e(\vec{\beta}, \vec{\gamma})\) for QAOA with \(p=1, 2\)
and \(3\). Moreover, we used 

\begin{itemize}
    \tightlist
    \item \(10\) different values for the penalty \(\varrho\), starting from the minimum that gives a feasible solution and advancing
    in steps of \(0.1\).
    \item For every combination of \(p\) and \(\varrho\) we ran \(50\) COBYLA minimizations with random initial guesses
    \(\vec{\beta}^{(0)} \in [0, \pi]^p\), \(\vec{\gamma}^{(0)} \in [0, 2\pi]^p\). The remaining parameters of COBYLA where
    left by their default values. 
    \item We used an exact state vector simulation to compute the expectation value.
\end{itemize}

We report the \textbf{expectation value
\(e(\vec{\beta^{\ast}}, \vec{\gamma^{\ast}})\)} (termed
\(\mathrm{cost}\) in the figures) below, where
\textbf{\(\vec{\beta^{\ast}}\) and \(\vec{\gamma^{\ast}}\)} are the
\textbf{result of the COBYLA minimization}: 
\[
\text{random } \vec{\beta}^{(0)}, \vec{\gamma}^{(0)} \stackrel{\text{COBYLA}}{\rightsquigarrow} \vec{\beta^{\ast}}, \vec{\gamma^{\ast}} \ .
\] 
The \textbf{dashed line} in the plots corresponds to the
\textbf{solution of the original minimization problem}, i.e.~this is the
value we try to reach with
\(e(\vec{\beta^{\ast}}, \vec{\gamma^{\ast}})\).

\emph{Remark:} In the subsequence we will often plot results as \textbf{box plots}. 
In these plots the ends of the box represent the lower and upper quartiles, while the median (second quartile) is marked by a line inside the box.

\begin{figure}
    \begin{center}
        \begin{subfigure}{0.95\textwidth}
            \centering
            \includegraphics{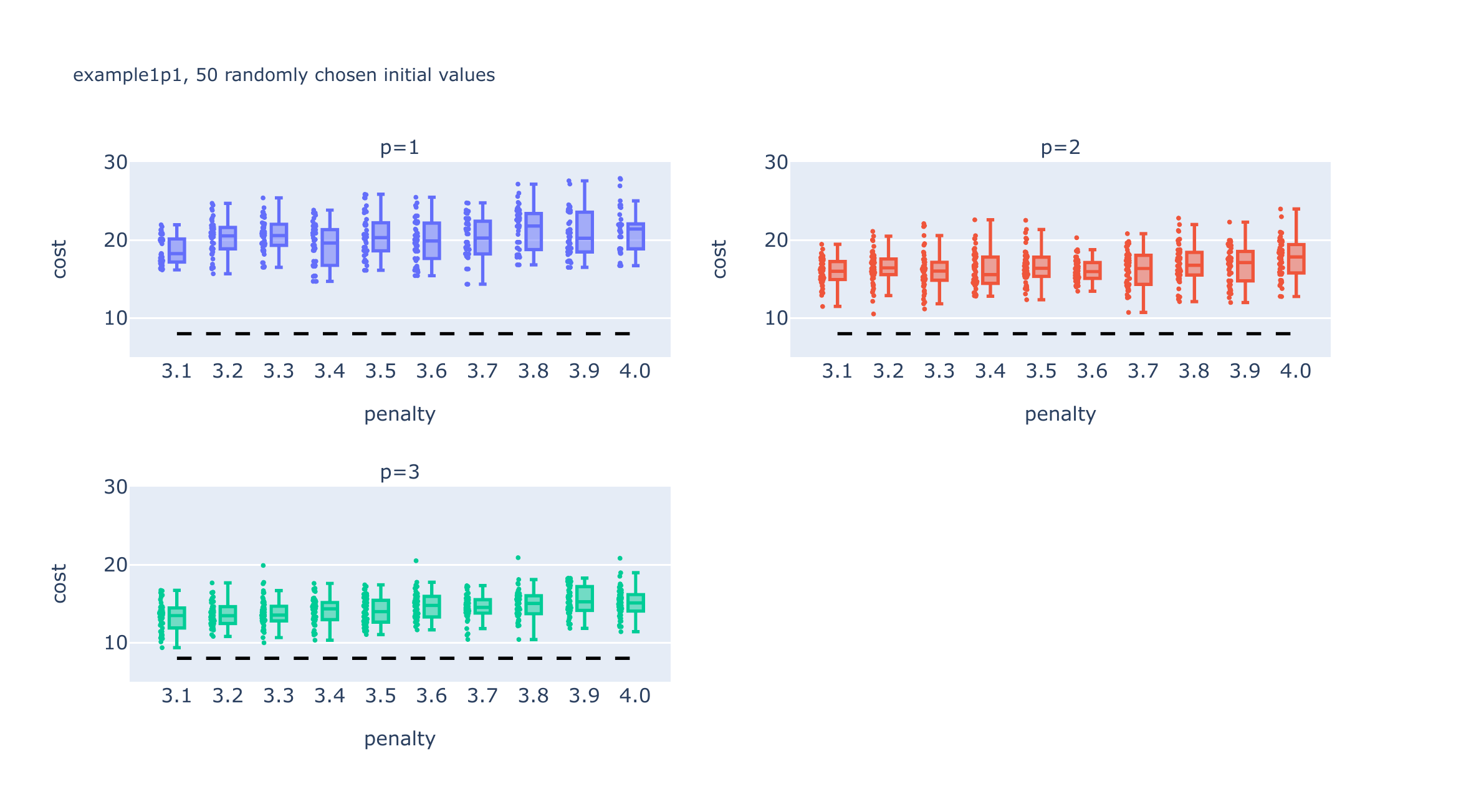}
            \caption{Results for example1p1.}
            \label{fig:notebook-4-cobyla-1p1}
        \end{subfigure}
        \begin{subfigure}{0.95\textwidth}
            \centering
            \includegraphics{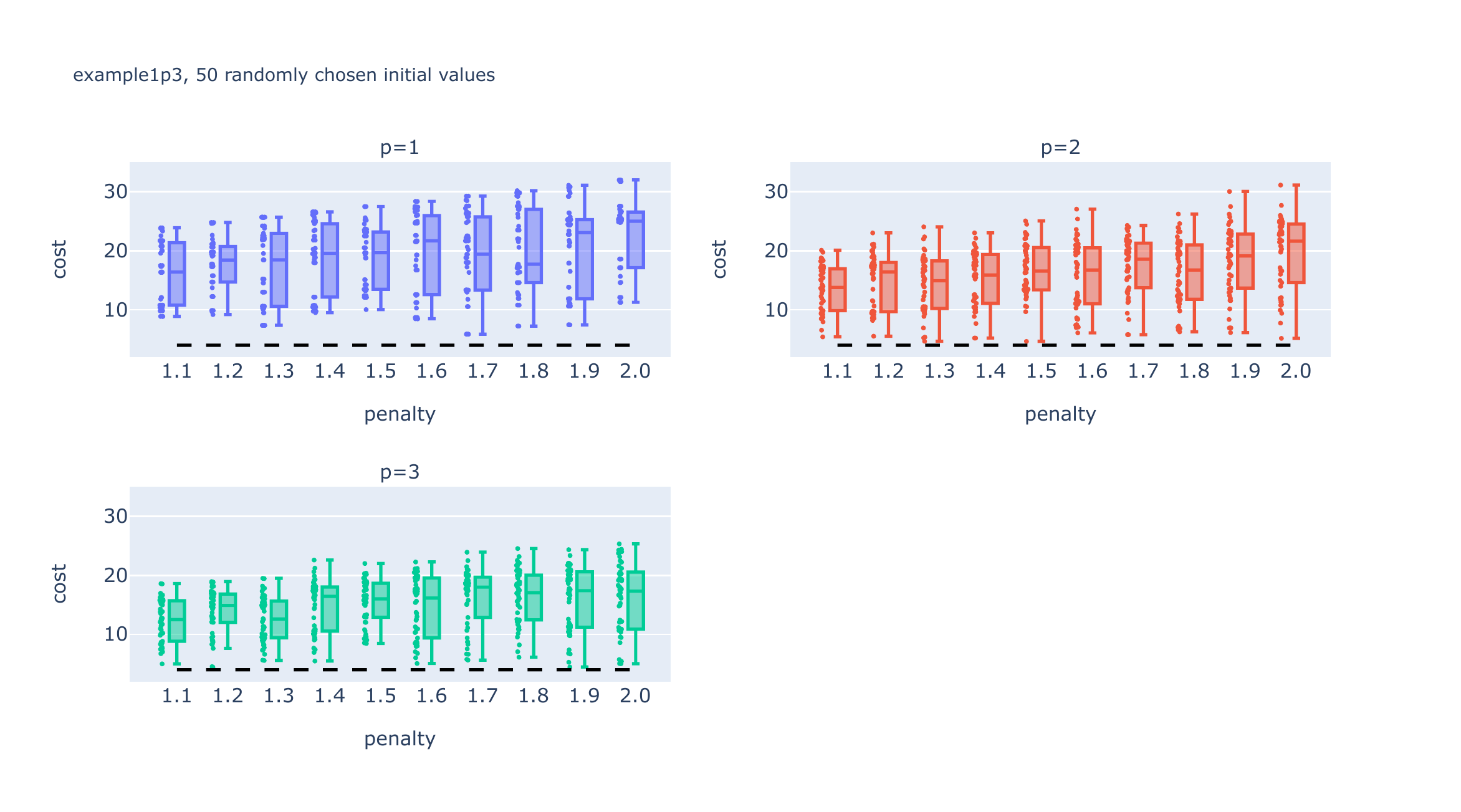}
            \caption{Results for example1p3.}
            \label{fig:notebook-4-cobyla-1p3}
        \end{subfigure}
        \caption{Results of COBYLA optimization for two subexamples of Example Series 1. We used $p=1, 2$, and $3$, and different penalties $\varrho$.
            Every dot corresponds to a randomly chosen initial guess $\vec{\beta}^{(0)}, \vec{\gamma}^{(0)}$. The dasehd line indicates the minimum cost, i.e.
            the exact solution.}
        \label{fig:notebook-4-cobyla-1}
    \end{center}
\end{figure}
    
% \begin{figure}
%     \begin{center}
%         \includegraphics{images/classical_optimization/example1p3.png}
%         \caption{example1p3}
%         \label{fig:notebook-4-cobyla-1p3}
%     \end{center}
% \end{figure}

\begin{figure}
    \begin{center}
        \begin{subfigure}{0.95\textwidth}
            \centering
            \includegraphics{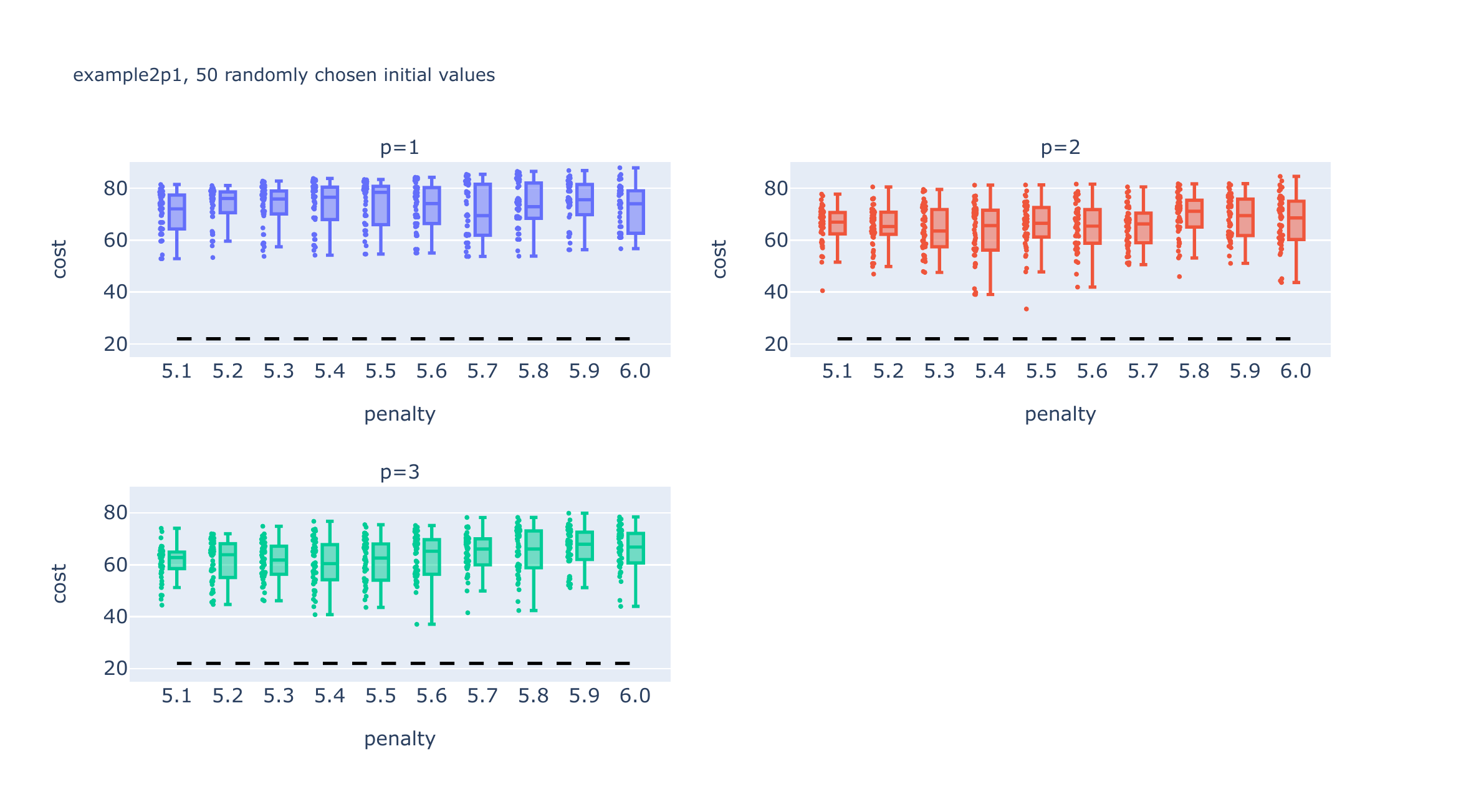}
            \caption{Results for example2p1.}
            \label{fig:notebook-4-cobyla-2p1}
        \end{subfigure}
        \begin{subfigure}{0.95\textwidth}
            \centering
            \includegraphics{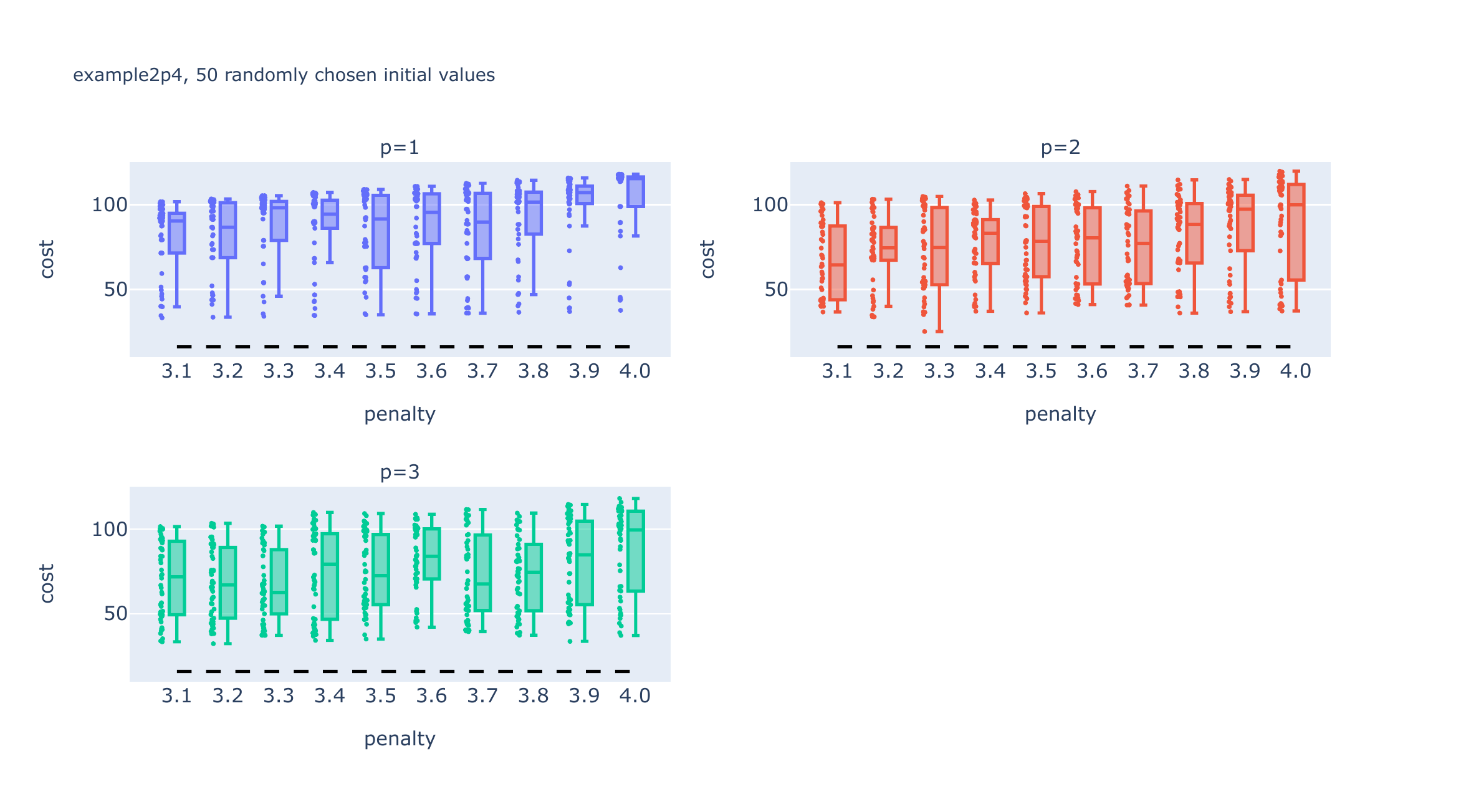}
            \caption{Reults for example2p4.}
            \label{fig:notebook-4-cobyla-2p4}
        \end{subfigure}
        \caption{Results of COBYLA optimization for two subexamples of Example Series 2. We used $p=1, 2$, and $3$, and different penalties $\varrho$.
            Every dot corresponds to a randomly chosen initial guess $\vec{\beta}^{(0)}, \vec{\gamma}^{(0)}$. The dasehd line indicates the minimum cost, i.e.
            the exact solution.}
        \label{fig:notebook-4-cobyla-2}
    \end{center}
\end{figure}
    
% \begin{figure}
%     \begin{center}
%         \includegraphics{images/classical_optimization/example2p1.png}
%         \caption{example2p1}
%         \label{fig:notebook-4-cobyla-2p1}
%     \end{center}
% \end{figure}

% \begin{figure}
%     \begin{center}
%         \includegraphics{images/classical_optimization/example2p4.png}
%         \caption{example2p4}
%         \label{fig:notebook-4-cobyla-2p4}
%     \end{center}
% \end{figure}

    In general, we see a \textbf{strong dependency} (and thus a \textbf{high
variance}) of the result of the optimization with the \textbf{choice of
the initial values} \(\vec{\beta}^{(0)}\) and \(\vec{\gamma}^{(0)}\)
(every dot in the upper figures corresponds to one random choice of the
initial values). Comparing the simpler examples with the more
complicated ones, i.e.~\(\mathrm{example1p1}\) with
\(\mathrm{example1p3}\) and \(\mathrm{example2p1}\) with
\(\mathrm{example2p4}\), we observe a higher variance and that QAOA with
\(p=1\) performs poorly for the more complicated problems (note the
accumulation of results at very high expectation values).

Another observation is that a \textbf{higher parameter \(p\)} (as
expected) \textbf{improves the quality of the QAOA solution}. In
particuar, for example1p1 we observe that QAOA with \(p=3\) yields
expectation values near to the exact solution whereas QAOA with \(p=1\)
is bounded away from the exact solution for every initial choice
\(\vec{\beta}^{(0)}\) and \(\vec{\gamma}^{(0)}\).

In summary, we have seen that optimizing the parameters \(\vec{\beta}\)
and \(\vec{\gamma}\) is indeed a difficult task. So, we can conclude
that different optimizers should be tested and at least a few different
initial values should be compared. We also observe that the choice of
penalty plays a role but can hardly conclude a recommendation from the
data.

    With this we end our experiments that solely used a simulator. The
\textbf{following sections} deal with \textbf{results from real quantum
computers}. An important \textbf{first step} here is to \textbf{analyze
the transpilation} of our QAOA circuits and the \textbf{number of basis
gates} (i.e.~the number of gates that the quantum computing device
natively supports) it requires in order to evaluate how well it can be
executed on NISQ computers.

    \hypertarget{analysis-of-transpiled-qaoa-circuit}{%
\section{Analysis of Transpiled QAOA
Circuits}\label{analysis-of-transpiled-qaoa-circuit}}

In Section \ref{sec:notebook-1-example-convert-to-qubo} we already saw a sparsity plot of the QUBO matrix. Now, we
explain how the sparsity pattern affects the transpilation of the QAOA
circuit associated with the QUBO.

    \hypertarget{sparsity-of-qubo-matrix-and-number-rzz-gates}{%
\subsection{\texorpdfstring{Sparsity of QUBO Matrix and Number of \(\RZZ\)
Gates}{Sparsity of QUBO Matrix and Number \textbackslash RZZ Gates}}\label{sparsity-of-qubo-matrix-and-number-rzz-gates}}

    Recall from Notebooks \ref{chap:notebook-1} and \ref{chap:notebook-2} the QUBO cost function

\begin{equation*}
f_3(\vec{b})
= \vec{b}^{\, t} A \vec{b} + L \vec{b} + c
= \sum_{i=0}^{n-1} \sum_{j>i}^{n-1} a_{ij} b_i b_j
+ \sum_{i=0}^{n-1} l_i b_i 
+ c \ ,
\end{equation*}

    and that the transformation
\(b_i \leftrightarrow \tfrac12 \bigl( I^{\otimes n} - \sigma_Z^{(i)} \bigr)\)
resulted in the cost Hamiltonian

\[
\HP = \sum_{i=0}^{n-1} \sum_{j>i}^{n-1} h_{ij} \sigma_Z^{(i)} \sigma_Z^{(j)}
+ \sum_{i=0}^{n-1} h_i' \sigma_Z^{(i)}
+ h'' I^{\otimes n} \ .
\]

    Moreover, recall the mixing and the phase operators

\[
\UM(\beta)
= \mathrm{exp}(- \mathrm{i} \beta \HM),
\qquad \text{and} \qquad
\UP(\gamma) 
=  \mathrm{exp}(- \mathrm{i} \gamma \HP) \ ,
\]

    respectively, as well as the gates needed to implement them:

    \[
\UM(\beta)
= \RX_0(2 \beta) \RX_1(2 \beta) \cdots \RX_{n-1}(2 \beta) \ ,
\]

    \[
\UP(\gamma) 
= \RZZ_{0,1}(2 \gamma h_{01}) \cdots \RZZ_{n-2,n-1}(2 \gamma h_{n-2,n-1})
\ 
\RZ_0(2\gamma h_0^\prime)% \RZ_1(2\gamma h_1^\prime)
\cdots \RZ_{n-1}(2\gamma h_{n-1}^\prime) \ .
\]

    Observe that we need the \textbf{one-qubit gates} \(\RX\) and \(\RZ\) as
well as the \textbf{two-qubit gate} \(\RZZ\). More precisely, we obtain
one \(\RZZ\) gate for every coefficient \(h_{ij} \neq 0\). It is easy to
see that we have \(h_{ij} \neq 0 \Leftrightarrow a_{ij} \neq 0\). This means: \textbf{the sparser} the matrix \(A\) (= the lesser the
number of non-zero entries in \(A\)) the \textbf{lesser \(\RZZ\) gates}
we have in the circuit:

\[
\text{number of non-zero entries of } A \quad 
\Longleftrightarrow \quad \text{number of } \RZZ \text{ gates in the circuit} \ .
\]

    \emph{Remark:} One can include \(L\) into \(A\) (in the QUBO cost function
\(f_3\)) because \(b_i^2 = b_i\). The resulting matrix, let's call it
\(\tilde{A}\), then can have \emph{non-zero} diagonal entries
\(\tilde{a}_{ii} = l_i\). However, these entries \emph{do not give rise}
to a \(\RZZ\) gate. In this case we have

\[
\text{number of non-zero (off-diagonal) entries of } \tilde{A} \quad
\Longleftrightarrow \quad \text{number of } \RZZ \text{ gates in the circuit} \ .
\]

    Clearly, the question arises why we should be mainly interested in the
number of \(\RZZ\) gates and can neglect (to a certain degree) the
number of the single qubit gates. In order to understand this we have to
analyze the \textbf{transpilations} of the gates.

    \hypertarget{transpilation-of-rx-rz-and-rzz}{%
\subsection{\texorpdfstring{Transpilation of \(\RX\), \(\RZ\) and
\(\RZZ\)}{Transpilation of \textbackslash RX, \textbackslash RZ and \textbackslash RZZ}}\label{transpilation-of-rx-rz-and-rzz}}

    Let us write a quantum circuit with one \(\RX\), one \(\RZ\), and one
\(\RZZ\) gate and transpile it to the \textbf{basis gates} of the
current IBM quantum computers (see Notebook \ref{chap:notebook-3}).

    \begin{tcolorbox}[breakable, size=fbox, boxrule=1pt, pad at break*=1mm,colback=cellbackground, colframe=cellborder]
\prompt{In}{incolor}{1}{\boxspacing}
\begin{Verbatim}[commandchars=\\\{\}]
\PY{k+kn}{from} \PY{n+nn}{qiskit}\PY{n+nn}{.}\PY{n+nn}{circuit} \PY{k+kn}{import} \PY{n}{QuantumCircuit}\PY{p}{,} \PY{n}{Parameter}
\PY{k+kn}{from} \PY{n+nn}{qiskit}\PY{n+nn}{.}\PY{n+nn}{compiler} \PY{k+kn}{import} \PY{n}{transpile}

\PY{n}{theta} \PY{o}{=} \PY{n}{Parameter}\PY{p}{(}\PY{l+s+s2}{\PYZdq{}}\PY{l+s+s2}{\PYZdl{}}\PY{l+s+se}{\PYZbs{}\PYZbs{}}\PY{l+s+s2}{theta\PYZdl{}}\PY{l+s+s2}{\PYZdq{}}\PY{p}{)}
\PY{n}{qc} \PY{o}{=} \PY{n}{QuantumCircuit}\PY{p}{(}\PY{l+m+mi}{2}\PY{p}{)}
\PY{n}{qc}\PY{o}{.}\PY{n}{rx}\PY{p}{(}\PY{n}{theta}\PY{p}{,} \PY{l+m+mi}{0}\PY{p}{)}
\PY{n}{qc}\PY{o}{.}\PY{n}{barrier}\PY{p}{(}\PY{p}{)}
\PY{n}{qc}\PY{o}{.}\PY{n}{rz}\PY{p}{(}\PY{n}{theta}\PY{p}{,} \PY{l+m+mi}{0}\PY{p}{)}
\PY{n}{qc}\PY{o}{.}\PY{n}{barrier}\PY{p}{(}\PY{p}{)}
\PY{n}{qc}\PY{o}{.}\PY{n}{rzz}\PY{p}{(}\PY{n}{theta}\PY{p}{,} \PY{l+m+mi}{0}\PY{p}{,} \PY{l+m+mi}{1}\PY{p}{)}
\PY{n}{qc}\PY{o}{.}\PY{n}{draw}\PY{p}{(}\PY{p}{)}
\end{Verbatim}
\end{tcolorbox}

\prompt{Out}{outcolor}{1}{}
    
    \begin{center}
    \adjustimage{max size={0.9\linewidth}{0.9\paperheight}}{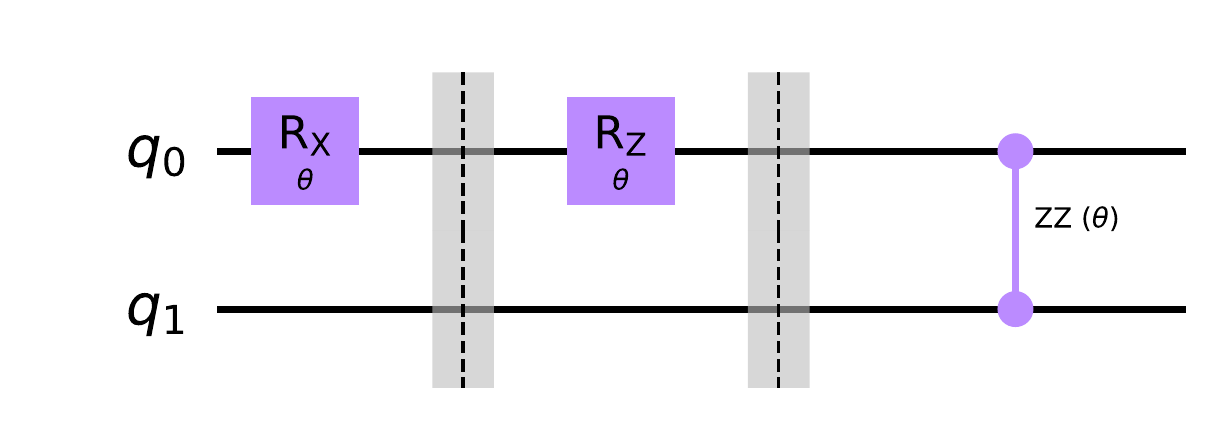}
    \end{center}
    { \hspace*{\fill} \\}

    \begin{tcolorbox}[breakable, size=fbox, boxrule=1pt, pad at break*=1mm,colback=cellbackground, colframe=cellborder]
\prompt{In}{incolor}{2}{\boxspacing}
\begin{Verbatim}[commandchars=\\\{\}]
\PY{c+c1}{\PYZsh{} See Notebook 3}
\PY{n}{basis\PYZus{}gates} \PY{o}{=} \PY{p}{[}\PY{l+s+s1}{\PYZsq{}}\PY{l+s+s1}{id}\PY{l+s+s1}{\PYZsq{}}\PY{p}{,} \PY{l+s+s1}{\PYZsq{}}\PY{l+s+s1}{rz}\PY{l+s+s1}{\PYZsq{}}\PY{p}{,} \PY{l+s+s1}{\PYZsq{}}\PY{l+s+s1}{sx}\PY{l+s+s1}{\PYZsq{}}\PY{p}{,} \PY{l+s+s1}{\PYZsq{}}\PY{l+s+s1}{x}\PY{l+s+s1}{\PYZsq{}}\PY{p}{,} \PY{l+s+s1}{\PYZsq{}}\PY{l+s+s1}{cx}\PY{l+s+s1}{\PYZsq{}}\PY{p}{,} \PY{l+s+s1}{\PYZsq{}}\PY{l+s+s1}{reset}\PY{l+s+s1}{\PYZsq{}}\PY{p}{]}

\PY{n}{qc\PYZus{}transpiled} \PY{o}{=} \PY{n}{transpile}\PY{p}{(}\PY{n}{qc}\PY{p}{,} \PY{n}{basis\PYZus{}gates}\PY{o}{=}\PY{n}{basis\PYZus{}gates}\PY{p}{)}
\PY{n}{qc\PYZus{}transpiled}\PY{o}{.}\PY{n}{draw}\PY{p}{(}\PY{p}{)}
\end{Verbatim}
\end{tcolorbox}

\prompt{Out}{outcolor}{2}{}
    
    \begin{center}
    \adjustimage{max size={0.9\linewidth}{0.9\paperheight}}{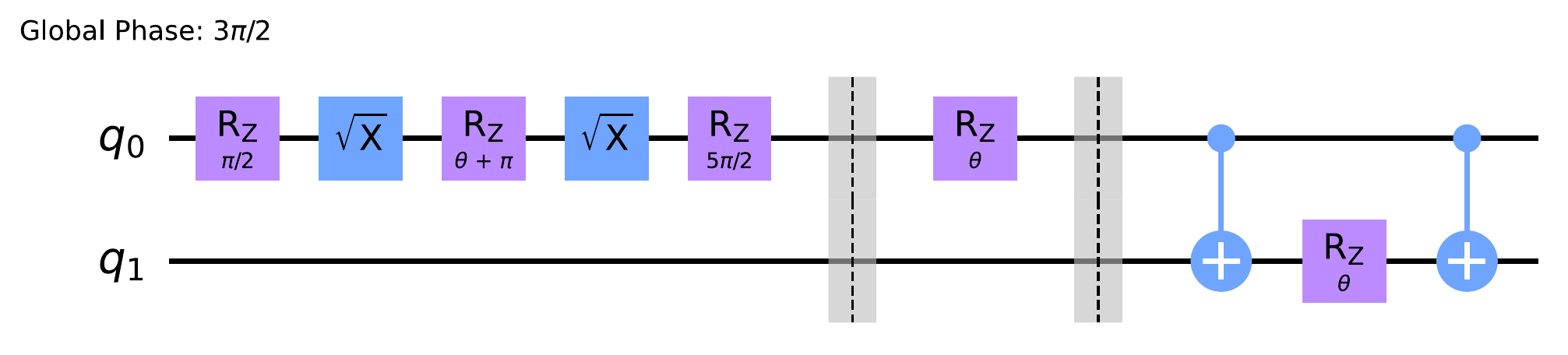}
    \end{center}
    { \hspace*{\fill} \\}

    Note that the \href{https://qiskit.org/documentation/stubs/qiskit.circuit.library.RZGate.html}{\(\RZ\) gate} is a virtual gate and let us thus call the
set of basis gates without \(\RZ\) the \textbf{hardware gates}. Then, we
see that \(\RX\) requires two single-qubit hardware gates, the \(\RZ\)
gate needs none, and \(\RZZ\) needs \textbf{two two-qubit hardware
gates} (i.e.~the two \(\CNOT\) gates). Recalling from Notebook \ref{chap:notebook-3} that
the \textbf{\(\CNOT\) gate} is the \textbf{most erroneous hardware gate}
(around one magnitude higher error rate than the two single qubit gates)
explains why we have to lay our focus on the \(\RZZ\) gates when
analyzing how well a QAOA circuit can be executed on real quantum
hardware.

    \hypertarget{cnot-gates-and-coupling-map}{%
\subsection{\texorpdfstring{\(\CNOT\) Gates and Coupling
Map}{\textbackslash CNOT Gates and Coupling Map}}\label{cnot-gates-and-coupling-map}}

\begin{figure}
    \begin{center}
        \includegraphics[width=7cm]{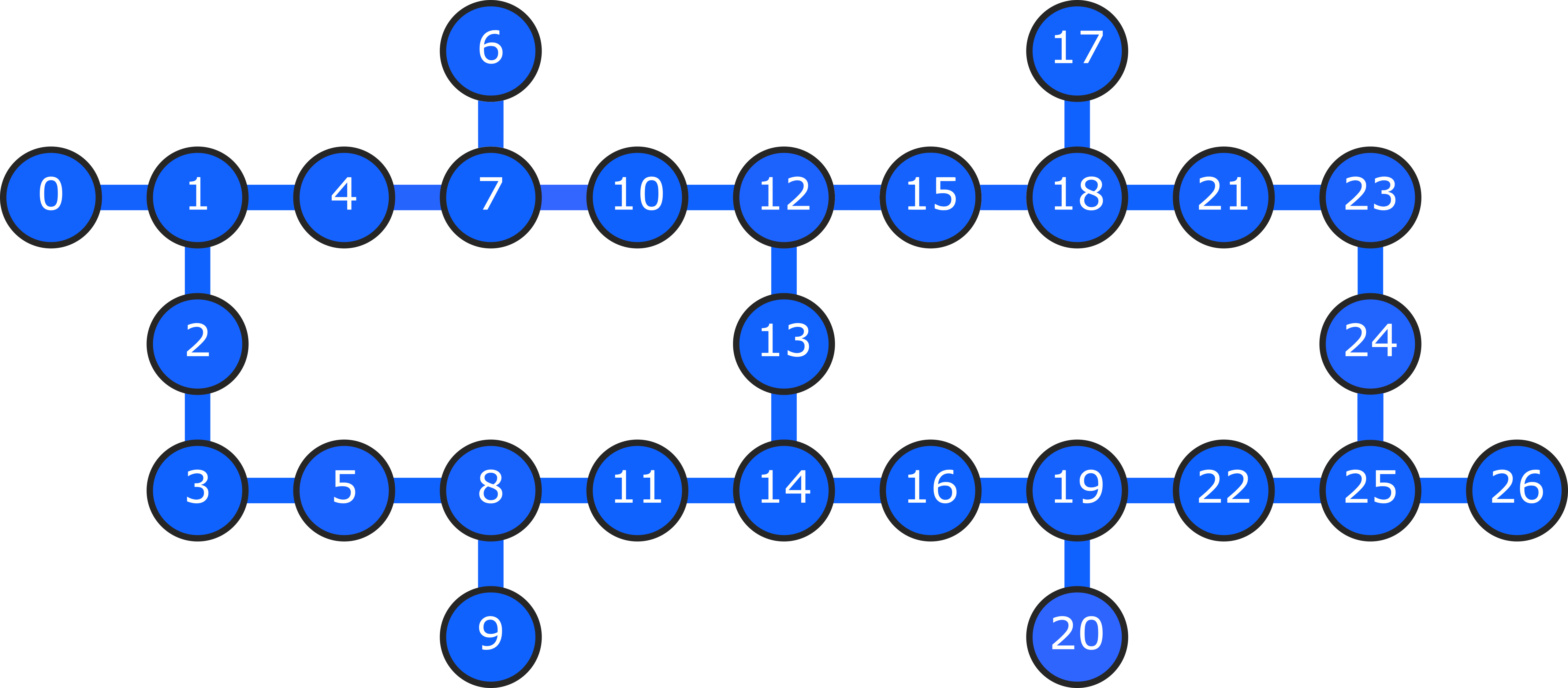}
        \caption{Coupling map of ibmq\_ehningen.}
        \label{fig:notebook-4-coupling-map-ehningen}
    \end{center}
\end{figure}

    Recall from Notebook \ref{chap:notebook-3} that not all qubits in a quantum computer are
connected with each other (see also the coupling map in Figure \ref{fig:notebook-4-coupling-map-ehningen}). So, let's see how a
\(\CNOT\) gate is transpiled for two qubits that are not connected.

    \begin{tcolorbox}[breakable, size=fbox, boxrule=1pt, pad at break*=1mm,colback=cellbackground, colframe=cellborder]
\prompt{In}{incolor}{3}{\boxspacing}
\begin{Verbatim}[commandchars=\\\{\}]
\PY{n}{qc} \PY{o}{=} \PY{n}{QuantumCircuit}\PY{p}{(}\PY{l+m+mi}{3}\PY{p}{)}
\PY{n}{qc}\PY{o}{.}\PY{n}{cx}\PY{p}{(}\PY{l+m+mi}{0}\PY{p}{,} \PY{l+m+mi}{2}\PY{p}{)}
\PY{n}{qc}\PY{o}{.}\PY{n}{draw}\PY{p}{(}\PY{p}{)}
\end{Verbatim}
\end{tcolorbox}

\prompt{Out}{outcolor}{3}{}
    
    \begin{center}
    \adjustimage{max size={0.9\linewidth}{0.9\paperheight}}{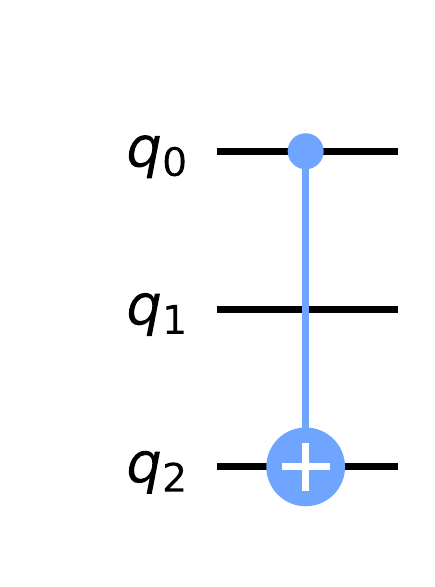}
    \end{center}
    { \hspace*{\fill} \\}

    \begin{tcolorbox}[breakable, size=fbox, boxrule=1pt, pad at break*=1mm,colback=cellbackground, colframe=cellborder]
\prompt{In}{incolor}{4}{\boxspacing}
\begin{Verbatim}[commandchars=\\\{\}]
\PY{c+c1}{\PYZsh{} Only qubits 0\PYZhy{}1, 1\PYZhy{}2 are connected.}
\PY{n}{coupling\PYZus{}map}\PY{o}{=}\PY{p}{[}\PY{p}{[}\PY{l+m+mi}{0}\PY{p}{,} \PY{l+m+mi}{1}\PY{p}{]}\PY{p}{,} \PY{p}{[}\PY{l+m+mi}{1}\PY{p}{,} \PY{l+m+mi}{0}\PY{p}{]}\PY{p}{,} \PY{p}{[}\PY{l+m+mi}{1}\PY{p}{,} \PY{l+m+mi}{2}\PY{p}{]}\PY{p}{,} \PY{p}{[}\PY{l+m+mi}{2}\PY{p}{,} \PY{l+m+mi}{1}\PY{p}{]}\PY{p}{]}
\PY{n}{qc\PYZus{}transpiled} \PY{o}{=} \PY{n}{transpile}\PY{p}{(}
    \PY{n}{qc}\PY{p}{,} 
    \PY{n}{coupling\PYZus{}map}\PY{o}{=}\PY{n}{coupling\PYZus{}map}\PY{p}{,} 
    \PY{n}{initial\PYZus{}layout}\PY{o}{=}\PY{p}{[}\PY{l+m+mi}{0}\PY{p}{,} \PY{l+m+mi}{1}\PY{p}{,} \PY{l+m+mi}{2}\PY{p}{]}\PY{p}{,}
    \PY{n}{seed\PYZus{}transpiler}\PY{o}{=}\PY{l+m+mi}{123}\PY{p}{)}
\PY{n}{qc\PYZus{}transpiled}\PY{o}{.}\PY{n}{draw}\PY{p}{(}\PY{p}{)}
\end{Verbatim}
\end{tcolorbox}

\prompt{Out}{outcolor}{4}{}
    
    \begin{center}
    \adjustimage{max size={0.9\linewidth}{0.9\paperheight}}{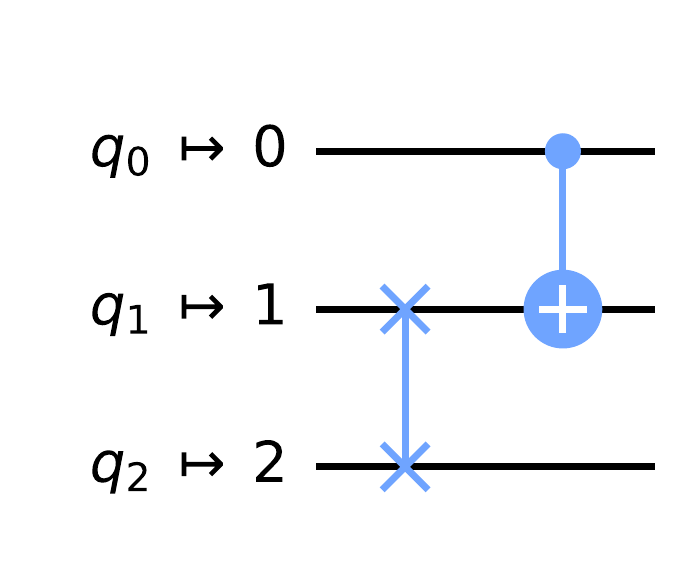}
    \end{center}
    { \hspace*{\fill} \\}

    We see that the transpiler needs to insert a \(\SWAP\) gate between
qubits 1 and 2 in order to realize the \(\CNOT\) gate between qubits 0
and 2. Clearly, this raises the question how a \(\SWAP\) gate is
transpiled to basis gates:

    \begin{tcolorbox}[breakable, size=fbox, boxrule=1pt, pad at break*=1mm,colback=cellbackground, colframe=cellborder]
\prompt{In}{incolor}{5}{\boxspacing}
\begin{Verbatim}[commandchars=\\\{\}]
\PY{n}{qc\PYZus{}transpiled} \PY{o}{=} \PY{n}{transpile}\PY{p}{(}
    \PY{n}{qc\PYZus{}transpiled}\PY{p}{,}
    \PY{n}{coupling\PYZus{}map}\PY{o}{=}\PY{n}{coupling\PYZus{}map}\PY{p}{,}
    \PY{n}{initial\PYZus{}layout}\PY{o}{=}\PY{p}{[}\PY{l+m+mi}{0}\PY{p}{,} \PY{l+m+mi}{1}\PY{p}{,} \PY{l+m+mi}{2}\PY{p}{]}\PY{p}{,}
    \PY{n}{basis\PYZus{}gates}\PY{o}{=}\PY{n}{basis\PYZus{}gates}\PY{p}{,}
    \PY{n}{seed\PYZus{}transpiler}\PY{o}{=}\PY{l+m+mi}{123}\PY{p}{)}
\PY{n}{qc\PYZus{}transpiled}\PY{o}{.}\PY{n}{draw}\PY{p}{(}\PY{p}{)}
\end{Verbatim}
\end{tcolorbox}

\prompt{Out}{outcolor}{5}{}
    
    \begin{center}
    \adjustimage{max size={0.9\linewidth}{0.9\paperheight}}{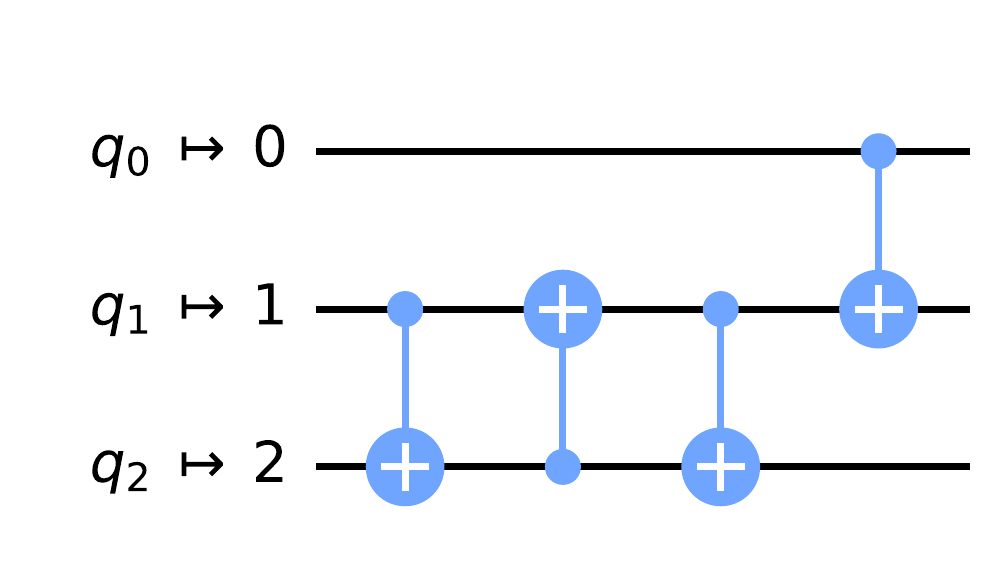}
    \end{center}
    { \hspace*{\fill} \\}

    We see that a \textbf{\(\mathrm{SWAP}\) gate} requires \textbf{three
\(\CNOT\) gates}! Having in mind the limited coupling map of current
(IBM) quantum computers it is easy to imagine that the denser the QUBO
matrix \(A\) is populated (i.e.~the more non-zero entries) the more
\(\mathrm{SWAP}\) gates (and therefore \(\CNOT\) gates) are needed
because many \(\RZZ\) gates between many different qubits have to be
transpiled.

    Now, let us observe these effects on our examples.

    \hypertarget{example-series-1}{%
\subsection{Example Series 1}\label{sec:results-example-series-1}}

    \hypertarget{sparsity-pattern-of-qubo-matrix}{%
\subsubsection{Sparsity Pattern of QUBO Matrix}
\label{sec:notebook-4-sparsity-pattern-qubo-matrix-1}}

    We begin by recalling Example Series 1:

\begin{figure}
    \begin{center}
        \includegraphics{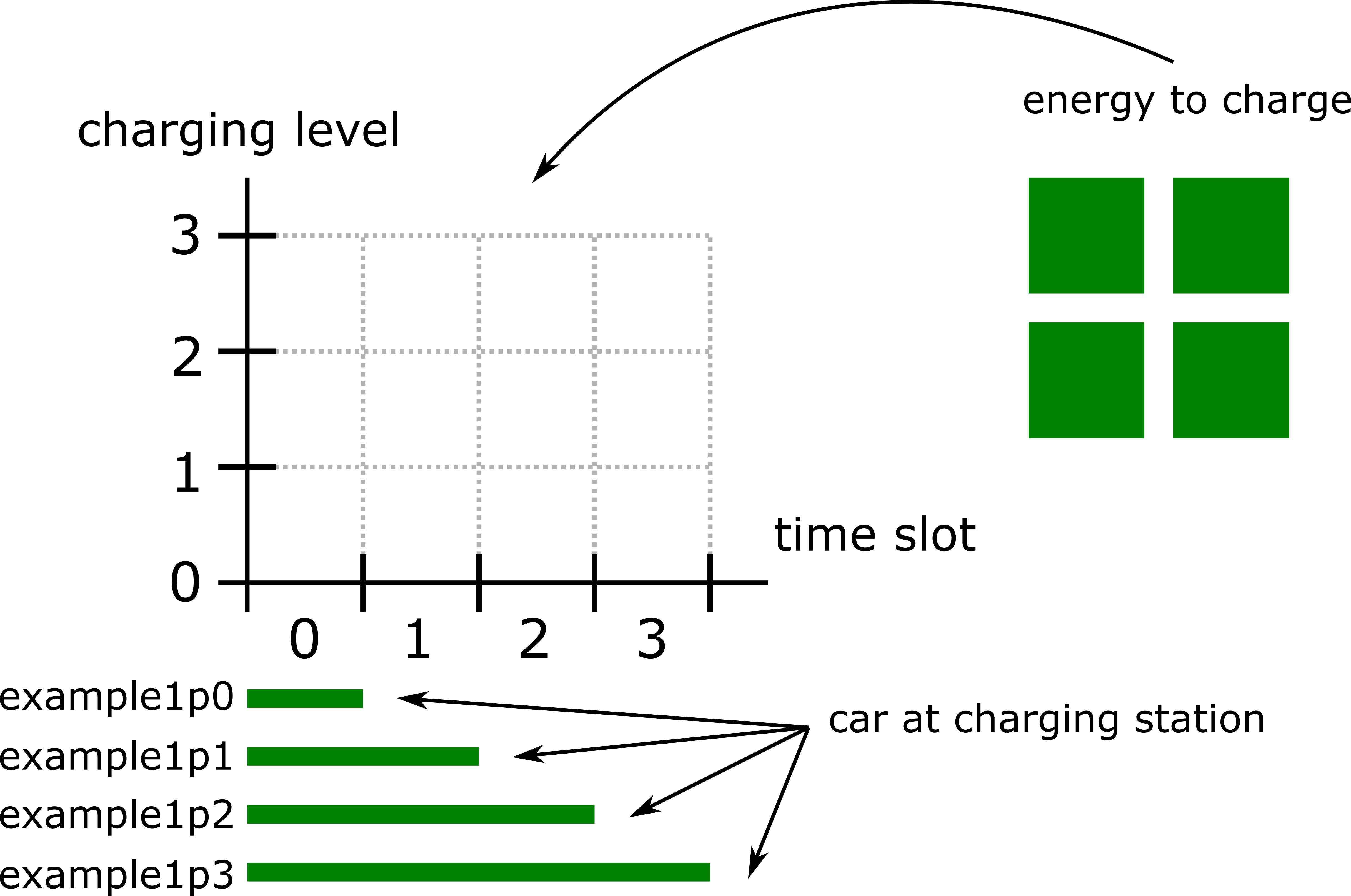}
        \caption{Reminder of Example series 1, see Section \ref{sec:example-series-1}.}
        % \label{}
    \end{center}
\end{figure}

    The next plot shows the sparsity pattern of the QUBO matrices.

\begin{figure}
    \begin{center}
        \includegraphics{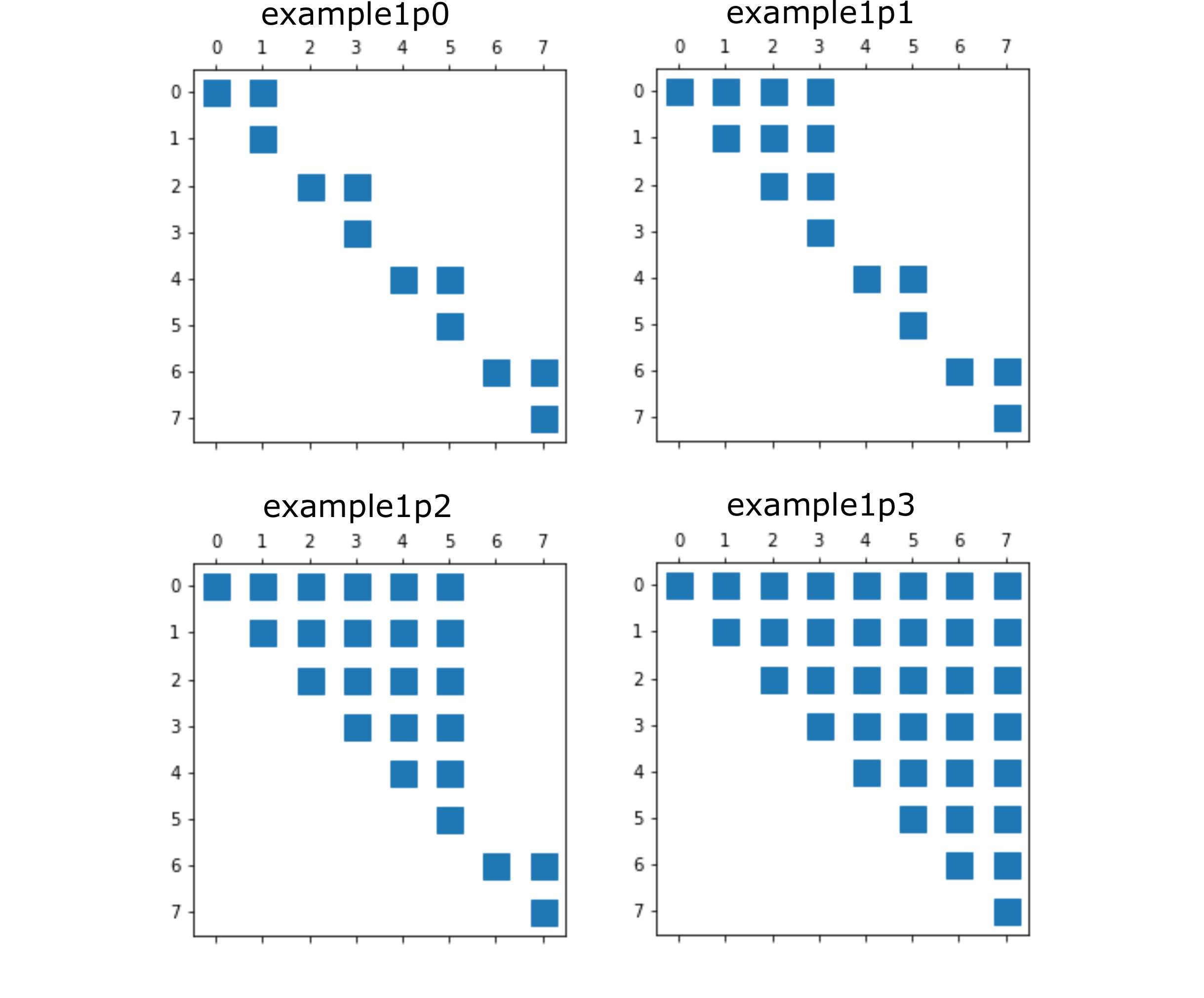}
        \caption{Sparsity pattern of QUBO matrices for Example Series 1.}
        \label{fig:notebook-4-sparsity-qubo-matrix-1}
    \end{center}
\end{figure}

    For these simple examples we can easily derive the structure of the QUBO
matrices: We need to represent the charging level of the car for every
time slot 0, 1, 2 and 3. Since we have 4 charging levels we need 2
qubits to represent the charging level for every time slot. So, we need
4 \(\cdot\) 2 = 8 qubits for every example in Example Series 1.

In the plot for example1p0 you can see that we only have a coupling
(off-diagonal element) between qubits 0 -- 1, 2 -- 3, 4 -- 5, and 6 -- 7.
Every two qubit pair represents the charging level for a time slot.
Since the car in this example is only at the charging station at time
slot 0 no coupling between the different time slots is necessary. This
changes for example1p1 where we see a coupling between qubits 0 -- 1 -- 2
-- 3. This stems from the fact that in this example the car is at the
charging station at time slots 0 and 1, and thus also the qubits involved
must be coupled. And so it goes on until example1p3, where the car is at
the charging station for all the time slots and we observe a full
coupling between all the qubits.

    Now, let us investigate how the sparsity pattern of the different
examples affect the count of basis gates for QAOA with \(p=1\).

    \hypertarget{gate-count-fully-connected-topology}{%
\subsubsection{Gate Count: Fully Connected
Topology}\label{sec:gate-count-fully-connected-topology}}

    First, let us assume that we have a \textbf{fully connected topology},
i.e.~each qubit is connected with all other qubits. In the following
figure we plot the number of hardware gates, i.e.~of \(\X\), \(\SX\) and
\(\CNOT\) (in Qiskit: \texttt{x}, \texttt{sx} and \texttt{cx}), and the
depth of the circuit for all subexamples of Example Series 1.

\begin{figure}
    \begin{center}
        \includegraphics{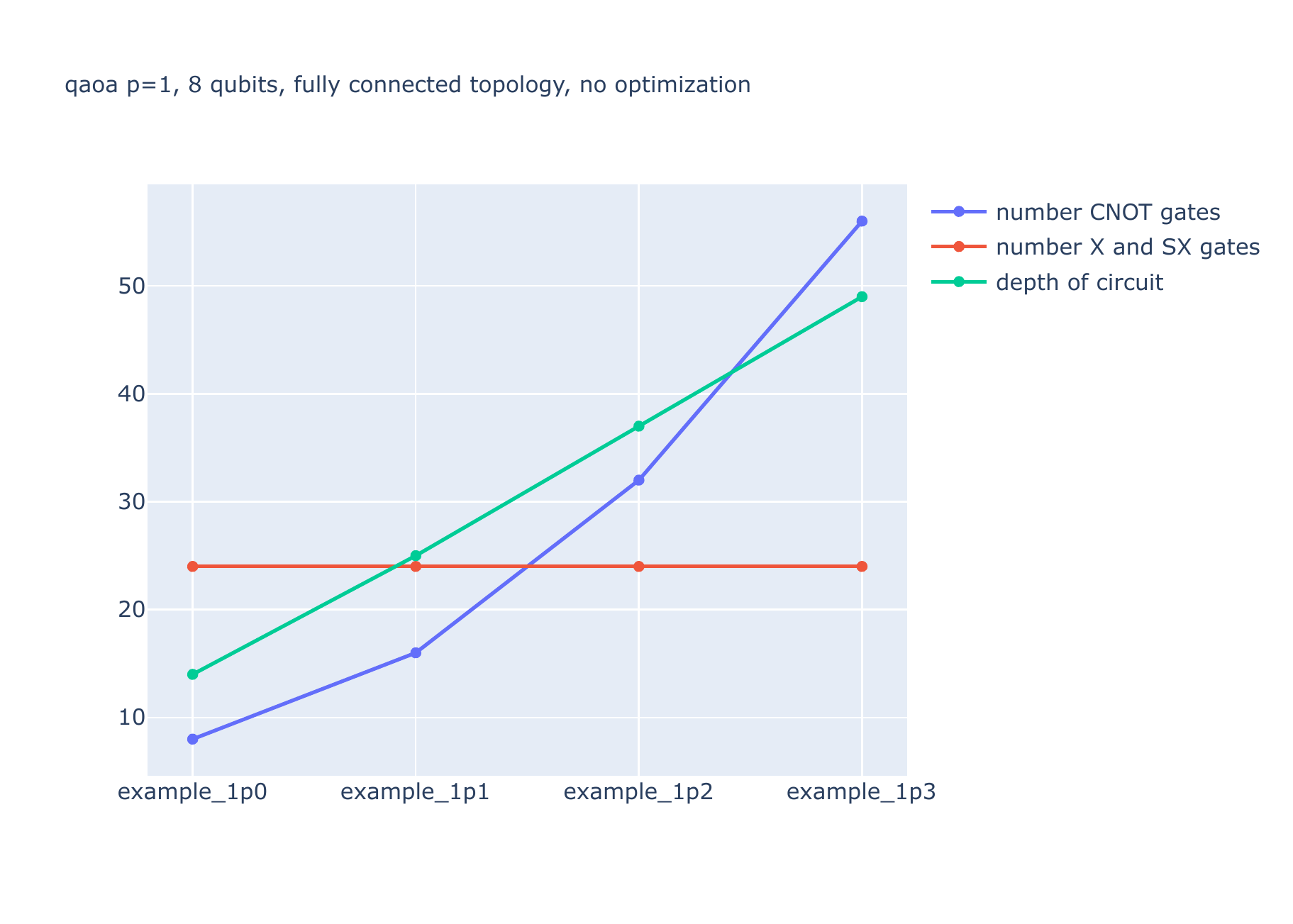}
        \caption{Number of $\X$, $\SX$, and $\CNOT$ gates and depth of the transpiled circuits
        for Example Series 1 for a fully connected topology.}
        \label{fig:notebook-4-gate-count-1}
    \end{center}
\end{figure}

    We see that the number of \(\CNOT\) gates grows whereas the number of
single qubit gates stays constant. The reason is that the subexamples
only differ in the number of time slots that the car is at the charging
station and (as we discussed above) this directly affects the number of
\(\RZZ\) gates which then directly translate to the number of \(\CNOT\)
gates.

\begin{center}\rule{0.5\linewidth}{0.5pt}\end{center}
Code snippet:

\begin{tcolorbox}[breakable, size=fbox, boxrule=1pt, pad at break*=1mm,colback=cellbackground, colframe=cellborder]
\begin{Verbatim}[commandchars=\\\{\}]
\PY{c+c1}{\PYZsh{} See Notebook 3.}
\PY{n}{backend\PYZus{}basis\PYZus{}gates} \PY{o}{=} \PY{p}{[}\PY{l+s+s2}{\PYZdq{}}\PY{l+s+s2}{id}\PY{l+s+s2}{\PYZdq{}}\PY{p}{,} \PY{l+s+s2}{\PYZdq{}}\PY{l+s+s2}{rz}\PY{l+s+s2}{\PYZdq{}}\PY{p}{,} \PY{l+s+s2}{\PYZdq{}}\PY{l+s+s2}{sx}\PY{l+s+s2}{\PYZdq{}}\PY{p}{,} \PY{l+s+s2}{\PYZdq{}}\PY{l+s+s2}{x}\PY{l+s+s2}{\PYZdq{}}\PY{p}{,} \PY{l+s+s2}{\PYZdq{}}\PY{l+s+s2}{cx}\PY{l+s+s2}{\PYZdq{}}\PY{p}{,} \PY{l+s+s2}{\PYZdq{}}\PY{l+s+s2}{reset}\PY{l+s+s2}{\PYZdq{}}\PY{p}{]}
\PY{n}{backend\PYZus{}single\PYZus{}qubit\PYZus{}basis\PYZus{}gates} \PY{o}{=} \PY{p}{[}\PY{l+s+s2}{\PYZdq{}}\PY{l+s+s2}{sx}\PY{l+s+s2}{\PYZdq{}}\PY{p}{,} \PY{l+s+s2}{\PYZdq{}}\PY{l+s+s2}{x}\PY{l+s+s2}{\PYZdq{}}\PY{p}{]}

\PY{n}{qaoa\PYZus{}ansatz\PYZus{}transpiled} \PY{o}{=} \PY{n}{transpile}\PY{p}{(}
    \PY{n}{qaoa\PYZus{}ansatz}\PY{p}{,}
    \PY{n}{basis\PYZus{}gates}\PY{o}{=}\PY{n}{backend\PYZus{}basis\PYZus{}gates}\PY{p}{,}
    \PY{n}{optimization\PYZus{}level}\PY{o}{=}\PY{l+m+mi}{0}\PY{p}{)} \PY{c+c1}{\PYZsh{} has no effect in this case}

\PY{n}{number\PYZus{}cnots} \PY{o}{=} \PY{n}{qaoa\PYZus{}ansatz\PYZus{}transpiled}\PY{o}{.}\PY{n}{count\PYZus{}ops}\PY{p}{(}\PY{p}{)}\PY{p}{[}\PY{l+s+s1}{\PYZsq{}}\PY{l+s+s1}{cx}\PY{l+s+s1}{\PYZsq{}}\PY{p}{]}
\PY{c+c1}{\PYZsh{} The function count\PYZus{}gates is provided in utils.}
\PY{n}{number\PYZus{}sx\PYZus{}x} \PY{o}{=} \PY{n}{count\PYZus{}gates}\PY{p}{(}
    \PY{n}{qaoa\PYZus{}ansatz\PYZus{}transpiled}\PY{p}{,} 
    \PY{n}{backend\PYZus{}single\PYZus{}qubit\PYZus{}basis\PYZus{}gates}\PY{p}{)}
\PY{n}{depth} \PY{o}{=} \PY{n}{qaoa\PYZus{}ansatz\PYZus{}transpiled}\PY{o}{.}\PY{n}{depth}\PY{p}{(}\PY{p}{)}
\end{Verbatim}
\end{tcolorbox}

\begin{center}\rule{0.5\linewidth}{0.5pt}\end{center}

    \hypertarget{gate-count-ibmq_ehningen}{%
\subsubsection{Gate Count:
ibmq\_ehningen}\label{sec:gate-count-ibmq_ehningen}}

    Next, we transpile the QAOA circuits for our example series to the
\textbf{ibmq\_ehningen backend}. The hardware gates are the same as
above but recall the \textbf{limited connectivity between the qubits}
(see Figure \ref{fig:notebook-4-coupling-map-ehningen}) that makes it necessary to introduce \(\SWAP\)
gates (which we have seen are transpiled to three \(\CNOT\) gates).

\begin{figure}
    \begin{center}
        \includegraphics{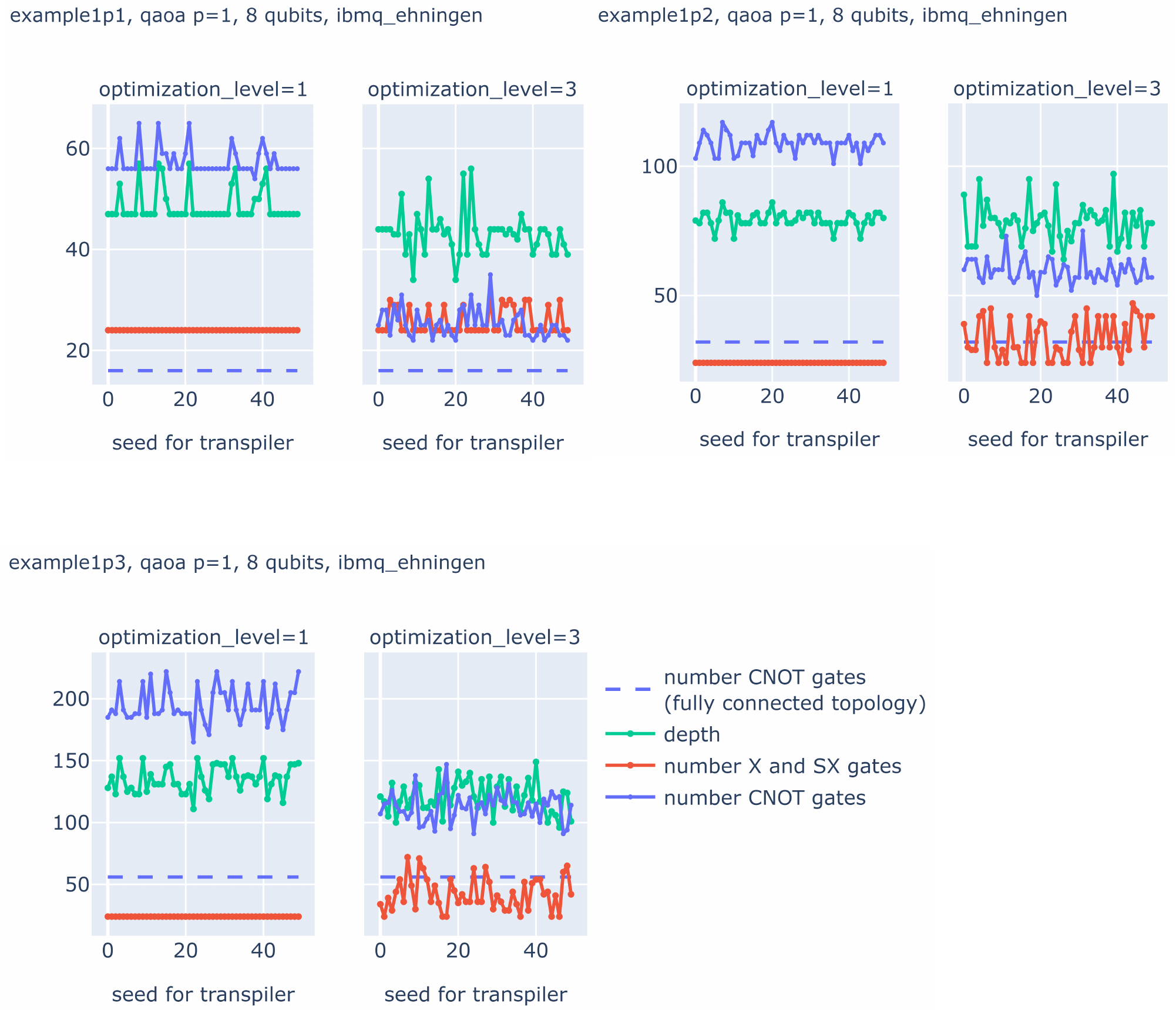}
        \caption{Number of $\X$, $\SX$, and $\CNOT$ gates and depth of transpiled circuits
        for Example Series 1 for ibmq\_ehningen. We used 50 different transpiler seeds and optimization levels 1 and 3.}
        \label{fig:notebook-4-gate-count-ehningen}
    \end{center}
\end{figure}

    Figure \ref{fig:notebook-4-gate-count-ehningen} shows the results for Example Series 1 transpiled to
ibmq\_ehningen.

    As discussed in Notebook \ref{chap:notebook-3} there are many parameters to \textbf{control
the transpilation process}. Two important ones are the
\textbf{\texttt{optimization\_level}} and the
\textbf{\texttt{seed\_transpiler}}. For the later we already explained
that it controls the stochastic part of the transpilation process.
Looking at the figures above we see that this can make a significant
difference. As demonstrated in Notebook \ref{chap:notebook-3} it is thus a \textbf{good
practice} to run a transpilation several times and \textbf{choose the
circuit with the least \(\CNOT\) gates}. Moreover, we see that a
transpilation can be done in many different ways and so there is much
potential for optimization. This is controlled with the
\texttt{optimization\_level}. The higher the number, the more optimized
the transpiled circuit is but at the expense of more computation time.
We refer to \cite{QiskitTranspilerWeb,QiskitTranspilerWeb2,QuantumEnablementWeb}
for further information. Looking at the severe restrictions of NISQ
hardware we would recommend \texttt{optimization\_level=3} in order to
obtain the best circuits and take the most of the current potential. However, keep in mind
that a higher optimization level comes at higher cost on the classical computer.

\begin{center}\rule{0.5\linewidth}{0.5pt}\end{center}
Code snippet:

\begin{tcolorbox}[breakable, size=fbox, boxrule=1pt, pad at break*=1mm,colback=cellbackground, colframe=cellborder]
\begin{Verbatim}[commandchars=\\\{\}]
\PY{n}{ibmq\PYZus{}ehningen} \PY{o}{=} \PY{n}{provider}\PY{o}{.}\PY{n}{get\PYZus{}backend}\PY{p}{(}\PY{l+s+s2}{\PYZdq{}}\PY{l+s+s2}{ibmq\PYZus{}ehningen}\PY{l+s+s2}{\PYZdq{}}\PY{p}{)}
\PY{n}{backend\PYZus{}coupling\PYZus{}map} \PY{o}{=} \PY{n}{ibmq\PYZus{}ehningen}\PY{o}{.}\PY{n}{configuration}\PY{p}{(}\PY{p}{)}\PY{o}{.}\PY{n}{coupling\PYZus{}map}
\PY{n}{backend\PYZus{}basis\PYZus{}gates} \PY{o}{=} \PY{n}{ibmq\PYZus{}ehningen}\PY{o}{.}\PY{n}{configuration}\PY{p}{(}\PY{p}{)}\PY{o}{.}\PY{n}{basis\PYZus{}gates}

\PY{n}{seeds\PYZus{}for\PYZus{}transpiler} \PY{o}{=} \PY{p}{[}\PY{n}{k} \PY{k}{for} \PY{n}{k} \PY{o+ow}{in} \PY{n+nb}{range}\PY{p}{(}\PY{l+m+mi}{50}\PY{p}{)}\PY{p}{]}
\PY{n}{number\PYZus{}seeds\PYZus{}for\PYZus{}transpiler} \PY{o}{=} \PY{n+nb}{len}\PY{p}{(}\PY{n}{seeds\PYZus{}for\PYZus{}transpiler}\PY{p}{)}

\PY{n}{quantum\PYZus{}circuit\PYZus{}transpilations} \PY{o}{=} \PY{n}{transpile}\PY{p}{(}
    \PY{p}{[}\PY{n}{qaoa\PYZus{}ansatz}\PY{p}{]}\PY{o}{*}\PY{n}{number\PYZus{}seeds\PYZus{}for\PYZus{}transpiler}\PY{p}{,}
    \PY{n}{coupling\PYZus{}map}\PY{o}{=}\PY{n}{backend\PYZus{}coupling\PYZus{}map}\PY{p}{,}
    \PY{n}{basis\PYZus{}gates}\PY{o}{=}\PY{n}{backend\PYZus{}basis\PYZus{}gates}\PY{p}{,}
    \PY{n}{optimization\PYZus{}level}\PY{o}{=}\PY{l+m+mi}{3}\PY{p}{,} \PY{c+c1}{\PYZsh{} 0, 1, 2 or 3}
    \PY{n}{seed\PYZus{}transpiler}\PY{o}{=}\PY{n}{seeds\PYZus{}for\PYZus{}transpiler}\PY{p}{)}
\end{Verbatim}
\end{tcolorbox}

\begin{center}\rule{0.5\linewidth}{0.5pt}\end{center}

    \hypertarget{gate-count-comparison}{%
\subsubsection{Gate Count: Comparison}\label{sec:gate-count-comparison}}

    As a last plot for Example Series 1 let us \textbf{compare} the
\textbf{number of \(\CNOT\) gates} between the \textbf{fully connected
topology} and \textbf{ibmq\_ehningen}:

\begin{figure}
    \begin{center}
        \includegraphics{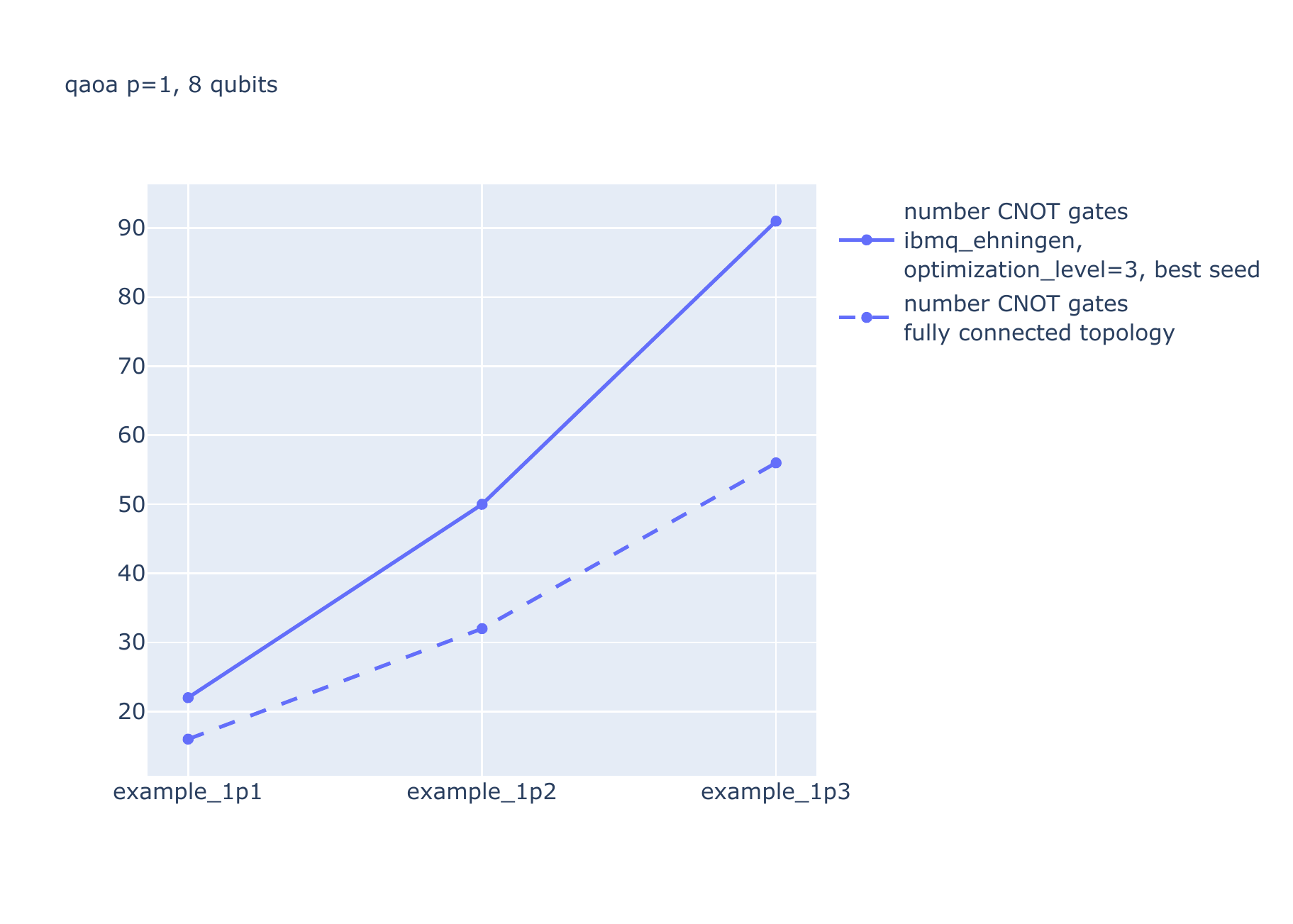}
        \caption{Comparison of number of $\CNOT$ gates for Example Series 1
        between a fully connected topology and the best transpilation for ibmq\_ehningen.}
        \label{fig:notebook-4-cnots-fully-connected-ehningen}
    \end{center}
\end{figure}

    We see that the difference in the number of \(\CNOT\) gates
\textbf{diverges} from example\_1p1 over example\_1p2 to example\_1p3.
This has the reason that the \textbf{coupling between the
variables} (i.e.~the non-zero entries in the QUBO matrix) increases from
example\_1p1 to example\_1p3 and that this coupling is realized via a \(\RZZ\) gates
between the corresponding qubits. The more qubits are connected via
\(\RZZ\) gates the more \(\SWAP\) gates have to be used in order to compensate
for the limited connectivity of ibmq\_ehningen and thus the divergence.

    \hypertarget{example-series-2}{%
\subsection{Example Series 2}\label{sec:results-example-series-2}}

    Now, we present the figures for Example Series 2 for the same experiments
as for Example Series 1. We will observe the same effects and thus will
only make a few comments.

    \hypertarget{sparsity-pattern-of-qubo-matrix}{%
\subsubsection{Sparsity Pattern of QUBO
Matrix}\label{sec:notebook-4-sparsity-pattern-of-qubo-matrix-2}}

As for Example Series 1 we start by recalling Example Series 2:

\begin{figure}
    \begin{center}
        \includegraphics{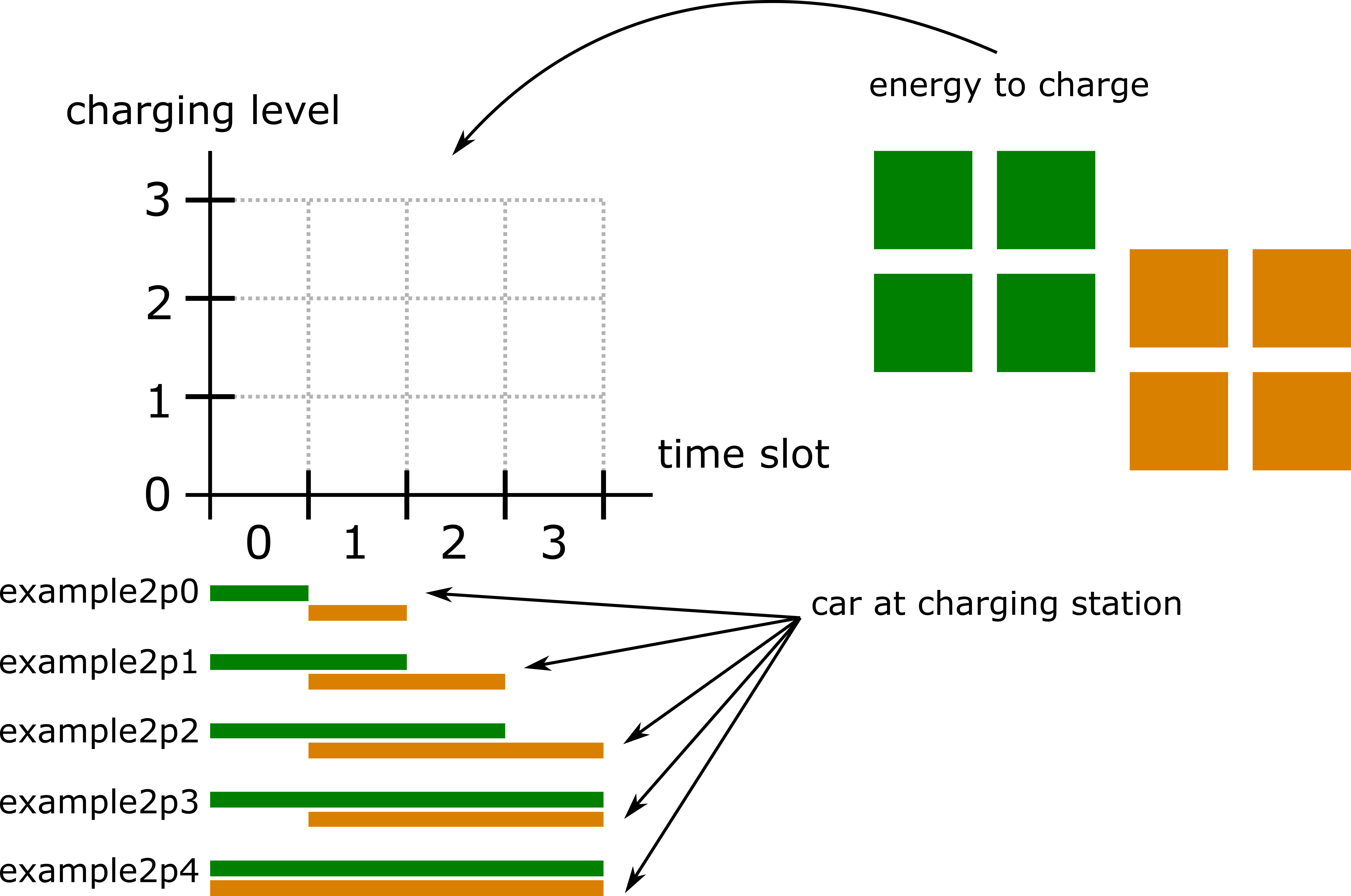}
        \caption{Reminder of Example Series 2, see Section \ref{sec:example-series-2}.}
        % \label{}
    \end{center}
\end{figure}

The sparsity pattern of the QUBO matrices is given in the next figure:
\begin{figure}
    \begin{center}
        \includegraphics{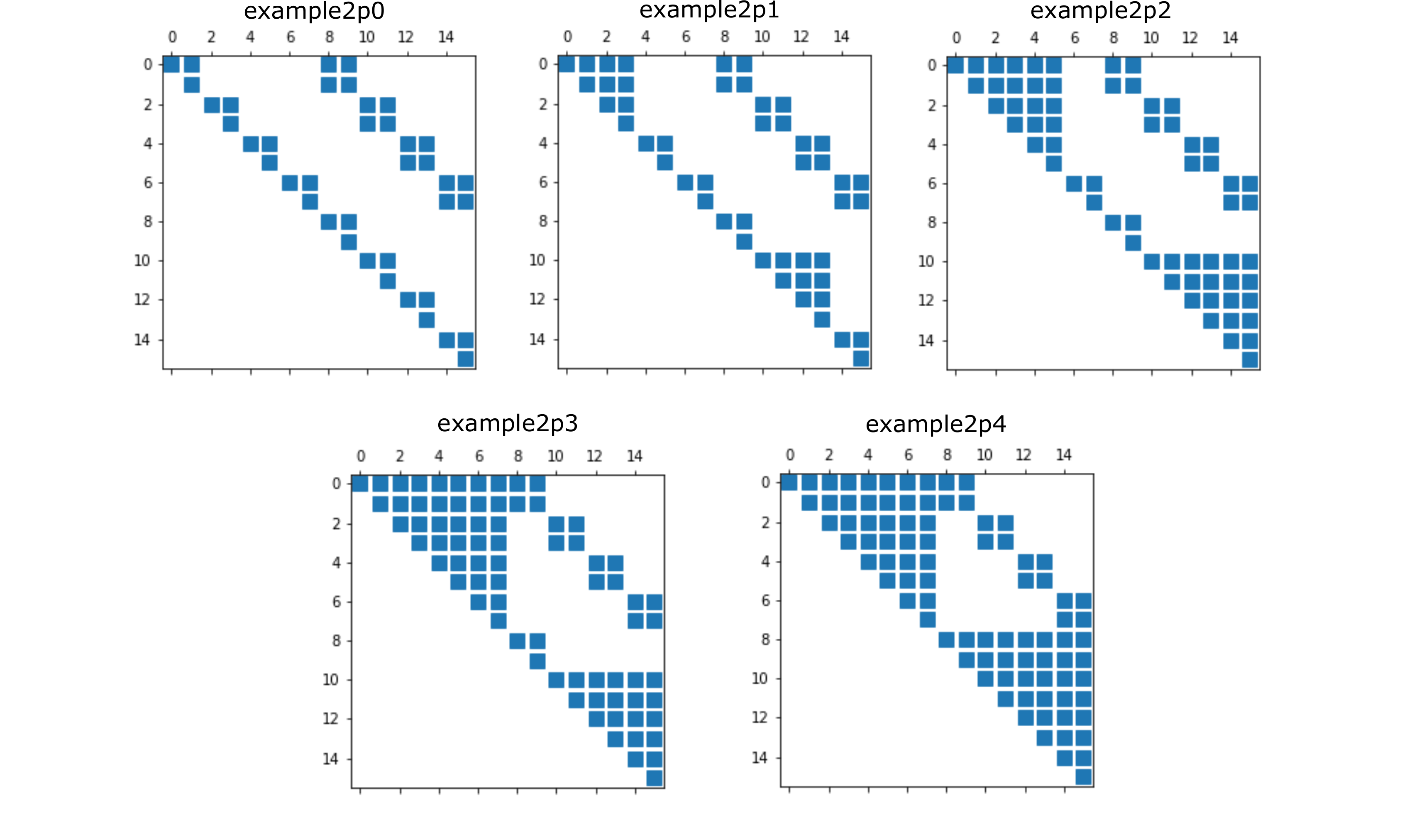}
        \caption{Sparsity pattern of the QUBO matrices for Example Series 2.}
        \label{fig:notebook-4-sparsity-pattern-2}
    \end{center}
\end{figure}

    Note the following differences to Examples Series 1: Clearly, we now
need 16 qubits -- 8 qubits for the green car and 8 qubits for the orange
car. Moreover, note that we have coupling of qubits 0 -- 1 -- 8 -- 9, 2 -- 3 -- 10 -- 11,
4 -- 5 -- 12 -- 13, 6 -- 7 -- 14 -- 15. These are needed for every time slot to add the
charging level for the green and the orange car. Altogether, we see a
more complicated sparsity pattern of the QUBO matrix and expect that
this will translate to deeper circuits with more gates and eventually a
poorer quality from results of real quantum computers.

% What we discussed for Example Series 1 also applies to the now following plots. So, we just present the plots and don't repeat the upper analysis.

    \hypertarget{gate-count-fully-connected-topology}{%
\subsubsection{Gate Count: Fully Connected
Topology}\label{sec:notebook-4-gate-count-fully-connected-topology-2}}

\begin{figure}
    \begin{center}
        \includegraphics{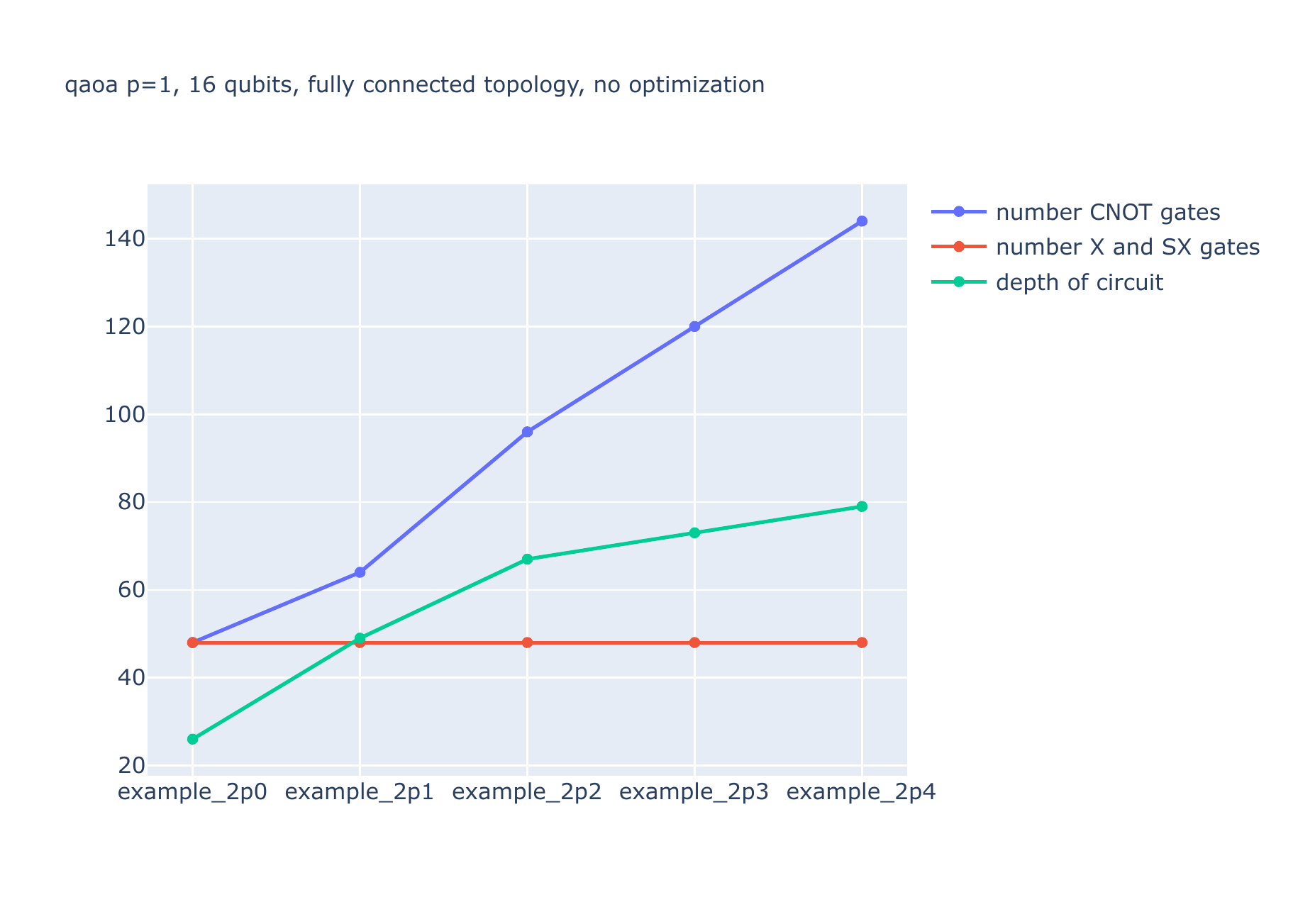}
        \caption{Number of $\X$, $\SX$, and $\CNOT$ gates and depth of transpiled circuits
        for Example Series 2 for a fully connected topology.}
        \label{fig:notebook-4-number-gates-fully-connected-2}
    \end{center}
\end{figure}

    \hypertarget{gate-count-ibmq_ehningen}{%
\subsubsection{Gate Count:
ibmq\_ehningen}\label{sec:notebook-4-gate-count-ibmq_ehningen-2}}

\begin{figure}
    \begin{center}
        \includegraphics{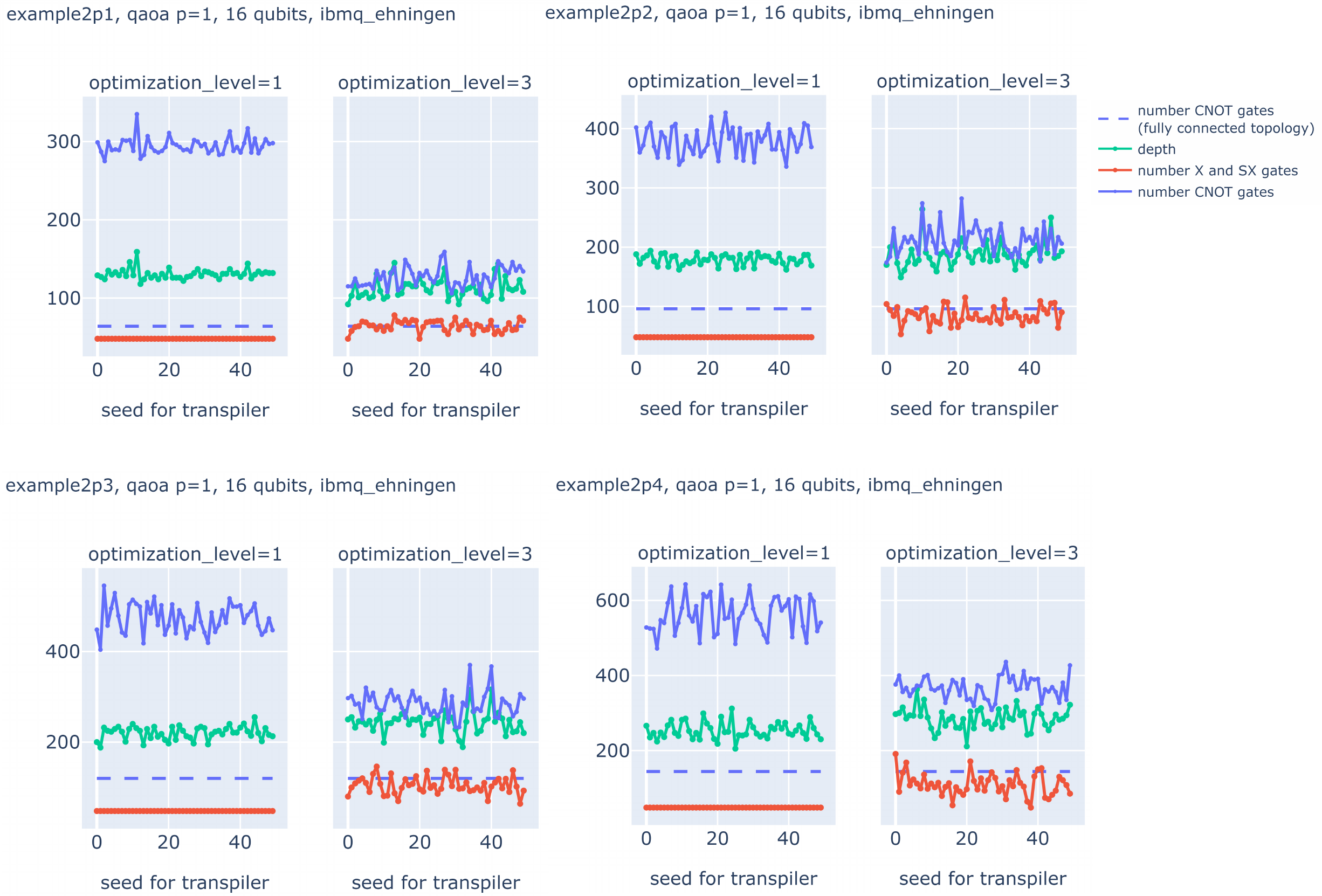}
        \caption{Number of $\X$, $\SX$, and $\CNOT$ gates and depth of transpiled circuits
        for Example Series 2 for ibmq\_ehningen. We used 50 different transpiler seeds and optimization levels 1 and 3.}
        \label{fig:notebook-4-number-gates-ehningen-2}
    \end{center}
\end{figure}

    \hypertarget{gate-count-comparison}{%
\subsubsection{Gate Count: Comparison}\label{sec:notebook-4-gate-count-comparison-2}}

\begin{figure}
    \begin{center}
        \includegraphics{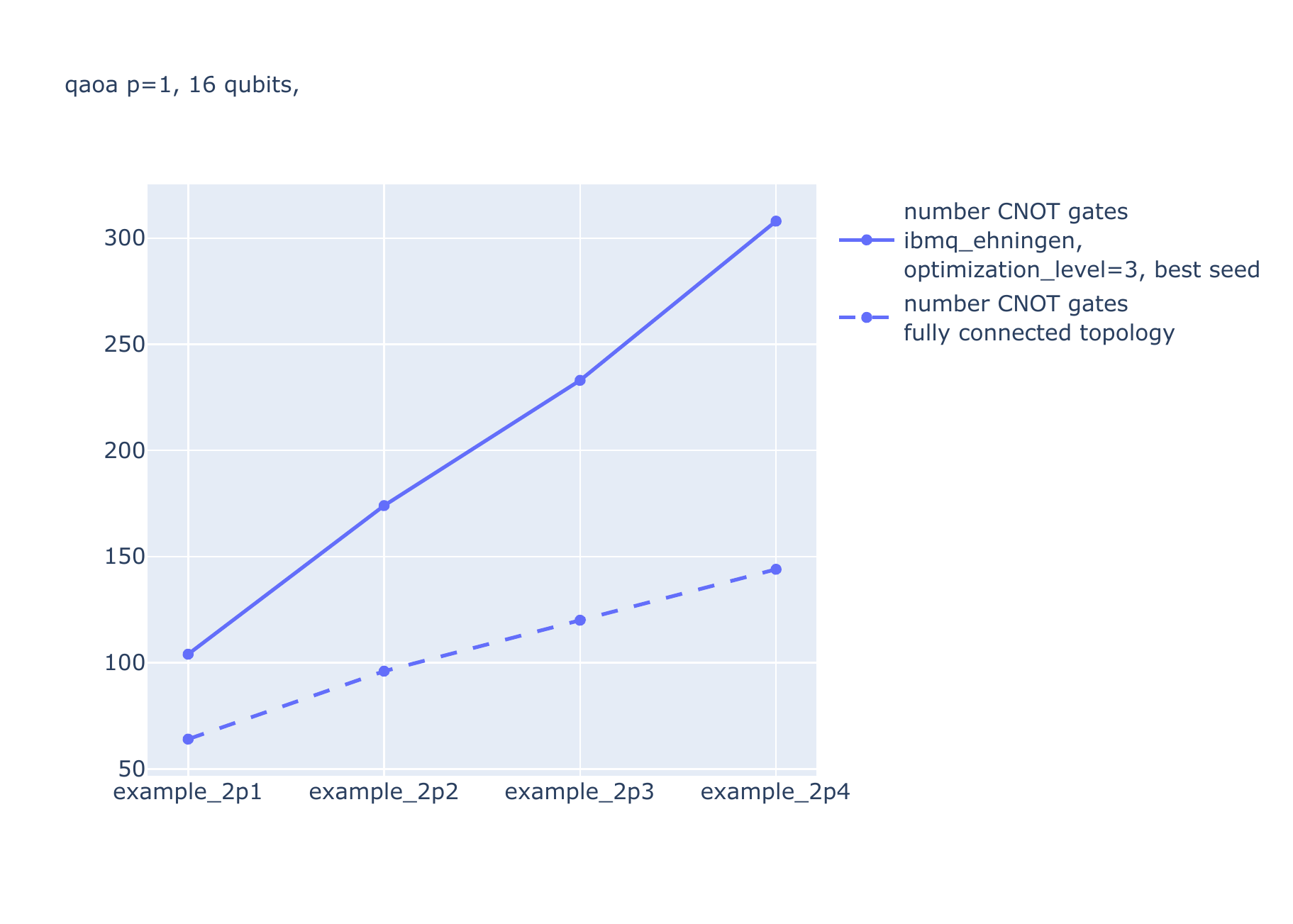}
        \caption{Comparison of number of $\CNOT$ gates for Example Series 2
        between a fully connected topology and the best transpilation for ibmq\_ehningen.
        Here we see the divergence in the number of \(\CNOT\) gates
        between a fully connected quantum computer and ibmq\_ehningen even
        stronger than in Example Series 1.}
        \label{fig:notebook-4-cnots-fully-connected-ehningen-2}
    \end{center}
\end{figure}

    In the last part of this notebook we present results from solving
Example Series 1 and 2 with QAOA on ibmq\_ehningen.

    \hypertarget{results-qaoa-on-ibmq_ehnigen}{%
\section{QAOA Results on
ibmq\_ehnigen}\label{sec:results-qaoa-on-ibmq_ehnigen}}

Our set up for the experiments reported below is the following:
\begin{itemize}
    \tightlist
        \item We use the \textbf{best parameters} \(\vec{\beta}\), \(\vec{\gamma}\), and 
        \(\varrho\) that we found in our simulator experiments in Section
        \ref{sec:notebook-4-classical-optimization}, see Table \ref{tab:paramters-ehningen-experiments}. Thus, we can analyze the \textbf{performance of
        ibmq\_ehningen} for the \textbf{best possible QAOA circuits}. 
        \item We \textbf{transpile} the QAOA circuits with \textbf{75 different seeds}
        and \textbf{optimization level 3}. This transpilation is abbreviated
        with \(\mathrm{std}\) in the figures below.
        \item We additionally add \textbf{dynamical decoupling} to the transpiled circuits (abbreviation
        then is \(\mathrm{dd}\)).
        \item We employ \textbf{measurement error mitigation} to the \(\mathrm{std}\) and \(\mathrm{dd}\) circuits via the
        package \texttt{mthree} (abbreviation then is \(\mathrm{xx\_mit}\), where
        \(\mathrm{xx}\) = \(\mathrm{std}\) or \(\mathrm{dd}\))
\end{itemize}

For more information on the error mitigation techniques see Notebook \ref{chap:notebook-3}.

    \hypertarget{quality-metric}{%
\subsection{Quality Metric}\label{sec:notebook-4-quality-metric}}

There are many ways how we can measure the quality of the result of a computation on a real quantum computer.
We will use the \textbf{fidelity} and the \textbf{expectation value}.

    \hypertarget{fidelity}{%
\subsubsection{Fidelity}\label{sec:notebook-4-fidelity}}

    In the following we will mainly measure the quality of the probability distribution \(\{q_i\}\)
stemming from running the QAOA circuit on \textbf{ibmq\_ehningen} by
computing the \textbf{fidelity} \(F\) with respect to the probability
distribution \(\{p_i\}\) from an \textbf{exact state vector simulation}. Thereby, the fidelity is defined as

\[
F( \{p_i\}, \{q_i\}) = \sum_i \sqrt{p_i q_i} \ ,
\]

see further \cite[Chapter 9]{NieC10}. We note that the fidelity is between
\(0\) and \(1\), where \(0\) is the worst case and \(1\) is the best
case. In \texttt{Qiskit} the fidelity can be computed via
\href{https://qiskit.org/documentation/stubs/qiskit.quantum_info.hellinger_fidelity.html}{\texttt{qiskit.quantum\_info.hellinger\_fidelity}}.

\begin{center}\rule{0.5\linewidth}{0.5pt}\end{center}
Code snippet:
\begin{tcolorbox}[breakable, size=fbox, boxrule=1pt, pad at break*=1mm,colback=cellbackground, colframe=cellborder]
\begin{Verbatim}[commandchars=\\\{\}]
\PY{k+kn}{from} \PY{n+nn}{qiskit}\PY{n+nn}{.}\PY{n+nn}{quantum\PYZus{}info} \PY{k+kn}{import} \PY{n}{hellinger\PYZus{}fidelity}

\PY{c+c1}{\PYZsh{} See the end of Notebook 3 for experiment\PYZus{}df}
\PY{n}{fidelity} \PY{o}{=} \PY{n}{hellinger\PYZus{}fidelity}\PY{p}{(}
    \PY{n}{experiment\PYZus{}df}\PY{p}{[}\PY{l+s+s2}{\PYZdq{}}\PY{l+s+s2}{probability\PYZus{}exact}\PY{l+s+s2}{\PYZdq{}}\PY{p}{]}\PY{o}{.}\PY{n}{to\PYZus{}dict}\PY{p}{(}\PY{p}{)}\PY{p}{,} 
    \PY{n}{experiment\PYZus{}df}\PY{p}{[}\PY{l+s+s2}{\PYZdq{}}\PY{l+s+s2}{probability}\PY{l+s+s2}{\PYZdq{}}\PY{p}{]}\PY{o}{.}\PY{n}{to\PYZus{}dict}\PY{p}{(}\PY{p}{)}\PY{p}{)}
\end{Verbatim}
\end{tcolorbox}
\begin{center}\rule{0.5\linewidth}{0.5pt}\end{center}

    \hypertarget{expectation-value}{%
\subsubsection{Expectation Value}\label{expectation-value}}

    In Notebook \ref{chap:notebook-2} we explained that the expectation value

\[
e(\vec{\beta}, \vec{\gamma})
=
\langle \psi_\mathrm{QAOA}(\vec{\beta}, \vec{\gamma}) | \HP |\psi_\mathrm{QAOA}(\vec{\beta}, \vec{\gamma}) \rangle
\]

is connected to the QUBO cost function \(f_3\) by

\[
e(\vec{\beta}, \vec{\gamma})
= \sum_{b \in \{0, 1\}^n}
|\lambda_{b}(\vec{\beta}, \vec{\gamma})|^2 f_3(\vec{b}) \ ,
\]

where the amplitudes \(\lambda_{b}(\vec{\beta}, \vec{\gamma})\) belong
to the state \(|\psi_\mathrm{QAOA}(\vec{\beta}, \vec{\gamma}) \rangle\),
i.e.

\[
|\psi_\mathrm{QAOA}(\vec{\beta}, \vec{\gamma}) \rangle 
= \sum_{\vec{b} \in \{0, 1\}^n}
\lambda_{b}(\vec{\beta}, \vec{\gamma}) |b\rangle \ .
\]

    This means a low expectation value \(e\) indicates that QAOA
generates a quantum state
\(|\psi_\mathrm{QAOA}(\vec{\beta}, \vec{\gamma}) \rangle\) with
\textbf{large amplitudes} for \textbf{bit strings} \(b\) with
\textbf{low cost} \(f_3(\vec{b})\). Thus, measuring \(e\) is a
meaningful quality metric, in particular if one compares it to the
minimum value of \(f_3\),
i.e.~\(\min_{\vec{b} \in \{0, 1\}^n} f_3(\vec{b})\).

In the following sections we first present results from experiments for Example Series 1 and then close this notebook by presenting results for Example Series 2.

    \hypertarget{fidelity-example-series-1-p1}{%
\subsection{\texorpdfstring{Fidelity, Example Series 1,
\(p=1\)}{Fidelity, Example Series 1, p=1}}\label{sec:fidelity-example-series-1-p1}}

\begin{figure}
    \begin{center}
        \includegraphics{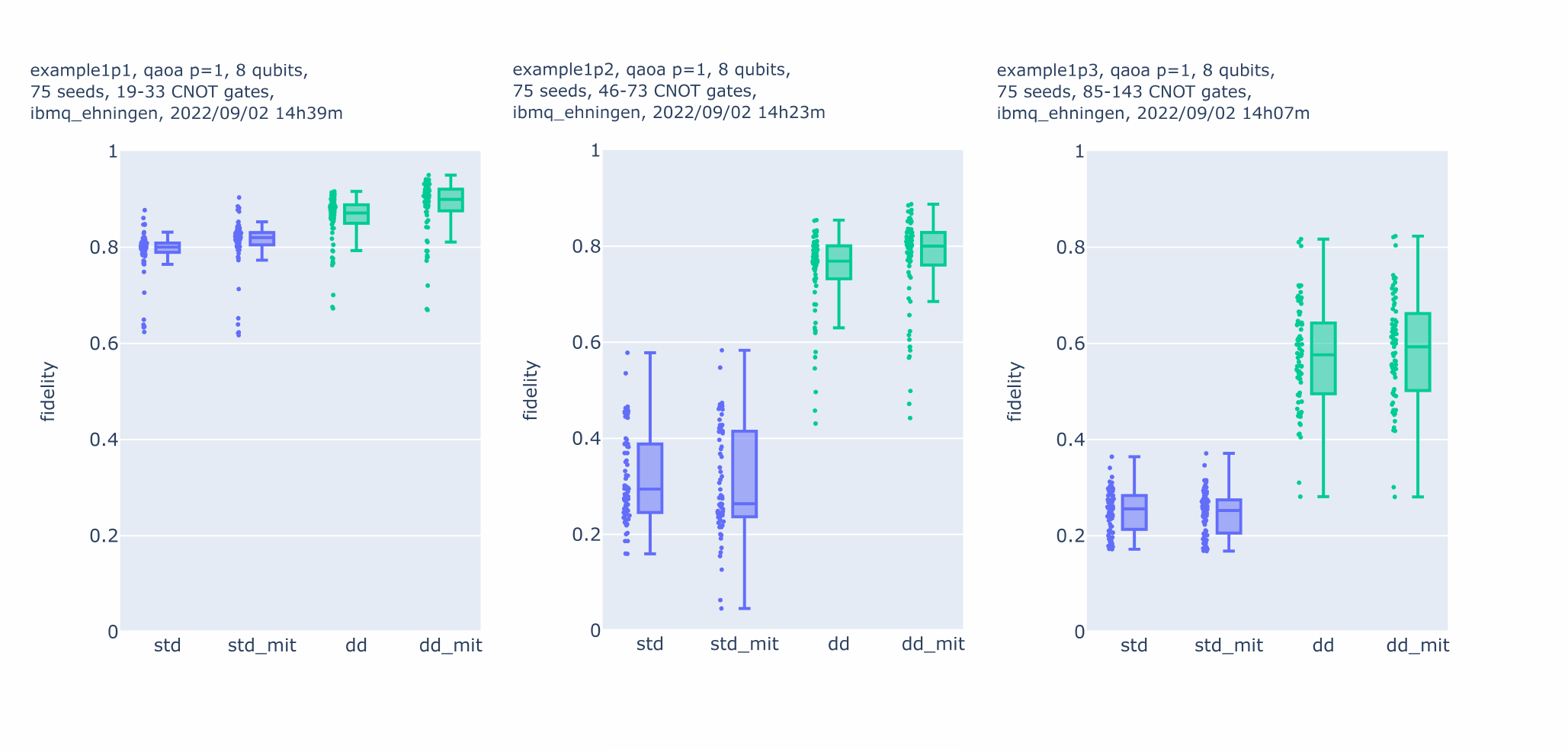}
        \caption{Fidelity vs. transpilation method for example1p1 (left), example1p2 (center), and example1p3 (right). Every dot stems from one of 75 transpilation seeds.}
        \label{fig:notebook-4-fidelity-ex1-p1-2022-09-02}
    \end{center}
\end{figure}

    In the upper three plots we can clearly see the influence of the number
of \(\CNOT\) gates. On the left, we see that a \textbf{moderate number
of \(\CNOT\)s} leads to a \textbf{fairly high fidelity} -- with and
without dynamical decoupling. We see some outliers but in general all
transpilations lead to a good quality (the variance between results is
small). In the middle we see that for circuits with \textbf{more
\(\CNOT\)s} the \textbf{fidelity} for the \textbf{standard transpilation
drops significantly}. There are still some circuits that lead to medium
fidelities but also many with poor performance, i.e.~the variance of the
results is very high. However, the \textbf{circuit depth} seems to be
\textbf{low enough} so that \textbf{dynamical decoupling can mitigate
many errors} and yields considerably better results. On the right plot
we see \textbf{\(\CNOT\) numbers} that are \textbf{definitely too high
for ibmq\_ehningen} so that without dynamical decoupling the
\textbf{fidelity is poor for all transpilations}. Adding dynamical
decoupling can in some cases give better results but looking at the
high variance we see that there is no guarantee that it works in general.

    \hypertarget{fidelity-p1-different-dates}{%
\subsection{\texorpdfstring{Fidelity, \(p=1\), Different
Dates}{Fidelity, p=1, Different Dates}}\label{fidelity-p1-different-dates}}

    The \textbf{error rates} of quantum computers are not static but
\textbf{significantly change over time}. In the following figures we
observe that this has a \textbf{drastic effect} on the \textbf{quality}
of the results obtained from ibmq\_ehningen.

\begin{figure}
    \begin{center}
        \begin{subfigure}{0.95\textwidth}
            \centering
            \includegraphics{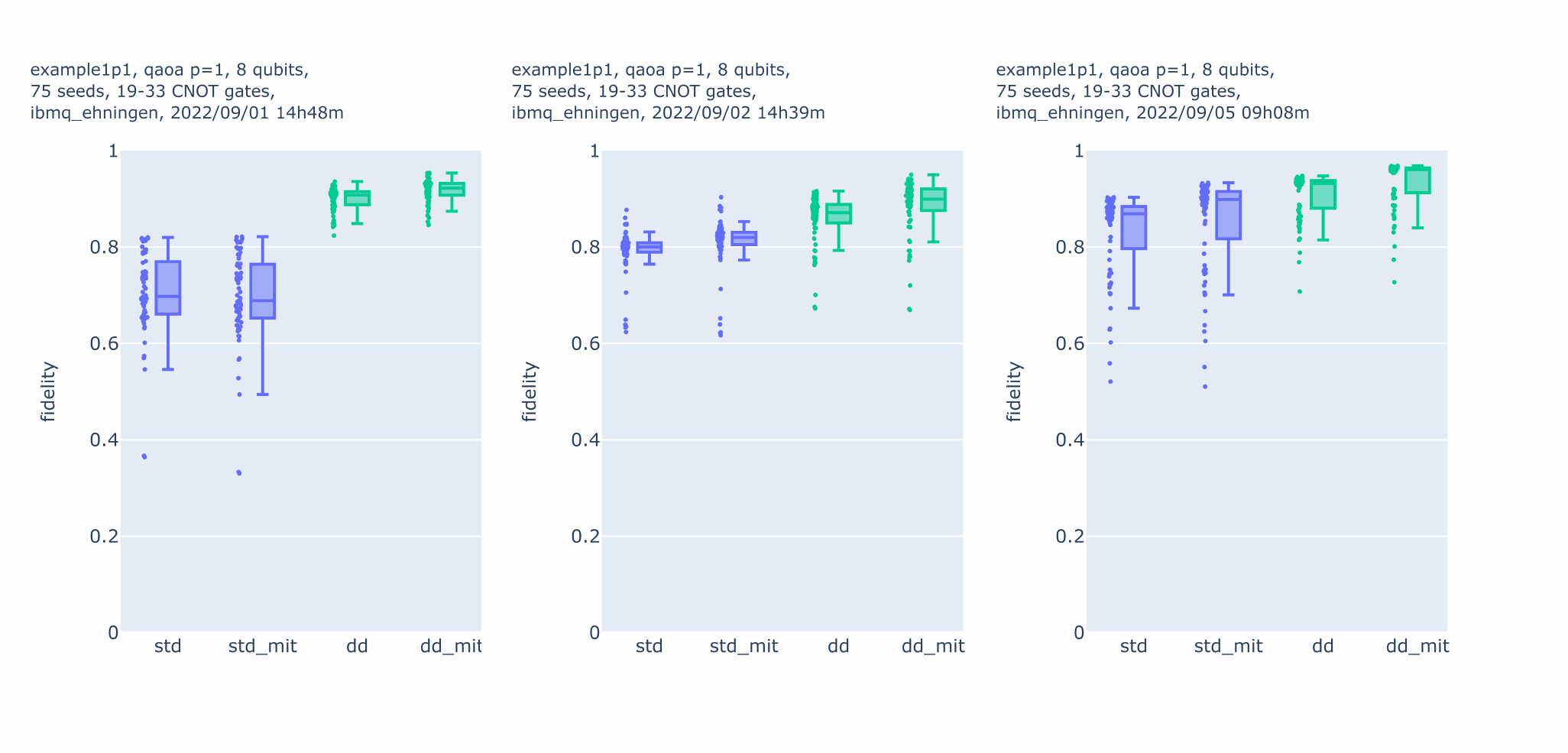}
            \caption{Example1p1.}
            \label{fig:notebook-4-fidelity-ex1p1-p1-2022-09-010205}
        \end{subfigure}
        \begin{subfigure}{0.95\textwidth}
            \centering
            \includegraphics{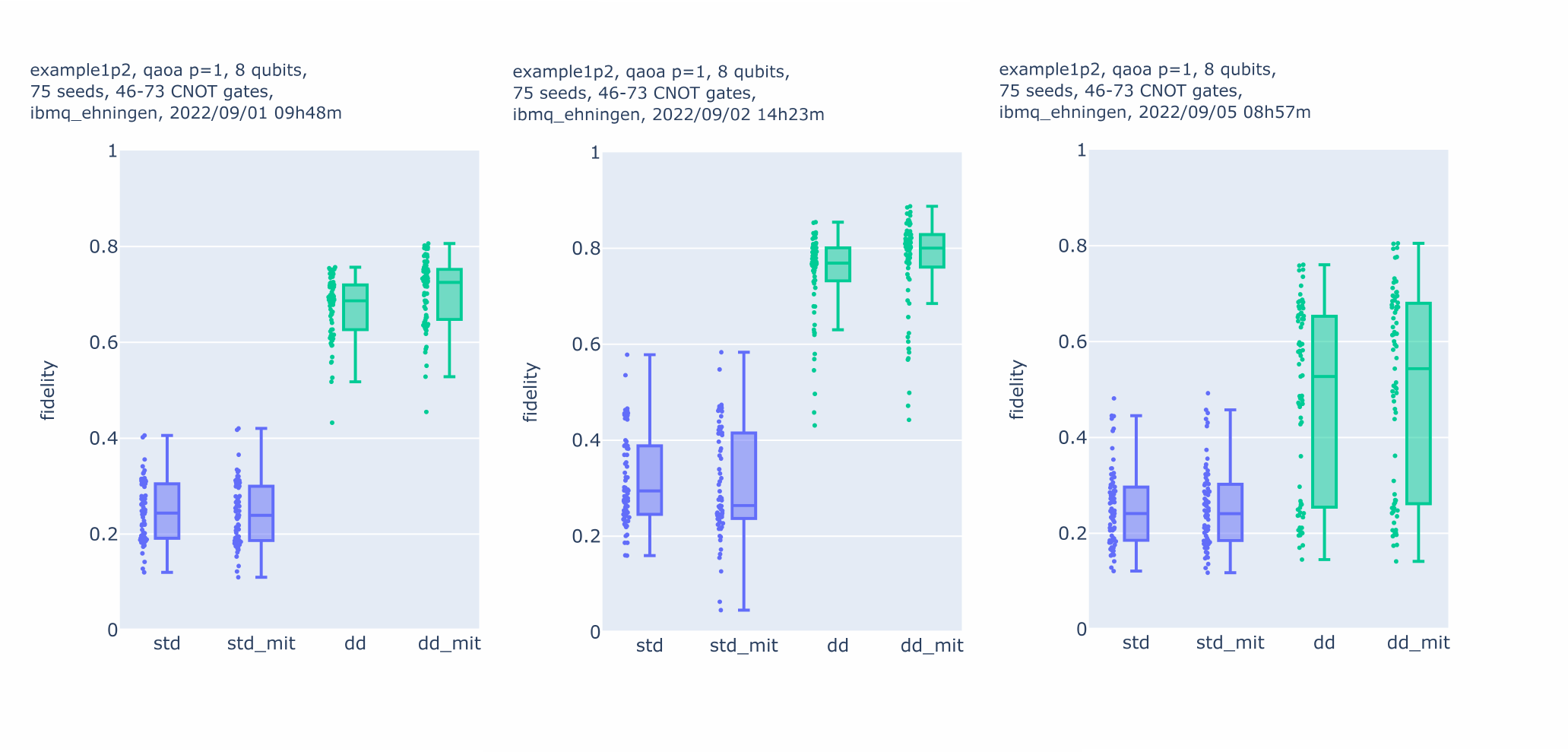}
            \caption{Example1p2.}
            \label{fig:notebook-4-fidelity-ex1p2-p1-2022-09-010205}
        \end{subfigure}
        \caption{Fidelity vs. transpilation method for example1p1 (upper plots) and example1p2 (lower plots) on three different dates. Every dot stems from one of 75 transpilation seeds.}
        \label{fig:notebook-4-fidelity-ex1-p1-2022-09-010205}
    \end{center}
\end{figure}

%     \hypertarget{exampe1p1}{%
% \subsubsection{Exampe1p1}}

% \begin{figure}
%     \begin{center}
%         \includegraphics{images/fidelities/example1p1_p1_2022_09_010205.png}
%         \caption{TODO}
%         \label{fig:notebook-4-fidelity-ex1p1-p1-2022-09-010205}
%     \end{center}
% \end{figure}

%     \hypertarget{exampe1p2}{%
% \subsubsection{Exampe1p2}}

% \begin{figure}
%     \begin{center}
%         \includegraphics{images/fidelities/example1p2_p1_2022_09_010205.png}
%         \caption{TODO}
%         \label{fig:notebook-4-fidelity-ex1p2-p1-2022-09-010205}
%     \end{center}
% \end{figure}

    Seeing how drastically the quality of our results changes from day to
day we recommend running experiments on a series of different dates and
(if possible) on different quantum computers.

    \hypertarget{fidelity-and-expectation-value-p1-and-p2}{%
\subsection{\texorpdfstring{Fidelity and Expectation Value, \(p=1\) and
\(p=2\)}{Fidelity and Expectation Value, p=1 and p=2}}
\label{sec:notebook-4-fidelity-and-expectation-value-p1-and-p2}}

    In this section we want to analyze the effect of the parameter \(p\). In
particular, we are interested in the \textbf{trade-off} between
\textbf{better approximation quality} but \textbf{longer circuits} that
come with higher values of \(p\).

%     \hypertarget{example1p1}{%
% \subsubsection{Example1p1}}

\begin{figure}
    \begin{center}
        \includegraphics{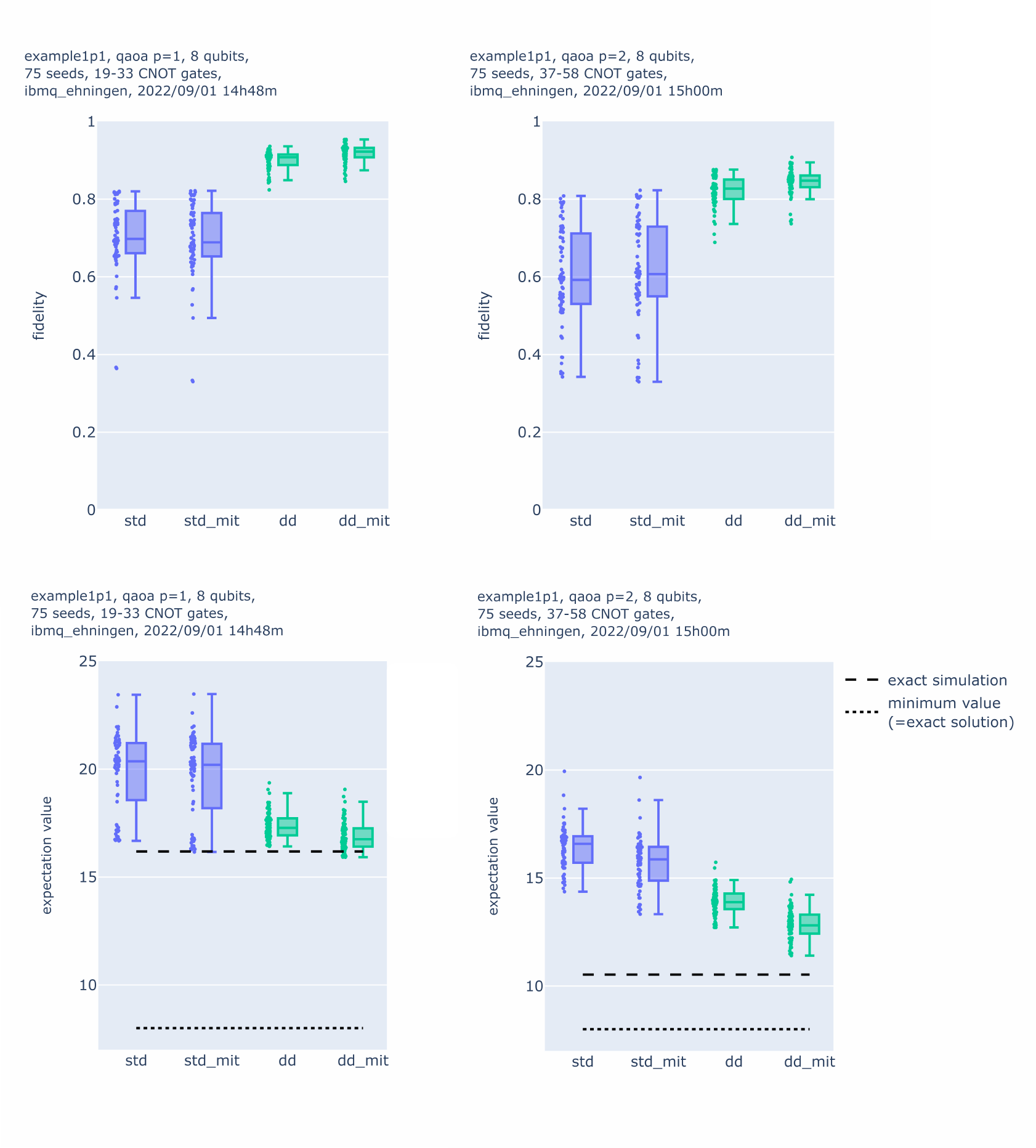}
        \caption{Fidelity (upper plots) and expectation value (lower plots) vs. transpilation method for example1p1.
        On the left we have $p=1$ and on the right $p=2$. Every dot stems from one of 75 transpilation seeds.}
        \label{fig:notebook-4-fidelity-ex1p1-p1-p2-2022-09-01}
    \end{center}
\end{figure}

    For the simplest example that we consider (i.e. example1p1) we see
in Figure \ref{fig:notebook-4-fidelity-ex1p1-p1-p2-2022-09-01} 
that choosing \(p=2\) gives a better result (in terms of a lower
expectation value). In theory this is expected, compare the dashed lines
in the two lower plots. However, in practice on real quantum computers the deeper circuits for
\(p=2\) could be a problem, but for the example at hand the circuits are shallow enough
so that we don't run into problems when executing them on ibmq\_ehningen.
This will change in the next examples.

%     \hypertarget{example1p2}{%
% \subsubsection{Example1p2}}

\begin{figure}
    \begin{center}
        \includegraphics{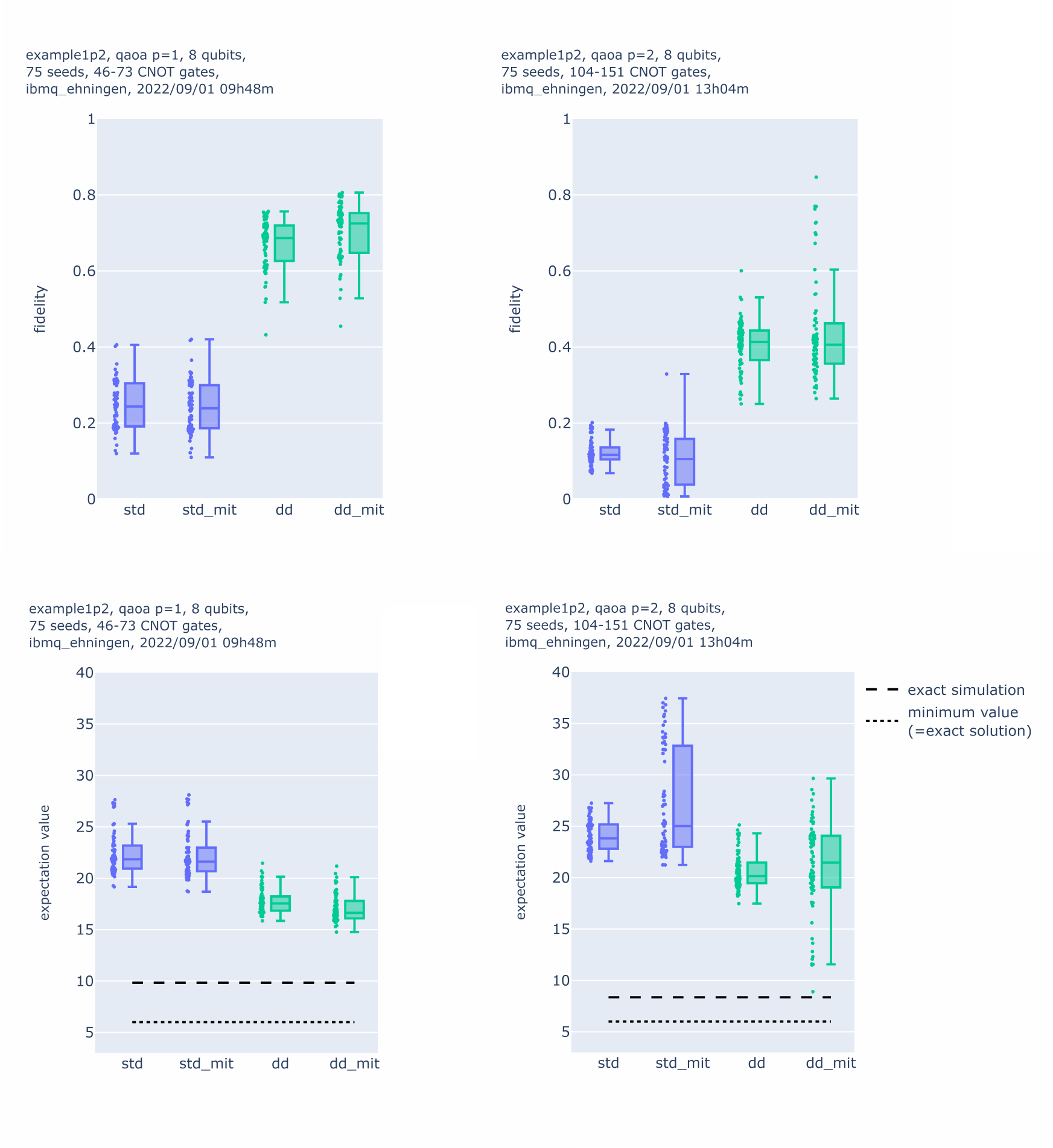}
        \caption{Fidelity (upper plots) and expectation value (lower plots) vs. transpilation method for example1p2.
        On the left we have $p=1$ and on the right $p=2$. Every dot stems from one of 75 transpilation seeds.}
        \label{fig:notebook-4-fidelity-ex1p2-p1-p2-2022-09-01}
    \end{center}
\end{figure}

    Already for example1p2 we see in Figure \ref{fig:notebook-4-fidelity-ex1p2-p1-p2-2022-09-01}
that the trade-off between better theoretical solution and practical result speak rather in favor of
\(p=1\). For this choice we get more certainly a solution with a good
expectation value. Choosing \(p=2\) might give a better solution but
looking at the high variance this is pretty uncertain.

%     \hypertarget{example1p3}{%
% \subsubsection{Example1p3}}

\begin{figure}
    \begin{center}
        \includegraphics{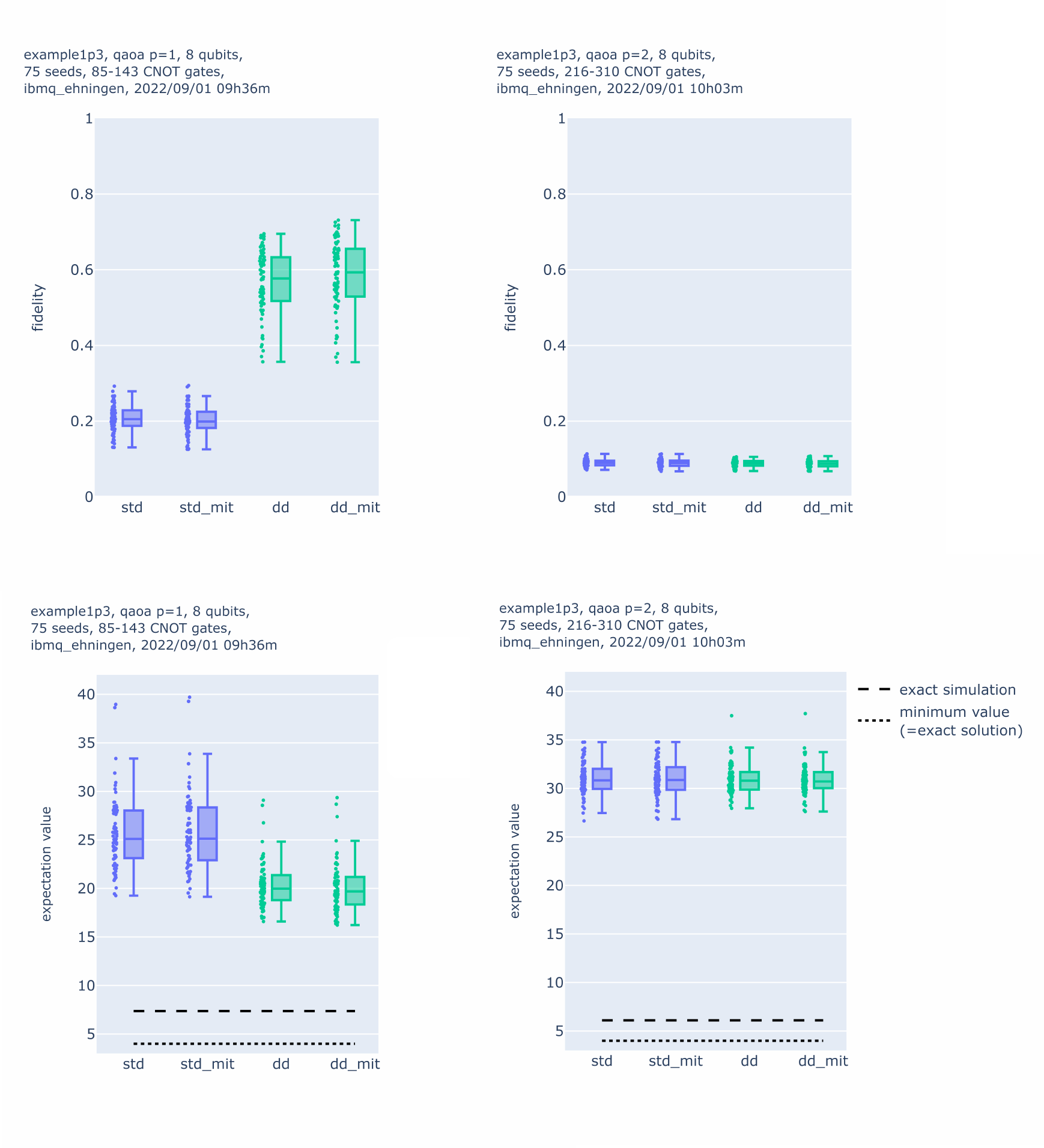}
        \caption{Fidelity (upper plots) and expectation value (lower plots) vs. transpilation method for example1p3.
        On the left we have $p=1$ and on the right $p=2$. Every dot stems from one of 75 transpilation seeds.}
        \label{fig:notebook-4-fidelity-ex1p3-p1-p2-2022-09-01}
    \end{center}
\end{figure}

    For example1p3 the situation is clear: The QAOA circuits for \(p=2\) are
too deep for ibmq\_ehningen so that no meaningful result can be
obtained, see Figure \ref{fig:notebook-4-fidelity-ex1p3-p1-p2-2022-09-01}.

    \hypertarget{number-cnot-gates-vs.fidelity}{%
\subsection{\texorpdfstring{Number \(\CNOT\) Gates
vs.~Fidelity}{Number \textbackslash CNOT Gates vs.~Fidelity}}
\label{sec:notebook-4-number-cnot-gates-vs-fidelity}}

    Next, we present figures showing the dependence of the fidelity that the circuit achieved
    with the number of \(\CNOT\) gates in the circuit.

%     \hypertarget{example1p1}{%
% \subsubsection{Example1p1}}

\begin{figure}
    \begin{center}
        \includegraphics{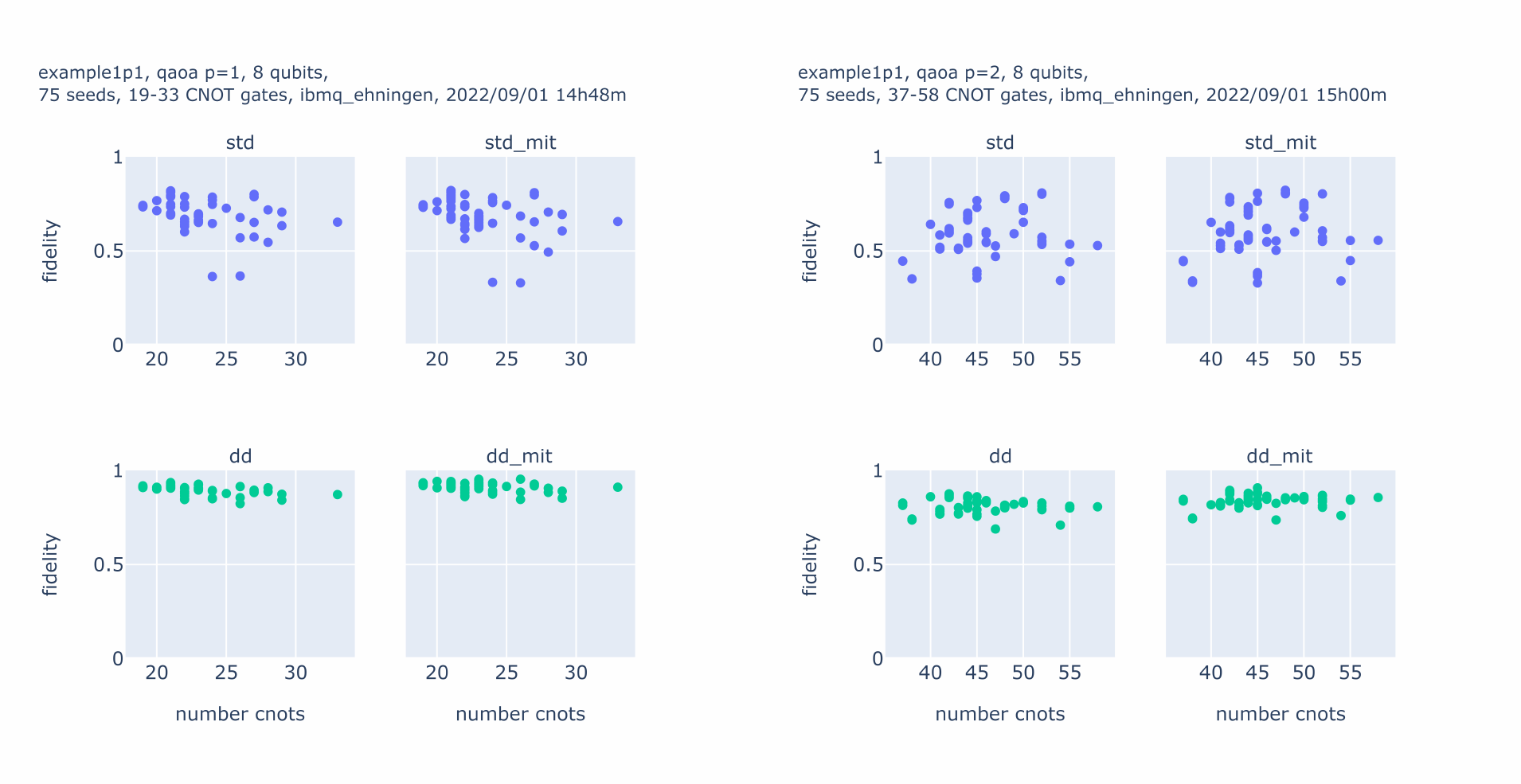}
        \caption{Fidelity vs. number of $\CNOT$ gates for example1p1 for different transpilations and $p=1$ (left plots) as well as $p=2$ (right plots). 
            Every dot stems from one of 75 transpilation seeds.}
        \label{fig:notebook-4-fidelity-vs-cnot-ex1p1-p1-p2-2022-09-01}
    \end{center}
\end{figure}

    Comparing the left and the right plots (i.e.~comparing \(p=1\) and
\(p=2\)) we see that circuits with more \(\CNOT\) gates give (in general) results
with a lower fidelity. Moreover, we see in the top plots that for the
standard transpilation we don't have the expected decrease in fidelity
when increasing the number of \(\CNOT\)s. In particular, in the top
right plot we see the best results for a medium number of \(\CNOT\)
gates and a very poor quality for the lowest number of \(\CNOT\)s. The
explanation is probably that the \textbf{standard transpilation method} is \textbf{not
aware of all kinds of errors} that appear in a quantum device and thus
did not choose the best qubits. A hint in this direction is also the
remarkable fact that for the circuits with dynamical decoupling the
fidelity stays nearly constant for the range of number of \(\CNOT\)
gates appearing in the examples here. It seems that \textbf{dynamical decoupling
removed the errors that spoiled the quality} for the top plots.

    For example1p2 and example1p3 below we make the same observations, but
additionally observe that dynamical decoupling and measurement error
mitigation are not sufficient to remove all appearing errors, see the plots
with \(\CNOT\) range between 40 and 150. (One reason might be cross-talk
between the qubits). For examples with more than 200 \(\CNOT\) gates
we don't get a meaningful solution.

%     \hypertarget{example1p2}{%
% \subsubsection{Example1p2}}

\begin{figure}
    \begin{center}
        \includegraphics{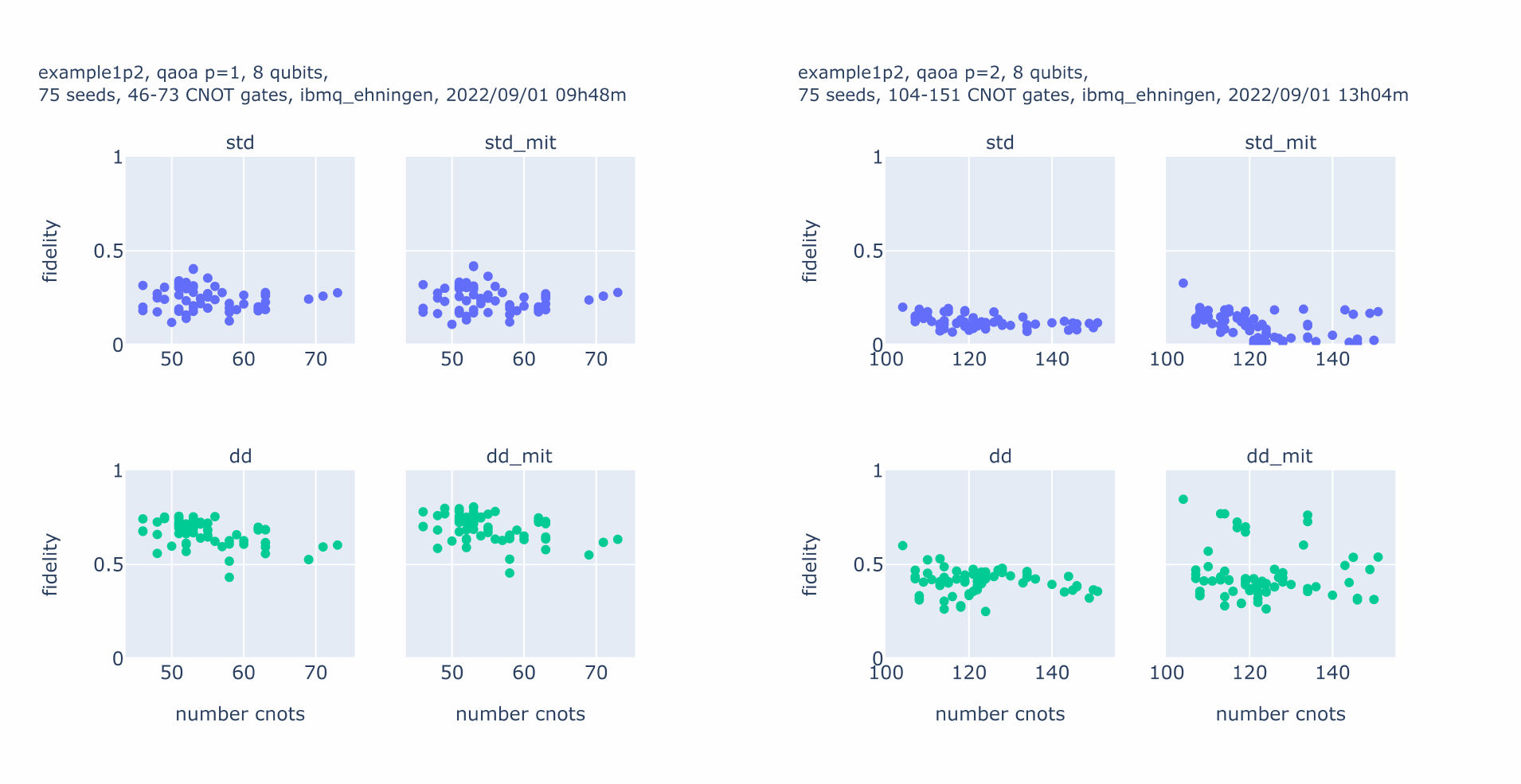}
        \caption{Fidelity vs. number of $\CNOT$ gates for example1p2 for different transpilations and $p=1$ (left plots) as well as $p=2$ (right plots). 
        Every dot stems from one of 75 transpilation seeds.}
        \label{fig:notebook-4-fidelity-vs-cnot-ex1p2-p1-p2-2022-09-01}
    \end{center}
\end{figure}

%     \hypertarget{example1p3}{%
% \subsubsection{Example1p3}}

\begin{figure}
    \begin{center}
        \includegraphics{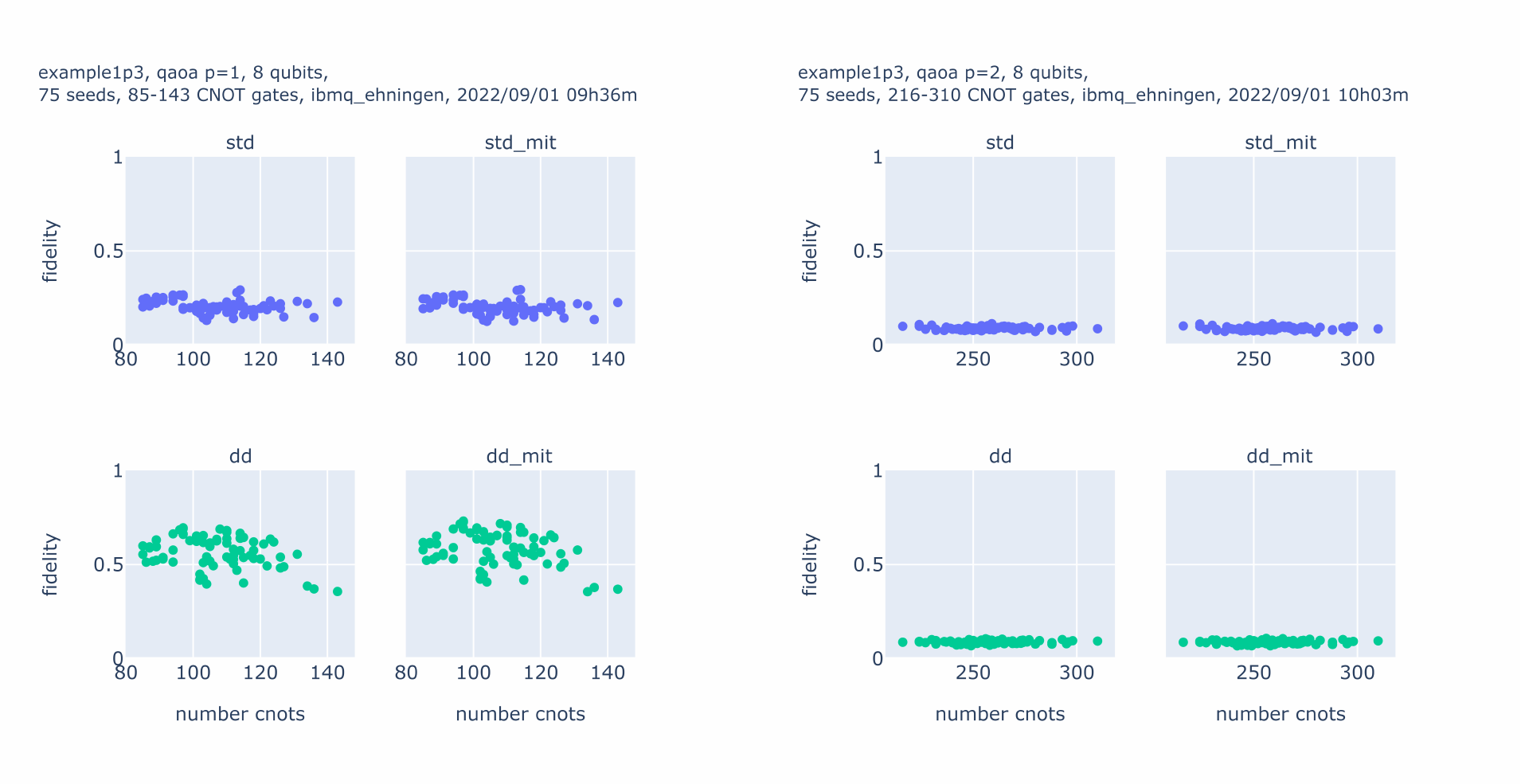}
        \caption{Fidelity vs. number of $\CNOT$ gates for example1p3 for different transpilations and $p=1$ (left plots) as well as $p=2$ (right plots). 
        Every dot stems from one of 75 transpilation seeds.}
        \label{fig:notebook-4-fidelity-vs-cnot-ex1p3-p1-p2-2022-09-01}
    \end{center}
\end{figure}

    \hypertarget{example1p3-different-dates}{%
\subsubsection{Example1p3: Different
Dates}}

\begin{figure}
    \begin{center}
        \includegraphics{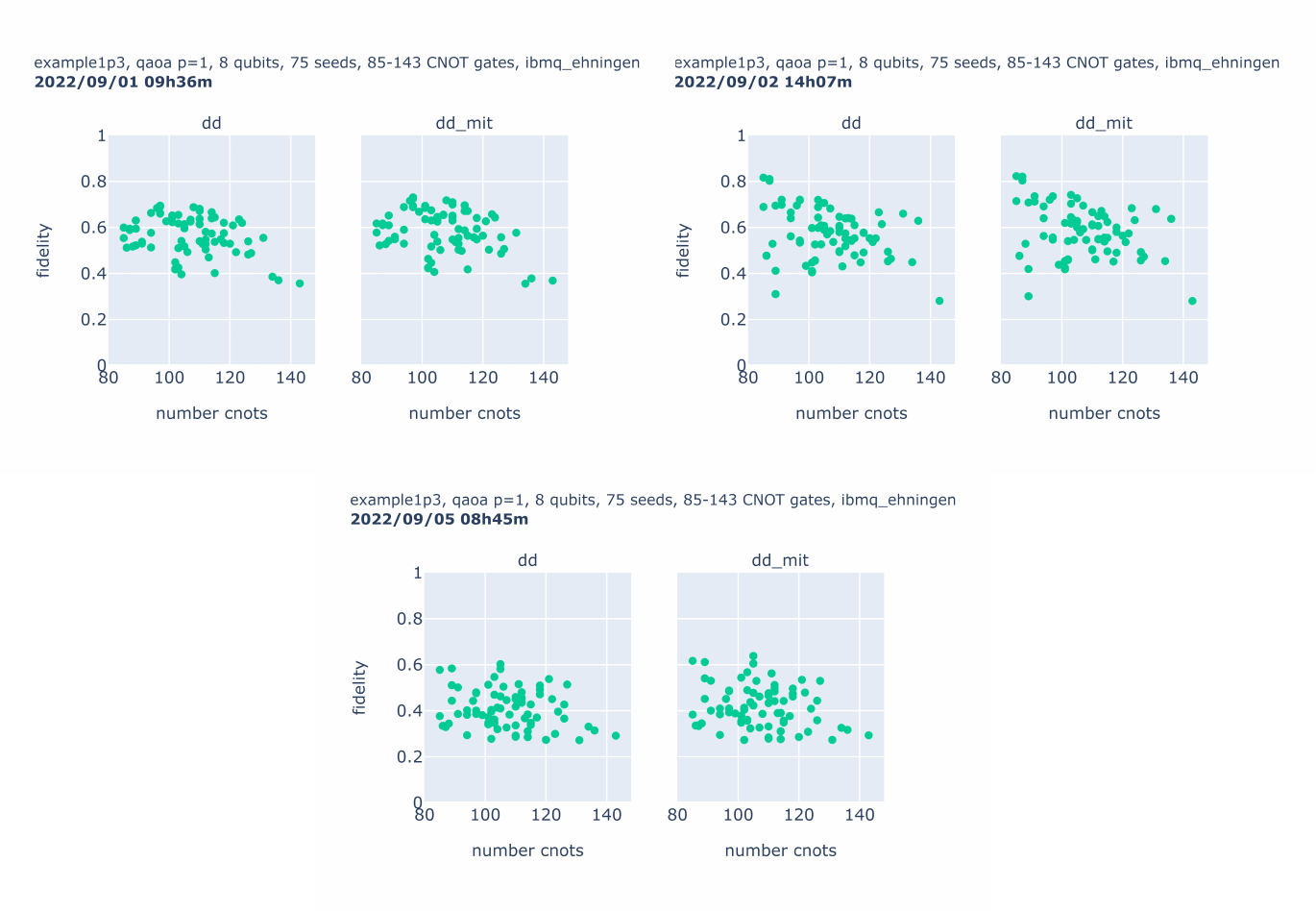}
        \caption{Fidelity vs. number of $\CNOT$ gates for example1p3 for $p=1$. The circuits for upper left plot were run on 2022/09/01, for the upper right on 2022/09/02 and for the lower plot on 2022/09/05.
        Every dot stems from one of 75 transpilation seeds.}
        \label{fig:notebook-4-fidelity-vs-cnot-ex1p3-p1-p2-2022-09-010205}
    \end{center}
\end{figure}

    As mentioned above the error rates of quantum computers change
significantly over time. The effect on the quality of our results is
clearly visible in Figure \ref{fig:notebook-4-fidelity-vs-cnot-ex1p3-p1-p2-2022-09-010205}.

We end this notebook with results from Example Series 2.

    \hypertarget{fidelity-and-expectation-value-p1-and-p2}{%
\subsection{\texorpdfstring{Fidelity and Expectation Value, \(p=1\) and
\(p=2\)}{Fidelity and Expectation Value, p=1 and p=2}}
\label{sec:notebook-4-fidelity-and-expectation-value-p1-and-p2-2}}

    For all examples of Example Series 2 we are in regimes of number of \(\CNOT\) gates that are
too high for current quantum computers. Thus, we see poor fidelities and
expectation values that stay bounded away from an exact simulation and
also from the exact solution.

    \hypertarget{example2p1}{%
\subsubsection{Example2p1}}

\begin{figure}
    \begin{center}
        \includegraphics{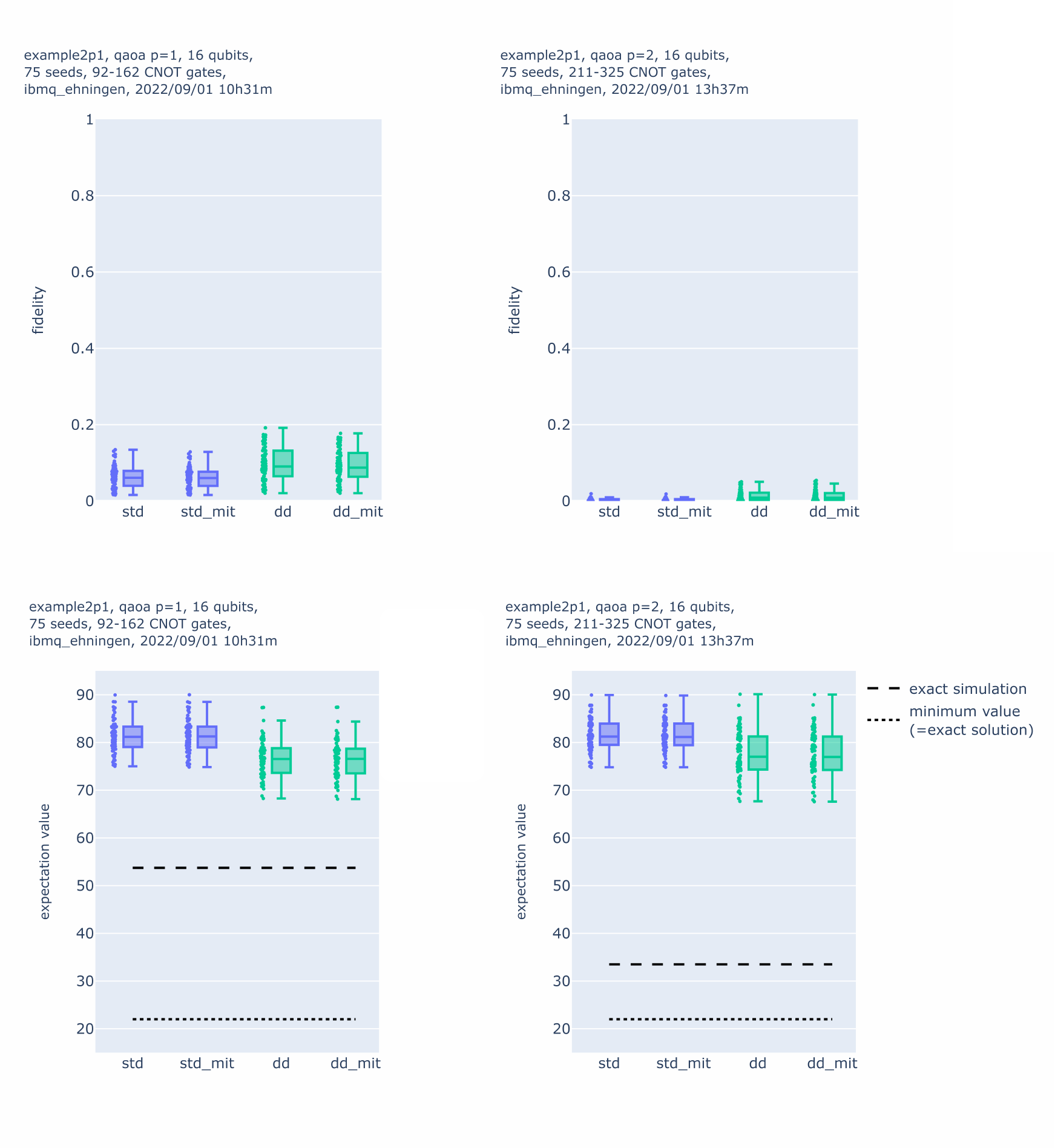}
        \caption{Fidelity (upper plots) and expectation value (lower plots) vs. transpilation method for example2p1.
        On the left we have $p=1$ and on the right $p=2$. Every dot stems from one of 75 transpilation seeds.}
        \label{fig:notebook-4-fidelity-ex2p1-p1-p2-2022-09-01}
    \end{center}
\end{figure}

    \hypertarget{example2p4}{%
\subsubsection{Example2p4}}

\begin{figure}
    \begin{center}
        \includegraphics{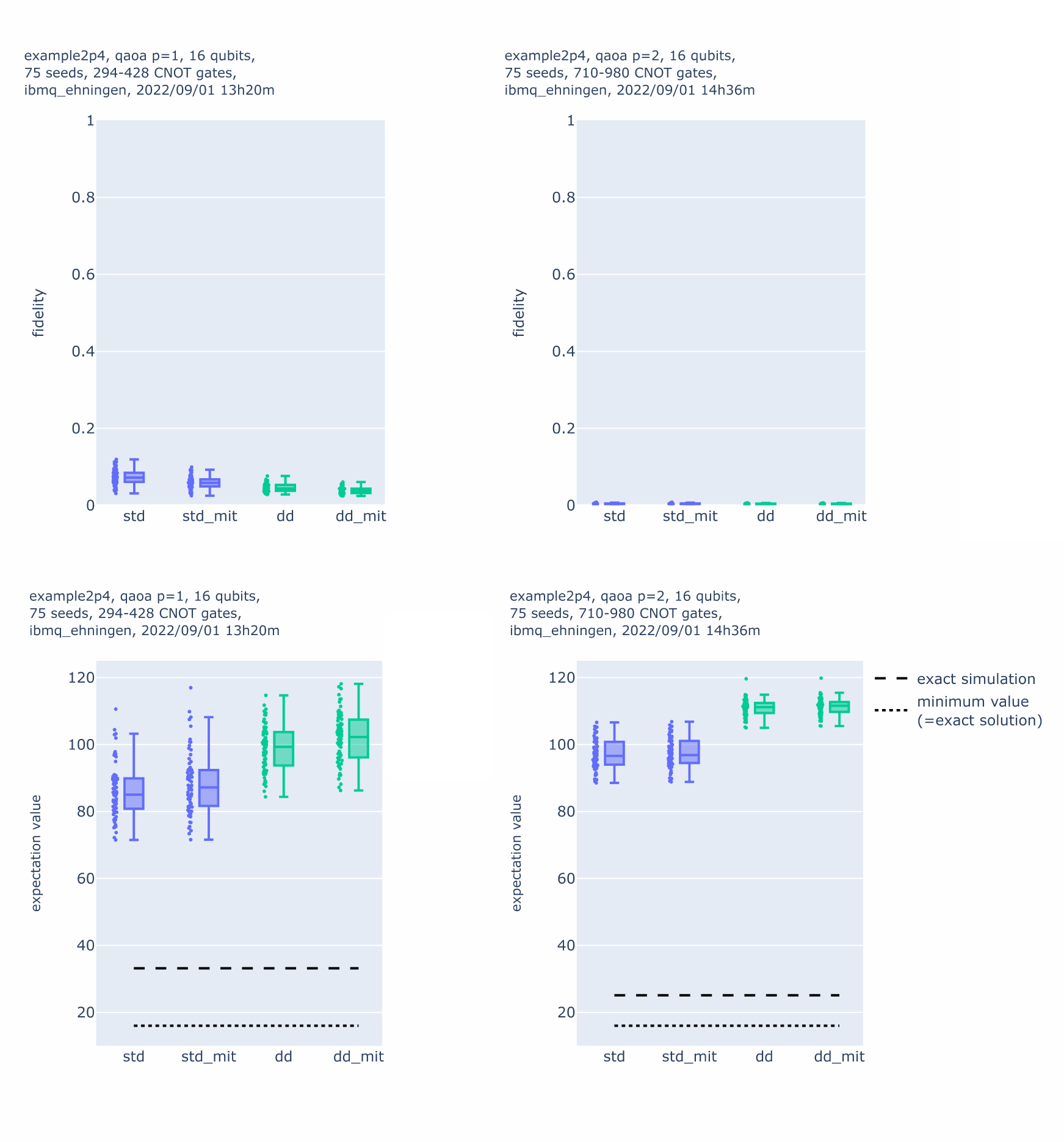}
        \caption{Fidelity (upper plots) and expectation value (lower plots) vs. transpilation method for example2p4.
        On the left we have $p=1$ and on the right $p=2$. Every dot stems from one of 75 transpilation seeds.}
        \label{fig:notebook-4-fidelity-ex2p4-p1-p2-2022-09-01}
    \end{center}
\end{figure}

\section{Remarks on Implementation} \label{sec:remarks-implementation}
All experiments in this notebook were run using the following package versions:
\begin{itemize}
    \tightlist
    \item \texttt{mthree==1.0}
    \item \texttt{qiskit==0.37}
    \item \texttt{qiskit\_aer==0.10.4}
    \item \texttt{qiskit\_terra==0.21.0}
\end{itemize}

For the experiments in Section \ref{sec:results-qaoa-on-ibmq_ehnigen} we used the parameters given in Table \ref{tab:paramters-ehningen-experiments}.
\begin{table}
\begin{center}
    \begin{tabular}{l|l|ll}
        Example           & penalty $\varrho$ & \multicolumn{2}{l}{QAOA parameters}
        \\[0.1cm]
        \hline \hline
        example1p1, $p=1$ & $\varrho = 3.1$   & $\beta_0 = 0.86470653$, &$\gamma_0 = -0.31959353$
        \\[0.05cm]
        example1p2, $p=1$ & $\varrho = 3.5$   & $\beta_0 = 0.70449479$, &$\gamma_0 = 4.48896374$
        \\[0.05cm]
        example1p3, $p=1$ & $\varrho = 1.3$   & $\beta_0 = 2.48199357$, &$\gamma_0 = 4.86664818$
        \\[0.1cm]
        \hline
        example1p1, $p=2$ & $\varrho = 3.2$   & $\beta_0 = 1.50891553$, &$\gamma_0 = 2.59570247$
        \\
                          &                   & $\beta_1 = 2.37353917$, &$\gamma_1 = 2.3102415$
        \\[0.05cm]
        example1p2, $p=2$ & $\varrho = 3.6$   & $\beta_0 = 3.99890724$, &$\gamma_0 = 6.11303759$
        \\
                          &                   & $\beta_1 = 2.72012026$, &$\gamma_1 = 1.75840967$
        \\[0.1cm]
        example1p3, $p=2$ & $\varrho = 1.6$   & $\beta_0 = 0.48954047$, &$\gamma_0 = 1.98033881$
        \\
                          &                   & $\beta_1 = 0.9098274$, &$\gamma_1 = 3.92848365$
        \\[0.1cm]
        \hline
        example2p1, $p=1$ & $\varrho = 5.7$   & $\beta_0 = 0.48207365$, &$\gamma_0 = 6.11153126$
        \\[0.05cm]
        example2p4, $p=1$ & $\varrho = 3.1$   & $\beta_0 = 2.77112559$, &$\gamma_0 = 0.02398761$
        \\[0.1cm]
        \hline
        example2p1, $p=2$ & $\varrho = 5.5$   & $\beta_0 = 1.60523831$, &$\gamma_0 = 0.08836239$
        \\
                          &                   & $\beta_1 = 3.88382278$, &$\gamma_1 = 6.19852592$
        \\[0.05cm]
        example2p4, $p=2$ & $\varrho = 3.3$   & $\beta_0 = 1.50645202$, &$\gamma_0 = 5.76044569$
        \\
                          &                   & $\beta_1 = 3.84358483$, &$\gamma_1 = 3.7570749$
    \end{tabular}
\end{center}
\caption{Parameters for the experiments in Section \ref{sec:results-qaoa-on-ibmq_ehnigen}.}
\label{tab:paramters-ehningen-experiments}
\end{table}

%% file: appendix_ehningen.tex
The following properties are from November, 11, 2022.

\begin{figure}
    \begin{center}
        \includegraphics[width=7cm]{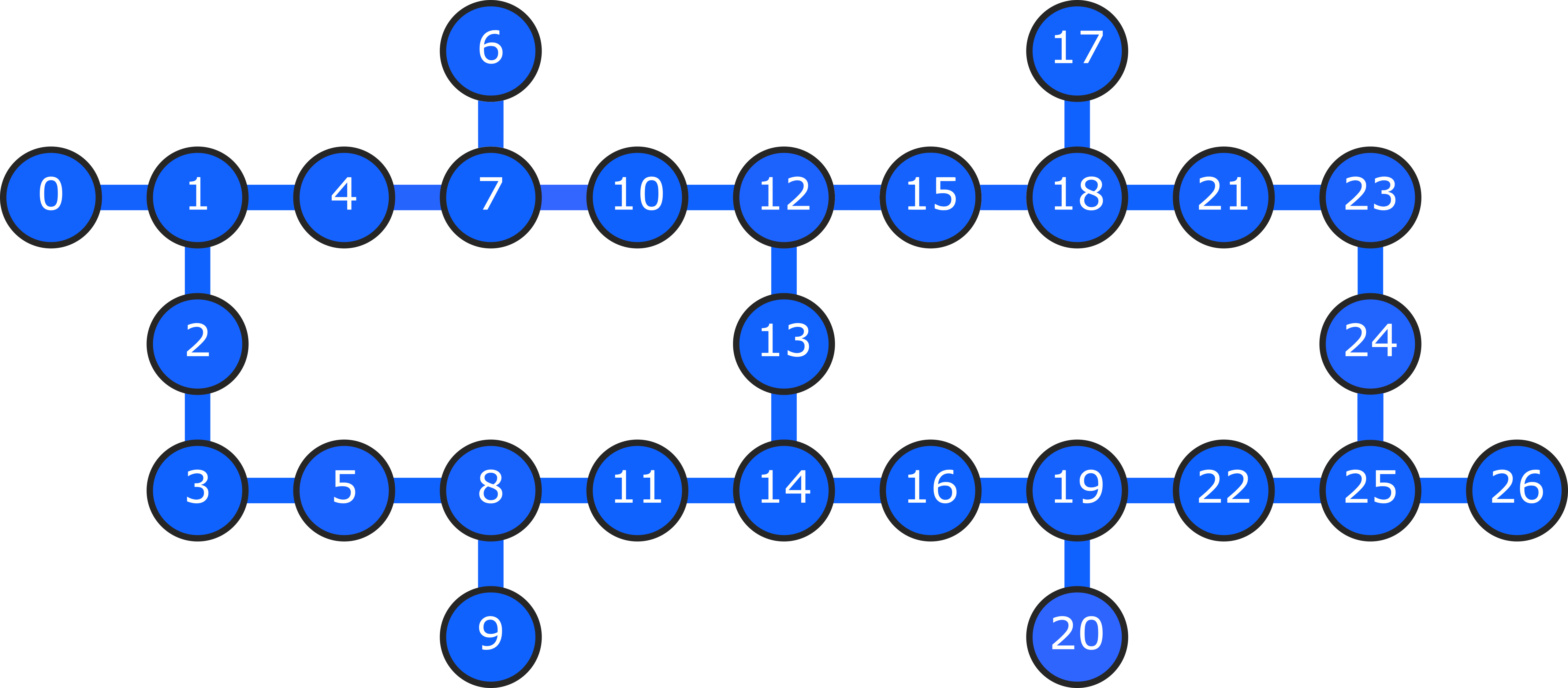}
        \caption{Coupling map of ibmq\_ehningen.}
    \end{center}
\end{figure}

\begin{figure}
    \begin{center}
        \includegraphics{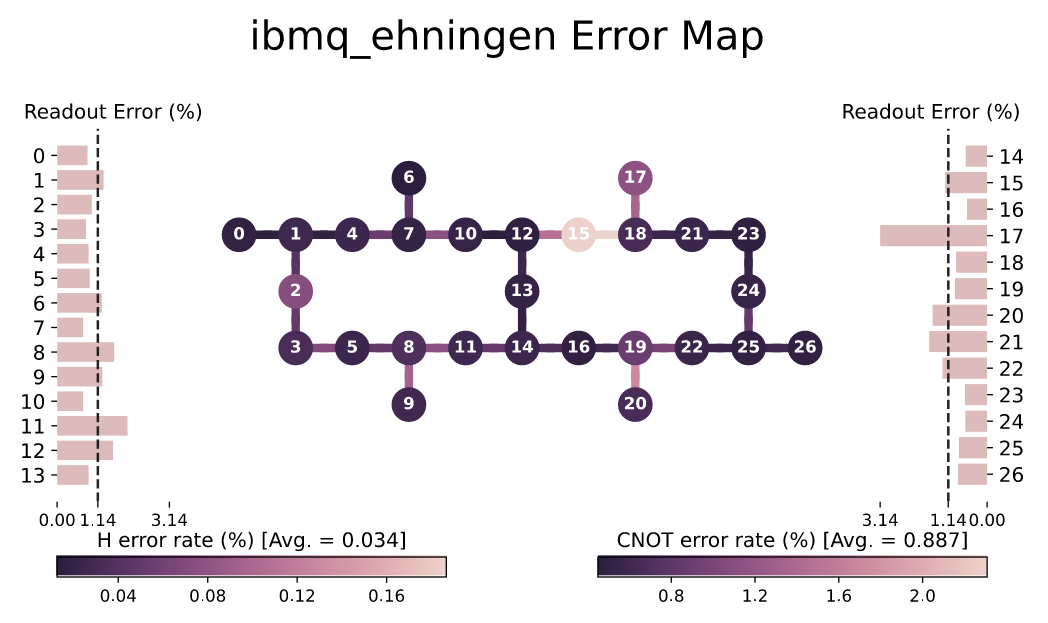}
        \caption{Error map of ibmq\_ehningen.}
    \end{center}
\end{figure}

\begin{figure}
    \begin{center}
        \includegraphics{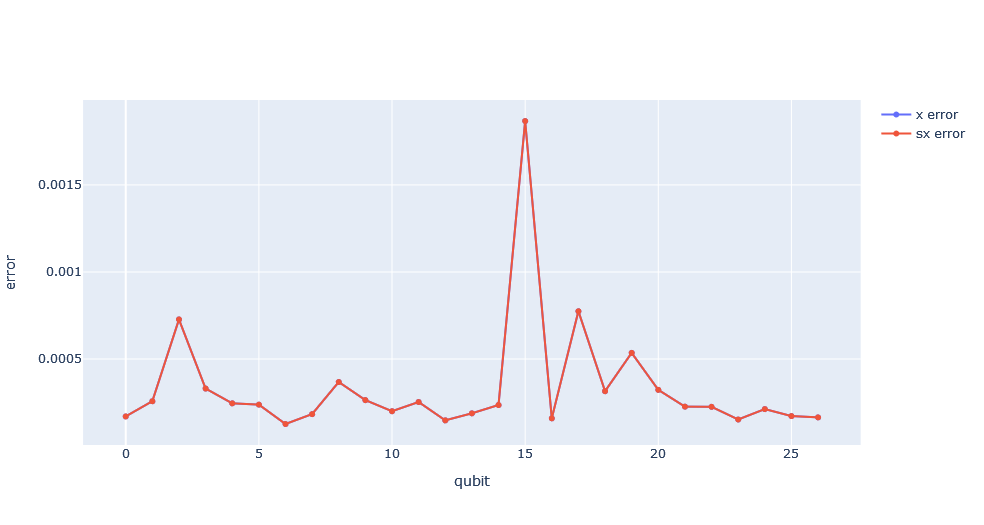}
        \caption{$\X$ and $\SX$ errors of ibmq\_ehningen.}
    \end{center}
\end{figure}

\begin{figure}
    \begin{center}
        \includegraphics{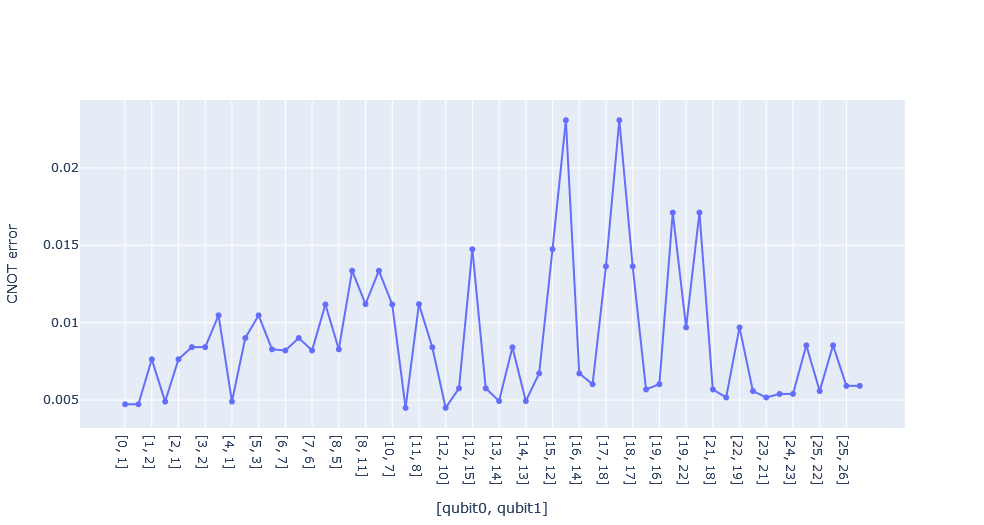}
        \caption{$\CNOT$ errors of ibmq\_ehningen.}
    \end{center}
\end{figure}

%% file: appendix_helper.tex
Code for \verb|codes_notebook_1.py|:

\begin{tcolorbox}[breakable, size=fbox, boxrule=1pt, pad at break*=1mm,colback=cellbackground, colframe=cellborder]
\begin{Verbatim}[commandchars=\\\{\}]
\PY{k+kn}{from} \PY{n+nn}{typing} \PY{k+kn}{import} \PY{n}{List}\PY{p}{,} \PY{n}{Union}
\PY{k+kn}{import} \PY{n+nn}{numpy} \PY{k}{as} \PY{n+nn}{np}
\PY{k+kn}{from} \PY{n+nn}{qiskit\PYZus{}optimization} \PY{k+kn}{import} \PY{n}{QuadraticProgram}
\PY{k+kn}{from} \PY{n+nn}{qiskit\PYZus{}optimization}\PY{n+nn}{.}\PY{n+nn}{converters} \PY{k+kn}{import} \PY{n}{QuadraticProgramConverter}\PY{p}{,} \PY{n}{LinearEqualityToPenalty}\PY{p}{,} \PY{n}{IntegerToBinary}
\PY{k+kn}{from} \PY{n+nn}{qiskit\PYZus{}optimization}\PY{n+nn}{.}\PY{n+nn}{algorithms} \PY{k+kn}{import} \PY{n}{CplexOptimizer}

\PY{c+c1}{\PYZsh{}\PYZsh{} \PYZhy{}\PYZhy{}\PYZhy{} Codes from Notebook 1 \PYZhy{}\PYZhy{}\PYZhy{} \PYZsh{}\PYZsh{}}

\PY{k}{class} \PY{n+nc}{Car}\PY{p}{:}
    \PY{k}{def} \PY{n+nf+fm}{\PYZus{}\PYZus{}init\PYZus{}\PYZus{}}\PY{p}{(}
        \PY{n+nb+bp}{self}\PY{p}{,}
        \PY{n}{car\PYZus{}id}\PY{p}{:} \PY{n+nb}{str}\PY{p}{,} \PY{c+c1}{\PYZsh{} an arbitrary name for the car}
        \PY{n}{time\PYZus{}slots\PYZus{}at\PYZus{}charging\PYZus{}unit}\PY{p}{:} \PY{n}{List}\PY{p}{[}\PY{n+nb}{int}\PY{p}{]}\PY{p}{,} \PY{c+c1}{\PYZsh{} time slots when the car is at the charging unit}
        \PY{n}{required\PYZus{}energy}\PY{p}{:} \PY{n+nb}{int} \PY{c+c1}{\PYZsh{} energy units that should be charged}
    \PY{p}{)} \PY{o}{\PYZhy{}}\PY{o}{\PYZgt{}} \PY{k+kc}{None}\PY{p}{:}
        \PY{n+nb+bp}{self}\PY{o}{.}\PY{n}{car\PYZus{}id} \PY{o}{=} \PY{n}{car\PYZus{}id}
        \PY{n+nb+bp}{self}\PY{o}{.}\PY{n}{time\PYZus{}slots\PYZus{}at\PYZus{}charging\PYZus{}unit} \PY{o}{=} \PY{n}{time\PYZus{}slots\PYZus{}at\PYZus{}charging\PYZus{}unit}
        \PY{n+nb+bp}{self}\PY{o}{.}\PY{n}{required\PYZus{}energy} \PY{o}{=} \PY{n}{required\PYZus{}energy}

    \PY{k}{def} \PY{n+nf+fm}{\PYZus{}\PYZus{}str\PYZus{}\PYZus{}}\PY{p}{(}\PY{n+nb+bp}{self}\PY{p}{)} \PY{o}{\PYZhy{}}\PY{o}{\PYZgt{}} \PY{n+nb}{str}\PY{p}{:}
        \PY{k}{return} \PY{l+s+sa}{f}\PY{l+s+s2}{\PYZdq{}}\PY{l+s+s2}{Car }\PY{l+s+s2}{\PYZsq{}}\PY{l+s+si}{\PYZob{}}\PY{n+nb+bp}{self}\PY{o}{.}\PY{n}{car\PYZus{}id}\PY{l+s+si}{\PYZcb{}}\PY{l+s+s2}{\PYZsq{}}\PY{l+s+s2}{:}\PY{l+s+se}{\PYZbs{}n}\PY{l+s+s2}{\PYZdq{}} \PYZbs{}
            \PY{l+s+sa}{f}\PY{l+s+s2}{\PYZdq{}}\PY{l+s+s2}{  at charging station at time slots }\PY{l+s+si}{\PYZob{}}\PY{n+nb+bp}{self}\PY{o}{.}\PY{n}{time\PYZus{}slots\PYZus{}at\PYZus{}charging\PYZus{}unit}\PY{l+s+si}{\PYZcb{}}\PY{l+s+se}{\PYZbs{}n}\PY{l+s+s2}{\PYZdq{}} \PYZbs{}
            \PY{l+s+sa}{f}\PY{l+s+s2}{\PYZdq{}}\PY{l+s+s2}{  requires }\PY{l+s+si}{\PYZob{}}\PY{n+nb+bp}{self}\PY{o}{.}\PY{n}{required\PYZus{}energy}\PY{l+s+si}{\PYZcb{}}\PY{l+s+s2}{ energy units}\PY{l+s+s2}{\PYZdq{}}
    
\PY{k}{class} \PY{n+nc}{ChargingUnit}\PY{p}{:}
    \PY{k}{def} \PY{n+nf+fm}{\PYZus{}\PYZus{}init\PYZus{}\PYZus{}}\PY{p}{(}
        \PY{n+nb+bp}{self}\PY{p}{,}
        \PY{n}{charging\PYZus{}unit\PYZus{}id}\PY{p}{:} \PY{n+nb}{str}\PY{p}{,} \PY{c+c1}{\PYZsh{} an arbitrary name for the charging unit}
        \PY{n}{number\PYZus{}charging\PYZus{}levels}\PY{p}{:} \PY{n+nb}{int}\PY{p}{,}
        \PY{n}{number\PYZus{}time\PYZus{}slots}\PY{p}{:} \PY{n+nb}{int}\PY{p}{,}
    \PY{p}{)} \PY{o}{\PYZhy{}}\PY{o}{\PYZgt{}} \PY{k+kc}{None}\PY{p}{:}
        \PY{n+nb+bp}{self}\PY{o}{.}\PY{n}{charging\PYZus{}unit\PYZus{}id} \PY{o}{=} \PY{n}{charging\PYZus{}unit\PYZus{}id}
        \PY{n+nb+bp}{self}\PY{o}{.}\PY{n}{number\PYZus{}charging\PYZus{}levels} \PY{o}{=} \PY{n}{number\PYZus{}charging\PYZus{}levels}
        \PY{n+nb+bp}{self}\PY{o}{.}\PY{n}{number\PYZus{}time\PYZus{}slots} \PY{o}{=} \PY{n}{number\PYZus{}time\PYZus{}slots}
        \PY{n+nb+bp}{self}\PY{o}{.}\PY{n}{cars\PYZus{}to\PYZus{}charge} \PY{o}{=} \PY{p}{[}\PY{p}{]}

    \PY{k}{def} \PY{n+nf+fm}{\PYZus{}\PYZus{}str\PYZus{}\PYZus{}}\PY{p}{(}\PY{n+nb+bp}{self}\PY{p}{)} \PY{o}{\PYZhy{}}\PY{o}{\PYZgt{}} \PY{n+nb}{str}\PY{p}{:}
        \PY{n}{info\PYZus{}cars\PYZus{}registered} \PY{o}{=} \PY{l+s+s2}{\PYZdq{}}\PY{l+s+s2}{\PYZdq{}}
        \PY{k}{for} \PY{n}{car} \PY{o+ow}{in} \PY{n+nb+bp}{self}\PY{o}{.}\PY{n}{cars\PYZus{}to\PYZus{}charge}\PY{p}{:}
            \PY{n}{info\PYZus{}cars\PYZus{}registered} \PY{o}{=} \PY{n}{info\PYZus{}cars\PYZus{}registered} \PY{o}{+} \PY{l+s+s2}{\PYZdq{}}\PY{l+s+s2}{ }\PY{l+s+s2}{\PYZdq{}} \PY{o}{+} \PY{n}{car}\PY{o}{.}\PY{n}{car\PYZus{}id}
        \PY{k}{return} \PY{l+s+s2}{\PYZdq{}}\PY{l+s+s2}{Charging unit with}\PY{l+s+se}{\PYZbs{}n}\PY{l+s+s2}{\PYZdq{}} \PYZbs{}
            \PY{l+s+s2}{\PYZdq{}}\PY{l+s+s2}{  charging levels: }\PY{l+s+s2}{\PYZdq{}} \PY{o}{+} \PY{n+nb}{str}\PY{p}{(}\PY{n+nb}{list}\PY{p}{(}\PY{n+nb}{range}\PY{p}{(}\PY{n+nb+bp}{self}\PY{o}{.}\PY{n}{number\PYZus{}charging\PYZus{}levels}\PY{p}{)}\PY{p}{)}\PY{p}{)}\PY{p}{[}\PY{l+m+mi}{1}\PY{p}{:}\PY{o}{\PYZhy{}}\PY{l+m+mi}{1}\PY{p}{]} \PY{o}{+} \PY{l+s+s2}{\PYZdq{}}\PY{l+s+se}{\PYZbs{}n}\PY{l+s+s2}{\PYZdq{}} \PYZbs{}
            \PY{l+s+s2}{\PYZdq{}}\PY{l+s+s2}{  time slots: }\PY{l+s+s2}{\PYZdq{}} \PY{o}{+} \PY{n+nb}{str}\PY{p}{(}\PY{n+nb}{list}\PY{p}{(}\PY{n+nb}{range}\PY{p}{(}\PY{n+nb+bp}{self}\PY{o}{.}\PY{n}{number\PYZus{}time\PYZus{}slots}\PY{p}{)}\PY{p}{)}\PY{p}{)}\PY{p}{[}\PY{l+m+mi}{1}\PY{p}{:}\PY{o}{\PYZhy{}}\PY{l+m+mi}{1}\PY{p}{]} \PY{o}{+} \PY{l+s+s2}{\PYZdq{}}\PY{l+s+se}{\PYZbs{}n}\PY{l+s+s2}{\PYZdq{}} \PYZbs{}
            \PY{l+s+s2}{\PYZdq{}}\PY{l+s+s2}{  cars to charge:}\PY{l+s+s2}{\PYZdq{}} \PY{o}{+} \PY{n}{info\PYZus{}cars\PYZus{}registered}

    \PY{k}{def} \PY{n+nf}{register\PYZus{}car\PYZus{}for\PYZus{}charging}\PY{p}{(}\PY{n+nb+bp}{self}\PY{p}{,} \PY{n}{car}\PY{p}{:} \PY{n}{Car}\PY{p}{)} \PY{o}{\PYZhy{}}\PY{o}{\PYZgt{}} \PY{k+kc}{None}\PY{p}{:}
        \PY{k}{if} \PY{n+nb}{max}\PY{p}{(}\PY{n}{car}\PY{o}{.}\PY{n}{time\PYZus{}slots\PYZus{}at\PYZus{}charging\PYZus{}unit}\PY{p}{)} \PY{o}{\PYZgt{}} \PY{n+nb+bp}{self}\PY{o}{.}\PY{n}{number\PYZus{}time\PYZus{}slots} \PY{o}{\PYZhy{}} \PY{l+m+mi}{1}\PY{p}{:}
            \PY{k}{raise} \PY{n+ne}{ValueError}\PY{p}{(}\PY{l+s+s2}{\PYZdq{}}\PY{l+s+s2}{From car required time slots not compatible with charging unit.}\PY{l+s+s2}{\PYZdq{}}\PY{p}{)}
        \PY{n+nb+bp}{self}\PY{o}{.}\PY{n}{cars\PYZus{}to\PYZus{}charge}\PY{o}{.}\PY{n}{append}\PY{p}{(}\PY{n}{car}\PY{p}{)}

    \PY{k}{def} \PY{n+nf}{reset\PYZus{}cars\PYZus{}for\PYZus{}charging}\PY{p}{(}\PY{n+nb+bp}{self}\PY{p}{)} \PY{o}{\PYZhy{}}\PY{o}{\PYZgt{}} \PY{k+kc}{None}\PY{p}{:}
        \PY{n+nb+bp}{self}\PY{o}{.}\PY{n}{cars\PYZus{}to\PYZus{}charge} \PY{o}{=} \PY{p}{[}\PY{p}{]}

    \PY{k}{def} \PY{n+nf}{generate\PYZus{}constraint\PYZus{}matrix}\PY{p}{(}\PY{n+nb+bp}{self}\PY{p}{)} \PY{o}{\PYZhy{}}\PY{o}{\PYZgt{}} \PY{n}{np}\PY{o}{.}\PY{n}{ndarray}\PY{p}{:}
        \PY{l+s+sd}{\PYZdq{}\PYZdq{}\PYZdq{}Matrix with ones for times when car is at charging station}
\PY{l+s+sd}{         and with zeros if car is not at charging station\PYZdq{}\PYZdq{}\PYZdq{}}
        \PY{n}{number\PYZus{}cars\PYZus{}to\PYZus{}charge} \PY{o}{=} \PY{n+nb}{len}\PY{p}{(}\PY{n+nb+bp}{self}\PY{o}{.}\PY{n}{cars\PYZus{}to\PYZus{}charge}\PY{p}{)}
        \PY{n}{constraint\PYZus{}matrix} \PY{o}{=} \PY{n}{np}\PY{o}{.}\PY{n}{zeros}\PY{p}{(}
            \PY{p}{(}\PY{n}{number\PYZus{}cars\PYZus{}to\PYZus{}charge}\PY{p}{,} \PY{n}{number\PYZus{}cars\PYZus{}to\PYZus{}charge}\PY{o}{*}\PY{n+nb+bp}{self}\PY{o}{.}\PY{n}{number\PYZus{}time\PYZus{}slots}\PY{p}{)}\PY{p}{)}
        \PY{k}{for} \PY{n}{row\PYZus{}index} \PY{o+ow}{in} \PY{n+nb}{range}\PY{p}{(}\PY{l+m+mi}{0}\PY{p}{,} \PY{n}{number\PYZus{}cars\PYZus{}to\PYZus{}charge}\PY{p}{)}\PY{p}{:}
            \PY{n}{offset} \PY{o}{=} \PY{n}{row\PYZus{}index}\PY{o}{*}\PY{n+nb+bp}{self}\PY{o}{.}\PY{n}{number\PYZus{}time\PYZus{}slots}
            \PY{n}{cols} \PY{o}{=} \PY{n}{np}\PY{o}{.}\PY{n}{array}\PY{p}{(}\PY{n+nb+bp}{self}\PY{o}{.}\PY{n}{cars\PYZus{}to\PYZus{}charge}\PY{p}{[}\PY{n}{row\PYZus{}index}\PY{p}{]}\PY{o}{.}\PY{n}{time\PYZus{}slots\PYZus{}at\PYZus{}charging\PYZus{}unit}\PY{p}{)}
            \PY{n}{constraint\PYZus{}matrix}\PY{p}{[}\PY{n}{row\PYZus{}index}\PY{p}{,} \PY{n}{offset}\PY{o}{+}\PY{n}{cols}\PY{p}{]} \PY{o}{=} \PY{l+m+mi}{1}
        \PY{k}{return} \PY{n}{constraint\PYZus{}matrix}

    \PY{k}{def} \PY{n+nf}{generate\PYZus{}constraint\PYZus{}rhs}\PY{p}{(}\PY{n+nb+bp}{self}\PY{p}{)} \PY{o}{\PYZhy{}}\PY{o}{\PYZgt{}} \PY{n}{np}\PY{o}{.}\PY{n}{ndarray}\PY{p}{:}
        \PY{l+s+sd}{\PYZdq{}\PYZdq{}\PYZdq{}Vector with required energy as entries\PYZdq{}\PYZdq{}\PYZdq{}}
        \PY{n}{number\PYZus{}cars\PYZus{}to\PYZus{}charge} \PY{o}{=} \PY{n+nb}{len}\PY{p}{(}\PY{n+nb+bp}{self}\PY{o}{.}\PY{n}{cars\PYZus{}to\PYZus{}charge}\PY{p}{)}
        \PY{n}{constraint\PYZus{}rhs} \PY{o}{=} \PY{n}{np}\PY{o}{.}\PY{n}{zeros}\PY{p}{(}\PY{p}{(}\PY{n}{number\PYZus{}cars\PYZus{}to\PYZus{}charge}\PY{p}{,} \PY{l+m+mi}{1}\PY{p}{)}\PY{p}{)}
        \PY{k}{for} \PY{n}{row\PYZus{}index} \PY{o+ow}{in} \PY{n+nb}{range}\PY{p}{(}\PY{l+m+mi}{0}\PY{p}{,} \PY{n}{number\PYZus{}cars\PYZus{}to\PYZus{}charge}\PY{p}{)}\PY{p}{:}
            \PY{n}{constraint\PYZus{}rhs}\PY{p}{[}\PY{n}{row\PYZus{}index}\PY{p}{]} \PY{o}{=} \PY{n+nb+bp}{self}\PY{o}{.}\PY{n}{cars\PYZus{}to\PYZus{}charge}\PY{p}{[}\PY{n}{row\PYZus{}index}\PY{p}{]}\PY{o}{.}\PY{n}{required\PYZus{}energy}
        \PY{k}{return} \PY{n}{constraint\PYZus{}rhs}

    \PY{k}{def} \PY{n+nf}{generate\PYZus{}cost\PYZus{}matrix}\PY{p}{(}\PY{n+nb+bp}{self}\PY{p}{)} \PY{o}{\PYZhy{}}\PY{o}{\PYZgt{}} \PY{n}{np}\PY{o}{.}\PY{n}{ndarray}\PY{p}{:}
        \PY{n}{number\PYZus{}cars\PYZus{}to\PYZus{}charge} \PY{o}{=} \PY{n+nb}{len}\PY{p}{(}\PY{n+nb+bp}{self}\PY{o}{.}\PY{n}{cars\PYZus{}to\PYZus{}charge}\PY{p}{)}
        \PY{k}{return} \PY{n}{np}\PY{o}{.}\PY{n}{kron}\PY{p}{(}
            \PY{n}{np}\PY{o}{.}\PY{n}{ones}\PY{p}{(}\PY{p}{(}\PY{n}{number\PYZus{}cars\PYZus{}to\PYZus{}charge}\PY{p}{,} \PY{l+m+mi}{1}\PY{p}{)}\PY{p}{)} \PY{o}{@} \PY{n}{np}\PY{o}{.}\PY{n}{ones}\PY{p}{(}\PY{p}{(}\PY{l+m+mi}{1}\PY{p}{,} \PY{n}{number\PYZus{}cars\PYZus{}to\PYZus{}charge}\PY{p}{)}\PY{p}{)}\PY{p}{,}
            \PY{n}{np}\PY{o}{.}\PY{n}{eye}\PY{p}{(}\PY{n+nb+bp}{self}\PY{o}{.}\PY{n}{number\PYZus{}time\PYZus{}slots}\PY{p}{)}\PY{p}{)}
    
\PY{k}{def} \PY{n+nf}{generate\PYZus{}qcio}\PY{p}{(}
    \PY{n}{charging\PYZus{}unit}\PY{p}{:} \PY{n}{ChargingUnit}\PY{p}{,}
    \PY{n}{name}\PY{p}{:} \PY{n+nb}{str}\PY{o}{=}\PY{k+kc}{None}
\PY{p}{)} \PY{o}{\PYZhy{}}\PY{o}{\PYZgt{}} \PY{n}{QuadraticProgram}\PY{p}{:}
    \PY{k}{if} \PY{n}{name} \PY{o+ow}{is} \PY{k+kc}{None}\PY{p}{:}
        \PY{n}{name} \PY{o}{=} \PY{l+s+s2}{\PYZdq{}}\PY{l+s+s2}{\PYZdq{}}
    \PY{n}{qcio} \PY{o}{=} \PY{n}{QuadraticProgram}\PY{p}{(}\PY{n}{name}\PY{p}{)}

    \PY{k}{for} \PY{n}{car} \PY{o+ow}{in} \PY{n}{charging\PYZus{}unit}\PY{o}{.}\PY{n}{cars\PYZus{}to\PYZus{}charge}\PY{p}{:}
        \PY{n}{qcio}\PY{o}{.}\PY{n}{integer\PYZus{}var\PYZus{}list}\PY{p}{(}
            \PY{n}{keys}\PY{o}{=}\PY{p}{[}\PY{l+s+sa}{f}\PY{l+s+s2}{\PYZdq{}}\PY{l+s+si}{\PYZob{}}\PY{n}{car}\PY{o}{.}\PY{n}{car\PYZus{}id}\PY{l+s+si}{\PYZcb{}}\PY{l+s+s2}{\PYZus{}t}\PY{l+s+si}{\PYZob{}}\PY{n}{t}\PY{l+s+si}{\PYZcb{}}\PY{l+s+s2}{\PYZdq{}} \PY{k}{for} \PY{n}{t} \PY{o+ow}{in} \PY{n+nb}{range}\PY{p}{(}\PY{l+m+mi}{0}\PY{p}{,} \PY{n}{charging\PYZus{}unit}\PY{o}{.}\PY{n}{number\PYZus{}time\PYZus{}slots}\PY{p}{)}\PY{p}{]}\PY{p}{,}
            \PY{n}{lowerbound}\PY{o}{=}\PY{l+m+mi}{0}\PY{p}{,}
            \PY{n}{upperbound}\PY{o}{=}\PY{n}{charging\PYZus{}unit}\PY{o}{.}\PY{n}{number\PYZus{}charging\PYZus{}levels}\PY{o}{\PYZhy{}}\PY{l+m+mi}{1}\PY{p}{,}
            \PY{n}{name}\PY{o}{=}\PY{l+s+s2}{\PYZdq{}}\PY{l+s+s2}{power.}\PY{l+s+s2}{\PYZdq{}}\PY{p}{)}

    \PY{n}{constraint\PYZus{}matrix} \PY{o}{=} \PY{n}{charging\PYZus{}unit}\PY{o}{.}\PY{n}{generate\PYZus{}constraint\PYZus{}matrix}\PY{p}{(}\PY{p}{)}
    \PY{n}{constraint\PYZus{}rhs} \PY{o}{=} \PY{n}{charging\PYZus{}unit}\PY{o}{.}\PY{n}{generate\PYZus{}constraint\PYZus{}rhs}\PY{p}{(}\PY{p}{)}
    \PY{k}{for} \PY{n}{row\PYZus{}index} \PY{o+ow}{in} \PY{n+nb}{range}\PY{p}{(}\PY{l+m+mi}{0}\PY{p}{,} \PY{n}{constraint\PYZus{}matrix}\PY{o}{.}\PY{n}{shape}\PY{p}{[}\PY{l+m+mi}{0}\PY{p}{]}\PY{p}{)}\PY{p}{:}
        \PY{n}{qcio}\PY{o}{.}\PY{n}{linear\PYZus{}constraint}\PY{p}{(}
            \PY{n}{linear}\PY{o}{=}\PY{n}{constraint\PYZus{}matrix}\PY{p}{[}\PY{n}{row\PYZus{}index}\PY{p}{,} \PY{p}{:}\PY{p}{]}\PY{p}{,}
            \PY{n}{rhs}\PY{o}{=}\PY{n}{constraint\PYZus{}rhs}\PY{p}{[}\PY{n}{row\PYZus{}index}\PY{p}{]}\PY{p}{[}\PY{l+m+mi}{0}\PY{p}{]}\PY{p}{,}
            \PY{n}{sense}\PY{o}{=}\PY{l+s+s2}{\PYZdq{}}\PY{l+s+s2}{==}\PY{l+s+s2}{\PYZdq{}}\PY{p}{,}
            \PY{n}{name}\PY{o}{=}\PY{l+s+sa}{f}\PY{l+s+s2}{\PYZdq{}}\PY{l+s+s2}{charge\PYZus{}correct\PYZus{}energy\PYZus{}for\PYZus{}}\PY{l+s+si}{\PYZob{}}\PY{n}{charging\PYZus{}unit}\PY{o}{.}\PY{n}{cars\PYZus{}to\PYZus{}charge}\PY{p}{[}\PY{n}{row\PYZus{}index}\PY{p}{]}\PY{o}{.}\PY{n}{car\PYZus{}id}\PY{l+s+si}{\PYZcb{}}\PY{l+s+s2}{\PYZdq{}}\PY{p}{)}

    \PY{n}{cost\PYZus{}matrix} \PY{o}{=} \PY{n}{charging\PYZus{}unit}\PY{o}{.}\PY{n}{generate\PYZus{}cost\PYZus{}matrix}\PY{p}{(}\PY{p}{)}
    \PY{n}{qcio}\PY{o}{.}\PY{n}{minimize}\PY{p}{(}\PY{n}{quadratic}\PY{o}{=}\PY{n}{cost\PYZus{}matrix}\PY{p}{)}

    \PY{k}{return} \PY{n}{qcio}

\PY{k}{class} \PY{n+nc}{Converter}\PY{p}{(}\PY{n}{QuadraticProgramConverter}\PY{p}{)}\PY{p}{:}
    \PY{k}{def} \PY{n+nf+fm}{\PYZus{}\PYZus{}init\PYZus{}\PYZus{}}\PY{p}{(}
        \PY{n+nb+bp}{self}\PY{p}{,} 
        \PY{n}{penalty}\PY{p}{:} \PY{n+nb}{float}\PY{o}{=}\PY{k+kc}{None} \PY{c+c1}{\PYZsh{} the penalty paramter for step 1}
    \PY{p}{)} \PY{o}{\PYZhy{}}\PY{o}{\PYZgt{}} \PY{k+kc}{None}\PY{p}{:}
        \PY{n+nb}{super}\PY{p}{(}\PY{p}{)}\PY{o}{.}\PY{n+nf+fm}{\PYZus{}\PYZus{}init\PYZus{}\PYZus{}}\PY{p}{(}\PY{p}{)}
        \PY{n+nb+bp}{self}\PY{o}{.}\PY{n}{\PYZus{}penalty} \PY{o}{=} \PY{n}{penalty}
        \PY{n+nb+bp}{self}\PY{o}{.}\PY{n}{linear\PYZus{}equality\PYZus{}to\PYZus{}penalty\PYZus{}converter} \PY{o}{=} \PY{n}{LinearEqualityToPenalty}\PY{p}{(}\PY{n}{penalty}\PY{p}{)}
        \PY{n+nb+bp}{self}\PY{o}{.}\PY{n}{integer\PYZus{}to\PYZus{}binary\PYZus{}converter} \PY{o}{=} \PY{n}{IntegerToBinary}\PY{p}{(}\PY{p}{)}

    \PY{k}{def} \PY{n+nf}{convert}\PY{p}{(}\PY{n+nb+bp}{self}\PY{p}{,} \PY{n}{quadratic\PYZus{}program}\PY{p}{:} \PY{n}{QuadraticProgram}\PY{p}{)} \PY{o}{\PYZhy{}}\PY{o}{\PYZgt{}} \PY{n}{QuadraticProgram}\PY{p}{:}
        \PY{k}{return} \PY{n+nb+bp}{self}\PY{o}{.}\PY{n}{integer\PYZus{}to\PYZus{}binary\PYZus{}converter}\PY{o}{.}\PY{n}{convert}\PY{p}{(}
            \PY{n+nb+bp}{self}\PY{o}{.}\PY{n}{linear\PYZus{}equality\PYZus{}to\PYZus{}penalty\PYZus{}converter}\PY{o}{.}\PY{n}{convert}\PY{p}{(}\PY{n}{quadratic\PYZus{}program}\PY{p}{)}\PY{p}{)}
    
    \PY{k}{def} \PY{n+nf}{interpret}\PY{p}{(}\PY{n+nb+bp}{self}\PY{p}{,} \PY{n}{x}\PY{p}{:} \PY{n}{Union}\PY{p}{[}\PY{n}{np}\PY{o}{.}\PY{n}{ndarray}\PY{p}{,} \PY{n}{List}\PY{p}{[}\PY{n+nb}{float}\PY{p}{]}\PY{p}{]}\PY{p}{)} \PY{o}{\PYZhy{}}\PY{o}{\PYZgt{}} \PY{n}{np}\PY{o}{.}\PY{n}{ndarray}\PY{p}{:}
        \PY{k}{return} \PY{n+nb+bp}{self}\PY{o}{.}\PY{n}{linear\PYZus{}equality\PYZus{}to\PYZus{}penalty\PYZus{}converter}\PY{o}{.}\PY{n}{interpret}\PY{p}{(}
            \PY{n+nb+bp}{self}\PY{o}{.}\PY{n}{integer\PYZus{}to\PYZus{}binary\PYZus{}converter}\PY{o}{.}\PY{n}{interpret}\PY{p}{(}\PY{n}{x}\PY{p}{)}\PY{p}{)}

\PY{c+c1}{\PYZsh{}\PYZsh{} \PYZhy{}\PYZhy{}\PYZhy{} Example for Notebook 2 \PYZhy{}\PYZhy{}\PYZhy{} \PYZsh{}\PYZsh{}}

\PY{k}{def} \PY{n+nf}{generate\PYZus{}example}\PY{p}{(}\PY{p}{)}\PY{p}{:}
    \PY{n}{charging\PYZus{}unit} \PY{o}{=} \PY{n}{ChargingUnit}\PY{p}{(}
        \PY{n}{charging\PYZus{}unit\PYZus{}id}\PY{o}{=}\PY{l+s+s2}{\PYZdq{}}\PY{l+s+s2}{charging\PYZus{}unit}\PY{l+s+s2}{\PYZdq{}}\PY{p}{,}
        \PY{n}{number\PYZus{}charging\PYZus{}levels}\PY{o}{=}\PY{l+m+mi}{4}\PY{p}{,}
        \PY{n}{number\PYZus{}time\PYZus{}slots}\PY{o}{=}\PY{l+m+mi}{4}\PY{p}{)}
    \PY{n}{car\PYZus{}green} \PY{o}{=} \PY{n}{Car}\PY{p}{(}
        \PY{n}{car\PYZus{}id}\PY{o}{=}\PY{l+s+s2}{\PYZdq{}}\PY{l+s+s2}{car\PYZus{}green}\PY{l+s+s2}{\PYZdq{}}\PY{p}{,}
        \PY{n}{time\PYZus{}slots\PYZus{}at\PYZus{}charging\PYZus{}unit}\PY{o}{=}\PY{p}{[}\PY{l+m+mi}{0}\PY{p}{,} \PY{l+m+mi}{1}\PY{p}{,} \PY{l+m+mi}{2}\PY{p}{]}\PY{p}{,}
        \PY{n}{required\PYZus{}energy}\PY{o}{=}\PY{l+m+mi}{4}\PY{p}{)}
    \PY{n}{charging\PYZus{}unit}\PY{o}{.}\PY{n}{register\PYZus{}car\PYZus{}for\PYZus{}charging}\PY{p}{(}\PY{n}{car\PYZus{}green}\PY{p}{)}

    \PY{n}{qcio} \PY{o}{=} \PY{n}{generate\PYZus{}qcio}\PY{p}{(}\PY{n}{charging\PYZus{}unit}\PY{p}{,} \PY{n}{name}\PY{o}{=}\PY{l+s+s2}{\PYZdq{}}\PY{l+s+s2}{QCIO}\PY{l+s+s2}{\PYZdq{}}\PY{p}{)}
    \PY{n}{converter} \PY{o}{=} \PY{n}{Converter}\PY{p}{(}\PY{n}{penalty}\PY{o}{=}\PY{l+m+mf}{3.6}\PY{p}{)}
    \PY{n}{qubo} \PY{o}{=} \PY{n}{converter}\PY{o}{.}\PY{n}{convert}\PY{p}{(}\PY{n}{qcio}\PY{p}{)}
    \PY{n}{number\PYZus{}binary\PYZus{}variables} \PY{o}{=} \PY{n}{qubo}\PY{o}{.}\PY{n}{get\PYZus{}num\PYZus{}binary\PYZus{}vars}\PY{p}{(}\PY{p}{)}

    \PY{n}{cplex\PYZus{}optimizer} \PY{o}{=} \PY{n}{CplexOptimizer}\PY{p}{(}\PY{p}{)}
    \PY{n}{qubo\PYZus{}minimization\PYZus{}result} \PY{o}{=} \PY{n}{cplex\PYZus{}optimizer}\PY{o}{.}\PY{n}{solve}\PY{p}{(}\PY{n}{qubo}\PY{p}{)}

    \PY{k}{return} \PY{n}{charging\PYZus{}unit}\PY{p}{,} \PY{n}{car\PYZus{}green}\PY{p}{,} \PY{n}{qcio}\PY{p}{,} \PY{n}{converter}\PY{p}{,} \PY{n}{qubo}\PY{p}{,} \PY{n}{number\PYZus{}binary\PYZus{}variables}\PY{p}{,} \PY{n}{qubo\PYZus{}minimization\PYZus{}result}
\end{Verbatim}
\end{tcolorbox}

\vspace{1.cm}
Code for \verb|codes_notebook_2.py|:

\begin{tcolorbox}[breakable, size=fbox, boxrule=1pt, pad at break*=1mm,colback=cellbackground, colframe=cellborder]
\begin{Verbatim}[commandchars=\\\{\}]
\PY{k+kn}{from} \PY{n+nn}{qiskit}\PY{n+nn}{.}\PY{n+nn}{circuit}\PY{n+nn}{.}\PY{n+nn}{library}\PY{n+nn}{.}\PY{n+nn}{n\PYZus{}local} \PY{k+kn}{import} \PY{n}{QAOAAnsatz}
\PY{k+kn}{from} \PY{n+nn}{codes\PYZus{}notebook\PYZus{}1} \PY{k+kn}{import} \PY{n}{generate\PYZus{}example} \PY{k}{as} \PY{n}{generate\PYZus{}example\PYZus{}notebook\PYZus{}1}

\PY{k}{def} \PY{n+nf}{generate\PYZus{}example}\PY{p}{(}\PY{p}{)}\PY{p}{:}
    \PY{n}{charging\PYZus{}unit}\PY{p}{,} \PY{n}{car\PYZus{}green}\PY{p}{,} \PY{n}{qcio}\PY{p}{,} \PY{n}{converter}\PY{p}{,} \PY{n}{qubo}\PY{p}{,} \PY{n}{number\PYZus{}binary\PYZus{}variables}\PY{p}{,} \PY{n}{qubo\PYZus{}minimization\PYZus{}result} \PY{o}{=} \PY{n}{generate\PYZus{}example\PYZus{}notebook\PYZus{}1}\PY{p}{(}\PY{p}{)}
    
    \PY{n}{ising}\PY{p}{,} \PY{n}{ising\PYZus{}offset} \PY{o}{=} \PY{n}{qubo}\PY{o}{.}\PY{n}{to\PYZus{}ising}\PY{p}{(}\PY{p}{)}
    
    \PY{n}{qaoa\PYZus{}reps} \PY{o}{=} \PY{l+m+mi}{2}
    \PY{n}{qaoa\PYZus{}circuit} \PY{o}{=} \PY{n}{QAOAAnsatz}\PY{p}{(}\PY{n}{cost\PYZus{}operator}\PY{o}{=}\PY{n}{ising}\PY{p}{,} \PY{n}{reps}\PY{o}{=}\PY{n}{qaoa\PYZus{}reps}\PY{p}{)}
    \PY{n}{qaoa\PYZus{}circuit}\PY{o}{.}\PY{n}{measure\PYZus{}all}\PY{p}{(}\PY{p}{)}

    \PY{k}{return} \PY{n}{charging\PYZus{}unit}\PY{p}{,} \PY{n}{car\PYZus{}green}\PY{p}{,} \PY{n}{qcio}\PY{p}{,} \PY{n}{converter}\PY{p}{,} \PY{n}{qubo}\PY{p}{,} \PY{n}{number\PYZus{}binary\PYZus{}variables}\PY{p}{,} \PY{n}{qubo\PYZus{}minimization\PYZus{}result}\PY{p}{,} \PY{n}{ising}\PY{p}{,} \PY{n}{ising\PYZus{}offset}\PY{p}{,} \PY{n}{qaoa\PYZus{}reps}\PY{p}{,} \PY{n}{qaoa\PYZus{}circuit}
\end{Verbatim}
\end{tcolorbox}

\vspace{1.cm}
Code for \verb|utils.py|:

\begin{tcolorbox}[breakable, size=fbox, boxrule=1pt, pad at break*=1mm,colback=cellbackground, colframe=cellborder]
    \begin{Verbatim}[commandchars=\\\{\}]
    \PY{k+kn}{from} \PY{n+nn}{typing} \PY{k+kn}{import} \PY{n}{Union}\PY{p}{,} \PY{n}{List}
    \PY{k+kn}{import} \PY{n+nn}{numpy} \PY{k}{as} \PY{n+nn}{np}
    \PY{k+kn}{from} \PY{n+nn}{datetime} \PY{k+kn}{import} \PY{n}{datetime}
    \PY{k+kn}{from} \PY{n+nn}{pathlib} \PY{k+kn}{import} \PY{n}{Path}
    \PY{k+kn}{import} \PY{n+nn}{pickle}
    \PY{k+kn}{import} \PY{n+nn}{plotly}\PY{n+nn}{.}\PY{n+nn}{graph\PYZus{}objects} \PY{k}{as} \PY{n+nn}{go}
    \PY{k+kn}{from} \PY{n+nn}{qiskit}\PY{n+nn}{.}\PY{n+nn}{circuit} \PY{k+kn}{import} \PY{n}{QuantumCircuit}
    
    \PY{k}{def} \PY{n+nf}{plot\PYZus{}charging\PYZus{}schedule}\PY{p}{(}
            \PY{n}{charging\PYZus{}unit}\PY{p}{,}
            \PY{n}{minimization\PYZus{}result\PYZus{}x}\PY{p}{,}
            \PY{n}{marker\PYZus{}size}\PY{o}{=}\PY{l+m+mi}{50}\PY{p}{,}
        \PY{p}{)} \PY{o}{\PYZhy{}}\PY{o}{\PYZgt{}} \PY{n}{go}\PY{o}{.}\PY{n}{Figure}\PY{p}{:}
        \PY{n}{marker\PYZus{}colors} \PY{o}{=} \PY{p}{[}\PY{l+s+s2}{\PYZdq{}}\PY{l+s+s2}{green}\PY{l+s+s2}{\PYZdq{}}\PY{p}{,} \PY{l+s+s2}{\PYZdq{}}\PY{l+s+s2}{orange}\PY{l+s+s2}{\PYZdq{}}\PY{p}{,} \PY{l+s+s2}{\PYZdq{}}\PY{l+s+s2}{blue}\PY{l+s+s2}{\PYZdq{}}\PY{p}{,} \PY{l+s+s2}{\PYZdq{}}\PY{l+s+s2}{red}\PY{l+s+s2}{\PYZdq{}}\PY{p}{,} \PY{l+s+s2}{\PYZdq{}}\PY{l+s+s2}{magenta}\PY{l+s+s2}{\PYZdq{}}\PY{p}{,} \PY{l+s+s2}{\PYZdq{}}\PY{l+s+s2}{goldenrod}\PY{l+s+s2}{\PYZdq{}}\PY{p}{]}
        \PY{n}{time\PYZus{}slots} \PY{o}{=} \PY{n}{np}\PY{o}{.}\PY{n}{arange}\PY{p}{(}\PY{l+m+mi}{0}\PY{p}{,} \PY{n}{charging\PYZus{}unit}\PY{o}{.}\PY{n}{number\PYZus{}time\PYZus{}slots}\PY{p}{)}
        \PY{n}{fig} \PY{o}{=} \PY{n}{go}\PY{o}{.}\PY{n}{Figure}\PY{p}{(}\PY{p}{)}
        \PY{n}{already\PYZus{}in\PYZus{}legend} \PY{o}{=} \PY{p}{[}\PY{p}{]}
        \PY{k}{for} \PY{n}{t} \PY{o+ow}{in} \PY{n}{time\PYZus{}slots}\PY{p}{:}
            \PY{n}{offset} \PY{o}{=} \PY{l+m+mi}{0}
            \PY{k}{for} \PY{n}{car\PYZus{}num} \PY{o+ow}{in} \PY{n}{np}\PY{o}{.}\PY{n}{arange}\PY{p}{(}\PY{l+m+mi}{0}\PY{p}{,} \PY{n+nb}{len}\PY{p}{(}\PY{n}{charging\PYZus{}unit}\PY{o}{.}\PY{n}{cars\PYZus{}to\PYZus{}charge}\PY{p}{)}\PY{p}{)}\PY{p}{:}
                \PY{n}{car\PYZus{}id\PYZus{}current\PYZus{}car} \PY{o}{=} \PY{n}{charging\PYZus{}unit}\PY{o}{.}\PY{n}{cars\PYZus{}to\PYZus{}charge}\PY{p}{[}\PY{n}{car\PYZus{}num}\PY{p}{]}\PY{o}{.}\PY{n}{car\PYZus{}id}
                \PY{n}{minimization\PYZus{}result\PYZus{}x\PYZus{}current\PYZus{}car} \PY{o}{=} \PY{n}{minimization\PYZus{}result\PYZus{}x}\PY{p}{[}
                    \PY{n}{car\PYZus{}num}\PY{o}{*}\PY{n}{charging\PYZus{}unit}\PY{o}{.}\PY{n}{number\PYZus{}time\PYZus{}slots}\PY{p}{:}\PY{p}{(}\PY{n}{car\PYZus{}num}\PY{o}{+}\PY{l+m+mi}{1}\PY{p}{)}\PY{o}{*}\PY{n}{charging\PYZus{}unit}\PY{o}{.}\PY{n}{number\PYZus{}time\PYZus{}slots}\PY{p}{]}
                \PY{n}{power\PYZus{}t} \PY{o}{=} \PY{n}{minimization\PYZus{}result\PYZus{}x\PYZus{}current\PYZus{}car}\PY{p}{[}\PY{n}{t}\PY{p}{]}
                \PY{n}{fig}\PY{o}{.}\PY{n}{add\PYZus{}trace}\PY{p}{(}\PY{n}{go}\PY{o}{.}\PY{n}{Scatter}\PY{p}{(}
                    \PY{n}{x}\PY{o}{=}\PY{p}{[}\PY{n}{t}\PY{o}{+}\PY{l+m+mf}{0.5}\PY{p}{]}\PY{o}{*}\PY{n+nb}{int}\PY{p}{(}\PY{n}{power\PYZus{}t}\PY{p}{)}\PY{p}{,}
                    \PY{n}{y}\PY{o}{=}\PY{n}{offset} \PY{o}{+} \PY{n}{np}\PY{o}{.}\PY{n}{arange}\PY{p}{(}\PY{l+m+mi}{0}\PY{p}{,} \PY{n}{power\PYZus{}t}\PY{p}{)}\PY{p}{,}
                    \PY{n}{mode}\PY{o}{=}\PY{l+s+s2}{\PYZdq{}}\PY{l+s+s2}{markers}\PY{l+s+s2}{\PYZdq{}}\PY{p}{,}
                    \PY{n}{marker\PYZus{}symbol}\PY{o}{=}\PY{l+s+s2}{\PYZdq{}}\PY{l+s+s2}{square}\PY{l+s+s2}{\PYZdq{}}\PY{p}{,}
                    \PY{n}{marker\PYZus{}size}\PY{o}{=}\PY{n}{marker\PYZus{}size}\PY{p}{,}
                    \PY{n}{marker\PYZus{}color}\PY{o}{=}\PY{n}{marker\PYZus{}colors}\PY{p}{[}\PY{n}{car\PYZus{}num}\PY{p}{]}\PY{p}{,}
                    \PY{n}{name}\PY{o}{=}\PY{n}{car\PYZus{}id\PYZus{}current\PYZus{}car}\PY{p}{,}
                    \PY{n}{showlegend}\PY{o}{=}\PY{k+kc}{False} \PY{k}{if} \PY{n}{car\PYZus{}id\PYZus{}current\PYZus{}car} \PY{o+ow}{in} \PY{n}{already\PYZus{}in\PYZus{}legend} \PY{k}{else} \PY{k+kc}{True}
                \PY{p}{)}\PY{p}{)}
                \PY{n}{offset} \PY{o}{+}\PY{o}{=} \PY{n}{power\PYZus{}t}
                \PY{k}{if} \PY{n}{power\PYZus{}t} \PY{o}{\PYZgt{}} \PY{l+m+mi}{0}\PY{p}{:}
                    \PY{n}{already\PYZus{}in\PYZus{}legend}\PY{o}{.}\PY{n}{append}\PY{p}{(}\PY{n}{car\PYZus{}id\PYZus{}current\PYZus{}car}\PY{p}{)}
        
        \PY{n}{fig}\PY{o}{.}\PY{n}{update\PYZus{}xaxes}\PY{p}{(}
            \PY{n}{tick0}\PY{o}{=}\PY{l+m+mi}{1}\PY{p}{,}
            \PY{n}{dtick}\PY{o}{=}\PY{l+m+mi}{1}\PY{p}{,}
            \PY{n+nb}{range}\PY{o}{=}\PY{p}{[}\PY{l+m+mf}{0.01}\PY{p}{,} \PY{n}{charging\PYZus{}unit}\PY{o}{.}\PY{n}{number\PYZus{}time\PYZus{}slots}\PY{p}{]}\PY{p}{,}
            \PY{n}{tickvals}\PY{o}{=}\PY{n}{np}\PY{o}{.}\PY{n}{arange}\PY{p}{(}\PY{l+m+mf}{0.5}\PY{p}{,} \PY{n}{charging\PYZus{}unit}\PY{o}{.}\PY{n}{number\PYZus{}time\PYZus{}slots}\PY{p}{)}\PY{p}{,}
            \PY{n}{ticktext}\PY{o}{=}\PY{n}{np}\PY{o}{.}\PY{n}{arange}\PY{p}{(}\PY{l+m+mi}{0}\PY{p}{,} \PY{n}{charging\PYZus{}unit}\PY{o}{.}\PY{n}{number\PYZus{}time\PYZus{}slots}\PY{p}{)}\PY{p}{,}
            \PY{n}{title}\PY{o}{=}\PY{l+s+s2}{\PYZdq{}}\PY{l+s+s2}{time slot}\PY{l+s+s2}{\PYZdq{}}\PY{p}{,}
            \PY{n}{title\PYZus{}font\PYZus{}size}\PY{o}{=}\PY{l+m+mi}{12}\PY{p}{,}
        \PY{p}{)}
        \PY{n}{fig}\PY{o}{.}\PY{n}{update\PYZus{}yaxes}\PY{p}{(}
            \PY{n+nb}{range}\PY{o}{=}\PY{p}{[}\PY{o}{\PYZhy{}}\PY{l+m+mf}{0.6}\PY{p}{,} \PY{n}{charging\PYZus{}unit}\PY{o}{.}\PY{n}{number\PYZus{}charging\PYZus{}levels}\PY{o}{\PYZhy{}}\PY{l+m+mi}{1}\PY{p}{]}\PY{p}{,}
            \PY{n}{tickvals}\PY{o}{=}\PY{n}{np}\PY{o}{.}\PY{n}{arange}\PY{p}{(}\PY{o}{\PYZhy{}}\PY{l+m+mf}{0.5}\PY{p}{,} \PY{n}{charging\PYZus{}unit}\PY{o}{.}\PY{n}{number\PYZus{}charging\PYZus{}levels}\PY{o}{\PYZhy{}}\PY{l+m+mf}{0.5}\PY{p}{)}\PY{p}{,}
            \PY{n}{ticktext}\PY{o}{=}\PY{n}{np}\PY{o}{.}\PY{n}{arange}\PY{p}{(}\PY{l+m+mi}{0}\PY{p}{,} \PY{n}{charging\PYZus{}unit}\PY{o}{.}\PY{n}{number\PYZus{}charging\PYZus{}levels}\PY{p}{)}\PY{p}{,}
            \PY{n}{title}\PY{o}{=}\PY{l+s+s2}{\PYZdq{}}\PY{l+s+s2}{charging level}\PY{l+s+s2}{\PYZdq{}}\PY{p}{,}
            \PY{n}{title\PYZus{}font\PYZus{}size}\PY{o}{=}\PY{l+m+mi}{12}\PY{p}{,}
            \PY{n}{zeroline}\PY{o}{=}\PY{k+kc}{False}
        \PY{p}{)}
        \PY{k}{return} \PY{n}{fig}
    
    \PY{k}{def} \PY{n+nf}{convert\PYZus{}to\PYZus{}date\PYZus{}and\PYZus{}time\PYZus{}string}\PY{p}{(}\PY{n}{time\PYZus{}stamp}\PY{p}{:} \PY{n}{Union}\PY{p}{[}\PY{n}{datetime}\PY{p}{,} \PY{n+nb}{str}\PY{p}{]}\PY{p}{)}\PY{p}{:}
        \PY{k}{if} \PY{n+nb}{isinstance}\PY{p}{(}\PY{n}{time\PYZus{}stamp}\PY{p}{,} \PY{n}{datetime}\PY{p}{)}\PY{p}{:}
            \PY{n}{output} \PY{o}{=} \PY{n+nb}{str}\PY{p}{(}\PY{n}{time\PYZus{}stamp}\PY{o}{.}\PY{n}{year}\PY{p}{)} \PY{o}{+} \PY{l+s+s2}{\PYZdq{}}\PY{l+s+s2}{\PYZus{}}\PY{l+s+s2}{\PYZdq{}} \PY{o}{+} \PYZbs{}
                \PY{n+nb}{str}\PY{p}{(}\PY{n}{time\PYZus{}stamp}\PY{o}{.}\PY{n}{month}\PY{p}{)}\PY{o}{.}\PY{n}{rjust}\PY{p}{(}\PY{l+m+mi}{2}\PY{p}{,} \PY{l+s+s1}{\PYZsq{}}\PY{l+s+s1}{0}\PY{l+s+s1}{\PYZsq{}}\PY{p}{)} \PY{o}{+} \PY{l+s+s2}{\PYZdq{}}\PY{l+s+s2}{\PYZus{}}\PY{l+s+s2}{\PYZdq{}} \PY{o}{+} \PYZbs{}
                \PY{n+nb}{str}\PY{p}{(}\PY{n}{time\PYZus{}stamp}\PY{o}{.}\PY{n}{day}\PY{p}{)}\PY{o}{.}\PY{n}{rjust}\PY{p}{(}\PY{l+m+mi}{2}\PY{p}{,} \PY{l+s+s1}{\PYZsq{}}\PY{l+s+s1}{0}\PY{l+s+s1}{\PYZsq{}}\PY{p}{)} \PY{o}{+} \PY{l+s+s2}{\PYZdq{}}\PY{l+s+s2}{\PYZhy{}}\PY{l+s+s2}{\PYZdq{}} \PY{o}{+} \PYZbs{}
                \PY{n+nb}{str}\PY{p}{(}\PY{n}{time\PYZus{}stamp}\PY{o}{.}\PY{n}{hour}\PY{p}{)}\PY{o}{.}\PY{n}{rjust}\PY{p}{(}\PY{l+m+mi}{2}\PY{p}{,} \PY{l+s+s1}{\PYZsq{}}\PY{l+s+s1}{0}\PY{l+s+s1}{\PYZsq{}}\PY{p}{)} \PY{o}{+} \PY{l+s+s2}{\PYZdq{}}\PY{l+s+s2}{h}\PY{l+s+s2}{\PYZdq{}} \PY{o}{+} \PYZbs{}
                \PY{n+nb}{str}\PY{p}{(}\PY{n}{time\PYZus{}stamp}\PY{o}{.}\PY{n}{minute}\PY{p}{)}\PY{o}{.}\PY{n}{rjust}\PY{p}{(}\PY{l+m+mi}{2}\PY{p}{,} \PY{l+s+s1}{\PYZsq{}}\PY{l+s+s1}{0}\PY{l+s+s1}{\PYZsq{}}\PY{p}{)} \PY{o}{+} \PY{l+s+s2}{\PYZdq{}}\PY{l+s+s2}{m}\PY{l+s+s2}{\PYZdq{}}
        \PY{k}{elif} \PY{n+nb}{isinstance}\PY{p}{(}\PY{n}{time\PYZus{}stamp}\PY{p}{,} \PY{n+nb}{str}\PY{p}{)}\PY{p}{:}
            \PY{n}{output} \PY{o}{=} \PY{n}{time\PYZus{}stamp}\PY{p}{[}\PY{l+m+mi}{0}\PY{p}{:}\PY{l+m+mi}{17}\PY{p}{]}\PY{o}{.}\PY{n}{replace}\PY{p}{(}\PY{l+s+s1}{\PYZsq{}}\PY{l+s+s1}{\PYZhy{}}\PY{l+s+s1}{\PYZsq{}}\PY{p}{,} \PY{l+s+s1}{\PYZsq{}}\PY{l+s+s1}{\PYZus{}}\PY{l+s+s1}{\PYZsq{}}\PY{p}{)}\PY{o}{.}\PY{n}{replace}\PY{p}{(}\PY{l+s+s1}{\PYZsq{}}\PY{l+s+s1}{T}\PY{l+s+s1}{\PYZsq{}}\PY{p}{,} \PY{l+s+s1}{\PYZsq{}}\PY{l+s+s1}{\PYZhy{}}\PY{l+s+s1}{\PYZsq{}}\PY{p}{)}\PY{o}{.}\PY{n}{replace}\PY{p}{(}\PY{l+s+s1}{\PYZsq{}}\PY{l+s+s1}{:}\PY{l+s+s1}{\PYZsq{}}\PY{p}{,} \PY{l+s+s1}{\PYZsq{}}\PY{l+s+s1}{h}\PY{l+s+s1}{\PYZsq{}}\PY{p}{,} \PY{l+m+mi}{1}\PY{p}{)}\PY{o}{.}\PY{n}{replace}\PY{p}{(}\PY{l+s+s1}{\PYZsq{}}\PY{l+s+s1}{:}\PY{l+s+s1}{\PYZsq{}}\PY{p}{,} \PY{l+s+s1}{\PYZsq{}}\PY{l+s+s1}{m}\PY{l+s+s1}{\PYZsq{}}\PY{p}{,} \PY{l+m+mi}{1}\PY{p}{)}
        \PY{k}{else}\PY{p}{:}
            \PY{k}{raise} \PY{n+ne}{ValueError}\PY{p}{(}\PY{l+s+s2}{\PYZdq{}}\PY{l+s+s2}{data type of }\PY{l+s+s2}{\PYZsq{}}\PY{l+s+s2}{time\PYZus{}stamp}\PY{l+s+s2}{\PYZsq{}}\PY{l+s+s2}{ not supported}\PY{l+s+s2}{\PYZdq{}}\PY{p}{)}
        \PY{k}{return} \PY{n}{output}
    
    \PY{k}{def} \PY{n+nf}{save\PYZus{}token}\PY{p}{(}\PY{n}{token}\PY{p}{:} \PY{n+nb}{str}\PY{p}{,} \PY{n}{file\PYZus{}name}\PY{p}{:} \PY{n+nb}{str}\PY{p}{)}\PY{p}{:}
        \PY{n}{path\PYZus{}token\PYZus{}file} \PY{o}{=} \PY{n}{Path}\PY{p}{(}\PY{n}{file\PYZus{}name}\PY{p}{)}\PY{o}{.}\PY{n}{with\PYZus{}suffix}\PY{p}{(}\PY{l+s+s2}{\PYZdq{}}\PY{l+s+s2}{.pickle}\PY{l+s+s2}{\PYZdq{}}\PY{p}{)}
        \PY{k}{if} \PY{n}{path\PYZus{}token\PYZus{}file}\PY{o}{.}\PY{n}{exists}\PY{p}{(}\PY{p}{)}\PY{p}{:}
            \PY{n+nb}{print}\PY{p}{(}\PY{l+s+s2}{\PYZdq{}}\PY{l+s+s2}{Token already saved.}\PY{l+s+s2}{\PYZdq{}}\PY{p}{)}
        \PY{k}{else}\PY{p}{:}
            \PY{k}{with} \PY{n+nb}{open}\PY{p}{(}\PY{n}{path\PYZus{}token\PYZus{}file}\PY{p}{,} \PY{l+s+s1}{\PYZsq{}}\PY{l+s+s1}{wb}\PY{l+s+s1}{\PYZsq{}}\PY{p}{)} \PY{k}{as} \PY{n}{file}\PY{p}{:}
                \PY{n}{pickle}\PY{o}{.}\PY{n}{dump}\PY{p}{(}\PY{n}{token}\PY{p}{,} \PY{n}{file}\PY{p}{)}
            \PY{n+nb}{print}\PY{p}{(}\PY{l+s+sa}{f}\PY{l+s+s2}{\PYZdq{}}\PY{l+s+s2}{Token has been saved in }\PY{l+s+s2}{\PYZsq{}}\PY{l+s+si}{\PYZob{}}\PY{n}{file\PYZus{}name}\PY{l+s+si}{\PYZcb{}}\PY{l+s+s2}{.pickle}\PY{l+s+s2}{\PYZsq{}}\PY{l+s+s2}{.}\PY{l+s+s2}{\PYZdq{}}\PY{p}{)}
            
    \PY{k}{def} \PY{n+nf}{load\PYZus{}token}\PY{p}{(}\PY{n}{file\PYZus{}name}\PY{p}{:} \PY{n+nb}{str}\PY{p}{)}\PY{p}{:}
        \PY{n}{path\PYZus{}token\PYZus{}file} \PY{o}{=} \PY{n}{Path}\PY{p}{(}\PY{n}{file\PYZus{}name}\PY{p}{)}\PY{o}{.}\PY{n}{with\PYZus{}suffix}\PY{p}{(}\PY{l+s+s2}{\PYZdq{}}\PY{l+s+s2}{.pickle}\PY{l+s+s2}{\PYZdq{}}\PY{p}{)}
        \PY{k}{try}\PY{p}{:}
            \PY{k}{with} \PY{n+nb}{open}\PY{p}{(}\PY{n}{path\PYZus{}token\PYZus{}file}\PY{p}{,} \PY{l+s+s1}{\PYZsq{}}\PY{l+s+s1}{rb}\PY{l+s+s1}{\PYZsq{}}\PY{p}{)} \PY{k}{as} \PY{n}{file}\PY{p}{:}
                \PY{n}{token} \PY{o}{=} \PY{n}{pickle}\PY{o}{.}\PY{n}{load}\PY{p}{(}\PY{n}{file}\PY{p}{)}
        \PY{k}{except} \PY{n+ne}{FileNotFoundError}\PY{p}{:}
            \PY{k}{raise} \PY{n+ne}{FileNotFoundError}\PY{p}{(}\PY{l+s+s2}{\PYZdq{}}\PY{l+s+s2}{Token has not been saved. Use the function save\PYZus{}token to to save your token.}\PY{l+s+s2}{\PYZdq{}}\PY{p}{)}
        \PY{n+nb}{print}\PY{p}{(}\PY{l+s+s2}{\PYZdq{}}\PY{l+s+s2}{Token loaded.}\PY{l+s+s2}{\PYZdq{}}\PY{p}{)}
        \PY{k}{return} \PY{n}{token}
    
    \PY{k}{def} \PY{n+nf}{count\PYZus{}gates}\PY{p}{(}
            \PY{n}{quantum\PYZus{}circuit}\PY{p}{:} \PY{n}{QuantumCircuit}\PY{p}{,}
            \PY{n}{gates\PYZus{}to\PYZus{}consider}\PY{p}{:} \PY{n}{List}\PY{p}{[}\PY{n+nb}{str}\PY{p}{]}
        \PY{p}{)} \PY{o}{\PYZhy{}}\PY{o}{\PYZgt{}} \PY{n+nb}{int}\PY{p}{:}
        \PY{n}{result} \PY{o}{=} \PY{l+m+mi}{0}
        \PY{k}{for} \PY{n}{gate} \PY{o+ow}{in} \PY{n}{gates\PYZus{}to\PYZus{}consider}\PY{p}{:}
            \PY{k}{try}\PY{p}{:}
                \PY{n}{count\PYZus{}gate} \PY{o}{=} \PY{n}{quantum\PYZus{}circuit}\PY{o}{.}\PY{n}{count\PYZus{}ops}\PY{p}{(}\PY{p}{)}\PY{p}{[}\PY{n}{gate}\PY{p}{]}
            \PY{k}{except} \PY{n+ne}{KeyError}\PY{p}{:}
                \PY{n}{count\PYZus{}gate} \PY{o}{=} \PY{l+m+mi}{0}
            \PY{n}{result} \PY{o}{=} \PY{n}{result} \PY{o}{+} \PY{n}{count\PYZus{}gate}
        \PY{k}{return} \PY{n}{result}
    \end{Verbatim}
    \end{tcolorbox}

%% file: paper.bbl
\begin{thebibliography}{22}
\providecommand{\natexlab}[1]{#1}
\providecommand{\url}[1]{\texttt{#1}}
\expandafter\ifx\csname urlstyle\endcsname\relax
  \providecommand{\doi}[1]{doi: #1}\else
  \providecommand{\doi}{doi: \begingroup \urlstyle{rm}\Url}\fi

\bibitem[Cerezo et~al.(2021)Cerezo, Arrasmith, Babbush, Benjamin, Endo, Fujii,
  McClean, Mitarai, Yuan, Cincio, and Coles]{CerEtAl21}
M.~Cerezo, A.~Arrasmith, R.~Babbush, S.~C. Benjamin, S.~Endo, K.~Fujii, J.~R.
  McClean, K.~Mitarai, X.~Yuan, L.~Cincio, and P.~J. Coles.
\newblock {Variational Quantum Algorithms}.
\newblock \emph{Nature Reviews Physics}, pages 1--20, 2021.
\newblock ISSN 2522-5820.
\newblock \url{https://www.nature.com/articles/s42254-021-00348-9}.

\bibitem[Farhi et~al.(2014)Farhi, Goldstone, and Gutmann]{FarGG14}
E.~Farhi, J.~Goldstone, and S.~Gutmann.
\newblock {A Quantum Approximate Optimization Algorithm}, 2014.
\newblock \url{http://arxiv.org/pdf/1411.4028v1}.

\bibitem[Guerra(2005)]{Gue05}
F.~Guerra.
\newblock {Spin Glasses}, 2005.
\newblock \url{https://arxiv.org/pdf/cond-mat/0507581}.

\bibitem[Hadfield et~al.(2019)Hadfield, Wang, O'Gorman, Rieffel, Venturelli,
  and Biswas]{HadWORVB19}
S.~Hadfield, Z.~Wang, B.~O'Gorman, E.~G. Rieffel, D.~Venturelli, and R.~Biswas.
\newblock {From the Quantum Approximate Optimization Algorithm to a Quantum
  Alternating Operator Ansatz}.
\newblock \emph{Algorithms}, 12\penalty0 (2):\penalty0 34, 2019.
\newblock ISSN 1999-4893.
\newblock \url{http://arxiv.org/pdf/1709.03489v2}.

\bibitem[IAO(2023 (online){\natexlab{a}})]{IaoWeb}
Fraunhofer IAO.
\newblock {LamA Laden am Arbeitsplatz}, January 20, 2023
  (online){\natexlab{a}}.
\newblock
  \url{https://www.iao.fraunhofer.de/de/forschung/smart-energy-and-mobility-solutions/lama-laden-am-arbeitsplatz.html}.

\bibitem[IAO(2023 (online){\natexlab{b}})]{LamaWeb}
Fraunhofer IAO.
\newblock {LamA - Laden am Arbeitsplatz}, January 20, 2023
  (online){\natexlab{b}}.
\newblock \url{https://www.lama.zone/}.

\bibitem[IBM(2023 (online){\natexlab{a}})]{QiskitOpflowWeb}
IBM.
\newblock {Operator Flow}, January 20, 2023 (online){\natexlab{a}}.
\newblock
  \url{https://github.com/Qiskit/qiskit-tutorials/blob/master/tutorials/operators/01_operator_flow.ipynb}.

\bibitem[IBM(2023 (online){\natexlab{b}})]{QiskitOptimizationWeb2}
IBM.
\newblock {Using Classical Optimization Solvers and Models with Qiskit
  Optimization}, January 20, 2023 (online){\natexlab{b}}.
\newblock
  \url{https://qiskit.org/documentation/optimization/tutorials/11_using_classical_optimization_solvers_and_models.html}.

\bibitem[IBM(2023 (online){\natexlab{c}})]{QiskitTranspilerWeb2}
IBM.
\newblock {Transpiler Passes and Pass Manager}, January 20, 2023
  (online){\natexlab{c}}.
\newblock
  \url{https://qiskit.org/documentation/tutorials/circuits_advanced/04_transpiler_passes_and_passmanager.html}.

\bibitem[IBM(2023 (online){\natexlab{d}})]{QuantumEnablementWeb}
IBM.
\newblock {Using the Qiskit Compiler}, January 20, 2023 (online){\natexlab{d}}.
\newblock \url{https://quantum-enablement.org/how-to/compiling/compiling.html}.

\bibitem[IBM(2023 (online){\natexlab{e}})]{QuantumEnablementWeb3}
IBM.
\newblock {Mesurement Error Mitigation Using M3}, January 20, 2023
  (online){\natexlab{e}}.
\newblock
  \url{https://quantum-enablement.org/how-to/mitigation/M3/m3_mitigation.html}.

\bibitem[IBM(2023 (online){\natexlab{f}})]{ShotNoiseWeb}
IBM.
\newblock {Noisy Real Hardware, Noise in Quantum Computers}, January 20, 2023
  (online){\natexlab{f}}.
\newblock
  \url{https://learn.qiskit.org/summer-school/2022/noisy-real-hardware-noise-quantum-computers}.

\bibitem[IEQ(2023 (online))]{IegWeb}
Fraunhofer IEQ.
\newblock {LamA - Laden am Arbeitsplatz (Charging at Work)}, January 20, 2023
  (online).
\newblock \url{https://www.ieg.fraunhofer.de/en/references/lama.html}.

\bibitem[Karimi and Ronagh(2019)]{KarP19}
S.~Karimi and P.~Ronagh.
\newblock {Practical Integer-to-Binary Mapping for Quantum Annealers}.
\newblock \emph{Quantum Information Processing}, 18\penalty0 (4):\penalty0
  042314, 2019.
\newblock ISSN 1570-0755.
\newblock \url{https://arxiv.org/pdf/1706.01945}.

\bibitem[Koßmann et~al.(2023)Koßmann, Binkowski, Tutschku, and
  Schwonnek]{KosBTS23}
G.~Koßmann, L.~Binkowski, C.~Tutschku, and R.~Schwonnek.
\newblock {Open-Shop Scheduling with Hard Constraints}, 2023.
\newblock \url{https://arxiv.org/abs/2211.05822}.

\bibitem[McClean et~al.(2018)McClean, Boixo, Smelyanskiy, Babbush, and
  Neven]{McCEtAl18}
J.~R. McClean, S.~Boixo, V.~N. Smelyanskiy, R.~Babbush, and H.~Neven.
\newblock {Barren Plateaus in Quantum Neural Network Training Landscapes}.
\newblock \emph{Nature Communications}, 9\penalty0 (1):\penalty0 4812, 2018.
\newblock ISSN 2041-1723.
\newblock \url{https://www.nature.com/articles/s41467-018-07090-4}.

\bibitem[Nielsen and Chuang(2010)]{NieC10}
M.~A. Nielsen and I.~L. Chuang.
\newblock \emph{{Quantum Computation and Quantum Information}}.
\newblock {Cambridge University Press}, Cambridge, 10th anniversary edition
  edition, 2010.
\newblock ISBN 9781107002173.

\bibitem[of~Mathematics(2023 (online))]{IsingWeb}
Encyclopedia of~Mathematics.
\newblock {Ising Model}, January 20, 2023 (online).
\newblock \url{https://encyclopediaofmath.org/index.php?title=Ising_model}.

\bibitem[Team(2023 (online){\natexlab{a}})]{QuantumEnablementWeb2}
IBM Quantum Enabling~Technologies Team.
\newblock Dynamical decoupling, January 20, 2023 (online){\natexlab{a}}.
\newblock
  \url{https://quantum-enablement.org/how-to/dynamical_decoupling/dynamical_decoupling.html}.

\bibitem[Team(2023 (online){\natexlab{b}})]{MthreeWeb}
Mthree Team.
\newblock {mthree}, January 20, 2023 (online){\natexlab{b}}.
\newblock \url{https://qiskit.org/documentation/partners/mthree/}.

\bibitem[Team(2023 (online){\natexlab{c}})]{QiskitTranspilerWeb}
Qiskit~Development Team.
\newblock Transpiler, January 20, 2023 (online){\natexlab{c}}.
\newblock \url{https://qiskit.org/documentation/apidoc/transpiler.html}.

\bibitem[Team(2023 (online){\natexlab{d}})]{QiskitOptimizationWeb}
Qiskit Optimization~Development Team.
\newblock {Optimization Tutorials}, January 20, 2023 (online){\natexlab{d}}.
\newblock
  \url{https://qiskit.org/documentation/optimization/tutorials/index.html}.

\end{thebibliography}
